\documentclass[3p,onecolumn]{elsarticle}
\biboptions{sort&compress}
\usepackage[usenames,dvipsnames,svgnames,table]{xcolor} 
\usepackage[utf8]{inputenc}
\usepackage{subfigure,lineno,wrapfig}

\usepackage[compat=1.1.0]{tikz-feynman}
\usepackage{graphicx}% Include figure files
\usepackage{dcolumn}% Align table columns on decimal point
\usepackage{bm}% bold math
\usepackage{hyperref}% add hypertext capabilities
\hypersetup{ 
    pdfnewwindow=true,      % links in new window
    colorlinks=true,       % false: boxed links; true: colored links
    allcolors=[RGB]{31 119 180}
} 
\usepackage{dsfont}
\usepackage{placeins}
\usepackage{todonotes}
\usepackage[utf8]{inputenc}
\usepackage{gensymb}
\usepackage{multirow}
\usepackage{amsmath}
\usepackage{amsfonts}
\usepackage{amssymb}
\usepackage{cleveref}
\usepackage{enumitem}
\usepackage{physics}
\usepackage{array}   % for \newcolumntype macro
\newcolumntype{L}{>{$}l<{$}} % math-mode version of "l" column type
\newcolumntype{R}{>{$}r<{$}}
\newcolumntype{C}{>{$}c<{$}}
\usepackage{xspace}
\usepackage{braket}
\usepackage{units}
\usepackage{overpic}
\usepackage{slashed}
\usepackage[normalem]{ulem}
\usepackage[format=plain,justification=RaggedRight,singlelinecheck=false]{caption}
\usepackage{graphbox}

\graphicspath{{figures/}}

\renewcommand{\abs}[1]{\ensuremath{|#1|}}
\let\Re\relax
\let\Im\relax

\DeclareMathOperator{\Re}{Re}
\DeclareMathOperator{\Im}{Im}

\newcommand{\mevnospace}{\ensuremath{{\mathrm{\,Me\kern -0.1em V}}}}
\newcommand{\gevnospace}{\ensuremath{{\mathrm{\,Ge\kern -0.1em V}}}}
\newcommand{\tevnospace}{\ensuremath{{\mathrm{\,Te\kern -0.1em V}}}}
\newcommand{\kev}{\ensuremath{{\mathrm{\,ke\kern -0.1em V}}}\xspace}
\newcommand{\mev}{\mevnospace\xspace}
\newcommand{\gev}{\gevnospace\xspace}

\newcommand{\mevp}{\ensuremath{{\mathrm{Me\kern -0.1em V}}}}
\newcommand{\gevp}{\ensuremath{{\mathrm{Ge\kern -0.1em V}}}}
\newcommand{\tevp}{\ensuremath{{\mathrm{Te\kern -0.1em V}}}}
\newcommand{\nb}{\ensuremath{\,\mathrm{nb}}}
\newcommand{\eV}{\ensuremath{{\mathrm{\,e\kern -0.1em V}}}\xspace}
\newcommand{\gevsq}{\ensuremath{{\mathrm{\,Ge\kern -0.1em V}^2}}\xspace}
\newcommand{\nsgev}{\ensuremath{{\mathrm{Ge\kern -0.1em V}}}\xspace}
\newcommand{\nsmev}{\ensuremath{{\mathrm{Me\kern -0.1em V}}}\xspace}
\newcommand{\nskev}{\ensuremath{{\mathrm{ke\kern -0.1em V}}}\xspace}

\newcommand{\babar}{BaBar\xspace}
\newcommand{\belle}{Belle\xspace}
\newcommand{\belletwo}{Belle~II\xspace}
\newcommand{\lhcb}{LHCb\xspace}
\newcommand{\bes}{BESIII\xspace}
\newcommand{\jlab}{JLab\xspace}
\newcommand{\aka}{{\it aka}\xspace}
\newcommand{\eg}{{\it e.g.}\xspace}
\newcommand{\cf}{{\it cf.}\xspace}

\newcommand{\ie}{{\it i.e.}\xspace}
\newcommand{\vs}{{\it vs.}\xspace}
\newcommand{\etc}{{etc.}\xspace}
\newcommand{\lhs}{{\it l.h.s.}\xspace}
\newcommand{\rhs}{{\it r.h.s.}\xspace}

\newcommand{\etapi}{\ensuremath{\eta^{(\prime)}\pi}\xspace}
\newcommand{\etapicompass}{\ensuremath{\eta^{(\prime)}\pi^{-}}\xspace}
\newcommand{\pione}{\ensuremath{\pi_1}\xspace}
\newcommand{\dof}{\ensuremath{\text{dof}}\xspace}

\newlist{todolist}{itemize}{2}
\setlist[todolist]{label=$\square$}
\usepackage{pifont}

\newcommand{\jpsi}{\ensuremath{J/\psi}\xspace}
\newcommand{\sig}{\ensuremath{\sigma/f_0(500)}\xspace}

\newcommand{\kap}{\ensuremath{\kappa/K_0^*(700)}\xspace}

\newcommand{\KSKS}{\ensuremath{K_S^0 K_S^0}\xspace}
\newcommand{\pizpiz}{\ensuremath{\pi^0 \pi^0}\xspace}
\newcommand{\h}{\ensuremath{h}\xspace}

\def\Dbar    {\kern 0.2em\bar{\kern -0.2em D}{}\xspace}

\def\Dz      {\ensuremath{D^0}\xspace}
\def\Dzb     {\ensuremath{\Dbar^0}\xspace}
\def\DzDzb   {\ensuremath{\Dz {\kern -0.16em \Dzb}}\xspace}
\def\Dp      {\ensuremath{D^+}\xspace}
\def\Dm      {\ensuremath{D^-}\xspace}

\def\DpDm    {\ensuremath{\Dp {\kern -0.16em \Dm}}\xspace}

\def\Bbar    {\kern 0.2em\bar{\kern -0.2em B}{}\xspace}

\def\Lbar     {\kern 0.2em\overline{\kern -0.2em\Lambda\kern 0.05em}\kern-0.05em{}\xspace}

\newcommand{\scenIII}{III\xspace}
\newcommand{\scenIIItr}{III+tr.\xspace}
\newcommand{\scenIV}{IV+tr.\xspace}
\newcommand{\scentr}{tr.\xspace}

\newcommand{\Pc}{\ensuremath{P_c(4312)}\xspace}
\newcommand{\SigmaD}{\ensuremath{\Sigma_c^+\bar{D}^0}\xspace}
\newcommand{\jpsip}{\ensuremath{J/\psi\,p}\xspace}

\newcommand{\Ac}{\mathcal{A}}

\newcommand{\2}{\bm{2}}
\newcommand{\3}{\bm{3}}
\renewcommand{\k}{\bm{k}}
\newcommand{\p}{\bm{p}}
\newcommand{\q}{\bm{q}}

\newcommand{\1}{\bm{1}}
\renewcommand{\2}{\bm{2}}
\renewcommand{\3}{\bm{3}}

\renewcommand{\Ac}{\mathcal{A}}

\newcommand{\Bc}{\mathcal{B}}
\newcommand{\Cc}{\mathcal{C}}
\newcommand{\Dc}{\mathcal{D}}
\newcommand{\Fc}{\mathcal{F}}
\newcommand{\Gc}{\mathcal{G}}

\newcommand{\Kc}{\mathcal{K}}
\newcommand{\Lc}{\mathcal{L}}
\newcommand{\Mc}{\mathcal{M}}
\newcommand{\Rc}{\mathcal{R}}
\newcommand{\Tc}{\mathcal{T}}
\newcommand{\Z}{\mathbb{Z}}

\newcommand{\XYZ}{\ensuremath{XY\!Z}\xspace}
\newcommand{\ALL}{\ensuremath{\textup{A}_{\textup{LL}}}\xspace}
\newcommand{\KLL}{\ensuremath{\textup{K}_{\textup{LL}}}\xspace}

\newcommand{\twototwo}{\ensuremath{\2\to\2}\xspace}
\newcommand{\onetothree}{\ensuremath{\1\to\3}\xspace}
\newcommand{\threetothree}{\ensuremath{\3\to\3}\xspace}

\newcommand{\diff}{\textrm{d}}

\newcommand{\m}{m}

\newcommand{\calB}{\ensuremath{\mathcal{B}}\xspace}

\newcommand{\calT}{\ensuremath{\mathcal{T}}\xspace}

\newcommand{\calP}{\ensuremath{\mathcal{P}}\xspace}
\newcommand{\calQ}{\ensuremath{\mathcal{Q}}\xspace}
\newcommand{\calE}{\ensuremath{\mathcal{E}}\xspace}

\journal{Progress in Particle and Nuclear Physics}

\newcommand{\catania}{INFN Sezione di Catania, I-95123 Catania, Italy}
\newcommand{\ceem}{Center for  Exploration  of  Energy  and  Matter, Indiana  University, Bloomington,  IN  47403,  USA}

\newcommand{\icn}{Instituto de Ciencias Nucleares, 
Universidad Nacional Aut\'onoma de M\'exico, Ciudad de M\'exico 04510, Mexico}
\newcommand{\icsup}{Pedagogical University of Krakow, 30-084 Krak\'ow, Poland}
\newcommand{\ific}{Instituto de F\'isica Corpuscular (IFIC), Centro Mixto CSIC-Universidad de Valencia, E-46071 Valencia, Spain}
\newcommand{\indiana}{Department of Physics,
Indiana  University, Bloomington,  IN  47405,  USA}
\newcommand{\jlabphys}{Theory Center and Physics Division, Thomas  Jefferson  National  Accelerator  Facility, Newport  News,  VA  23606,  USA}
\newcommand{\ekut}{Institute for Theoretical Physics, T\"ubingen University, Auf der Morgenstelle 14, 72076 T\"ubingen, Germany}
\newcommand{\ur}{
Institute for Theoretical Physics, Regensburg University, D-93040 Regensburg, Germany}
\newcommand{\lanl}{Theoretical Division, Los Alamos National Laboratory, Los Alamos, NM 87545, USA}
\newcommand{\lmu}{Ludwig-Maximilian University of Munich, Germany}
\newcommand{\messina}{Dipartimento di Scienze Matematiche e Informatiche, Scienze Fisiche e Scienze della Terra, 
Universit\`a degli Studi di Messina, I-98122 Messina, Italy}
\newcommand{\odu}{Department of Physics, Old Dominion University, Norfolk, VA 23529, USA}
\newcommand{\origins}{ORIGINS Excellence Cluster, 80939 Munich, Germany}
\newcommand{\scnuIQM}{Guangdong Provincial Key Laboratory of Nuclear Science, Institute of Quantum Matter, South China Normal University, Guangzhou 510006, China}
\newcommand{\scnuJLQM}{Guangdong-Hong Kong Joint Laboratory of Quantum Matter, Southern Nuclear Science Computing Center, South China Normal University, Guangzhou 510006, China}
\newcommand{\ub}{Departament de F\'isica Qu\`antica i Astrof\'isica and Institut de Ci\`encies del Cosmos, Universitat de Barcelona, E-08028, Spain}
\newcommand{\ucm}{Departamento de F\'isica Te\'orica, Universidad Complutense de Madrid and IPARCOS, E-28040 Madrid, Spain}
\newcommand{\uned}{Departamento de F\'isica Interdisciplinar, Universidad Nacional de Educaci\'on a Distancia (UNED), Madrid E-28040, Spain}
\newcommand{\wm}{College of William \& Mary, Williamsburg, VA 23187, USA}

\begin{document}
\hypersetup{allcolors=[RGB]{31 119 180}}
\begin{frontmatter}
\begin{flushright}
~\\
\vspace{-2cm}
\includegraphics[width=1.5in]{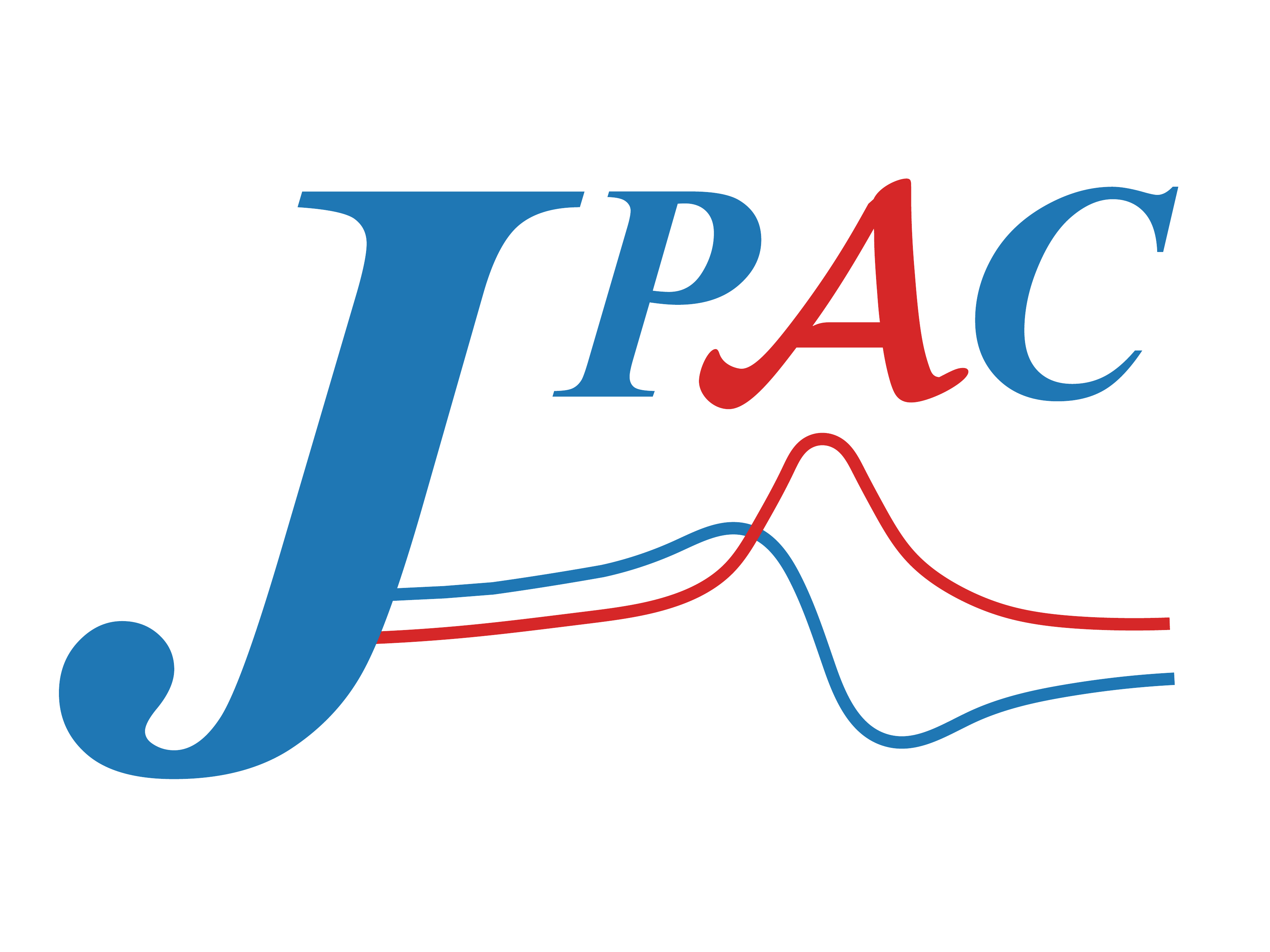}
\end{flushright}
\title{Novel approaches in Hadron Spectroscopy}
\author[jlab,ific]{Miguel~Albaladejo}
\author[icsup]{{\L}ukasz~Bibrzycki}
\author[indiana,ceem]{Sebastian~M.~Dawid}
\author[icn,uned]{C\'esar~Fern\'andez-Ram\'irez}
\author[indiana,ceem,lanl]{Sergi~Gonz\`alez-Sol\'is}
\author[jlab,ekut,ur]{Astrid~N.~Hiller~Blin}
\author[jlab,odu]{Andrew~W.~Jackura}
\author[ub,ucm]{Vincent~Mathieu}
\author[origins,lmu]{Mikhail~Mikhasenko}
\author[jlab]{Victor~I.~Mokeev}
\author[jlab,indiana,ceem]{Emilie~Passemar}
\author[messina,catania]{Alessandro~Pilloni\corref{mycorrespondingauthor}}
\ead{alessandro.pilloni@unime.it}
\author[jlab,wm]{Arkaitz~Rodas}
\author[icn]{Jorge~A.~Silva-Castro}
\author[indiana]{Wyatt~A.~Smith}
\author[jlab,indiana,ceem]{Adam~P.~Szczepaniak}
\author[indiana,ceem,scnuIQM,scnuJLQM]{Daniel~Winney}
\author{\\\vspace{.3cm}(Joint Physics Analysis Center)}
%addresses
\address[jlab]{\jlabphys}
\address[ific]{\ific}
\address[icsup]{\icsup}
\address[indiana]{\indiana}
\address[ceem]{\ceem}
\address[icn]{\icn}
\address[uned]{\uned}
\address[lanl]{\lanl}
\address[odu]{\odu}
\address[ub]{\ub}
\address[ucm]{\ucm}
\address[ekut]{\ekut}
\address[ur]{\ur}
\address[origins]{\origins}
\address[lmu]{\lmu}
\address[messina]{\messina}
\address[catania]{\catania}
\address[wm]{\wm}
\address[scnuIQM]{\scnuIQM}
\address[scnuJLQM]{\scnuJLQM}

\cortext[mycorrespondingauthor]{Corresponding author}

\begin{abstract}
The last two decades have witnessed the discovery of a myriad of new and unexpected
hadrons. The future holds more surprises for us, thanks to new-generation experiments.
Understanding the signals and determining the properties
of the states requires a parallel theoretical effort.
To make full use of available and forthcoming data, a careful amplitude modeling is required, together with a sound treatment of the statistical uncertainties, 
and a systematic survey of the model dependencies.
We review the contributions made by the Joint Physics Analysis Center 
to the field of hadron spectroscopy.
\end{abstract}

\begin{keyword}
Hadron spectroscopy \sep 
Exotic hadrons \sep 
Three-body scattering \sep 
Resonance production\\
{\it Preprint numbers: LA-UR-21-31664, JLAB-THY-22-3459}
\end{keyword}

\end{frontmatter}

\clearpage
\tableofcontents

\clearpage
%============================================
% Introduction
%============================================
\section{Introduction}
In the last two decades the quark model lore of baryons with three quarks and mesons with a quark-antiquark
pair has been challenged by the many unexpected exotic
hadron resonances found in high-energy experiments.
Many tetraquarks, pentaquarks, molecules, hybrids and glueball candidates have 
sprung forth and revitalized the field of hadron spectroscopy~\cite{pdg,Esposito:2016noz,Olsen:2017bmm,Guo:2017jvc,Lebed:2016hpi,Karliner:2017qhf,Guo:2019twa,ali2019multiquark,Brambilla:2019esw}.
Discovering and characterizing an exotic resonance is not a goal in and of itself,
but is a necessary step in identifying complete multiplets
and studying the emerging patterns and properties of the spectrum.
We still lack a comprehensive and consistent picture of this
sector of the QCD spectrum, as many of the candidates have been
identified in a single production or decay channel.
Nevertheless, this research needs to be pursued
as this kind of knowledge would provide
insight not only into the nature of said exotics,
but also into the inner workings of the nonperturbative regime of QCD, especially
since an analytic solution of QCD in this
regime will not be available in the foreseeable future. 

Current experiments such as LHCb, COMPASS, BESIII, GlueX, and CLAS, are providing datasets with unprecedented statistics. With the forthcoming data from~\belletwo~\cite{Belle-II:2018jsg}, we expect to have enough data to properly analyze some exotic channels at lepton colliders, which were previously limited by insufficient statistics. 
However, the measurements often depict 
multibody final states, which make it a challenge to perform
a model-independent determination of an exotic candidate.

On the theoretical front, 
Lattice QCD provides the most rigorous, albeit computationally expensive,
tool to calculate observables from first principles~\cite{Shepherd:2016dni,Briceno:2017max}.
However, it cannot explain
the emergence of confinement and mass generation,
or \emph{why} quarks and gluons organize themselves in the observed
hadron spectrum.
Functional methods~\cite{Polonyi:2001se,Maris:2003vk}, sum rules~\cite{Lucha:2021mwx}, models of QCD (as the quark model~\cite{DeRujula:1975qlm,Godfrey:1985xj,Capstick:1985xss}, Hamiltonian formalisms~\cite{Szczepaniak:2001rg,Szczepaniak:2006nx}, holography-inspired descriptions~\cite{Plessas:2015mpa,Brodsky:2014yha}), or quark-level Effective Field Theories~\cite{Brambilla:1999xf,Braaten:2014qka,Brambilla:2018pyn}
are employed to fill in that gap.

Together with these top-down approaches, bottom-up strategies are also feasible: one can  write ans\"atze for the amplitudes that respect the fundamental principles as much as possible, at least in a given kinematical domain, and fit to data. 
If the amplitude model space is large enough, the resonance properties obtained will be as unbiased as possible.  

Once the theoretical modeling of a reaction amplitude has been achieved,
it can be combined with a sound statistical data analysis. 
For example, one can use clustering methods to separate the physical resonances from the artifacts of the amplitude parametrizations.
All these studies allow one to give a robust determination of resonances and of their properties.
These analyses are computationally expensive and require high-performance computing resources, but they will become mandatory for the interpretation of present and future high-precision data. 

One can gain additional information by studying the production rates and the underlying mechanisms of resonances. In particular, the dual role of resonances as particles and forces implies that one can probe their properties in both regimes. These ideas motivate a program to impose duality constraints to the standard amplitude analysis.

In this review we summarize the contributions of the Joint Physics Analysis Center (JPAC) to the field. 
JPAC started in 2013, impulsed by Mike Pennington, originally to provide theory support to the most delicate spectroscopy analyses at Jefferson Lab (\jlab). In the following years, it has become a model example of collaboration between theorists and experimentalists, developing amplitude analysis tools and 
best practices for hadron spectroscopy.
The methods require complementary sets of skills in QCD, reaction theory, computer science, and experimental data analysis.
The group has a strong record of interactions with experiments:
JPAC members have contributed to analyses by \babar, \bes, CLAS, COMPASS, GlueX, and \lhcb, and to several proposals of future spectroscopy experiments and facilities. The tools implemented are publicly available~\cite{jpacweb}.
The review is organized as follows.
In Section~\ref{sec:resonances} we discuss the physics of resonances: the generalities of the QCD spectrum, the methods, and some practical applications, both to the light and heavy sectors. Section~\ref{sec:three} is devoted to the study of three-body physics,
a quickly developing topic within Lattice QCD with important applications in the experimental analyses. The production of resonances is discussed in Section~\ref{sec:production}. A brief summary is presented in Section~\ref{sec:conclusions}.

%============================================
% Resonance studies
%============================================
\section{Resonance studies}
\label{sec:resonances}

\subsection{\texorpdfstring{The $S$-matrix and amplitude parametrizations}{The S-matrix and amplitude parametrizations}}
\label{sec:smatrix}
The excited spectrum of QCD is composed of states with lifetimes $\lesssim 10^{-21}\text{~s}$, which need to be reconstructed from the energy and angular dependence of their decay products. 
The measured rates are proportional to the modulus squared of the {\em reaction amplitude}, which encodes the information about these states at the quantum level.
While the reaction amplitude's angular dependence is determined by the spin of the particles involved, the energy behavior is dynamical. 

The $S$-matrix theory traces back to the late 50s, as a possible formalism that circumvented the apparent inconsistencies of perturbative quantum field theory. The idea was that, even if no theory of strong interactions was available, the underlying $S$-matrix must satisfy certain properties. Lorentz invariance requires that the $S$-matrix elements, and therefore amplitudes, depend on particle momenta only through Mandelstam invariants.
In particular, Landau argued that causality of 
the interaction implies that the amplitudes must be analytic functions of the invariants~\cite{Landau:1959fi}. Similar analyticity requirements were argued by Regge, studying the Schr\"odinger equation for complex values of angular momentum~\cite{Regge:1959mz}.  
Analyticity, unitarity, and crossing symmetry, constitute the so-called $S$-matrix principles. Here, unitarity stems from probability conservation, and crossing symmetry relates particles and anti-particles and is proper of relativistic quantum theories. The hope was that these principles were sufficient to uniquely determine the strong
interaction $S$-matrix, once proper additional assumptions and initial conditions were given. The main additional hypothesis was the maximal analyticity principle, \ie that the only singularities appearing in an amplitude are the ones required by unitarity and crossing symmetry. This was verified at all orders in perturbative field theory, but has not yet been proven to hold nonperturbatively. Chew led the so-called bootstrap program, which, using an input model for resonances exchanged in the cross-channel, allowed one to recover self-consistently the same resonances in the direct channel. The main obstacle to this was that the dispersion relations one derives suffer from Castillejo-Dalitz-Dyson (CDD) ambiguities~\cite{Castillejo:1955ed}, and the solution cannot be determined uniquely. 
In modern terms, the $S$-matrix principles are not specific to the strong interactions, and the information about what theory they are applied to must be encoded in these ambiguities. 
When QCD was established as the underlying theory of strong interactions, it was proposed that CDD poles reflect the presence of bound states of quarks, about which the $S$-matrix theory knows nothing {\it a priori}, and must be imposed from data. With the discovery of $J/\psi$ and the triumph of quantum field theory, the bootstrap program was abandoned. Fifty years later, we still do not have a constructive solution of QCD. There is no simple connection between the interaction at the quark- and hadron-level, so there is a renewed interest in what one can learn from amplitude properties alone, and if possible, to constrain the space of feasible solutions rather than to look for a unique one. 
The new program is thus to postulate {\it ans\"atze} for the amplitudes that depend on a finite number of parameters and fit them to data.
Ideally, one requires the amplitudes to fulfill the constraints given by the $S$-matrix principles, to obtain physical results as sound as possible. It should be stressed, however, that implementing all the constraints simultaneously is extremely difficult, and the problem has to be approached on a case-by-case basis in order to enforce the constraints that are most relevant for the physics at hand.

\begin{figure}\begin{center}
\includegraphics[height=5cm,keepaspectratio]{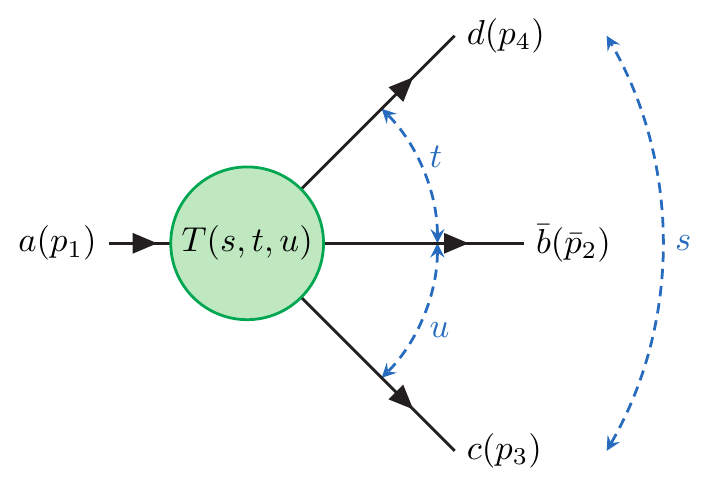} \hspace{1.cm}
\includegraphics[height=5cm,keepaspectratio]{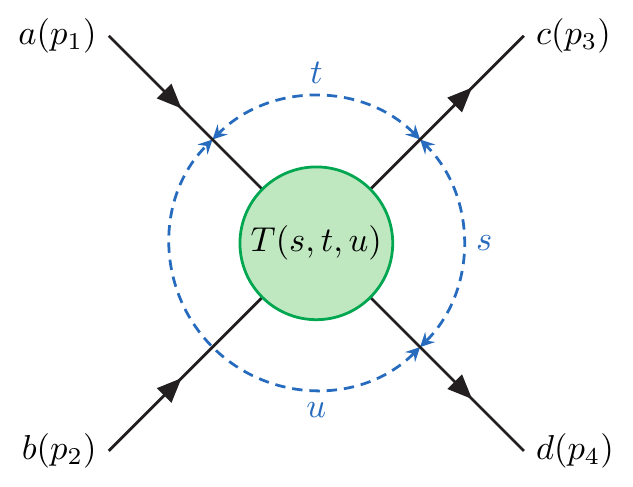}
\caption{ (left) Representation of the decay $a(p_1) \to \bar{b}(\bar{p}_2) \, c(p_3) \, d(p_4)$ and the kinematical variables involved. (right) Representation of the scattering process $a(p_1) \, b(p_2) \to c(p_3) \, d(p_4)$ and its kinematical variables.
\label{fig:scattdecaydiagrams}}
\end{center}\end{figure}

We review here the basics of \twototwo scattering and \onetothree decay, that will be used in the rest of the paper.
Consider a \twototwo scattering process of scalar particles $a(p_1)\, b(p_2)\to c(p_3)\, d(p_4)$, and the $\onetothree$ decay $a(p_1)\to \bar b(\bar p_2)\, c(p_3)\, d(p_4)$ one obtains by crossing particle $b$, as depicted in \figurename{~\ref{fig:scattdecaydiagrams}}. The Mandelstam variables, for scattering, are defined through

\begin{subequations}
\begin{align}
s  &= (p_1+p_2)^2  = (p_3+p_4)^2~,  \\
t  &= (p_1-p_3)^2  = (p_2-p_4)^2~,  \\
u  &= (p_1-p_4)^2  = (p_2-p_3)^2~,  \\
s + t + u &= m_1^2 + m_2^2 + m_3^3 + m_4^2 \equiv \Sigma~, \label{eq:stuscatt}
\end{align}
\end{subequations}
where for the $\onetothree$ decay, $\bar p_2 = -p_2$. In order to describe the $s$-channel center of mass frame it is convenient to introduce the initial- and final-state 3-momenta,  
\begin{align}
p(s) &=\frac{\lambda^\frac{1}{2}(s,m_1^2,m_2^2)}{2 \sqrt{s}} & q(s) &= \frac{\lambda^\frac{1}{2}(s,m_3^2,m_4^2)}{2\sqrt{s}}~,
\end{align}
where $\lambda(x,y,z) = x^2 + y^2 + z^2-2(x\,y+y\,z+z\,x)$ is the triangle or K\"all\'en  function~\cite{Kallen:1964lxa}. The four-momenta in the $s$-channel center-of-mass frame read
\begin{subequations}
\begin{align}
p_1 &=\left(\frac{s+m_1^2-m_2^2}{2\sqrt{s}},p(s)\hat{z}\right),  & p_3 &=\left(\frac{s+m_3^2-m_4^2}{2\sqrt{s}},q(s)\hat{n}\right), \\
p_2 &= \left(\frac{s+m_2^2-m_1^2}{2\sqrt{s}},-p(s)\hat{z}\right), & p_4 &=\left(\frac{s+m_4^2-m_3^2}{2\sqrt{s}},-q(s)\hat{n}\right)\, ,
\label{eq:momenscatt}
\end{align}
\end{subequations}
where $\hat z$ is the direction of particle $a$, usually taken as the $z$-axis, and $\hat n$ the direction of particle $c$, usually taken in the half $xz$-plane containing the positive $x$-axis. 
The cosine of the scattering angle $z_s$ is defined by $\hat{z}\cdot \hat{n} \equiv z_s$. It is given as a function of the Mandelstam variables as
\begin{equation}
z_s = \frac{s(t-u)+\left(m_1^{2}-m_2^{2}\right)\left(m_3^{2}-m_4^{2}\right)}{4 s\,p(s)\,q(s)}.
\end{equation}
Similar expressions can be obtained for the scattering angle $z_t$ and $z_u$ of the crossed processes in the respective center-of-mass frame.
 We will omit its $s$-, $t$-, and $u$-dependence when no ambiguity can arise. For simplicity, we consider the spinless case. The customary partial wave expansion of the amplitude $T(s,t,u)$ is
\begin{equation}\label{eq:pwdecomp1}
T(s,t,u) = \sum_{\ell=0}^{\infty} (2\ell+1) P_\ell(z_s) \,t_\ell(s)~,
\end{equation}
where $P_\ell(z_s)$ are the Legendre polynomials, $t_\ell(s)$ are the partial waves of angular momentum $\ell$. 
Note that for symmetry and compactness reasons we have left the explicit $u$-dependence in Eq.~\eqref{eq:pwdecomp1}, although the sum of the three Mandelstam variables is constrained, and as a result the full amplitude $T(s,t,u)$ depends only on two of them. The partial waves $t_\ell(s)$ can be obtained from the full amplitude by inversion of the aforementioned equation
\begin{align}
t_\ell(s) 
& = \frac{1}{2} \int_{-1}^{+1} \mathrm{d}z\, P_\ell(z)\, T\!\left( s,t(s,z),u(s,z) \right)~. \label{eq:pwprojection}
\end{align}

The physics of the unstable states we are interested in is encoded in these partial waves. A resonance of spin $\ell$ appears in $t_\ell(s)$ as a pole located at complex values of energy, the real and imaginary parts being the mass and half-width of the resonance, respectively. 
It is thus necessary to consider amplitudes that can be analytically continued from the physical real axis---where data exist---to the complex plane. The partial waves diagonalize unitarity, \ie they transform an integral equation into a tower of uncoupled algebraic equations,
\begin{equation}
\Im t_\ell(s)= t_\ell(s)\,\rho(s)\,t_\ell(s)^\dagger\,,
\end{equation}
where $\rho(s)_{ij}\propto \delta_{ij}\,q_j(s)/\sqrt{s} $ is the diagonal matrix of the phase space of all possible two-body channels.\footnote{For the specific normalization of partial waves given in Eq.~\eqref{eq:pwprojection}, $\rho(s)_{ij} = \delta_{ij}\,q_j(s)/8\pi\sqrt{s}$ for distinguishable particles. However, in most of the applications discussed further, the normalization of $\rho(s)$ can be reabsorbed in other model parameters, and different values will be used.}
The physical sheet is protected by analyticity: No dynamical singularity can appear in the complex plane, although poles on the real axis associated with bound states may appear.\footnote{In partial waves, a kinematical circular cut appears in the complex plane for the unequal masses case~\cite{Martin:1970}. While this important for the most precise amplitude determinations~\cite{Buettiker:2003pp,Descotes-Genon:2006sdr,Pelaez:2020gnd}, it will be largely ignored in the following.}  Unitarity provides us with the means to continue our partial waves into the next contiguous Riemann sheets, where resonances live. The existence of a non-zero imaginary part due to this principle produces a multi-valued complex function, which has a physical branch cut produced by $s$-channel unitarity. A parametrization that fulfills this principle is the customary $K$-matrix formalism~\cite{Rosenfeld:1970ve,Aitchison:1972ay},
\begin{equation}
    t_\ell(s)=K_{\ell}(s)\,\left[1-i\rho(s)K_{\ell}(s)\right]^{-1}\, ,
    \label{eq:kmat}
\end{equation}
where $K_\ell(s)$ is a real symmetric matrix. The simplest parametrizations for $K_{\ell}^{i j}(s)=g^{i}g^{j}/(M^2-s)$ contain the ``bare'' information about the resonance: Formally Eq.~\eqref{eq:kmat} can be expanded as
\begin{equation}
    t_\ell(s)\simeq K_{\ell}(s)+K_{\ell}(s)i\rho(s)K_{\ell}(s)+K_{\ell}(s)i\rho(s)K_{\ell}(s)i\rho(s)K_{\ell}(s)+\dots\, .
\end{equation}
In this limit, the resonance basically behaves like a quasi-stable particle of mass $M$ propagated between different initial $(i)$ and final $(j)$ states, and acquires a width due to the couplings with the continuum. However, the physical objects are the poles of $t_\ell(s)$, not of $K_\ell(s)$. As we will see in the following sections there is no one-to-one correspondence between the two, and so this interpretation of $K_\ell(s)$ must be taken with a grain of salt.

 Finally, in order to describe the data by means of analytic functions, we will make use of the {\it Chew-Mandelstam formalism}~\cite{Chew:1960iv}, in which the ordinary phase space $\rho(s)$ is replaced by its dispersive form.
 
In addition to right hand cuts produced by unitarity, the partial waves can exhibit more complicated structures, like left-hand cuts as a result of crossing symmetry and unitarity in the crossed channels (we refer the reader to~\cite{Martin:1970} for a reference textbook on the topic). 
The $N/D$ formalism~\cite{Chew:1960iv,Bjorken:1960zz,Oller:1998hw,Oller:1998zr} makes the splitting between these left and right hand cuts explicit. The partial wave can be recast as
\begin{equation}
    t_\ell(s)=N_\ell(s)\,D^{-1}_\ell(s),
\end{equation}
which is designed to separate the constraints coming from unitarity in the direct ($D$) and cross-channels ($N$). The latter can be often interpreted in terms of the intensity of the produced resonances. Although the two functions should be related through complicated dispersive equations, in practice we will use a simple functional form for $N$. The main reason for this is that we mostly study energy regions far from crossed channel cuts.

\begin{figure}
    \centering
    \includegraphics[width=0.4\textwidth]{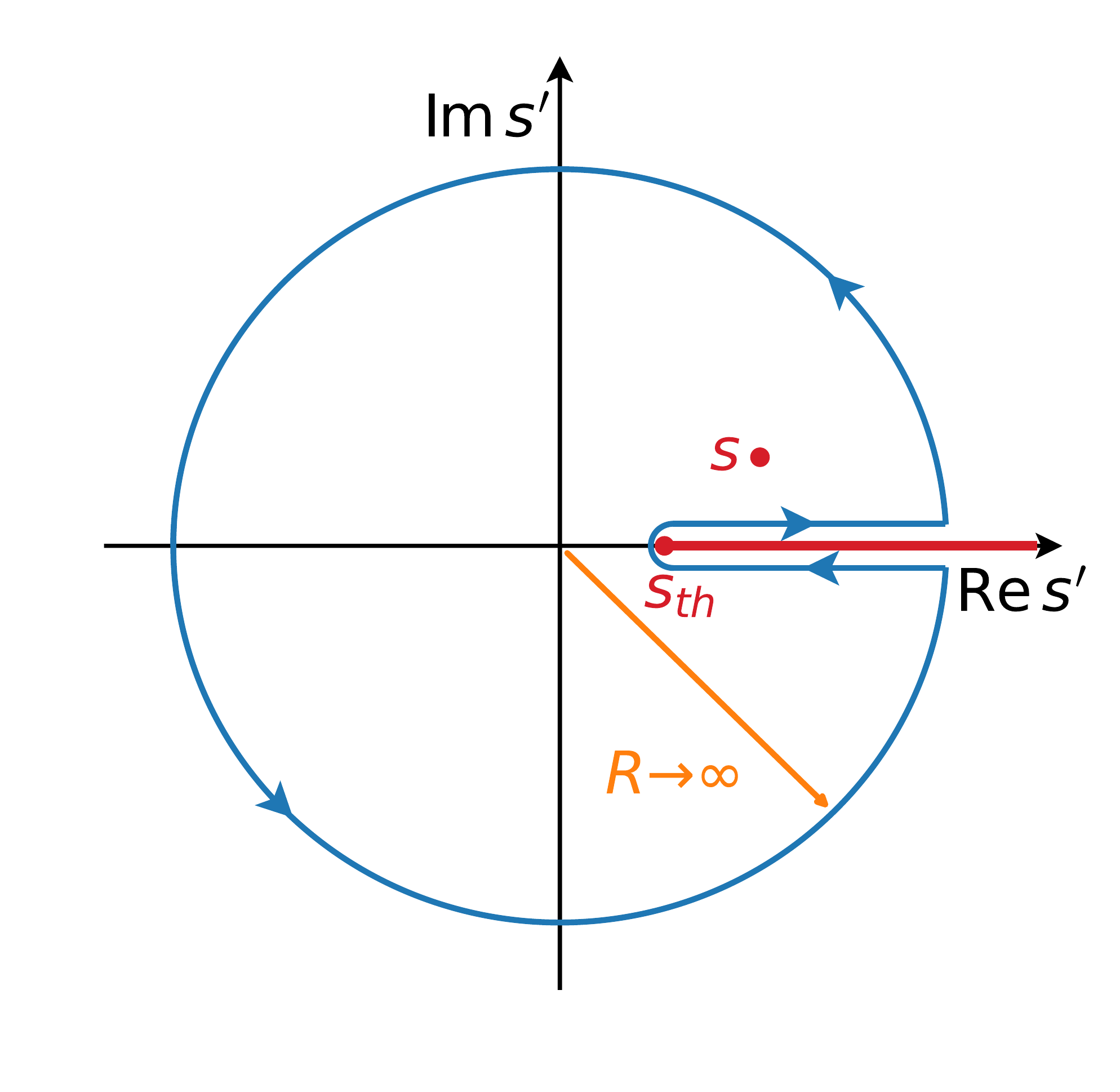}
      \caption{Typical application of Cauchy theorem and dispersion relations. The integral across the cut can be given in terms of the imaginary part. The integral over the circle at infinity vanishes if the integrand goes to zero fast enough. }\label{fig:cauchy}
\end{figure}

A fundamental tool to impose analytic properties is given by dispersion relations. Let us call $F(s)$ a function that is analytic everywhere in the complex plane, except for a right-hand cut, and fulfills Schwartz reflection principle, $F(s^*) = F^*(s)$. Its value at any point in the complex plane is given by Cauchy theorem, $F(s) = \frac{1}{2\pi i}\oint ds' \frac{F(s')}{s'-s}$, where the path can be any closed curve that does not cross the cut. By choosing the path represented in \figurename{~\ref{fig:cauchy}}, the integral can be decomposed as
\begin{equation}
    F(s) = \frac{1}{2\pi i}\int_{s_\text{th}}^\infty ds' \frac{F(s'+i\epsilon) - F(s'-i\epsilon) }{s'-s} + \int_R ds' \frac{F(s) }{s'-s}\,,
\end{equation}
where the second term is the integral over a circle of infinite radius ($R \to \infty$) in the complex plane. The first term is $\frac{1}{\pi}\int_{s_\text{th}}^\infty ds' \Im F(s') /(s'-s)$, and 
for example one can use unitarity to relate the imaginary part to other quantities. For the theorem to be useful, the integral over the contour at infinity must vanish, which happens only if the integrand goes to zero fast enough. If not, one can perform subtractions in $s=s_0$ to the dispersion relation, \ie apply Cauchy theorem to  $\left(F(s) - \sum_{k=0}^{N-1} F^{(k)}(s_0) (s-s_0)^k/k!\right)\Big/(s-s_0)^{N}$, and get
\begin{equation}
    F(s) = \sum_{k=0}^{N-1} F^{(k)}(s_0) \frac{(s-s_0)^k}{k!} + \frac{(s-s_0)^{N}}{\pi}\int_{s_\text{th}}^\infty ds' \frac{\Im F(s') }{(s'-s_0)^{N}(s'-s)}\,.
\end{equation}
By choosing $N$ large enough, one can always make the integral converge if the function has no essential singularities. The price to pay is that there are $N$ undetermined constants.\footnote{In Quantum Field Theories, this can be considered as a renormalization procedure, where a number of observables must be sacrificed to reabsorb the divergences.} 
More detailed introductions to dispersion relations can be found \eg in Refs.~\cite{Martin:1970,Oller:2019rej}.

When looking at the final state of $\onetothree$ decay products, one must consider that there are several processes that could produce a local enhancement in the cross section. We first focus on the study of Dalitz plot distributions, fixing the mass of the decaying particle. A resonance in a two-body subchannel generally produces broad structures when projected onto another channel, so no confusion usually arises. However, besides poles, if the kinematics overlap, a resonance of the crossed channel could rescatter into these final products, enhancing the cross section and mimicking a resonance in the direct channel. At leading order these rescattering processes are called triangle singularities, as named by Landau~\cite{Landau:1959fi}. These singularities appear in the integrand of the corresponding Feynman diagram and `pinch' the integration domain, producing a logarithmic branch point as a result. The implications of this on phenomenology have been widely discussed in the literature, see for example~\cite{Guo:2019twa}.
We will discuss examples of these processes in Sections~\ref{sec:zc3900} and~\ref{sec:pc4337}. When one focuses on the dependence on total energy, for example to study the line shape of a resonance decaying into a three-body final state, different complications arise. We will discuss examples of these processes in Sections~\ref{sec:3bodyApp}. 

We recall that more complications arise from particle spin, even though they are purely kinematic in nature.  For example, consider a \onetothree decay of particles with spin. The Legendre polynomials discussed above will be promoted to Wigner $D$-matrices. Writing helicity amplitudes for the three two-body subchannel in the different resonance frames requires boosts that do not conserve the helicity of the various particles. To add the various contributions coherently, one has to take into account the so-called Wigner rotations (crossing matrices) associated with precession of particle spins when moving from one frame to another. 
There is a recent interest in this, motivated by the fact that the practical implementation of such rotations is highly nontrivial~\cite{Wang:2020giv,Marangotto:2019ucc}. A proposal to write the \onetothree reaction as a sum over subchannels in the same reaction plane, and then rotate the whole sum together, was given in~\cite{JPAC:2019ufm}, and is referred to as ``Dalitz plot decomposition''. In this way, the dependence of the Wigner rotations on the relevant Mandelstam variables is apparent.
\begin{figure}
    \centering
    \includegraphics[width=\textwidth]{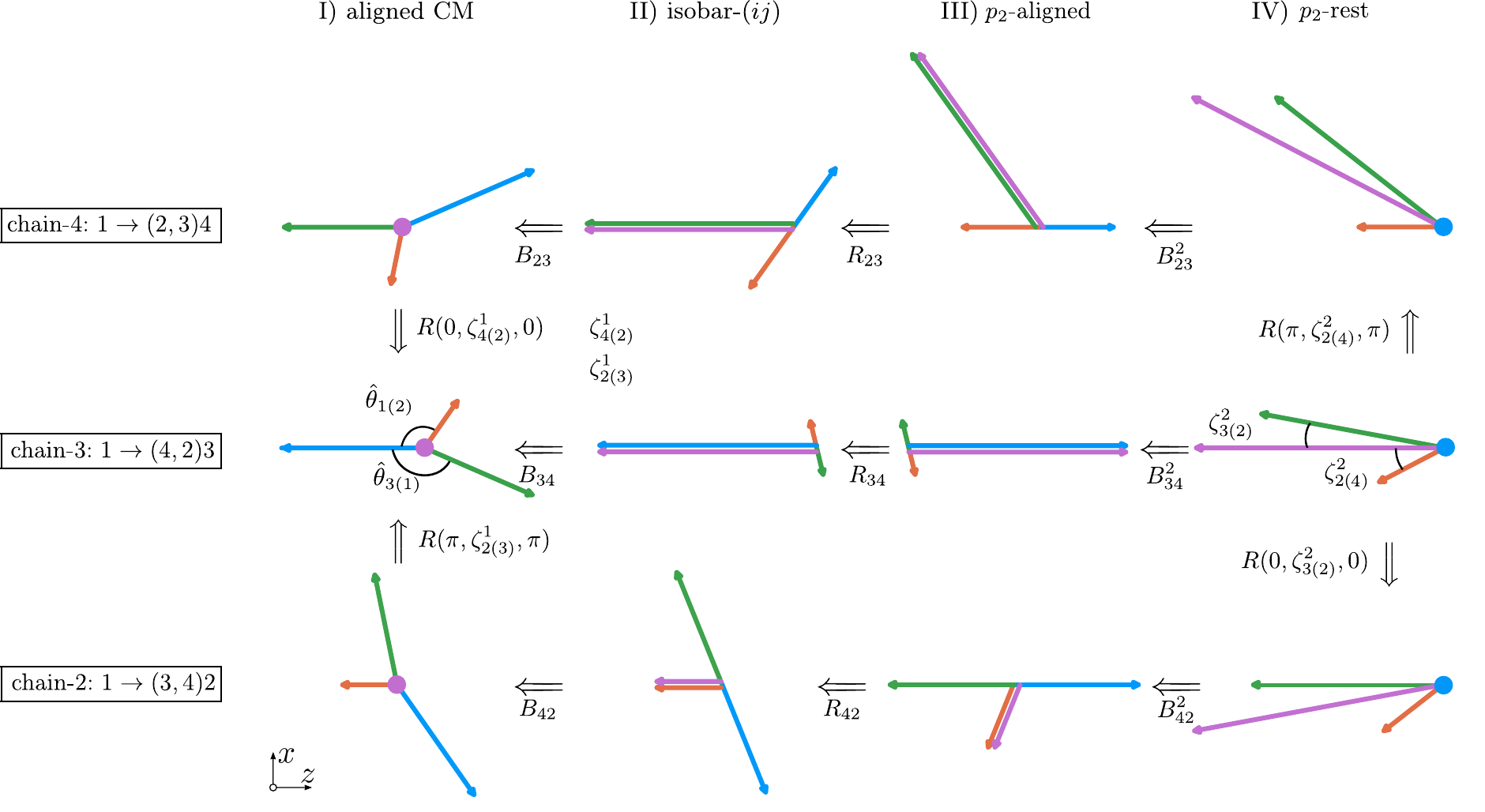}
    \caption{Demonstration of the spin alignment problem for the decay $1\to 2,3,4$ with the particles $1$ and $2$ having a non-trivial spin.
    The three rows show the angles in the amplitude construction
    in different decay chains. The double arrows stand for active transformations (boosts and rotations). Figure adapted from~\cite{JPAC:2019ufm}. }\label{fig:DPD}
\end{figure}
Explicitly, the amplitude factorizes in
\begin{align}
 T^{\Lambda}_{\lambda_2,\lambda_3,\lambda_4} = \sum_{\lambda_1} D^{j_1*}_{\Lambda,\lambda_1}\!\left(\alpha,\beta,\gamma\right)\, A_{\lambda_1,\lambda_2,\lambda_3,\lambda_4}(s,t,u)\,,
\end{align}
where $D^{j_1*}_{\Lambda,\lambda_1}\!\left(\alpha,\beta,\gamma\right)$ is the Wigner $D$-matrix that takes into account the alignment of the reaction plane in the laboratory frame in terms of the Euler angles $\alpha$, $\beta$, $\gamma$, and
\begin{align} \label{eq:dpd.isobar}
    A_{\lambda_1,\lambda_2,\lambda_3,\lambda_4}\!\left(s,t,u\right) =
        \sum_{j} A_{\lambda_1,\lambda_2,\lambda_3,\lambda_4}^{(23),j}\!\left(s,t,u\right) +
        \sum_{j} A_{\lambda_1,\lambda_2,\lambda_3,\lambda_4}^{(34),j}\!\left(s,t,u\right) +
        \sum_{j} A_{\lambda_1,\lambda_2,\lambda_3,\lambda_4}^{(42),j}\!\left(s,t,u\right)\,.
\end{align}
This decomposition looks like a partial-wave expansion, but is performed over all the two-body subchannels, and is known as {\it isobar representation}. This will be discussed further in Section~\ref{sec:KT}. The isobar amplitude $A_{\lambda_1,\lambda_2,\lambda_3,\lambda_4}^{(xy),j}$ contains resonances in the $(xy)$-channel of spin $j$, with particle $z$ as spectator. The situation in the reaction plane is represented in \figurename{~\ref{fig:DPD}}.

The unaligned isobar amplitude reads:
\begin{align} \label{eq:dpd.hdh}
    T_{\lambda_1',\lambda_2',\lambda_3',\lambda_4'}^{(xy),j} = 
    h^{(z),j}_{\lambda_z'+\lambda_1',\lambda_z'}(\sigma_{xy})\,
    (-1)^{j_{z}-\lambda'_{z'}}\,
    d^j_{\lambda_1'+\lambda'_{z},\lambda'_{x}-\lambda'_{y}}(\theta_{xy}) \,
    h^{(xy),j}_{\lambda_x',\lambda_y'}(\sigma_{xy})\,
    (-1)^{j_{y}-\lambda'_{y}}\,\mathcal{R}_j(\sigma_{xy})\,.
\end{align}
where $\theta_{xy}$ is the decay angle in the subchannel $(xy)$ rest frame, precisely,
the angle between $\vec{p}_x$ and the $z$ axis set by $-p_z$. 
Assuming the cascade process for the three-body decay, 
the energy-dependent part of the isobar amplitude is factorized into a product of two vertex functions,
$h^{(z),j}_{\lambda_z'+\lambda_1',\lambda_z'}$, and $h^{(xy),j}_{\lambda_x',\lambda_y'}$,
corresponding to $1\to (xy), z$, and $(xy)\to x,y$ decays, respectively,
and the isobar lineshape function, common to all helicity combinations.
Here, we have made explicit the phases $(-1)^{j-\lambda}$ due to the Jacob-Wick particle-2 convention, that leads to the natural matching with the $LS$ decomposition~\cite{Jacob:1959at,JPAC:2019ufm}. 
Equation~\eqref{eq:dpd.hdh} for different chains can be added together once they are all aligned to the common definition of the helicity indices as in Eq.~\eqref{eq:dpd.isobar}.
The difference in the definition of the helicity states for different chains is evident from \figurename{~\ref{fig:DPD}}.
\begin{align} \label{eq:dpd.master}
    A_{\lambda_1,\lambda_2,\lambda_3,\lambda_4}^{(xy),j} &= \sum_{\lambda_1',\lambda_2',\lambda_3', \lambda_4'}  T_{\lambda_1',\lambda_2',\lambda_3',\lambda_4'}^{(xy),j}\,
    \,\,
    d^{j_1}_{\lambda_1,\lambda_1'}(\zeta_{z(r_1)}^1)
    \, d^{j_2}_{\lambda_2',\lambda_2}(\zeta_{z(r_2)}^2)
    \, d^{j_3}_{\lambda_3',\lambda_3}(\zeta_{z(r_3)}^3)
    \, d^{j_4}_{\lambda_4',\lambda_4}(\zeta_{z(r_4)}^4)\,,
\end{align}
where the index $r$ for every particle indicates the frame where the unprimed helicities of this particle is defined. The alignment angles $\zeta_{z(r)}$ depend on $s,t,u$ and they are trivial if $r=z$. The explicit expressions for the general case are found in~\cite{JPAC:2019ufm}.
A practical method to validate the spin alignment is suggested in Ref.~\cite{Wang:2020giv}.

Another consequence of spin is the presence of kinematical singularities, that must be removed before studying dispersion relations. One can argue what the simplest factors needed to control these singularities  are, and what the minimal energy dependence is that one therefore expects. This was done in the context of $\bar B^0 \to \jpsi \pi^+ K^-$ and $\Lambda_b^0 \to \jpsi p K^-$ in~\cite{JPAC:2017vtd,JPAC:2018dfc}.

As a final remark, when dealing with all these different reactions, exploring a number of possible parametrizations helps to reduce the model dependence. It allows one to assess systematic uncertainties in one's results. Furthermore, exploring these parametrizations, combined with a proper statistical analysis, allows one to distinguish the poles corresponding to physical resonances from model artifacts. This will be shown in detail in Section~\ref{sec:spurious}. Hence, we will adopt this approach for our analyses in this review. Alternatively to this procedure, other model-independent analytic continuation methods have been pursued. We mention here Pad\'e approximants~\cite{Masjuan:2013jha,Masjuan:2014psa}, Laurent-Pietarinen expansion~\cite{Svarc:2014sqa}, the Schlessinger point method~\cite{Tripolt:2016cya,Binosi:2022ydc}, or Machine-Learning techniques~\cite{Yoon:2018,Fournier:2018}.

\subsection{Statistics tools}
\label{sec:statistics_tools}
The determination of the existence of each resonance and its properties 
relies on fitting experimental data accompanied by an uncertainty analysis.
We review the general strategy and some of the techniques employed by JPAC. 
This is particularly relevant for pole extraction, where the error propagation through standard means is complicated.
In doing so we mostly take a frequentist point of view~\cite{Cowan:2010js,Cowan:2018lhq}.
It is also possible to perform similar analyses from a Bayesian perspective. 
For an introduction to Bayesian statistics we refer the reader to~\cite{Sivia} and for the specific case of its application to high energy physics to~\cite{DAgostini:1999gfj}.

\subsubsection{Fitting data}

\begin{figure}
    \centering
    \includegraphics[width=.75\textwidth]{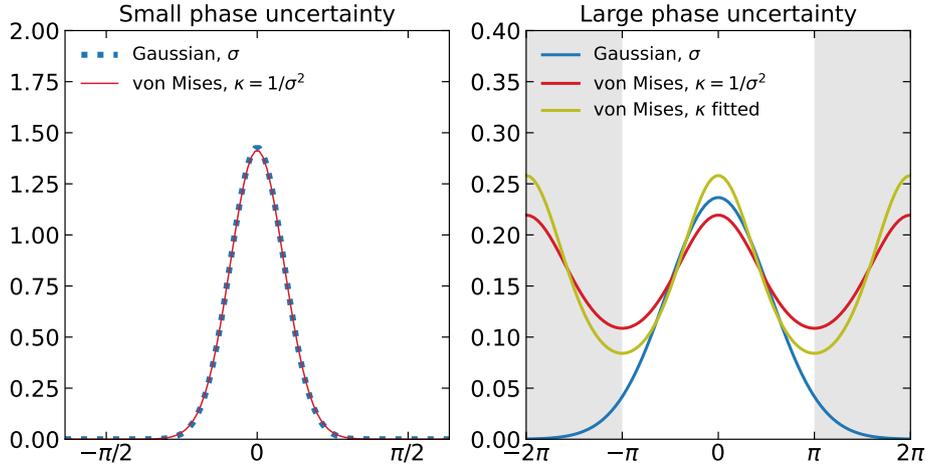} 
          \caption{Comparison between the Gaussian and von Mises distributions for small (left)
      and large (right) phase shift uncertainties, centered in $\mu = 0$. 
      In the upper plot the Gaussian distribution has $\sigma=0.28$ (dotted blue), 
      and von Mises has $\kappa = 1/\sigma^2 = 12.82$ (solid red).
      In the lower plot the Gaussian has $\sigma=1.69$ (solid blue), 
      von Mises has $\kappa = 1/\sigma^2=0.35$ (solid red),
      and another von Mises has $\kappa=0.56$ (solid gold) obtained by
      fitting to the Gaussian distribution.
      The gray bands
      hide the region outside the $[-\pi,\pi]$ range.
      Figures from~\cite{Bibrzycki:2021rwh}.}
    \label{fig:vonmises}
\end{figure}

The standard approach to fitting data is through maximizing the likelihood,
\begin{align}
L(\{ \theta\}|\{ y \})= \prod_i^N P_i(y_i|\theta_i) \, , \label{eq:likelihood}
\end{align}
where $P_i(y_i|\theta_i)$ stands for the probability density function
at fixed parameter $\theta_i$ and
$y_i$ is the experimental datapoint. 
If chosen as Gaussian
for binned data,
\begin{align}
P_i(y_i|\theta_i)  = \frac{1}{\sqrt{2\pi}\sigma_i}
\exp\left[ -\frac{1}{2}\left(\frac{f_i(\{\theta\})-y_i}{\sigma_i}\right)^2\right] \, ,
\end{align}
where $y_i$ is the binned experimental datapoint value and $\sigma_i$ its uncertainty. 
$f_i(\{\theta\})$ is the objective function to be fitted.
This expression assumes each bin to be statistically independent, as is customary.
Maximizing the likelihood is equivalent to 
minimizing the $\chi^2$ function,
\begin{align}
\chi^2(\{\theta\}) = \sum_i^N \left( \frac{f_i(\{\theta\})-y_i}{\sigma_i}\right)^2 . \label{eq:chi2}
\end{align}
The choice of a Gaussian distribution 
is standard in many physical problems, and is adequate
if the experimental uncertainties are of statistical origin.
However, there are situations where other probability densities
should be chosen. For example, in the case that the $y$
observable
is positively defined (\eg an intensity) 
and, because of the values of $y_i$ and  
$\sigma_i$ there is a significant overlap with unphysical negative values, a Gamma distribution may be more appropriate, 
\begin{align}
H(y_i|\theta_i) = \left( \frac{y_i\theta_i}{\sigma_i^2} \right)^\frac{\theta_i^2}{\sigma_i^2}\frac{\exp\left(-y_i\theta_i/\sigma_i^2\right)}{\theta_i\Gamma\left( y_i^2/\sigma_i^2\right)}\,. \label{eq:gammadistribution}
\end{align}
This was used for example in~\cite{HillerBlin:2016odx,Rodas:2021tyb}.
Another common issue occurs when the observable is periodic, for example when dealing with a relative phase. 
A simple solution is to redefine the difference in Eq.~\eqref{eq:chi2} to take the periodicity into account,
\begin{equation}
    \chi^2(\{\theta\}) = \sum_i^N \min_{k \in \Z} \left(\frac{f_i(\{\theta\}) - y_i - 2k\pi}{\sigma_i}\right)^2\,,
\end{equation}
as is done in~\cite{JPAC:2018zyd,Rodas:2021tyb}.
However, using a von Mises distribution 
may be more rigorous,
\begin{align}
M(y_i|\theta_i) = \frac{1}{2\pi I_0(\kappa_i)}\exp\left[\kappa_i \cos \left( f_i(\{\theta\}) - y_i\right) \right] \, , \label{eq:vonmises}
\end{align}
where $I_0(\kappa_i)$ is the modified Bessel function. 
The concentration parameter $\kappa_i$ is the reciprocal measurement of the dispersion. 
If the uncertainty is small, a Gaussian distribution with $\sigma_i$ equal to the experimental uncertainty is almost equivalent to a von Mises distribution with $\kappa_i = 1/\sigma_i^2$. For larger values of the uncertainty, 
Gaussian and the von Mises distribution with $\kappa_i = 1/\sigma_i^2$ are quite different, and it is better to refit the concentration parameter to the $(y_i,\sigma_i)$ Gaussian distribution as done in~\cite{Bibrzycki:2021rwh}
and shown in \figurename{~\ref{fig:vonmises}}.
Of course, in such a case $\chi^2$ is no longer the correct estimator
to maximize the likelihood. 
The best strategy here would be to compute the 
logarithm of the
likelihood from Eq.~\eqref{eq:likelihood} using the appropriate
distributions, and maximize the obtained function.
For example, if a partial wave is to be fitted with
experimentalists providing $N_{\text{pw}}$ phase shifts and intensities, some of them compatible with zero within uncertainties, 
the likelihood in~\eqref{eq:likelihood}
can be computed by defining the probability distributions $P_i(y_i| \theta_i)$ for the phase shifts through the 
von Mises
distribution in~\eqref{eq:vonmises} and the intensities
through the Gamma distribution in~\eqref{eq:gammadistribution}.
In this case the likelihood for a single energy bin reads:
\begin{align}
L(\{ \theta\}|\{ y \})= 
\prod_i^{N_{\text{pw}}} M(y^\phi_i|\theta_i)  \prod_i^{N_{\text{pw}}} H(y^I_i|\theta_i)\, ,
\end{align}
where the $y^\phi$ and $y^I_i$ stand for the experimental
phase shifts and intensities, respectively.
and the quantity to minimize would be
$-\log L(\{ \theta\}|\{ y \})$.

The standard likelihood function as presented in Eq.~\eqref{eq:likelihood} suits many data analysis. However, a
special situation happens when we need to fix the 
normalization in a fit to events. In that case
the likelihood function needs to be modified to incorporate such constraints, producing the so-called extended maximum likelihood method~\cite{Lyons:1985vx,Barlow:1990vc,James:2006zz}:
\begin{align}
{\mathcal L}(\{ \theta\}|\{ y \}) 
= \sum_{i}^N \left[
     f_i(\{\theta\}) -
     y_i \log f_i(\{\theta\}) \right]\, , \label{eq:enllh}
\end{align}
as was employed in~\cite{Bibrzycki:2021rwh}.
In the extended maximum likelihood formulation, the normalization of the probability distribution function is allowed to vary, and, thus it becomes applicable to problems in which the number of samples obtained is itself a relevant measurement.
As normalization correlates all the datapoints, one has to be careful
on how the D'Agostini bias might impact the fit~\cite{DAgostini:1993arp}.

The minimization is usually performed using
a gradient-based optimization method such as
{\sc MINUIT}~\cite{minuit}
or Levenberg-Marquardt~\cite{Levenberg:1944,Marquardt:1963}.
Unfortunately, multiple
local minima can appear, preventing the
optimizer from finding the physically sensible
minimum. The typical strategy is to
try many initial values for the parameters
at the beginning of the optimization process
and then compare all the minima obtained.
Another approach is to explore the
parameter space using a genetic algorithm~\cite{Ireland:2004kp,Fernandez-Ramirez:2008ixe,Fernandez-Ramirez:2015tfa}, and then improve the result with a 
gradient-based method.
Knowing about the existence of nearby local minima is necessary to have a better interpretation of the results.
Also, a good set of initial parameters 
is required to apply the bootstrap method 
 detailed below.

\subsubsection{Uncertainties estimation with bootstrap}
\label{sec:stat_bootstrap}

\begin{figure}
    \centering
    \includegraphics[width=.55\textwidth]{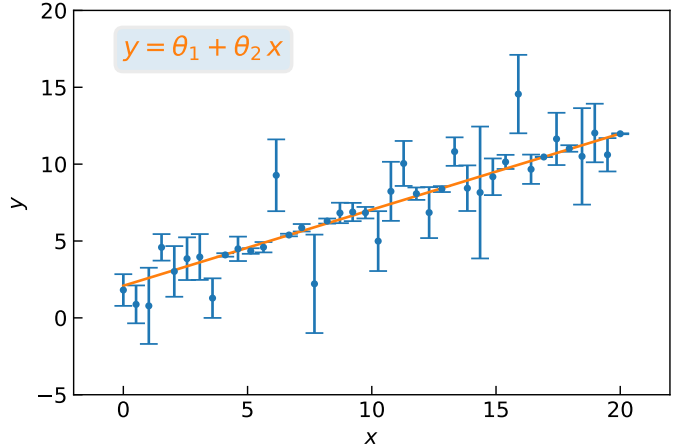}     
    \caption{Datapoints and BFF for the linear fit example.}
    \label{fig:linear_fit_stat}
\end{figure}

The fit needs to be accompanied by an
error analysis, as well as a method to propagate
the uncertainties from the fit parameters
to the physical observables, whose relationship may be highly nontrivial.
A standard approach is to
use the covariance matrix obtained
from the Hessian
of the likelihood
as given by, for example, {\sc MIGRAD}~\cite{minuit}.
This relies on the parabolic approximation of the 
likelihood function around the minimum, which always provides symmetric uncertainties for the fitted parameters. The main advantage of this is that this process is computationally cheap, and
in many circumstances this approximation is good enough.
For more refined determinations of the uncertainties, the high-energy physics community usually relies
on {\sc MINOS}, which samples
the likelihood in the neighborhood of the minimum and is able to provide asymmetric uncertainties for the fit parameters. 
However, propagating errors from parameters to observables using {\sc MINOS}
is unattainable for nontrivial functions like pole extractions.
To overcome this, we can use the method of bootstrapping, a Monte Carlo based method~\cite{EfroTibs93,Landay:2016cjw}.
Although computationally expensive,
its results are robust and rigorous.

For pedagogical reasons we explain the technique
through a linear fit example, and benchmark with the results
of {\sc MIGRAD} and {\sc MINOS}.\footnote{The Python code for this example and a simplified version of the analysis in~\cite{Fernandez-Ramirez:2019koa} can be downloaded from~\cite{Fernandez-Ramirez:2021github}.}
We consider a model $y=0.5 + 2\, x$. We generate $N=40$ datapoints 
 uniformly in $x \in \left[0,20 \right]$, and for each of them
generate an uncertainty $\Delta y_i$ extracted from a Gaussian distribution with zero mean
and $\sigma=1.5$. Then, we compute the noise $\nu_i = \hat{\nu}\times \Delta y_i$ where $\hat{\nu}$ is generated from a Normal
distribution.
Finally, the datapoint is $y_i = 0.5 + 2\, x_i + \nu_i$
with associated error $\Delta y_i$. 
Figure~\ref{fig:linear_fit_stat} shows the computed datapoints.
We use {\sc MINUIT} $\chi^2$ minimization to fit these data 
to a linear model $y=\theta_1+\theta_2\, x$. The best fit found (BFF)
has $\chi^2_\text{BFF}/\dof=36.75/\left(40-2\right)=0.967$,
$\theta_1=0.495\pm0.004$, and $\theta_2=2.09\pm 0.07$. 
The error is computed using
{\sc MIGRAD}, but {\sc MINOS} gets the same results, as the likelihood is symmetric by construction.

\begin{figure}
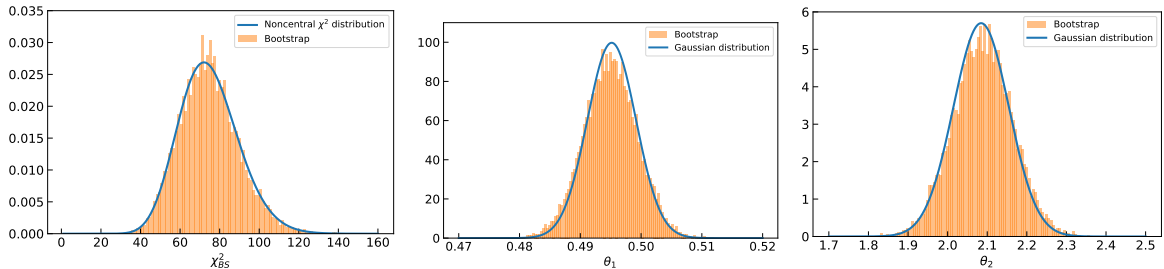

    \centering
    \includegraphics[width=.32\textwidth]{{../figures/statistics/noncentral_chi2}.pdf}     
    \includegraphics[width=.3\textwidth]{{../figures/statistics/theta1}.pdf} 
    \includegraphics[width=.3\textwidth]{{../figures/statistics/theta2}.pdf} 
    \caption{(left) Theoretical noncentral $\chi^2$ distribution \vs the histogram from the bootstrap fits for $\chi^2_\text{BS}$. 
    (center) Theoretical Gaussian distribution \vs the histogram from the bootstrap fits for $\theta_1$ parameter. 
    (right) Theoretical Gaussian distribution \vs the histogram from the bootstrap fits for $\theta_2$ parameter.}
    \label{fig:linear_example_distributions}
\end{figure}
We repeat the fit with bootstrap.
We find the best fit by minimizing the $\chi^2$.
We can resample each datapoint, generating a new one from a Gaussian distribution having by mean the original $y_i$ 
value and $\sigma_i = \Delta y_i$. 
In this way we generate a new pseudodata set $\{\tilde{y}\}$
that is compatible with the experimental measurement. 
The uncertainties of the new pseudodata set are
fixed to original ones $\{\Delta y \}$.
The new pseudodata set can be refitted with the original model, obtaining a set of
parameters $\{ \theta \}_1$
and the associated $\left[\chi^2_\text{BS}\right]_1$.
Then, we repeat the procedure until we have acquired the desired statistical significance. 
We call each fit to one pseudodata set a {\it bootstrap} (BS) {\it fit}.
The results of the process are the histograms of the
parameters $\{ \theta \}$ and the $\{ \chi^2_\text{BS} \}$.
Since $\Delta y_i$ is assumed Gaussian, the $\{\chi^2_\text{BS}\}$ follow a 
noncentral $\chi^2$ distribution,
\begin{align}
\chi^2_{nc}(x|k,\lambda)= \frac{1}{2}\exp\left[ -\frac{\lambda+x}{2}\right] 
\left(\frac{x}{\lambda}\right)^{\left(k-2\right)/4}
 I_{\left(k-2\right)/2}\left(\sqrt{\lambda x} \right) \, , \label{eq:noncentralchi2}
\end{align}
where $\lambda=\chi^2_\text{BFF}$, $k$ the number of degrees of freedom. Figure~\ref{fig:linear_example_distributions} shows the comparison
between Eq.~\eqref{eq:noncentralchi2} and the $\{\chi^2_\text{BS}\}$
distribution from the $M=10^4$ BS fits, which approximately peaks at 
$\sim 2\,\chi^2_\text{BFF}$.
Figure~\ref{fig:linear_example_distributions} also shows the $\{\theta_1\}$ and $\{\theta_2\}$ histograms,
which are Gaussian and give
$\theta_1=0.495\pm0.004$, and $\theta_2=2.09\pm 0.07$, the same result as {\sc MIGRAD}.
The expected value of the parameters is computed as the
mean of the $\{\theta_1\}$ and $\{\theta_2\}$ histograms,
and the $1\sigma$ uncertainties (68\% confidence level) from the 16th and 84th quantiles.
Any desired confidence level can be computed
selecting the appropriate quantiles, given that enough BS fits are computed, since the accuracy 
scales as $1\slash\sqrt{M}$.
For example, if $M=10^3$ BS fits are performed, the 
accuracy of our results would be $3.2\%$; not good enough to claim a $2\sigma$ (95.5\%) confidence level.

The covariance and correlation matrices are straightforward 
to compute from the BS fits,
\begin{align}
\text{cov}(\theta_i,\theta_j) = \sum_{k=1}^M \frac{\left(\left[\theta_i\right]_k-\left\langle \theta_i \right\rangle \right)\left( \left[\theta_j\right]_k-\left\langle \theta_j \right\rangle\right)  }{M} \, ; \,
\text{corr}(\theta_i,\theta_j) = \frac{\text{cov}(\theta_i,\theta_j)}{\sqrt{\text{cov}(\theta_i,\theta_i)}\sqrt{\text{cov}(\theta_j,\theta_j)}} \, .
\end{align}
The covariance and correlation matrices are very similar to the ones obtained with {\sc MIGRAD}.
\begin{align}
\text{cov}(\theta_0,\theta_1)_\text{Hessian}&=\begin{bmatrix}
55.7 & -3.27 \\
-3.27 & 0.19
\end{bmatrix} \times 10^{-4}\, ;& \,
\text{corr}(\theta_0,\theta_1)_\text{Hessian}&= \begin{bmatrix}
1 & -0.996  \\
-0.0996 & 1 
\end{bmatrix} \, ; \notag \\
\text{cov}(\theta_0,\theta_1)_\text{Bootstrap}&=\begin{bmatrix}
54.8 & -3.22 \\
 -3.22 & 0.19 
\end{bmatrix}\times 10^{-4}\, ;& \,
\text{corr}(\theta_0,\theta_1)_\text{Bootstrap}&= \begin{bmatrix}
1 & -0.996  \\
-0.0996 & 1 
\end{bmatrix} \, ; \notag
\end{align}
as expected in this simple example. 
Hence, we showed how bootstrap and a standard Hessian method
are equivalent, given enough BS fits are computed.

\begin{figure}
    \centering
    \includegraphics[width=.7\textwidth]{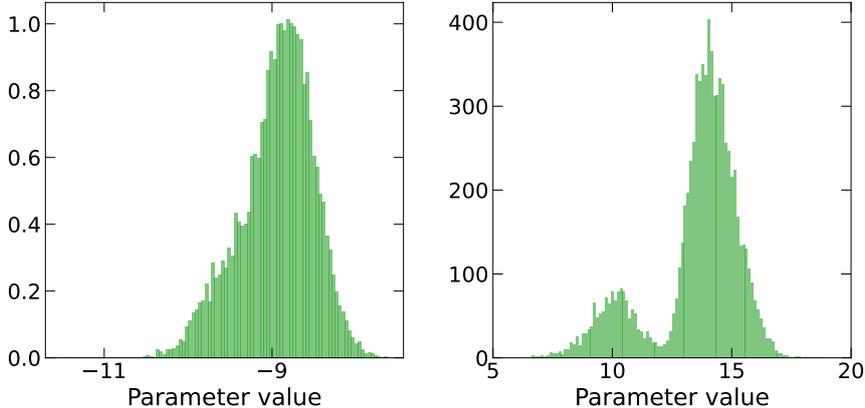} 
    \caption{Examples of a realistic case from~\cite{Bibrzycki:2021rwh} where one parameter histogram is well behaved (left) and follows a 
    Gaussian-like behavior while another 
    has two nearby minima (right).}
    \label{fig:parameter_distribution}
\end{figure}
The calculation of any observable $g(\{\theta\})$ and the propagation of the uncertainties 
is straightforward. For
each set of parameters $\left[ \{ \theta \}\right]_i$ 
obtained from a BS fit, we compute the observable 
$g_i=g(\left[ \{ \theta \}\right]_i)$, obtaining
$M$ values of $g_i$.
From the histogram we can compute the expected value $\left\langle g \right\rangle$
and the uncertainties as done for the parameters $\{ \theta \}$.
This procedure is independent of the functional form of $g$ 
and fully propagates the uncertainties in the parameters
and their correlations to the derived observable.

For simplicity we explained the method using data from a linear model 
who are statistically independent and whose
uncertainties follow Gaussian distributions.
Hence, there was only one minimum for the BFF
and the histograms of  $\{ \theta \}$ and the $\{ \chi^2_\text{BS} \}$
were Gaussian and noncentral $\chi^2$-distributed, respectively.
Extending the method to any other distribution is straightforward, both at the level of the likelihood function
and at the generation of the pseudodata sets.
If the experimental datapoints 
are correlated,
one can generate the pseudodata according to the correlation matrix.
Similarly, one can incorporate correlated errors, as systematic uncertainties.
The only disadvantage is that systematic and statistical uncertainties
propagate together,
so they cannot be disentangled in the observables.

If the BFF has a local minimum nearby,
it is possible
for the bootstrap to jump from the global minimum to the 
local one. In that case, the parameter distribution
can follow a two peak structure (see \figurename{~\ref{fig:parameter_distribution}}) and the expected value,
the uncertainties, and any other computed quantities
have to be taken with a grain of salt.
It is possible to analyze and study each minimum separately by computing uncertainties and comparing the solutions, but choosing one minimum over others and the resulting conclusions would depend on the separations of the peaks in the parameter histograms, on the correlations of the parameters of the model, and on whether the fits leading to the different minima have systematically different likelihoods.  
There is no simple general recipe to follow and each case must be studied independently.

\begin{figure}
    \centering
    \includegraphics[width=.7\textwidth]{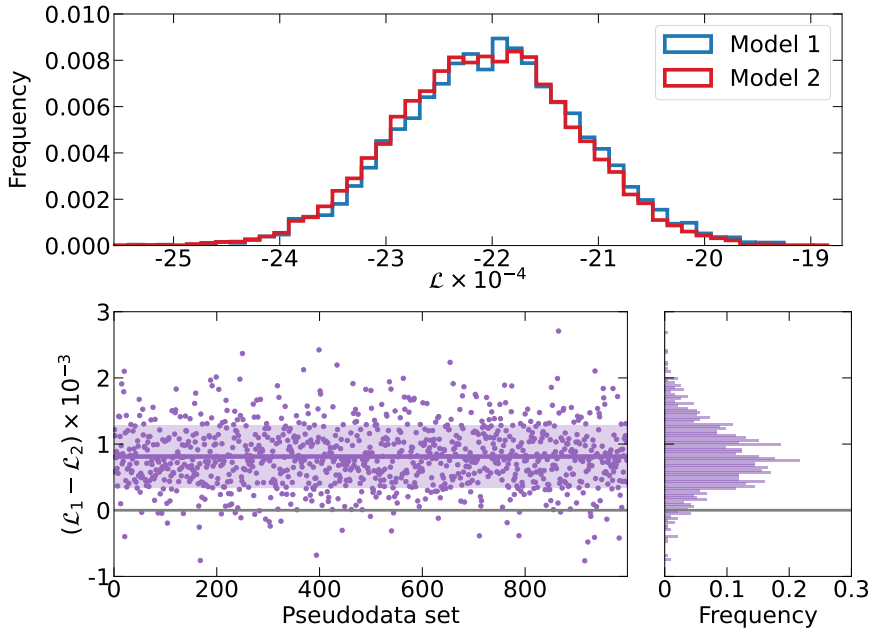} 
    \caption{Example from~\cite{Bibrzycki:2021rwh}.
    (top plot) Extended negative log-likelihood for two models. (bottom row) Difference in negative log-likelihood at every one of the first 1000 pseudodata sets (left) and frequency for the $10^4$ computed pseudodata sets (right). The purple line represents the mean, the band the $68\%$ confidence level. The distribution sits mostly above the zero difference (gray line).}
    \label{fig:likelihood}
\end{figure}

Bootstrap results can be exploited to compare two 
models of apparently similar quality in terms of the $\{\chi^2_\text{BFF}\}$
and $\{ \chi^2_\text{BS}\}$ distributions.
Given the two models,
for each pseudodata set we can fit both and compare
them for each BS fit. If one model systematically
outperforms the other, it is of better quality.
This was exploited in~\cite{Bibrzycki:2021rwh} 
for the case of extended negative 
loglikelihood~[Eq.~\eqref{eq:enllh}] fits to 
$\eta^{(\prime)}\pi$ data from the COMPASS collaboration.
The results for $\eta \pi$ are shown in \figurename{~\ref{fig:likelihood}},
where two models with the same amount of parameters
provided similar best likelihoods 
and likelihood distributions, but when bootstrap fits
were individually compared, a systematic pattern emerged
favoring a particular model.
In this case, for each bootstrap fit one of the models provides a better likelihood more of $90\%$ of the instances. Hence, we can state that this minimum is favored.

\subsubsection{Physical and spurious poles}
\label{sec:spurious}

\begin{figure}[t]
\centering
\includegraphics[width=.5\textwidth]{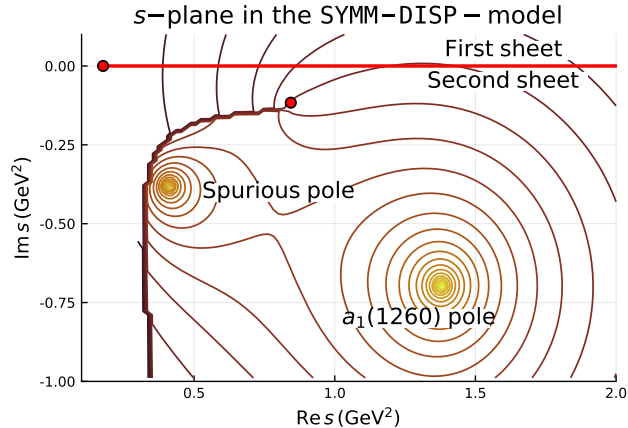} 
\caption{\label{fig:spurious_pole} 
Example of a spurious pole in the pole extraction of the $a_1(1260)$
resonance form the $\tau^-\to\pi^-\pi^+\pi^-\nu_\tau$ decay. 
The brown curve represents the $\rho\pi$ wholly cut.
Figure from~\cite{Mikhasenko:2019vhk}.
}
\end{figure}

Once the data have been fitted and the poles extracted,
the question of whether the found poles are truly 
physical resonances or artifacts of the
parametrization must be considered before attempting any physical interpretation.
This is because: (a) We fit a given energy range and poles can appear far away from the fitting region; (b)
Data have statistical noise, so an apparent signal can be compatible with
statistical fluctuations; and (c) The amplitude models
are incomplete, 
\ie they do not encompass the full 
physics of QCD, and sometimes are unwillingly biased.
If data are cut in a certain energy range, poles whose real part is outside
or at the edge of
the fitting region 
can allow the model to reproduce
the behavior of the data at the edge, acting as an effective background,
and their physical meaning is highly debatable.
Other poles can be forced by features of the model. 
For example, the unitarization of left-hand singularities can create poles in the unphysical Riemann sheets close to threshold. Without a careful examination of the model, of the data
and of the uncertainites, these poles can be mistakenly hailed
as new resonances,
when they are not really demanded by the data.
\figurename{~\ref{fig:spurious_pole}} shows an example of a spurious
pole in the extraction of the 
$a_1(1260)$ resonance parameters from the $\tau^-\to\pi^-\pi^+\pi^-\nu_\tau$ decay~\cite{JPAC:2018zwp}, discussed in Section~\ref{sec:a1}.
First of all, we note the presence of a branch cut starting at the complex $\rho\pi$ threshold~\cite{Frazer:1964zz,Holman:1965mxx,Ceci:2011ae}. While the position of the branch point is fixed by the $\rho$ mass and width,
the cut location and shape is determined by the integration path one chooses for the three-body phase-space. The choice in~\cite{JPAC:2018zwp} allows one to rotate the cut as shown in \figurename{~\ref{fig:spurious_pole}}, to discover another pole. This pole cannot affect sizeably the real axis, as it is hidden behind the branch cut. It is likely that such poles are not required by data, but rather artifacts required by the model. While in general distinguishing the two can be complicated, in this case the model is simple enough that this second pole can be related to the functional form of the phase space function, which by construction contains an extra singularity in the second sheet, rather than a resonance pole required by data.

\begin{figure}[t]
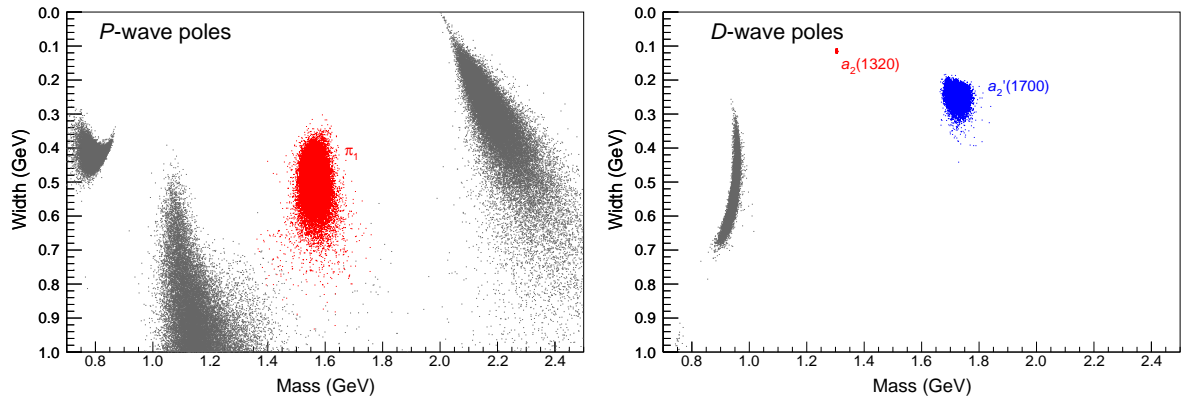

    \includegraphics[width=.47\textwidth]{{../figures/statistics/general_poleP}.pdf}
    \includegraphics[width=.47\textwidth]{{../figures/statistics/general_poleD}.pdf} 
    \caption{\label{fig:general_poles} 
Poles clusters generated by bootstrap for the $P$- (left) and $D$-waves (right). The $\pi_1$, $a_2(1320)$, and $a_2'(1700)$ resonances
are labeled.
}
\end{figure}

We have found that the error analysis based on the bootstrap
method, besides providing a proper uncertainty analysis,
often helps discerning true resonant poles from 
those which are artifacts of the parametrization 
or due to statistical noise, \aka spurious poles,
reducing the possibility of
signal misinterpretation.
Moreover, it also helps to assess the reliability
of the extracted pole, \ie if it truly
represents a resonance or it is an spurious effect.
As an example, in Section~\ref{sec:etapicompass}
we describe an actual physics example from the 
analysis of COMPASS $\eta^{(\prime)}\pi$ $P$- and $D$-waves
in the resonance region~\cite{JPAC:2018zyd}.
The BFF in this analysis has four poles in the $P$-wave
and three in the $D$-wave. \figurename{~\ref{fig:general_poles}}
shows the pole positions of the $\mathcal{O}(10^5)$ BS fits.
The three clusters that appear in the $D$-wave
are associated to each one of the three poles found in the BFF.
The two higher mass clusters are Gaussian, and stable against 
statistical fluctuations.
However, the lowest mass cluster has a 
nongaussian shape, with the mass close to threshold and the width as deep as $1\gev$. 
Given their position, it is intuitive that these poles cannot have a direct influence on data. Nevertheless, the cluster is relatively narrow and well accumulated, as if the data were actually constraining it. 
The reason for this is that this pole is actually built in the model, for similar reasons as were discussed above.
This pole is, therefore, not demanded by the data. 
For the single channel analysis in~\cite{JPAC:2017dbi}, we could track down the pole's origin by turning off the imaginary
part of the amplitude, finding that it arises from a left-hand pole in the model that mimics these effects due to the left-hand cut. Hence, it is an artifact consequence of the model.

In the $P$-wave we find four clusters. 
Only the one labeled as \pione has the correct behavior and 
is the one we associate to an exotic resonance.
We deem the other three spurious.
The heavier cluster is at higher masses than the fitted data 
(2\gev). 
This pole appears right above the fitted region, and
the model tries to overfit the last few data points by placing a pole. 
When the bootstrap is performed, the pole position is completely unstable, showing its unphysical origin as an artifact of the 
data selection.
The lowest mass one is equivalent to the left-hand pole found
in the $D$-wave, as its mass is very close to threshold.
The remaining cluster at $\sim 1.2\gev$, that did not appear in the BFF is unstable against bootstrap as it often escapes deep in the complex plane, so it is associated with statistical fluctuations. However, it could have happened that the BFF found such a pole, say at a width of $500\mev$, and could have misidentified that pole as a new state. This type of analyses allows us to distinguish such artifacts from physical states.
Additional examples can be found in Section~\ref{sec:scalars}
and Ref.~\cite{Rodas:2021tyb} for the $\jpsi$ radiative decays.
The procedure sketched here is not a rigid algorithm, and has to be adjusted to the physics problem at hand.
While on one hand a proper algorithmic definition could be given with $k$-means clustering~\cite{1056489}, especially in its unsupervised version~\cite{9072123}, it is still important to study the clustering on a case by case basis, in particular to compare the results of different systematic studies.

\subsection{Machine Learning for hadron spectroscopy}
\label{sec:ai}

There are two factors which have allowed Machine Learning (ML) to thrive in recent years: The first is an enormous progress in hardware, related mostly to the use of massive parallel GPU processors~\cite{Krizhevsky20121097}. This development makes it significantly easier to tackle problems involving ``big data". The second factor is related to rapid and multifaceted development of architectures, algorithms and computational techniques like convolutional neural networks~\cite{cccfa4f7238441b0a9021bb9f917e8ed}, rectified linear unit (ReLU)~\cite{10.5555/3104322.3104425,pmlr-v15-glorot11a}, improved stochastic gradient optimization~\cite{Kingma:2014vow} or batch normalization~\cite{Ioffe:2015ovl}. ML in nuclear and high-energy physics has already quite a long history: Applications to experimental studies include event selection~\cite{Gligorov:2012qt,Baldi:2014kfa,Santos:2016kno}, jet classification~\cite{Peterson:1993nk,ATLAS:2015thz}, track reconstruction~\cite{Adam:2003kg,ATLAS:2017kyn}, and event generation~\cite{Lazzarin:2020uvv,Buckley:2019wov}.
 On the theory front, ML has been extensively used for fitting~\cite{Ireland:2004kp,Fernandez-Ramirez:2008ixe,NNPDF:2014otw,Forte:2002fg,Rojo:2004iq}
 and to provide model-independent parametrizations of structure and spectral functions~\cite{Forte:2002fg,Rojo:2004iq,Ball:2013lla,DelDebbio:2007ee}, as well as of solutions of Schr\"odinger equations~\cite{Keeble:2019bkv,Adams:2020aax}.
Several reviews cover these applications extensively~\cite{Abdughani:2019wuv,Guest:2018yhq}.

In hadron spectroscopy, ML methods have not been explored so thoroughly. Among the many techniques available, 
we focus on classifiers, which are a kind of discriminative models. These are designed to capture the differences between groups of data 
(\eg canonical ``cat \vs dog'' classification) 
and estimate a conditional probability of the output to belong to a class given the input.
Recently, the use of classifiers has been proposed as a method to identify the nature of a given hadron state~\cite{Sombillo:2020ccg,Sombillo:2021rxv,Sombillo:2021yxe,Ng:2021ibr}.
The idea is that different natures of the states reflect into different lineshapes, 
and ML can be used to discriminate the interpretation which is most favored by data. 
In particular, this has been applied to the $P_c(4312)$ pentaquark candidate
in~\cite{Ng:2021ibr}: Since the peak appears very close to a two-body threshold, the amplitude can be expanded model-independently, and the resulting simple form permits a direct characterization of the $P_c$. Details about the physics that determines the different classes will be presented in Section~\ref{sec:pc4312}, while here we focus on the methodology.

\begin{figure}[t]
    \centering
    \includegraphics[width=0.6\textwidth,clip]{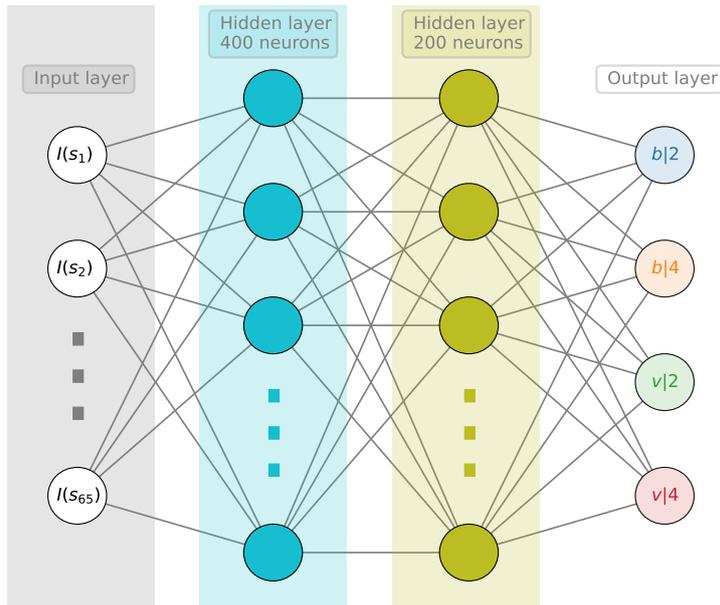}
    \caption{The neural network architecture of the classifier used in~\cite{Ng:2021ibr}. In the input layer, the feature vectors contain the intensity of the amplitude in each energy bin. The output layer is composed of the four classes which correspond to the various interpretations of the $P_c(4312)$. Figure from the Supplemental Material of~\cite{Ng:2021ibr}.}
    \label{fig:ann.classifier}
\end{figure}

Formally, the classification problem can be stated as seeking the function $f$ which maps the space of input data $x\in X^N$ (\aka feature vectors)  into a set of target classes $t\in T$.~\footnote{With  this definition, the only difference between classification and regression tasks is that the target set is finite for classification and continuous for regression.} Here the feature vectors consist of 65 intensity values in bins of energy, and labels were four possible interpretations of the $P_c(4312)$.
The problem lies in the choice of $f$~\cite{10.5555/1162264,ripley_1996,StatisticalLearning}.
In what follows we focus on the simplest version of a neural network classifier, a dense feed-forward network in which the trained parameters are encoded in weights of
the network node (neuron) connections. The typical architecture of such a network is depicted in \figurename{~\ref{fig:ann.classifier}}.
The values read in the output layer are obtained by passing the values of the feature vectors through consecutive hidden layers. For this process to be more than mere matrix multiplication (thus allowing us to model arbitrary nonlinear input-output dependencies), the output value of each neuron is obtained by subjecting the weighted values from all nodes of the preceding layer to an activation function $\sigma$. Thus the  output value of the $m$-th neuron of the $n$-th layer can be expressed as
\begin{equation}
    x_{m,n}=\sigma\!\left(\sum_{i=1}^N w_{im}x_{i,n-1}+b_{m,n}\right) \, ,
\end{equation}
where $N$ is the dimension of the $(n-1)$-th layer and $b$ is known as the bias vector of the $n$-th layer. Nonlinearities of the $\sigma$ function can be modeled in many different ways (step function, sigmoid, $\tanh$, \etc), but ReLU was employed in~\cite{Ng:2021ibr}. By continuing this procedure for all nodes in all layers, we finally obtain the values at the output layer that can be compared with ground truth labels (classification) or values (regression). To measure the quality of this comparison one uses a cost function, \eg a cross entropy or a $\chi^2$ (\aka mean squared error). 
Thus, `learning' is basically the minimization of the cost function by varying the model parameters, in this case the weights of the neural network. This is a difficult optimization problem and, along with the discussion of what the proper choice of the cost function is, it has a large body of literature devoted to it~\cite{Goodfellow-et-al-2016,10.5555/3283445,deisenroth_faisal_ong_2020}. For our example,
 the network must be trained on the line shapes one gets from the four classes. To do so, we calculate the line shapes for $10^5$ uniformly sampled model parameter sets,  described in Section~\ref{sec:pc4312} in detail. In this way we effectively scan the space of line shapes. For each sample one can calculate the pole position, and thus what class the line shape belongs to.
The feature vectors can be matched with these ground truth labels. It is clear that the stability of the optimization process and the precision of eventual inference are largely impacted by the size of the training dataset.

\begin{figure}[t]
    \centering
    \includegraphics[width=.8\textwidth]{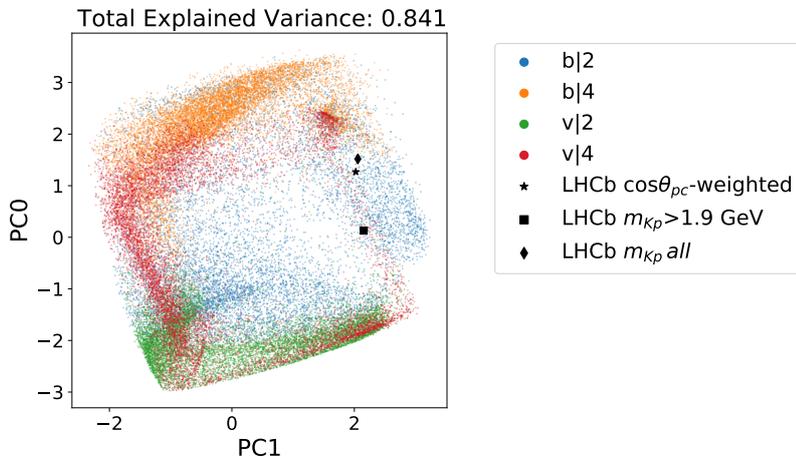}
    \caption{Training sets projected onto 2D space using PCA, compared with the values obtained from the \lhcb datasets. Experimental data are located in a region well represented in the training set. 
    In the $b|2$, $b|4$, $v|2$, and $v|4$ labels, the $b$ stands for bound, the $v$ for virtual, and the number for the Riemann sheet where the pole appears (see Section~\ref{sec:pc4312}).
    Figure from the Supplemental Material of~\cite{Ng:2021ibr}.}
    \label{fig:pca}
\end{figure}

As said, the feature vectors in this example have 65 elements. This is less than in typical ML problems and far less than the amount of data input to Convolutional Neural Networks.
But even here one expects substantial correlation which is related to information redundancy. Moreover, the system is governed mainly by threshold dynamics, so there should only be a few relevant features. Last but not least, working in a smaller dimensional space allows us to plot 2D projections to represent the data and to acquire a better intuition on the properties of the training set and the data we might want to classify. To isolate the relevant features, one customarily employs the Principal Component Analysis (PCA)~\cite{hastie2009elements}, which boils down to extracting the eigenvectors and eigenvalues of the covariance matrix built from the standardized feature vectors (see  \figurename{~\ref{fig:pca}}).
The covariance matrix expressed in terms of eigenfeatures is diagonal with diagonal elements summing up to the total variance. So, in the PCA, one retains those diagonal elements (and associated features) which ``explain most of the variance".

Tracing the path that leads the classifier to assign the input to a particular class is impossible, except for in the simplest of models. This makes the ML tools function as black boxes, whose decisions we are bound to trust rather than understand. This is uncomfortable not only in physics where we aim at understanding the dynamics that leads to a given choice, but also in other fields, like medicine or economics. 
However, we can at least select the features used for the classifier to make its class assignment. Partially, PCA already address this task by selecting ``principal directions," expressed in terms of combinations of the underlying features. These in turn may be difficult to interpret, and may be better to use SHapley Additive exPlanations (SHAP) values~\cite{NIPS2017_8a20a862}. This approach originates in game theory, and it shows whether an individual feature favors (positive) or disfavors (negative) a certain classification.

\begin{figure}[t]
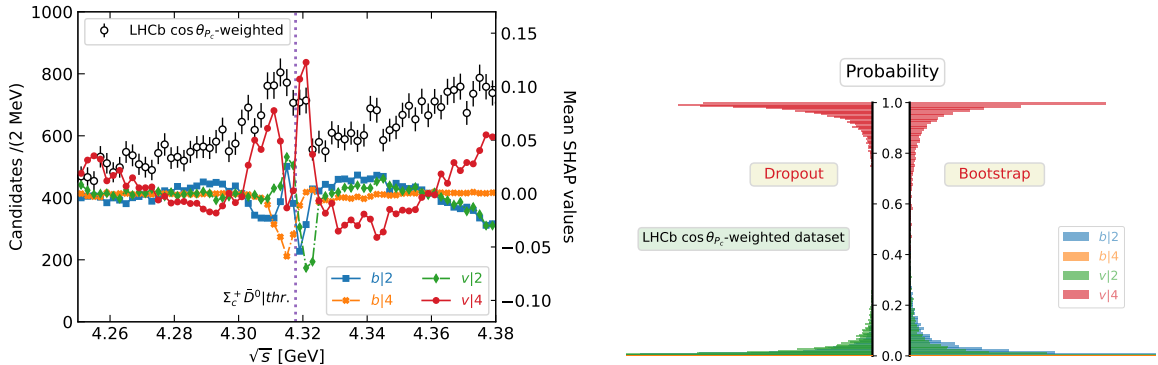

    \centering
    \includegraphics[width=0.47\textwidth]{{../figures/AI/ai.fig5}.pdf} 
    \includegraphics[width=0.47\textwidth]{{../figures/AI/ai.fig4}.pdf} 
    \caption{(left) SHAP values for the four classes overlaid on the experimental intensity plot. It is evident that the threshold region is the one impacting the decision. (right) Dropout and bootstrap classification probability densities for the predictions on one of the \lhcb datasets for each of the four classes.
    The $x$ axes are equally cut for the purpose of visibility and comparison.
    Class notation as in \figurename{~\ref{fig:pca}}.  Figures from~\cite{Ng:2021ibr}.}
    \label{fig:shap.violin}
\end{figure}

The SHAP values in \figurename{~\ref{fig:shap.violin}} show that most of the class assignment explanation comes from the $v|4$ class in the near-threshold region. This conclusion is both expected and valuable: It confirms the conjectured dominance of the threshold effects in shaping the experimental signal, thus providing an \emph{ex post} justification of the assumed scattering length approximation. Also it narrows the interval of energies relevant for the analysis to those neighboring the resonance peak.

Having established that our training set covers the region of the feature space where the experimental data are situated, and having identified the energy region of importance for the class assignment, we are ready to infer the nature of the $P_c(4312)$ from the experimental data. 
A probabilistic interpretation of the classification can be obtained by subjecting the signal $t$ produced by the neurons of the output layer to a softmax function,
\begin{equation}
    \text{softmax}(t_i)=\frac{\exp(t_i)}{\sum_{j=1}^4 \exp(t_j)},
\end{equation}
where $i,j$ run over the four classes.
Obviously, this function is positive definite and normalized to 1. We obtain a probability by resampling the data with the bootstrap procedure explained in Section~\ref{sec:stat_bootstrap}.
Alternatively, one can apply the Monte Carlo dropout to the trained layers~\cite{gal2016dropout}, which approximates the Bayesian inference in the deep Gaussian process. The results of these two procedures are shown in \figurename{~\ref{fig:shap.violin}}. Both the dropout and bootstrap distributions show the clear dominance of the $v|4$ class (virtual pole on the IV Riemann sheet, as explained later in Section~\ref{sec:pc4312}). In other words, based on the experimental data, the neural network assigns the highest probability for the $P_c(4312)$ to have a $v|4$ nature. As said, the meaning of the different classes will be given Section~\ref{sec:pc4312}.

One of the benefits of using Machine Learning and Artificial Intelligence at large is that of generalization: One hopes that, using ML or AI, it is possible to classify features the network was not explicitly trained on. 
Realistically speaking, this generalization ability is rather modest. Still, in the context of hadron spectroscopy, one may ask what would be the recognition rate for the classifier trained on the $P_c(4312)$ if applied to other resonances. One can speculate that, for resonances emerging due to similar threshold dynamics, this recognition ability may still persist.  This brings us to the concept of transfer learning, which is widely used in Convolutional Neural Networks as applied to image recognition~\cite{5288526,7333916,7493483}. One typical application of this is the transfer of ImageNet pretrained convolutional layers to a model one is interested in~\cite{s21061963}. 
In hadron physics,  there is no such pretrained ``amplitude database.'' Still, the potential to generate the training sets from amplitudes describing typical situations like a near-threshold peak is practically unlimited. Details of the resulting line shape would depend on the production mechanism, particle masses, detector resolution, \etc, but most of the information enabling the translation of the line shape into class assignment comes from a small region, as discussed above. Therefore, identifying layers of the CNN which extract this region and transfer them to other models may result in satisfactory classifier performance. Of course, the transferred layers will have to be supplemented with additional trainable layers (either dense or convolutional), to account for non-transferable properties.

Finally, other categories of ML methods can also find applications to spectroscopy: Generative models, as Variational Autoencoders (VAE)~\cite{Kingma:2013hel}, Restricted Boltzmann Machines (RBM)~\cite{Hinton2006ReducingTD} or Generative Adversarial Networks (GAN)~\cite{10.1145/3422622} can be used as well. 
Technically the difference between discriminative and generative models is that the former are designed to capture the differences between groups of data, while the latter compute the joint probability of input and output.
In particular GANs recently found application as an alternative to Monte Carlo Event Generators like PYTHIA~\cite{Sjostrand:2007gs}, Herwig~\cite{Bellm:2015jjp} or SHERPA~\cite{Gleisberg:2008ta}. Contrary to these conventional event generators, which are biased by the underlying physical models, the GAN-based generator might learn directly from experimental data. They seem effective in generating inclusive electron-proton scattering events and operated in the range of energies, even beyond those they were trained on~\cite{Velasco:2020nqr}. The application of these to exclusive channels of interest for spectroscopy is presently ongoing~\cite{Alanazi:2020jod,Alanazi:2021grv}.

\subsection{Light hadron spectroscopy}
\label{sec:lighthadron}
The light hadron sector
has been subject to fierce debate for many decades. Resonances are generally broad and overlap each other; experimental analyses were limited by statistics, and often implemented simplistic methods. All these issues hindered the extraction of reliable information. The natures, and in some cases even the existences, of some states are still under debate.

Quark models play a crucial role in guiding analysis, and in predicting the number and properties of states to search for~\cite{Godfrey:1985xj,Capstick:1985xss}. However, since we are entering an era of high-statistics experiments, we are now facing the limits of such models. A complementary path was followed with effective field theories having hadrons as degrees of freedom, in particular Chiral Perturbation Theory ($\chi$PT)~\cite{Weinberg:1978kz,Gasser:1983yg,Gasser:1984gg,Jenkins:1990jv,Bernard:1990kw,Ordonez:1992xp,Ordonez:1993tn,Bijnens:1995yn, Bijnens:1997vq, Bijnens:1999sh, Bijnens:2014lea,Bernard:1995dp,Epelbaum:2008ga}. The low energy constants at a given order can be fixed from experimental~\cite{Gasser:1983yg, Gasser:1984gg, Bernard:1990kw} or lattice QCD~\cite{Ecker:2010nc, Aoki:2016frl} data. However, fixed order effective theories respect unitarity only perturbatively, and cannot produce resonance poles if not explicitly incorporated. This problem was circumvented by various unitarization methods (U$\chi$PT)~\cite{Dobado:1989qm, Truong:1991gv, Dobado:1996ps, Oller:1997ng, Oller:1998hw, Oller:1998zr, GomezNicola:2001as,Oller:2019rej,Bolton:2015psa,Molina:2020qpw,Niehus:2020gmf}, at least in the low-energy region. 
Nevertheless, these methods still suffer from several model dependencies and approximations. This becomes particularly clear when dealing with light scalars, where all the $S$-matrix principles play a significant role. This is the main reason why dispersive approaches have been gaining attention in recent years~\cite{Roy:1971tc,Steiner:1970fv,Steiner:1971ms,Hite:1973pm,Johannesson:1976qp,Ananthanarayan:2000ht,Colangelo:2001df,Descotes-Genon:2006sdr,Garcia-Martin:2011iqs,Hoferichter:2011wk,Ditsche:2012fv,Hoferichter:2015hva,Pelaez:2016tgi,Pelaez:2018qny,Hoferichter:2019nlq,Pelaez:2019eqa,Pelaez:2020gnd,Pelaez:2021dak}. The combination of dispersion relations with experimental data is able to provide us the most robust information about the lightest mesons~\cite{Caprini:2005zr,GarciaMartin:2011nna,Moussallam:2011zg, Pelaez:2020uiw}. In particular, both the \sig and \kap mesons showcase a successful implementations of such approaches, achieving very high accuracy. These results triggered their acceptance by the Particle Data Group (PDG, for recent reviews we refer the reader to~\cite{Pelaez:2015qba,Pelaez:2020gnd}). Unfortunately, partial wave dispersive analyses are usually applicable only up $\sim 1\gev$. At a practical level, most of the data at higher energies come from photo-, electro- and hadroproduction, heavy meson decays, peripheral production, or $e^+e^-$ annihilations. Furthermore, the large number of open channels available make the rigorous application of unitarity unfeasible.
For these reasons, loosening the $S$-matrix constraints, and studying a number of phenomenological amplitudes to assess the systematic uncertainties and reduce the model bias seems the appropriate path to follow.

There are several interesting topics in the light sector. The most fundamental questions concern the existence of resonances where gluons play the role of constituents, as glueball or hybrid mesons~\cite{Mathieu:2008me,Llanes-Estrada:2021evz,Meyer:2015eta}. The analysis of isoscalar scalar and tensor mesons in the $1$--$2.5\gev$ region---where the lightest glueball is expected---is presented in Section~\ref{sec:scalars}. The $\eta^{(\prime)}\pi$ channel, where the $a_2^{(\prime)}$ and the exotic $\pi_1$ are seen, is discussed in Section~\ref{sec:etapicompass}. 
The $a_1$ and $\pi_2$ states, for which the three-body dynamics plays a major role, will be discussed later in Sections~\ref{sec:a1} and~\ref{sec:pi2}.
In the list of states that have received lots of attention in the past, we recall the $\eta(1405)$ as a pseudoscalar glueball candidate~\cite{Gutsche:2009jh}, the $X(1835)$ that appears at the $p\bar p$ threshold~\cite{BES:2005ega,Kochelev:2005vd,Zhu:2005ns}, and the poorly known strangeonium sector. 

The baryon sector is even more difficult, despite the efforts by a larger community. We will just mention the longstanding puzzles about the Roper and the $\Lambda(1405)$~\cite{Burkert:2017djo,Jido:2003cb}. A collective discussion of several baryon resonances [including the $\Lambda(1405)$] is performed by identifying the Regge trajectory they belong to, in Section~\ref{sec:baryons_and_hyperons}.

\subsubsection{\texorpdfstring{\jpsi radiative decays}{J/psi radiative decays}}
\label{sec:scalars}
As mentioned, the isoscalar-scalar mesons, and -tensor mesons to some extent, have played a central role in spectroscopy. 
They can mix with the lightest glueball with the same quantum numbers. In pure Yang-Mills, the spectrum is populated by glueballs, the lightest one expected to be around $1.5$--$2\gev$~\cite{Bali:1993fb,Patel:1986vv,Albanese:1987ds,Michael:1988jr,Sexton:1995kd,Morningstar:1999rf,Szczepaniak:2003mr,Chen:2005mg,Athenodorou:2020ani}. In nature, glueball production is expected to be enhanced in processes where quarks annihilate into gluons, like $p\bar p$ collisions or \jpsi radiative decays. 

Most of the literature traces the existence of a significant glueball component with the emergence of a supernumerary state with respect to how many are predicted by the quark model~\cite{Mathieu:2008me,Ochs:2013gi,Llanes-Estrada:2021evz}. The PDG lists seven inelastic scalar-isoscalars. In particular, among the $f_0(1370)$, $f_0(1500)$, $f_0(1710)$ in the $1.2$--$2\gev$ there is one more resonance than is expected by the quark model, which stimulated intense efforts to identify one of them as the glueball~\cite{Chanowitz:1980gu,Amsler:1995td,Amsler:1995tu,Lee:1999kv,Giacosa:2005zt,Giacosa:2005qr,Albaladejo:2008qa,Janowski:2014ppa}. The $f_0(1710)$ couples mostly to kaon pairs~\cite{Barberis:1999am,Uehara:2013mbo,BESIII:2018ubj}. Since photons do not couple directly to gluons, the poor production of $f_0(1500)$ in $\gamma\gamma$ suggests it may be mainly a glueball. On the other hand,
the chiral suppression of the matrix element of a scalar glueball to a $q\bar q$ pair point to the $f_0(1710)$ as a better candidate~\cite{Chanowitz:2005du,Albaladejo:2008qa}. Although this result is model-dependent, it is supported by a 
quenched Lattice QCD calculation~\cite{Sexton:1995kd}.

The tensor resonances are better understood. The $f_2(1270)$ and $f_2'(1525)$  are 
identified as $u\bar u + d\bar d$ and $s\bar s$ mesons, respectively.
 Indeed, the former couples mostly to $\pi \pi$, and  the latter to $K\bar  K$~\cite{Garcia-Martin:2011iqs,pdg}. Both resonances are narrow and 
  have also been extracted from lattice QCD~\cite{Briceno:2017qmb}. 

In this section we summarize our efforts to determine these inelastic scalar and tensor resonances from \jpsi radiative decays~\cite{Rodas:2021tyb}. We consider the data from the nominal solutions of the $\jpsi \to \gamma\pi^0\pi^0$~\cite{BESIII:2015rug} and $\to \gamma \KSKS$~\cite{BESIII:2018ubj} mass-independent analyses by \bes. Bose symmetry requires $J^{PC}=\text{(even)}^{++}$; and the isospin zero amplitude is dominant for both channels. We fit the intensities and relative phases of the $0^{++},\,2^{++}$ E1 multipoles between $1$--$2.5\gev$. 

\begin{figure}[t]
\centering
\includegraphics[width=0.32\textwidth]{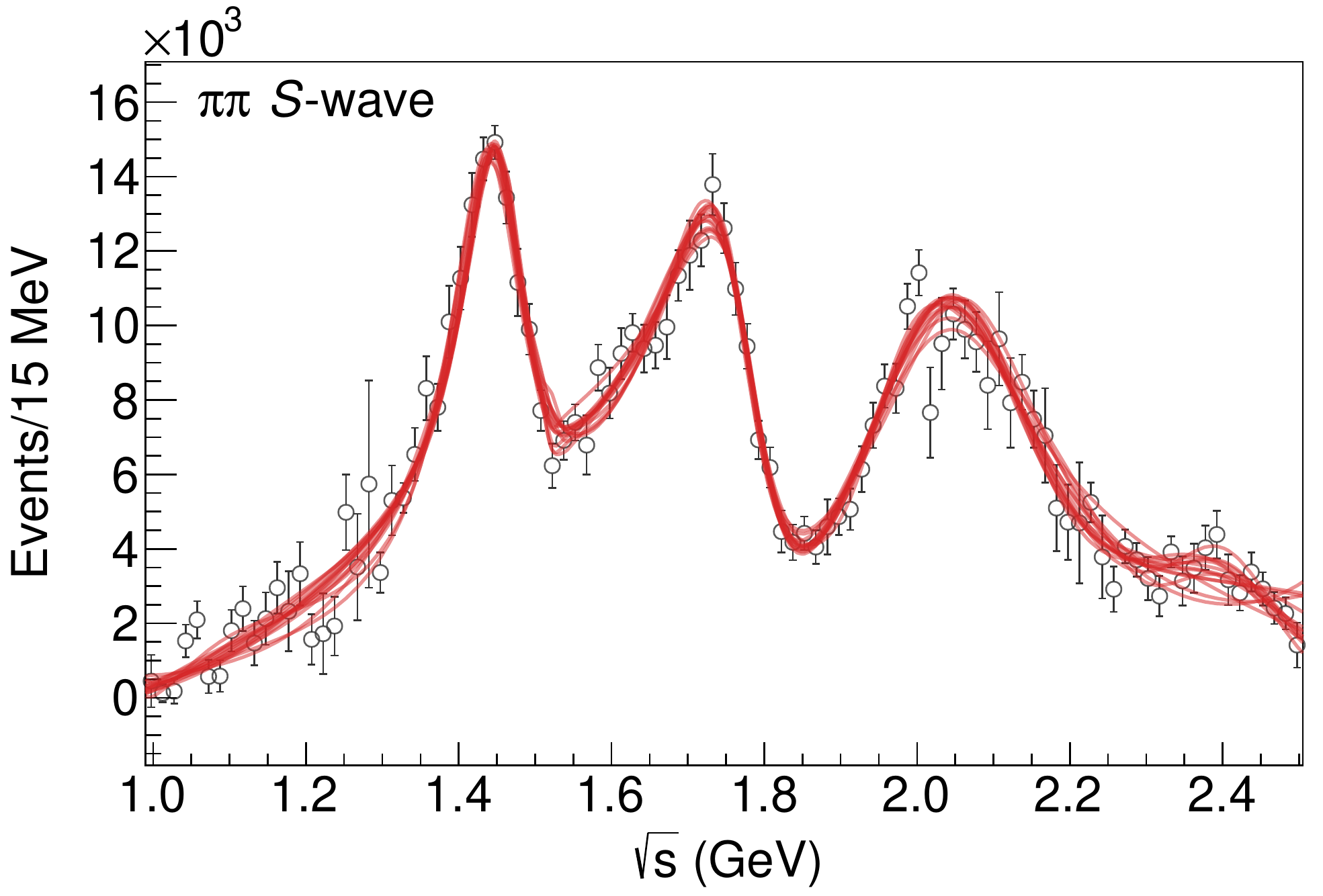} \includegraphics[width=0.32\textwidth]{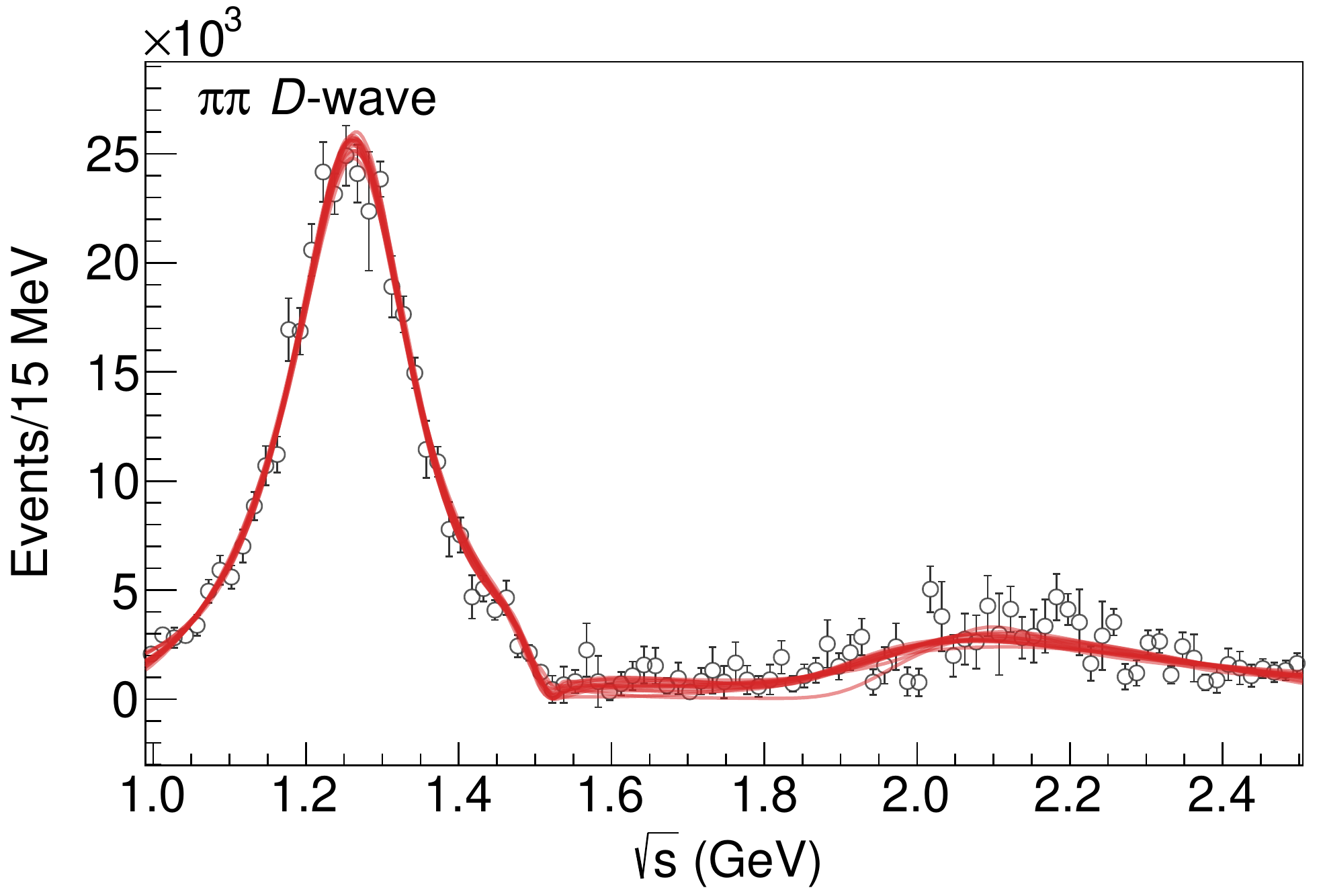} \includegraphics[width=0.32\textwidth]{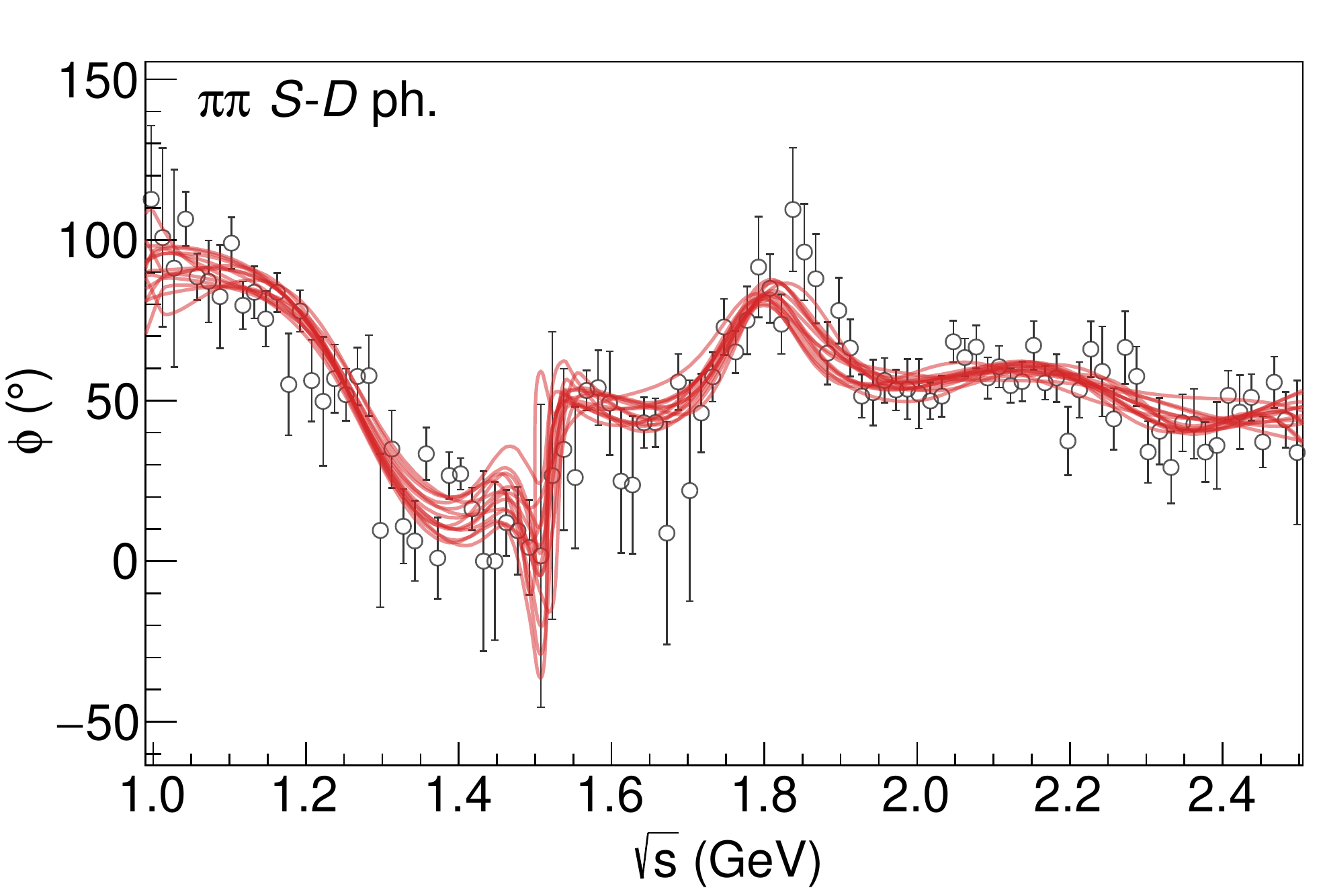} \\
\includegraphics[width=0.32\textwidth]{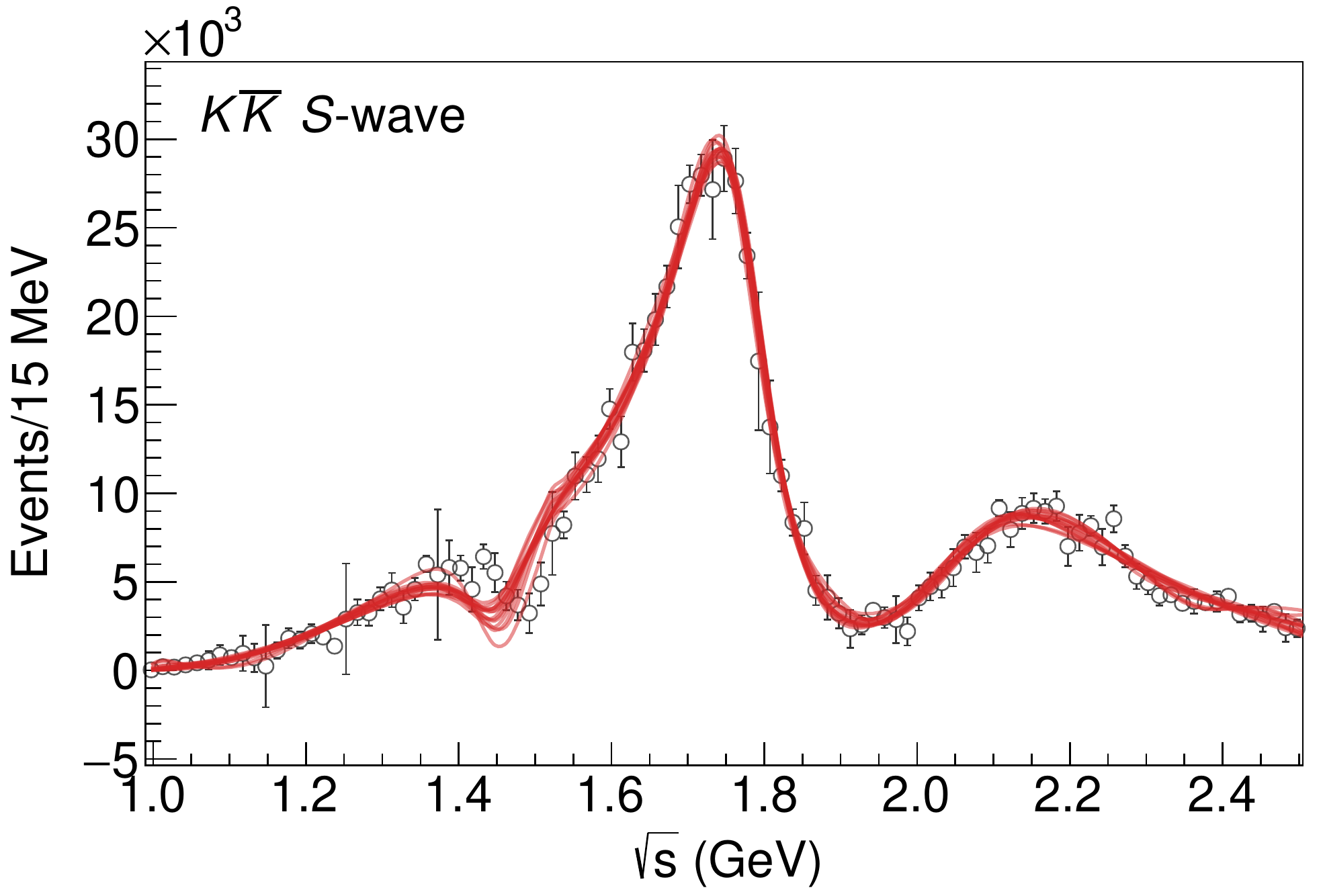} \includegraphics[width=0.32\textwidth]{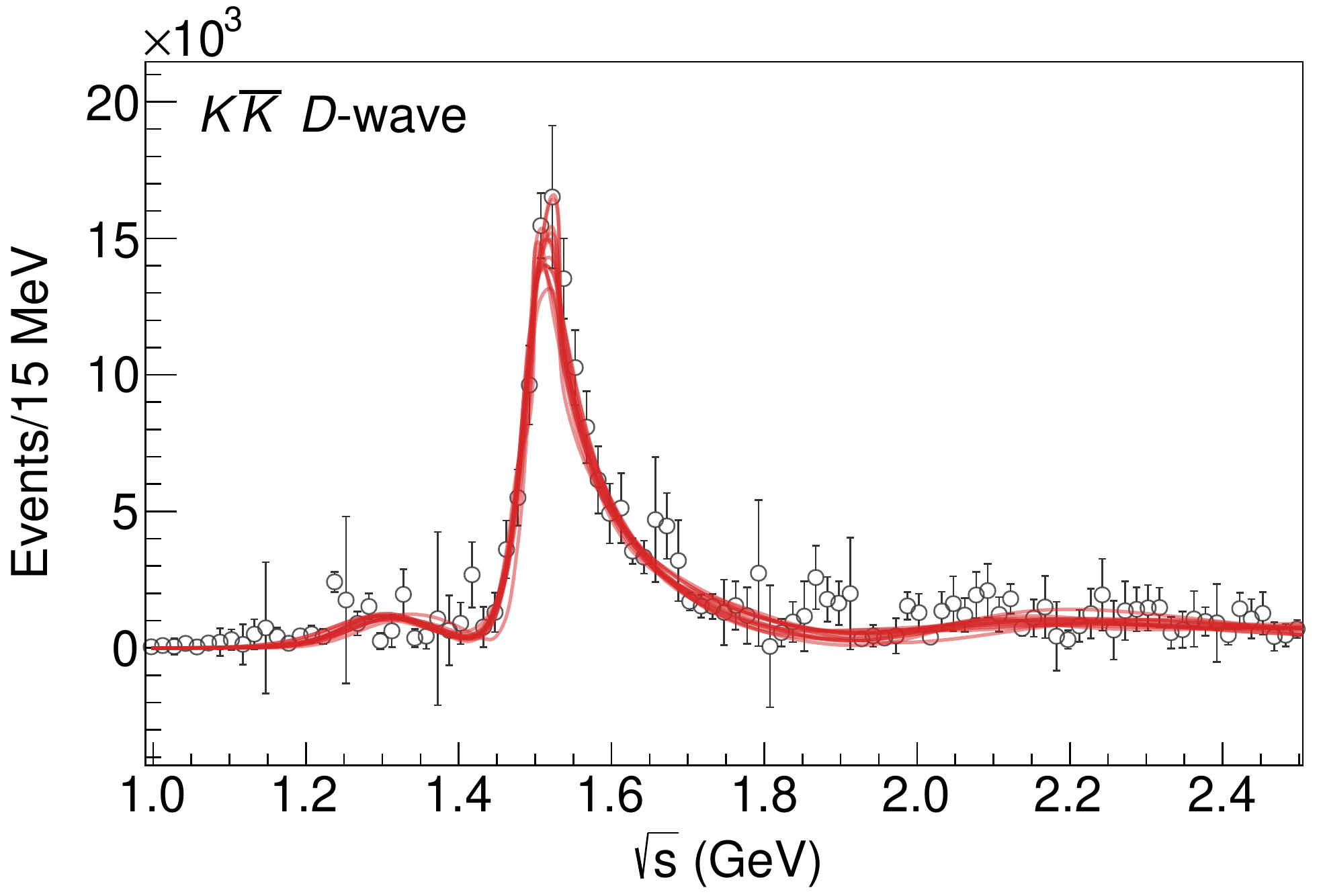} \includegraphics[width=0.32\textwidth]{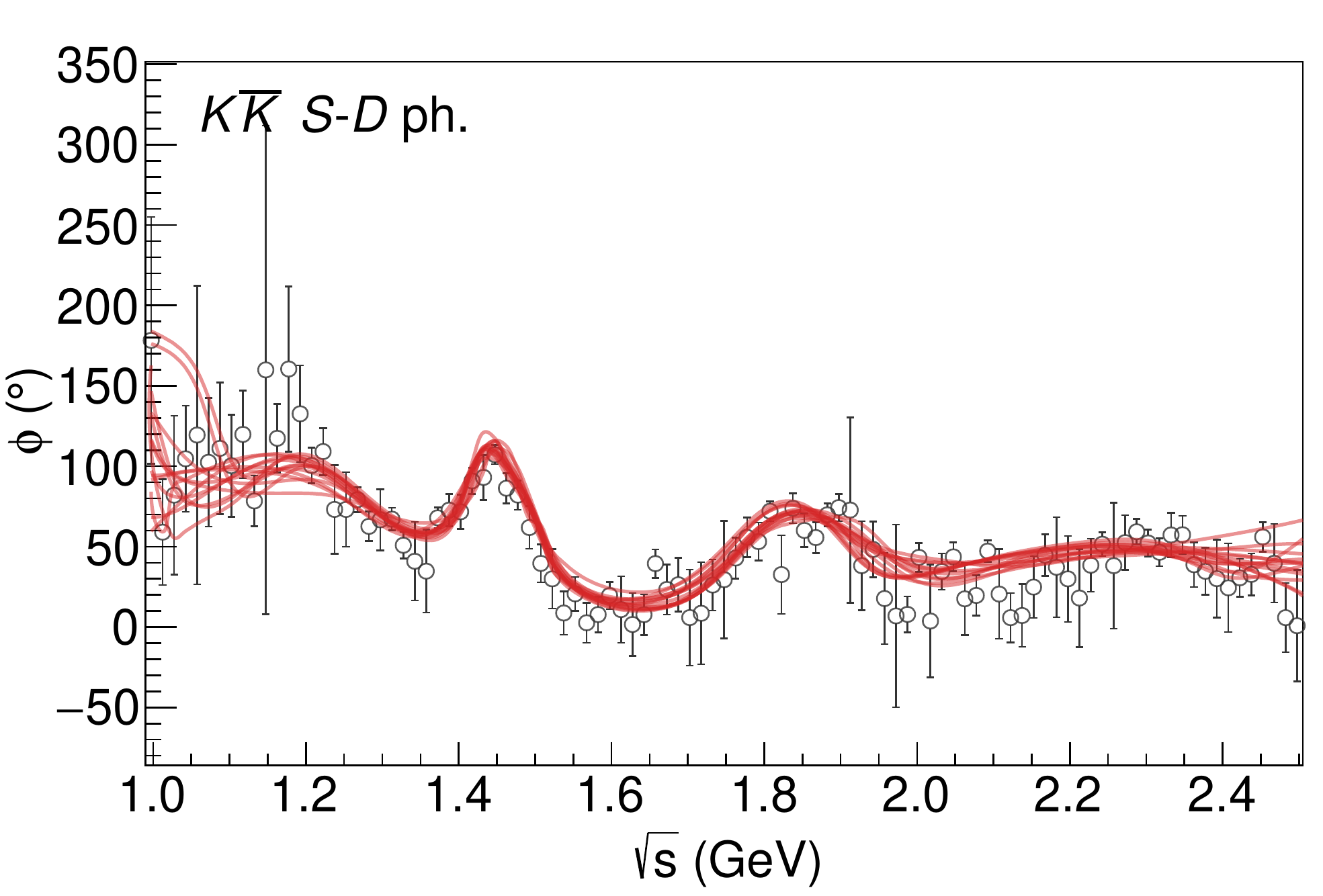} 
\caption{Best 3-channel fits to $\pi \pi$ (top) and $K \bar K$ (bottom) final states. The intensities for the $S$- (left), $D$-wave (center), and their relative phase (right) are shown. The red lines correspond to the central value of each one of the different fits. All these fits have $\chi^2/\dof\sim 1.1$--$1.2$. Figures from~\cite{Rodas:2021tyb}.
 }
\label{fig:3channelfits}
\end{figure}

As mentioned, we use a variety of parametrizations that fulfill as many $S$-matrix principles as possible in order to keep the model dependencies under control. We follow the coupled-channel $N/D$
formalism~\cite{Chew:1960iv,Bjorken:1960zz,Aitchison:1972ay,Oller:1998hw},
\begin{equation}
 a^J_i(s) = E_\gamma\, p_i^{J} \, \sum_k n^J_k(s) \left[ {D^J(s)}^{-1} \right]_{ki}\,,
\end{equation}
with $i=\h\bar \h$ the hadron index,$s$ the $\h\bar \h$ invariant mass squared and
$p_i$ the breakup momentum in the $h\bar h$ rest frame. Gauge invariance requires the inclusion of one power of photon energy
$E_\gamma$. 

The $n^J_k(s)$ incorporate exchange forces in the production process
and are smooth functions of $s$ in the physical region. It is parametrized by an effective polynomial expansion, possibly including background poles. The matrix 
$D^J(s)$  represents the $\h \bar \h \to \h \bar \h$ final state
interactions, and encodes the resonant content of the $\etapi$ system. 
A customary parametrization is given by~\cite{Aitchison:1972ay}
\begin{equation}
\label{eq:Dsol}
D^J_{ki}(s) =  \left[ {K^J(s)}^{-1}\right]_{ki} - \frac{s}{\pi}\int_{4m_{k}^2}^{\infty}ds'\frac{\rho N^J_{ki}(s') }{s'(s'-s - i\epsilon)}, 
\end{equation}
where $\rho N^J_{ki}(s')$ is smooth in the physical region, and describes the crossed-channel contribution to the scattering process. 
For the $K$-matrix, we consider 
\begin{subequations}
\begin{align}
K^J_{ki}(s)_\text{nominal} &= \sum_R \frac{g^{J,R}_k g^{J,R}_i}{m_R^2 - s} + c^J_{ki} + d^J_{ki} \,s,\label{eq:Kmatrix}
\end{align}
with $c^J_{ki}= c^J_{ik}$ and $d^J_{ki}= d^J_{ik}$.
Alternatively, we may parametrize the inverse $K$-matrix as a sum of CDD poles~\cite{Castillejo:1955ed,JPAC:2017dbi},
\begin{equation}
\left[K^J(s)^{-1}\right]_{ki}^\text{CDD} = c^J_{ki} - d^J_{ki} \,s - \sum_R \frac{g^{J,R}_k g^{J,R}_i}{m_R^2 - s} \, , \label{eq:CDD}
\end{equation}
\end{subequations}
where $c^J_{ki} = c^J_{ik}$ and $d^J_{ki} = d^J_{ik}$ are constrained to be positive. These coefficients are also referred to as ``the CDD pole at infinity''.
For a single channel, this choice ensures that no poles can appear on the physical Riemann sheet.\footnote{This is ensured by the fact that $D(s)$ satisfies the Herglotz-Nevanlinna representation~\cite{Nussenzveig:1972ca}. A direct check can be made: we write the CDD parametrization for single channel in the unsubtracted form (a single subtraction can still be done, as it just shifts the real parameter $c$).
\begin{equation}
D(s) = c - d \,s - \sum_R \frac{(g^{R})^2}{m_R^2 - s} \, - \frac{1}{\pi}\int_{s_\text{th}}^\infty ds' \frac{\rho N(s')}{s'-s}\nonumber
\end{equation}
We calculate the imaginary part in the upper $s$ plane. We thus write $s = x + iy$ with $y>0$.
\begin{equation}
\Im D(s) = - d \,y - y\,\sum_R \frac{(g^{R})^2}{\left|m_R^2 - s\right|^2} \, - y \frac{1}{\pi}\int_{s_\text{th}}^\infty ds'\frac{\rho N(s')}{\left|s'-s\right|^2}\,,\nonumber
\end{equation}
so it is a sum of negative terms, provided that $c \in \mathbb{R}$, $d > 0$, and $\rho N(s') \ge 0$ for $s'\ge s_\text{th}$. That implies that $D(s)$ can never vanish in the upper plane. By the Schwartz reflection principle, it cannot vanish in the lower plane either.} Even in the case of coupled channels their occurrence is scarce, and when they do occur they are deep in the complex plane, far from the physical region. Moreover, they can be mapped one into the other if a suitable background polynomial of the $K$-matrix is considered. 

Initially, we performed a 2-channel analysis using only data on the $\pizpiz$ and $\KSKS$ final states. However these models are too rigid and cannot fully reproduce the local features of the data. Indeed, many of these resonances couple substantially to $4\pi$. For this reason, we extend our model with an unconstrained $\rho\rho$ channel, using the previous results as starting point for the new fits. 
We identify 14 best fits with different parametrizations, shown in \figurename{~\ref{fig:3channelfits}}.
We perform the bootstrap analysis as discussed in Section~\ref{sec:stat_bootstrap}, generating  $\mathcal{O}(10^4)$ pseudodata sets per parametrization. 

\begin{table}[b] 
\caption{List of final pole position and uncertainties resulting from the combination of the different fits to the data. The errors correspond to the variance of the full samples, by assuming that the spread of results, shown in \figurename{~\ref{fig:finalpoles}}, resembles a Gaussian distribution.
Table from~\cite{Rodas:2021tyb}.}
\centering 
\begin{tabular}{c c c c}
\hline
\hline
$S$-wave& $\sqrt{s_p}$ (\mevp) & $D$-wave& $\sqrt{s_p}$ (\mevp) \\ \hline
\rule[-0.2cm]{-0.1cm}{.55cm} $f_0 (1500)$ &  $(1450 \pm 10) - i (106 \pm 16)/2$  &  $f_2 (1270)$ &  $(1268 \pm 8) - i (201 \pm 11)/2$ \\
\rule[-0.2cm]{-0.1cm}{.55cm} $f_0 (1710)$ &  $(1769 \pm 8) - i (156 \pm 12)/2$  &  $f_2 (1525)$ &  $(1503 \pm 11) - i (84 \pm 15)/2$ \\
\rule[-0.2cm]{-0.1cm}{.55cm} $f_0 (2020)$ &  $(2038 \pm 48) - i (312 \pm 82)/2$  &  $f_2 (1950)$ &  $(1955 \pm 75) - i (350 \pm 113)/2$ \\
\rule[-0.2cm]{-0.1cm}{.55cm} $f_0 (2330)$ &  $(2419 \pm 64) - i (274 \pm 94)/2$  &   &  \\
\hline
\hline
\end{tabular}
\label{tab:polesfinal}
\end{table}

\begin{figure}[t]
\centering
\includegraphics[width=0.9\textwidth]{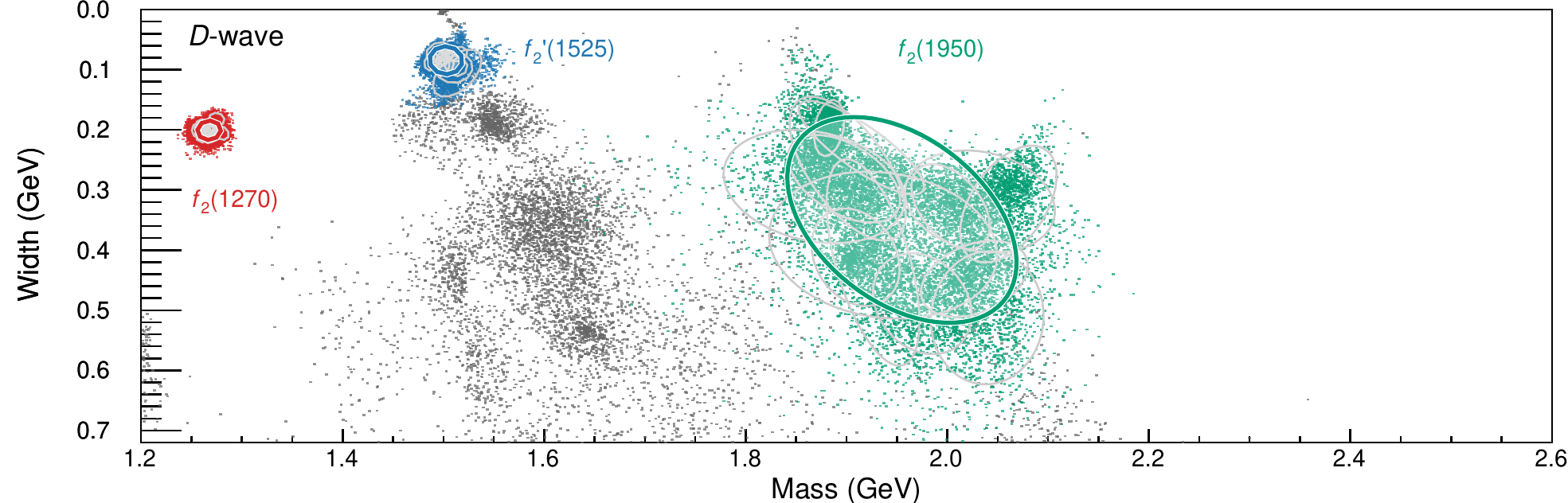} \\ \includegraphics[width=0.9\textwidth]{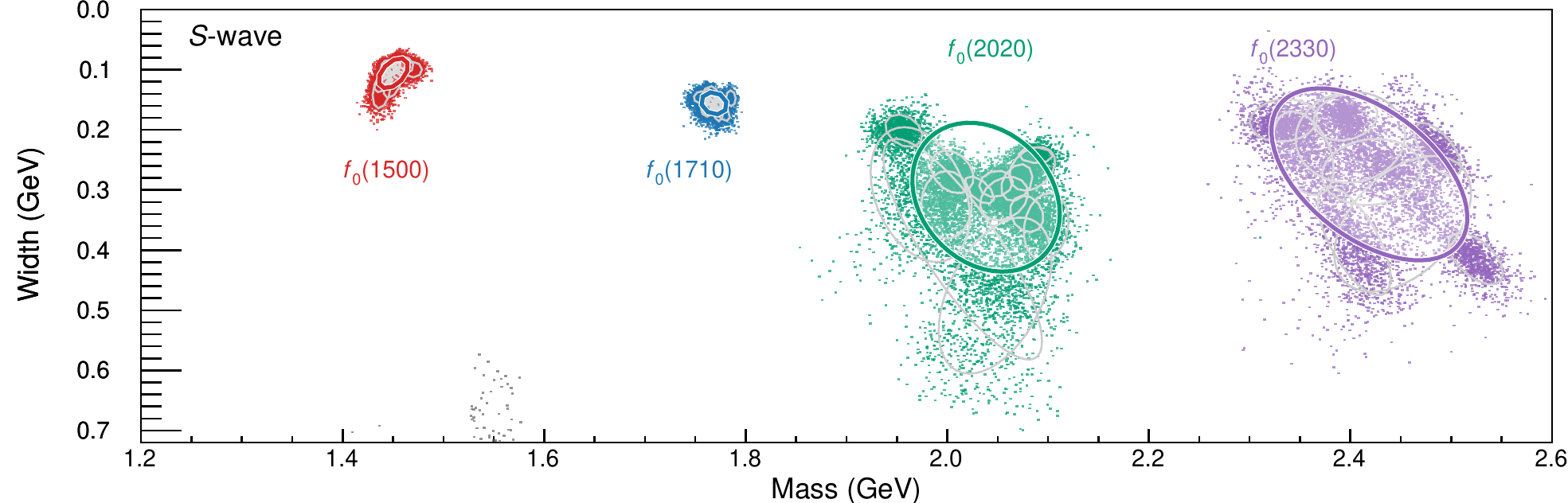} 
\caption{Shown in different colors are the final results for the pole positions, superimposed for the 14 models used in the analysis. A point is drawn for each pole found in each one of the $\mathcal{O}(10^4)$ bootstrap resamples. Gray points are identified as spurious resonances. For each physical resonance and systematic, gray ellipses show the $68\%$ confidence region. Colored ellipses show the final average of all systematics. Figures from~\cite{Rodas:2021tyb}.
}
\label{fig:finalpoles}
\end{figure}

Since our amplitudes respect analyticity and unitarity, we can look for resonant poles in the complex energy plane. 
Considering the 14 models and the bootstrap resampling, we have $\mathcal{O}(10^5)$ ``points'' per pole, as shown in \figurename{~\ref{fig:finalpoles}}. 
As discussed in Section~\ref{sec:spurious}, this analysis allows us to distinguish the spurious model artifacts from the physical ones. We found a total of 4 scalar and 3 tensor stable clusters that can be identified with physical resonances. The four lighter states produce a reasonably Gaussian spread, while the heavier ones have more complicated structures. The results are summarized in \tablename{~\ref{tab:polesfinal}}. The scalar resonances are compatible with other recent extractions~\cite{Ropertz:2018stk,Sarantsev:2021ein}. We found no evidence for a $f_0(1370)$ in these processes. However, it is customarily accepted that this is a $q\bar q$ state that couples mostly to $4\pi$, so that our findings do not challenge its existence.

Finally, we also studied the production and scattering couplings of these resonances. The  tensor sector look fairly simple, with the $f_2(1270)$ and $f_2'(1525)$ coupling almost entirely to $\pi \pi$ and $K\bar K$, respectively. This result is roughly compatible to those listed by the PDG~\cite{pdg}.\footnote{One must be cautious when comparing our results with those of the PDG: While we quote amplitude poles, the results listed by the PDG combine both amplitude poles and Breit-Wigner parameters. However, for narrow isolated resonances, the difference is not large.} The results for the scalar resonances are more involved. Although the scattering couplings are not well constrained, the couplings to the whole radiative process show that the coupling of the $f_0(1710)$ is larger than the $f_0(1500)$, in particular for the $K \bar K$ channel, where it becomes almost one order of magnitude greater.  As mentioned above, this favors the interpretation for the $f_0(1710)$ to have a sizeable glueball component. One might ask if this could instead be explained by an $s\bar s$ component. However, the $f_0(1710)$ is not seen in $B_s^0 \to J/\psi\,K^+K^-$ decay, where the $s\bar s$ pair is produced by the weak vertex, and enhances the production of $f_0(980)$~\cite{Ropertz:2018stk}. This non-observation also supports a glueball assignment.

\subsubsection{\texorpdfstring{$\etapicompass$ spectroscopy at COMPASS}{eta pi- spectroscopy at COMPASS}}
\label{sec:etapicompass}
%%%%%%%%%%%%%%%%%%%%%%%%%%%%%%%%%%%%
%	Figure :: Intensities
%%%%%%%%%%%%%%%%%%%%%%%%%%%%%%%%%%%%
\begin{figure}[t]
\centering
\subfigure[CDD$_{\infty}$ pole only.]{
\includegraphics[width=0.46\textwidth]{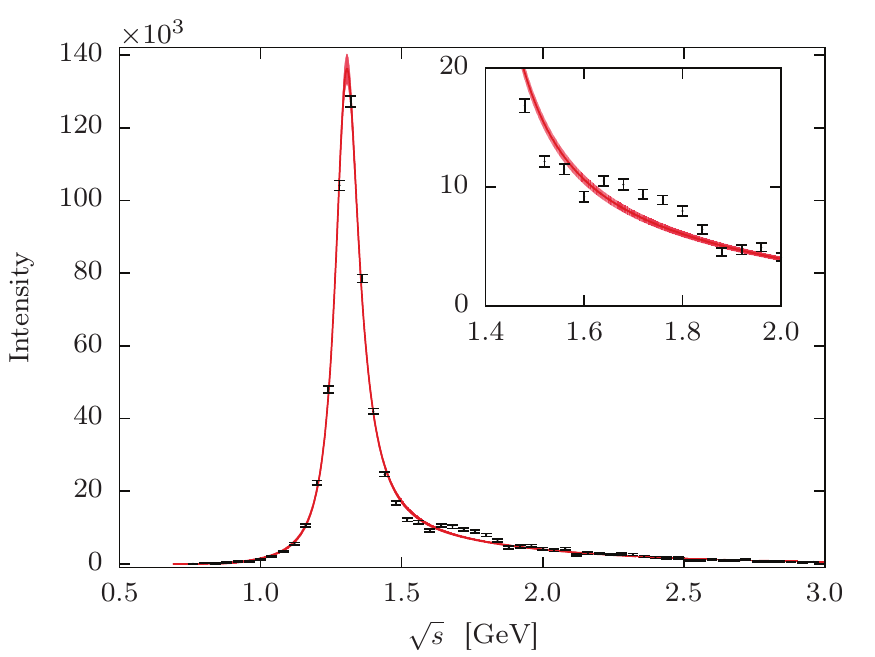}
\label{fig:1CDD}}
\subfigure[Two CDD poles.]{
\includegraphics[width=0.46\textwidth]{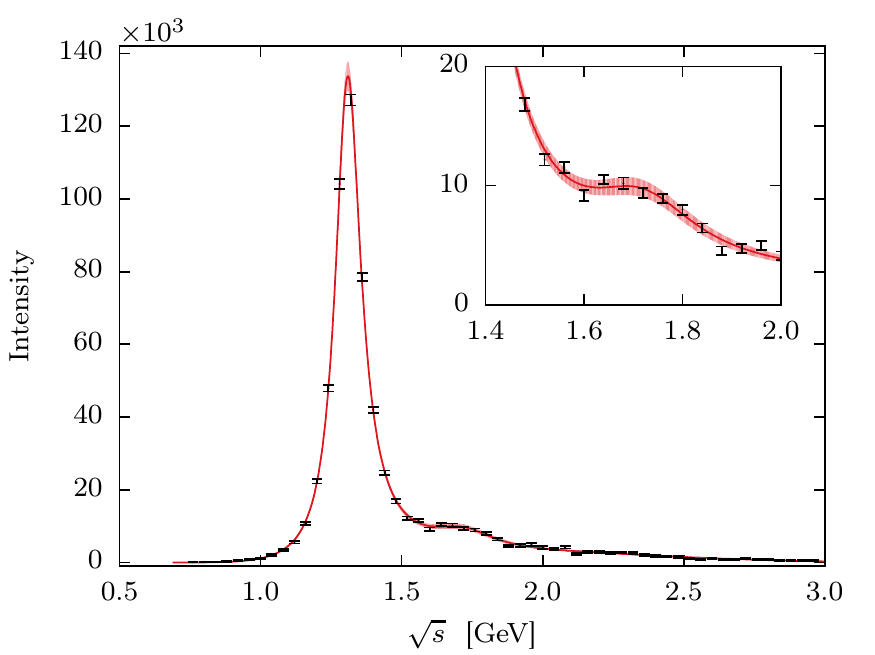}
\label{fig:2CDD}}
\caption{Intensity distribution and fits to the $J^{PC}=2^{++}$ wave  
for different number of CDD poles, (a) using only CDD$_{\infty}$  and (b) 
 using CDD$_{\infty}$ and the CDD pole at $s=c_3$. Red lines show the fit results.
 Data is taken from Ref.~\cite{COMPASS:2014vkj}. The inset shows the $a'_2$ region. 
The error bands correspond to the $3\sigma$ ($99.7\%$) confidence level. Figures from the single-channel analysis of~\cite{JPAC:2017dbi}}
\label{fig:intensities}
\end{figure}
%%%%%%%%%%%%%%%%%%%%%%%%%%%%%%%%%%%%
%	Figure : Poles
%%%%%%%%%%%%%%%%%%%%%%%%%%%%%%%%%%%%
\begin{figure}[t]
\includegraphics[width=\textwidth]{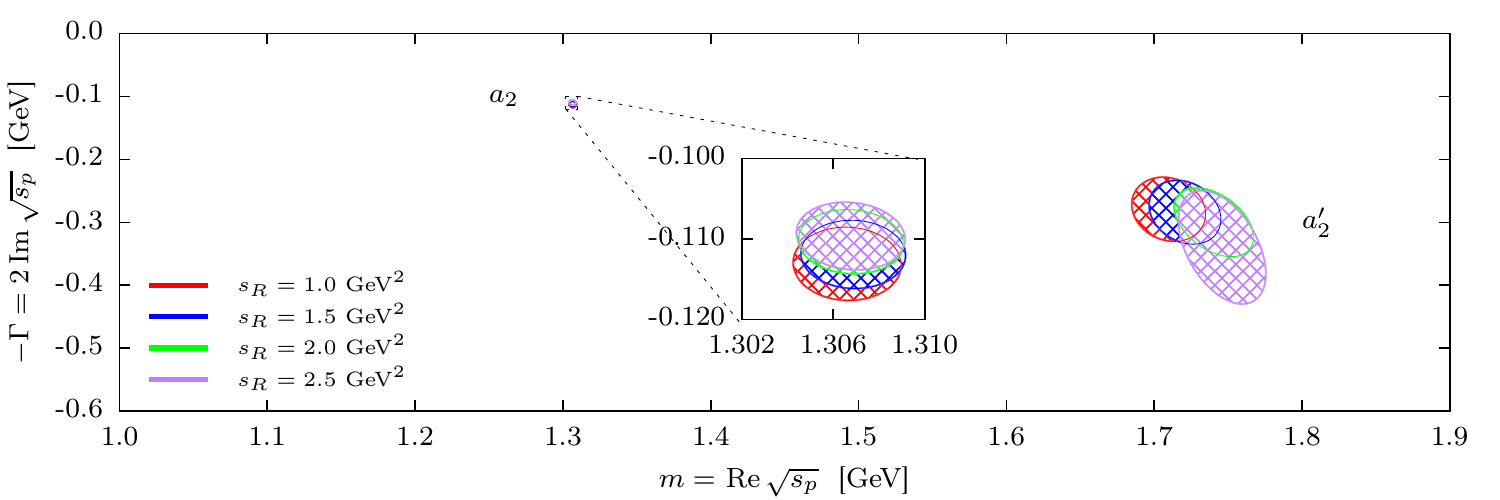}
\caption{Location of pole positions with two CDD poles. The various ellipses represent the $2\sigma$ confidence level for the several model variations  for $s_R$. The fixed parameter $s_R$ describes the initial positions of the left-hand cut. Figure from the single-channel analysis of~\cite{JPAC:2017dbi}
}
\label{fig:polepositions}  
\end{figure}

Although the $\eta^{(\prime)}$ belong to the same pseudoscalar nonet as the pion, it is too short-lived to permit measurement in any scattering experiment. The information about its interactions comes solely from production experiments, which stimulated an intense theoretical effort to obtain information about their interaction (see \eg~\cite{Albaladejo:2015aca,Albaladejo:2016mad,Isken:2017dkw}). 
The $\eta^{(\prime)} \pi$ system is particularly interesting, as its odd waves have exotic quantum numbers, and could be populated by hybrid mesons. Furthermore, its $D$-waves could contain the $a_2'(1700)$ resonance, which does not seem to often decay to two-body final states, making its precise determination challenging.

The first reported hybrid candidate was the $\pi_1(1400)$ in the $\eta \pi$ final state~\cite{Thompson:1997bs,Chung:1999we,Adams:2006sa,Abele:1998gn,Abele:1999tf}. Another state, the $\pi_1(1600)$,
was claimed to appear $\sim200\mev$ heavier in the $\rho \pi$ and $\eta'\pi$ channels~\cite{Ivanov:2001rv,Khokhlov:2000tk}. 
More refined extractions were not conclusive~\cite{Szczepaniak:2003vg}.
Both peaks were confirmed by COMPASS~\cite{COMPASS:2009xrl,COMPASS:2018uzl}. 
While the $\pi_1(1600)$ is closer to the theoretical expectations, having two nearby $1^{-+}$ hybrids below 2\gev is problematic~\cite{Close:1987aw,Chung:2002fz,Dudek:2011bn}. 
Establishing whether there exists one or two exotic states in this mass region 
is thus a stringent test for our understanding of QCD in the nonperturbative regime.

A high statistics dataset of diffractive production $\pi p\to \etapicompass p$, with $p_{\text{beam}} = 190\gev$, has been measured by the COMPASS collaboration~\cite{COMPASS:2014vkj}.
In this section we will focus on the analyses of the lowest waves in the resonance region by~\cite{JPAC:2017dbi,JPAC:2018zyd}, while we will discuss the high energy region in Section~\ref{sec:doubleregge}.

High-energy diffractive production is dominated by an effective Pomeron exchange ($\mathbb{P}$), which allows us to factorize this process into the nuclear target/recoil vertex and the $\pi \mathbb{P}\to \eta^{(\prime)}\pi$ process.\footnote{As we will discuss in Section~\ref{sec:doubleregge}, the contribution of the $f_2$ is also needed. For the sake of extracting the resonances, the details of the exchanges are not relevant: For example, no information on the trajectory enters the analysis. We will consider here a ``Pomeron'' exchange that effectively includes all the other ones, whose effective spin one dominates the $\phi$ distribution. } In the Gottfried-Jackson (GJ) frame~\cite{Gottfried:1964nx}, the $\mathbb{P}$ helicity equals the $\etapi$ total angular momentum projection $M$, and the corresponding helicity amplitude $a_M(s,t,t_1)$ can be expanded into partial waves $a_{JM}(s,t)$. Here, $s$ is the invariant mass squared of the $\eta^{(\prime)}\pi$ system, $t_1$ is the invariant momentum transfer squared between the $\eta^{(\prime)}$ and $\pi$, $t$ is the invariant momentum transfer from the $\pi$ beam to the nuclear target, and $J$ is the total angular momentum of the $\eta^{(\prime)}\pi$ system. 

At low transferred momentum, the $\mathbb{P}$ has a predominant coupling to $\lvert M\rvert = 1$ waves, indicating an effective vector coupling to the nuclear vertex. Since the COMPASS analysis integrates over $t \in [-1.0,-0.1]\gevsq$, we will consider a fixed effective $t_{\mathrm{eff}} = -0.1\gevsq$ value in our analysis, and we will vary it to assess systematic uncertainties. The production amplitude can be parametrized following the $N/D$ formalism,
\begin{equation}
\label{eq:amplitude}
 a^J_i(s) = q^{J-1} p_i^J \, \sum_k n^J_k(s) \left[ {D^J(s)}^{-1}  \right]_{ki}\, ,
\end{equation}
where the kinematic prefactors assure proper angular momentum barrier suppression. The $\pi\mathbb{P}$ momentum is represented by $q$, with the $(J-1)$ power coming from an additional momentum factor from the nuclear vertex as explained in~\cite{Jackura:2019ccv}. The parametrizations of $n^J(s)$ and $D^J(s)$ are similar to the ones discussed the previous Section~\ref{sec:scalars}.

%%%%%%%%%%%%%%%%%%%%%%%%%%%%%%%%%%%%
%	Figure :: Amplitudes
%%%%%%%%%%%%%%%%%%%%%%%%%%%%%%%%%%%%
\begin{figure}[t]
\centering
\includegraphics[width=.32\textwidth]{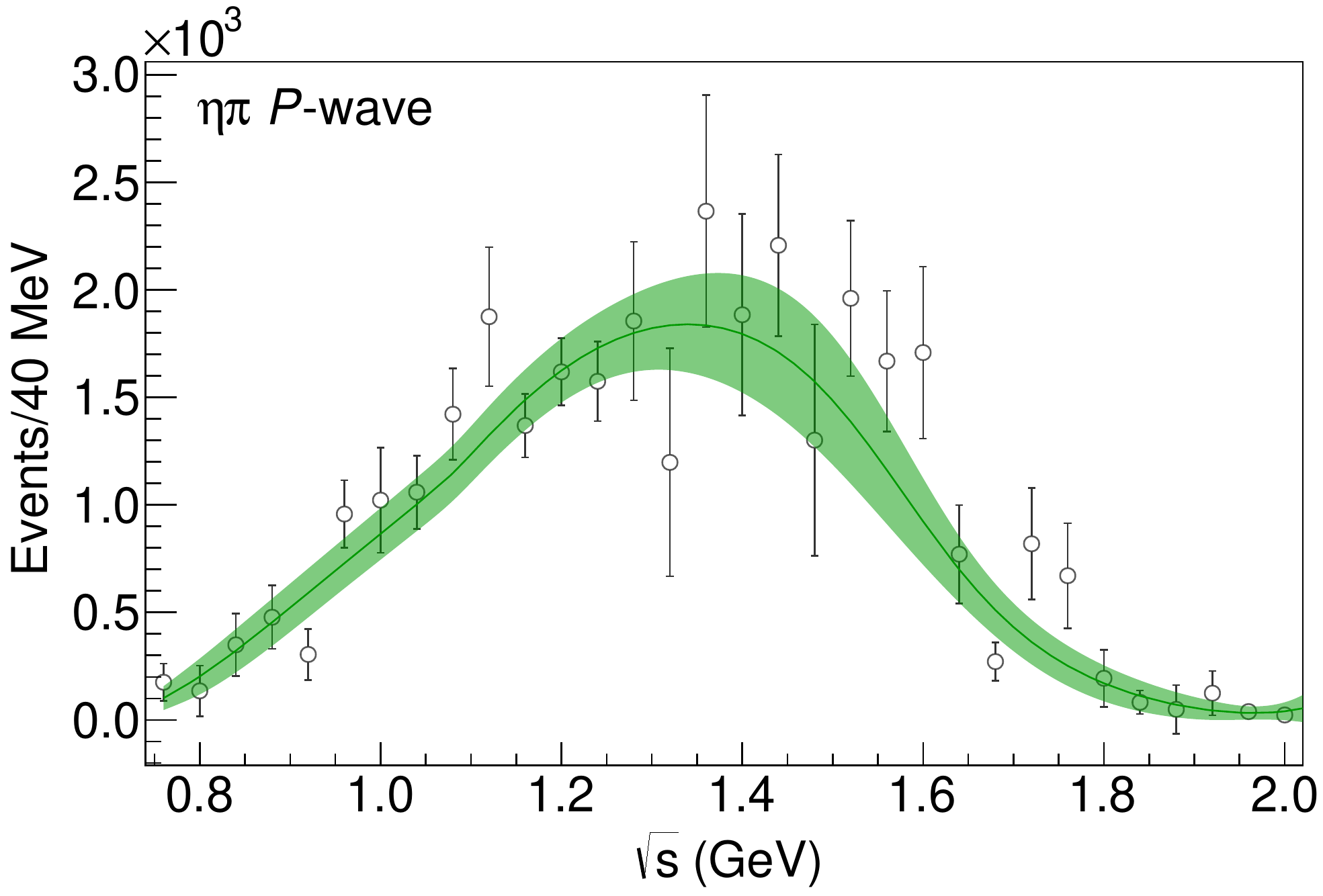}
\includegraphics[width=.32\textwidth]{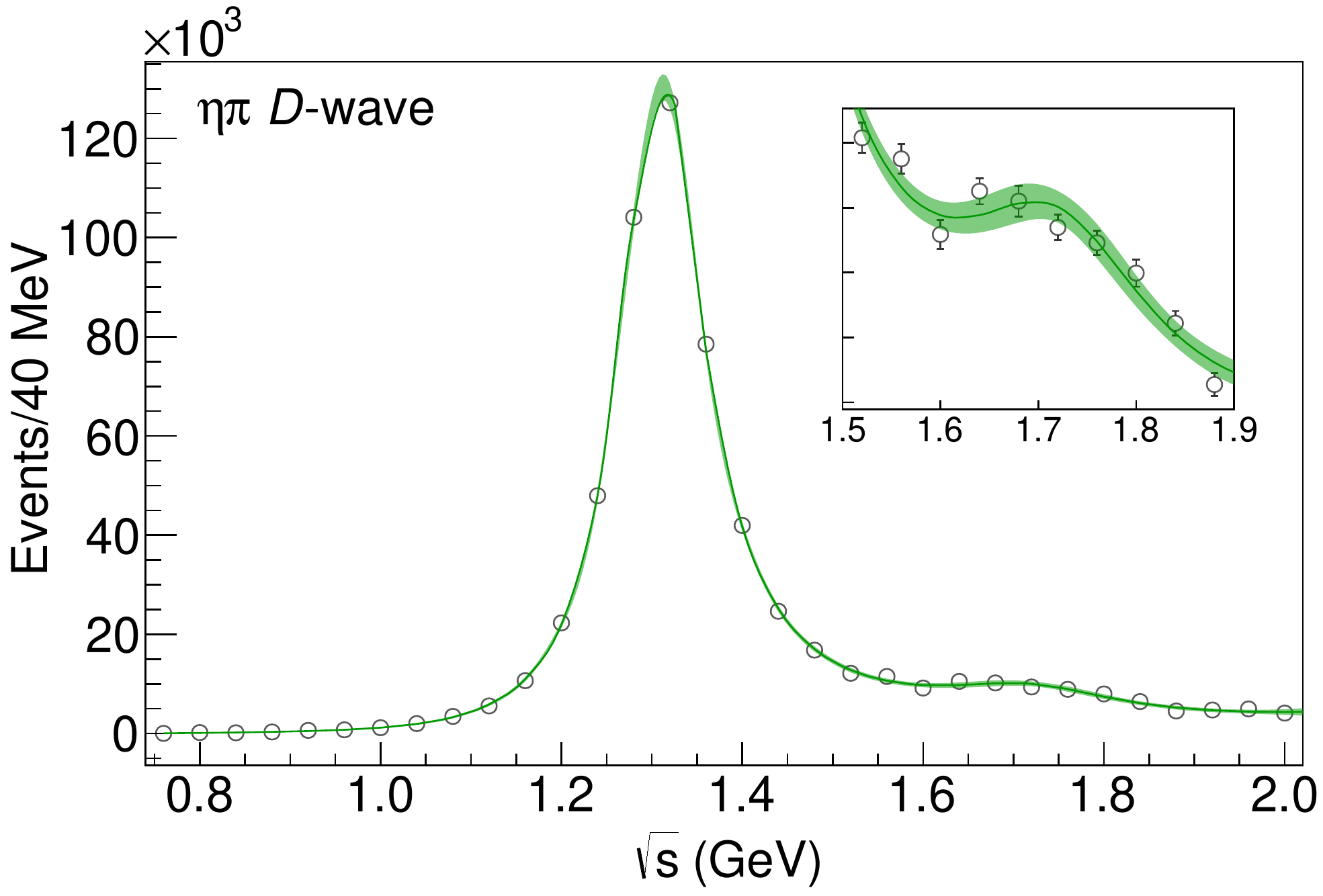}
\includegraphics[width=.32\textwidth]{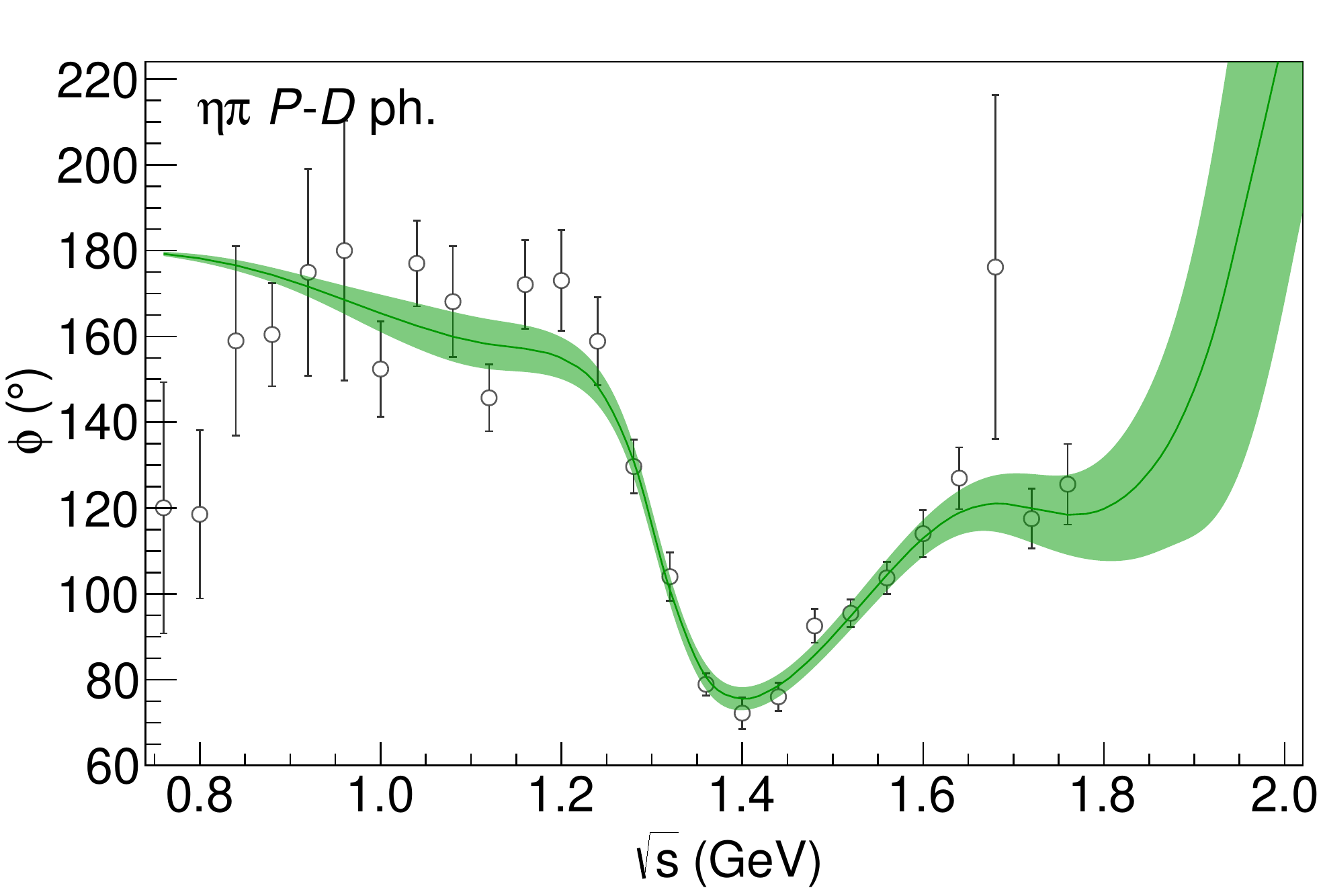}\\
\includegraphics[width=.32\textwidth]{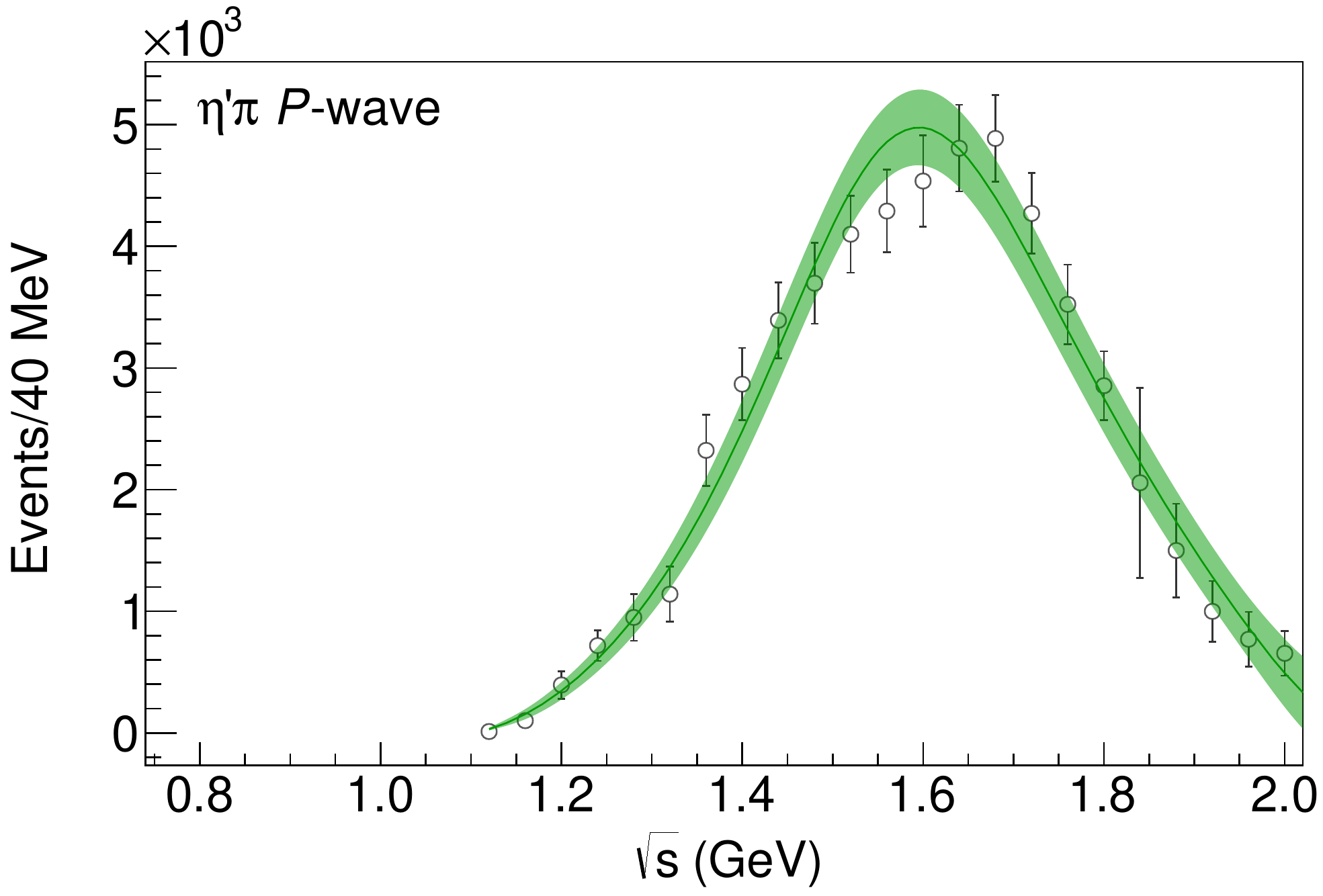}
\includegraphics[width=.32\textwidth]{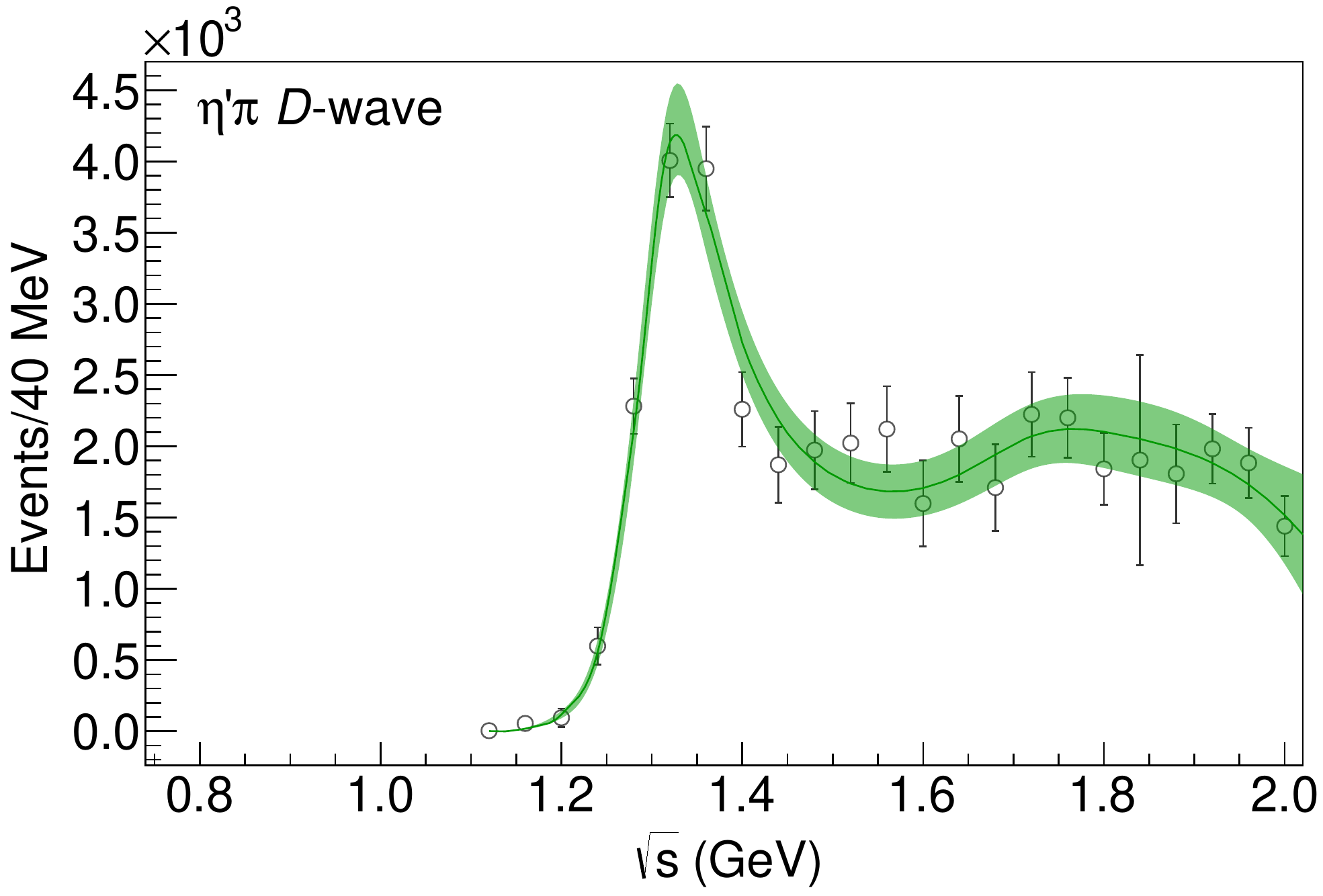}
\includegraphics[width=.32\textwidth]{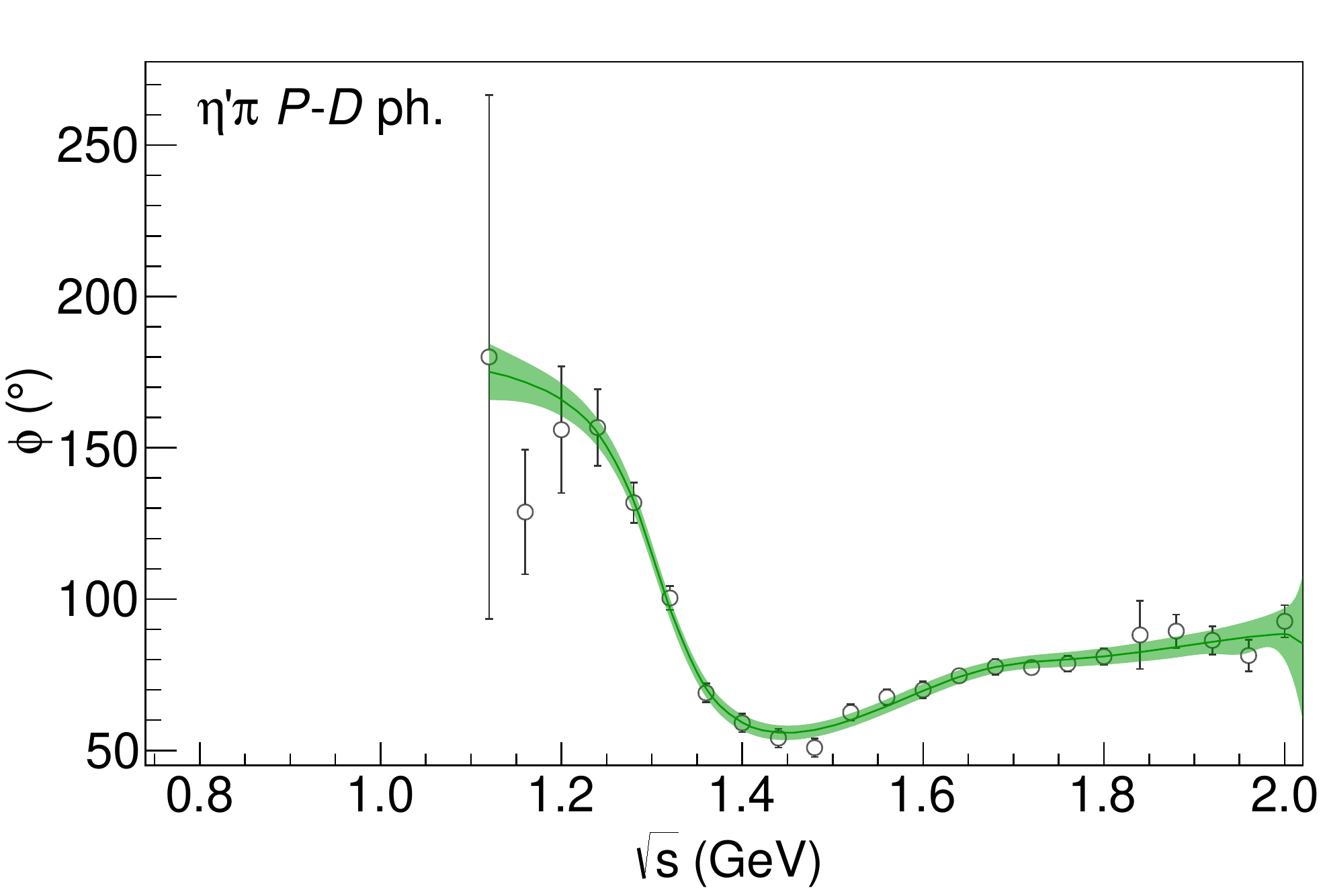}
\caption{Fits to the $\eta \pi$ (upper row) and $\eta' \pi$ (lower row) data from COMPASS~\cite{COMPASS:2014vkj}.
The intensities of $P$- (left), $D$-wave (center), and their relative phase
(right) are shown. The inset zooms into the region of the $a_2'(1700)$. The solid line and green band show the result of the fit and the $2\sigma$ confidence level provided by the bootstrap analysis, respectively. The best fit has $\chi^2/\dof = 162/122=1.3$.
Figures from the coupled-channel analysis of~\cite{JPAC:2018zyd}. 
 } 
  \label{fig:fits}
\end{figure}
As a first study, we perform a single-channel analysis of $\eta\pi$ in the $J^{PC} = 2^{++}$ wave and determine the spectral content of this channel~\cite{JPAC:2017dbi}. This serves both as a test for the model, since the tensor wave is the strongest, and an opportunity to investigate radial excitations of the $a_2$. For $D^J(s)$ the reference parametrization is CDD, which forces a zero in the amplitude that must be divided out from $n(s)$.

For the elastic case at hand, we choose CDD parametrizations as a reference as they automatically enforce that no poles can occur in the first sheet. $K$-matrix parametrizations were used for the coupled-channel fits as a systematic check. One can show that the two parametrizations are equivalent up to smooth background terms.

As seen in \figurename{~\ref{fig:intensities}}, the COMPASS data shows a dominant peak around $\sqrt{s}\sim 1.2$--$1.3\gev$, and a small enhancement around $\sqrt{s}\sim 1.7\gev$. To assess whether this is actually due to a resonance, we try and fit with just the CDD pole at infinity, and also by adding a second one. 
In the former case the fit captures the dominant $a_2$ peak, but for reasonable descriptions of the production model the secondary bump cannot be described (see \figurename{~\ref{fig:1CDD}}). In contrast, if we allow both CDD poles, we can resolve both the dominant and subdominant peaks in the spectrum, providing a good description of the data with a $\chi^2/\dof = 1.91$. The results for the pole positions of this single-channel analysis are shown in \figurename{~\ref{fig:polepositions}}.

%%%%%%%%%%%%%%%%%%%%%%%%%%%%%%%%%%%%
%	Figure : Poles-coupled-channel
%%%%%%%%%%%%%%%%%%%%%%%%%%%%%%%%%%%%
\begin{figure}[t]
\includegraphics[width=\textwidth]{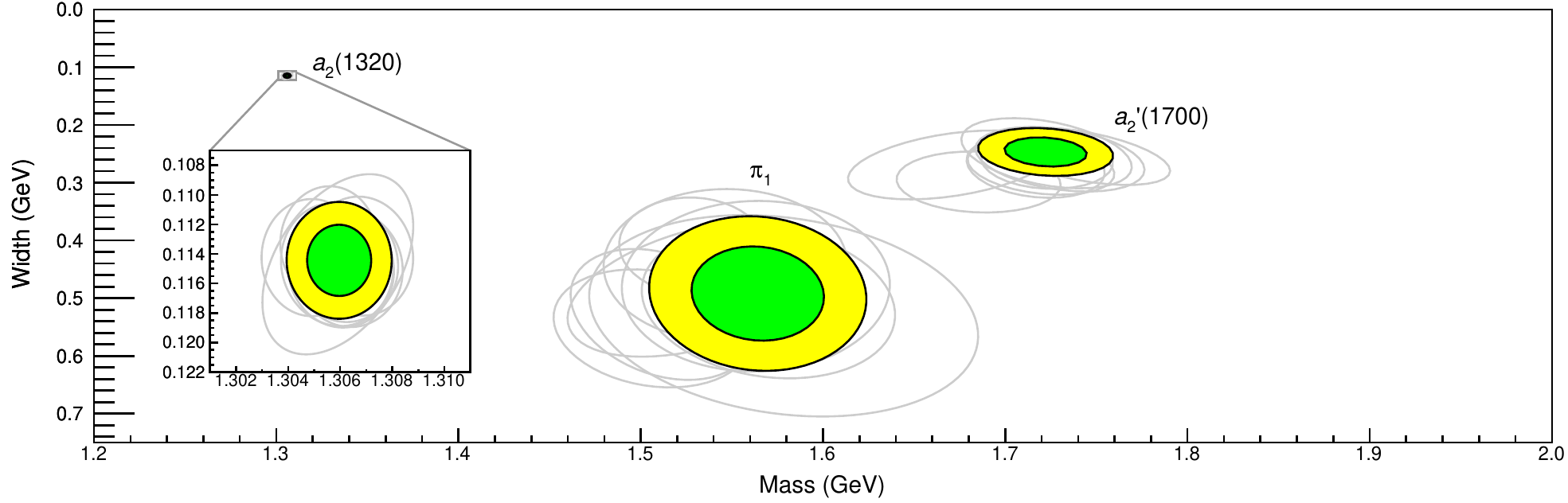}
\caption{Positions of the poles identified as the $a_2(1320)$, \pione, and $a_2'(1700)$. The inset shows the position of the $a_2(1320)$. The green and yellow ellipses show the $1\sigma$ and $2\sigma$ confidence levels, respectively. The gray ellipses in the background show, 
 within $2\sigma$, the different pole positions produced by each of the model variations as explained in~\cite{JPAC:2018zyd}. 
 }
 \label{fig:pionepoles}
\end{figure}
\begin{table}[b] 
\caption{Pole position from the $\etapi$ analysis of~\cite{JPAC:2018zyd}. The first error is statistical, the second systematic.}
\centering
\begin{tabular}{c| c c} 
\hline 
\hline
Poles & Mass (\mevp) & Width (\mevp) \\ \hline 
$a_2(1320)$ & $1306.0 \pm 0.8 \pm 1.3 $ & $114.4 \pm 1.6 \pm 0.0$ \\ 
$a_2'(1700)$ & $1722 \pm 15 \pm 67 $ & $247 \pm 17 \pm 63$ \\ 
$\pione$ & $1564 \pm 24 \pm 86 $ & $492 \pm 54 \pm 102$ \\ 
\hline 
\hline
\end{tabular}
\label{tab:pionepoles}
\end{table}

With this $2^{++}$ channel under control, we extend the analysis to a coupled-channel study where we investigate the $\pi_1$ hybrid candidate with the $\eta^{(\prime)}\pi$ data from COMPASS. To use the information on the relative phase, we have to fit the $P$- and $D$-wave data simultaneously. We also include the $\eta'\pi$ data in order to understand both of the $\pi_1(1400)$ and $\pi_1(1600)$ peaks. We will neglect any other possible decay channels other than the two at hand, even though these are not expected to be the dominant ones~\cite{Woss:2020ayi}. Our reasoning for this, as explained above, is that adding new channels should not produce a significant displacement of the pole positions in this analysis, as the missing imaginary parts will likely be absorbed by the existing channels. However, residues and thus couplings are clearly affected by this assumption, so we do not study them.

The best fit for our nominal model makes use of only a single $P$-wave $K$-matrix pole, and it is shown in \figurename{~\ref{fig:fits}}, where the global $\chi^2/\dof = 162/122=1.3$. The statistical uncertainties shown correspond to the $2\sigma$ confidence level associated to bootstrapping the sample data. This result is remarkable, considering the high precision data on the $D$-wave, and all the degrees of freedom exhibited on the data. As seen in the figure, all local features are nicely described by the fit. In particular both $P$-wave peaks, and even the elusive $a_2'(1700)$ peak are neatly captured.

For systematic checks, fits with different numbers of $K$-matrix or CDD poles in $P$-wave have been implemented. The ones with no poles are unable to describe the data, and the ones with more than one do not produce any noticeable difference in the goodness of the fit. In the complex plane extra poles can appear, but they are spread all over the place, are generally very broad, and behave erratically when the fit parameters are changed even slightly.

Once again, after obtaining a faithful description of the data, we can make use of our analytic parametrizations to search for poles in the complex plane. The statistical uncertainties are determined via bootstrap.
We perform 12 systematic variations of the nominal model and of its parameters. When continuing to the complex plane they all produce an isolated cluster in $P$-wave, that we identify with the \pione, together with two poles on the $D$-wave corresponding to the $a_2(1320)$ and $a_2'(1700)$ resonances. Their pole positions are listed in \tablename{~\ref{tab:pionepoles}}, and their spread of results is plotted in \figurename{~\ref{fig:pionepoles}}. We conclude that there is no more than one $J^{PC}=1^{-+}$ hybrid meson decaying to both \etapicompass channels. This picture reconciles experimental evidences with phenomenological and Lattice QCD expectations.

\subsubsection{Regge phenomenology of light baryons}
\label{sec:baryons_and_hyperons}
\begin{figure}[p]
\centering
\includegraphics[width=0.8\textwidth]{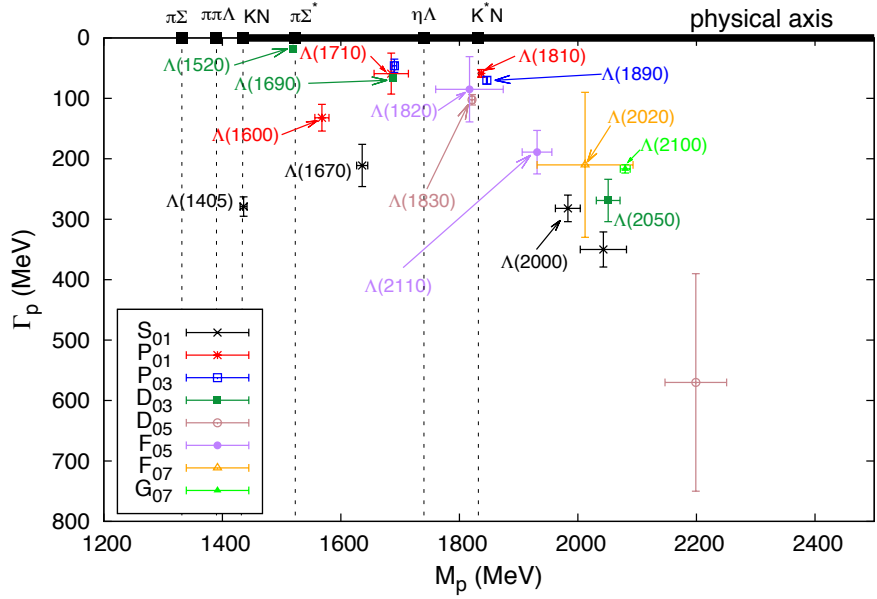} \\
\includegraphics[width=0.8\textwidth]{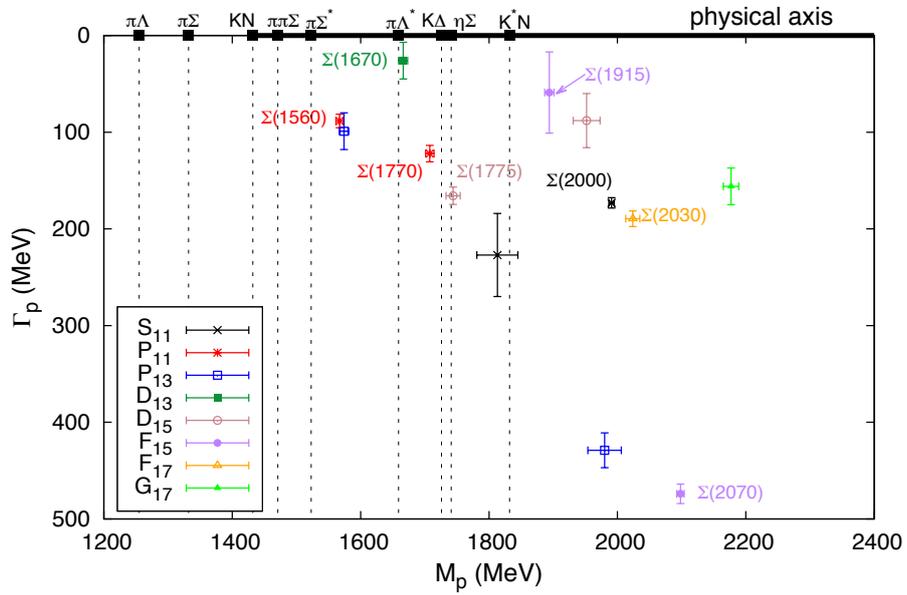}
\caption{Masses and widths from the $\Lambda$ (top) and $\Sigma$ (bottom) resonances. Figures from~\cite{Fernandez-Ramirez:2015tfa}. 
}\label{fig:hyperonpoles}
\end{figure}

The low-lying $N^*$, $\Delta$, $\Lambda$, and $\Sigma$
resonances, accessible 
in pion-nucleon and antikaon-nucleon scattering 
and in photoproduction
experiments, are a source of insights 
into the quark model and the inner works of
nonperturbative QCD phenomena~\cite{Burkert:2017djo}. 
One of the many goals of light baryon spectroscopy 
is to understand the origin and structure 
of resonances. In particular, one hopes to identify whether a 
compact three-quark interpretation holds
for these states or
if other components should be considered.

For example, the nature of the $\Lambda(1405)$ has been controversial, being a primary candidate for a $\bar{K}N$ and $\Sigma\pi$ molecule.
This interpretation has traditionally been favored
by chiral unitary approaches~\cite{Jido:2003cb,Hyodo:2007jq,Roca:2013av,Mai:2014xna}, which generally finds two poles that explain the experimental signal,
while a compact interpretation has been favored by quark models~\cite{Loring:2001ky,Melde:2008yr,Santopinto:2014opa,Faustov:2015eba} and large-$N_c$ calculations~\cite{Schat:2001xr,Goity:2002pu}.
Lattice QCD computations are inconclusive as the resonant nature of the
$\Lambda(1405)$ has not been accounted for~\cite{Engel:2012qp,Engel:2013ig,Hall:2014uca,Gubler:2016viv,Pavao:2020zle}.

Often the parameters of these resonances are extracted
from experimental data
through a partial wave analysis, assuming each partial wave independent of the others.
Such an approach does not take into account the fact that amplitudes are also analytic functions of the angular momentum, as described by Regge theory~\cite{Collins:1977jy,Gribov:2003nw,Gribov:2009zz}.
Therefore, resonances of increasing spin 
must lie on a so-called Regge trajectory,
whose shape can be
 used to gain insight on the microscopic mechanisms 
responsible for the formation of the resonance~\cite{Rossi:1977cy,Montanet:1980te,Londergan:2013dza,Carrasco:2015fva,Fernandez-Ramirez:2015fbq,Pelaez:2017sit}.
In QCD, Regge trajectories are approximately 
linear, as first shown by Chew and Frautschi~\cite{Chew:1962eu} by plotting
the spin of resonances $J_p$ versus their mass squared $M^2$,
which is one the strongest phenomenological indications of 
confinement~\cite{Greensite:2011zz}.
Constituent quark model predictions 
for baryons fit nicely in
the approximately linear behavior~\cite{Nakkagawa:1972az,Bijker:1994yr,Bijker:2000gq,Ortiz-Pacheco:2018ccl,Capstick:1985xss,Loring:2001kx,Inopin:1999nf,Tang:2000tb,Godfrey:1985xj,Koll:2000ke,Tang:2000tb,Ebert:2009ub} 
and so do flux tube models of baryons~\cite{Isgur:1984bm,Olsson:1995mv,Semay:2007cv}.
The emerging pattern can be used to
guide a partial wave analyses, for example 
gaps in the trajectories are usually due to missing states.

\begin{figure}[t]
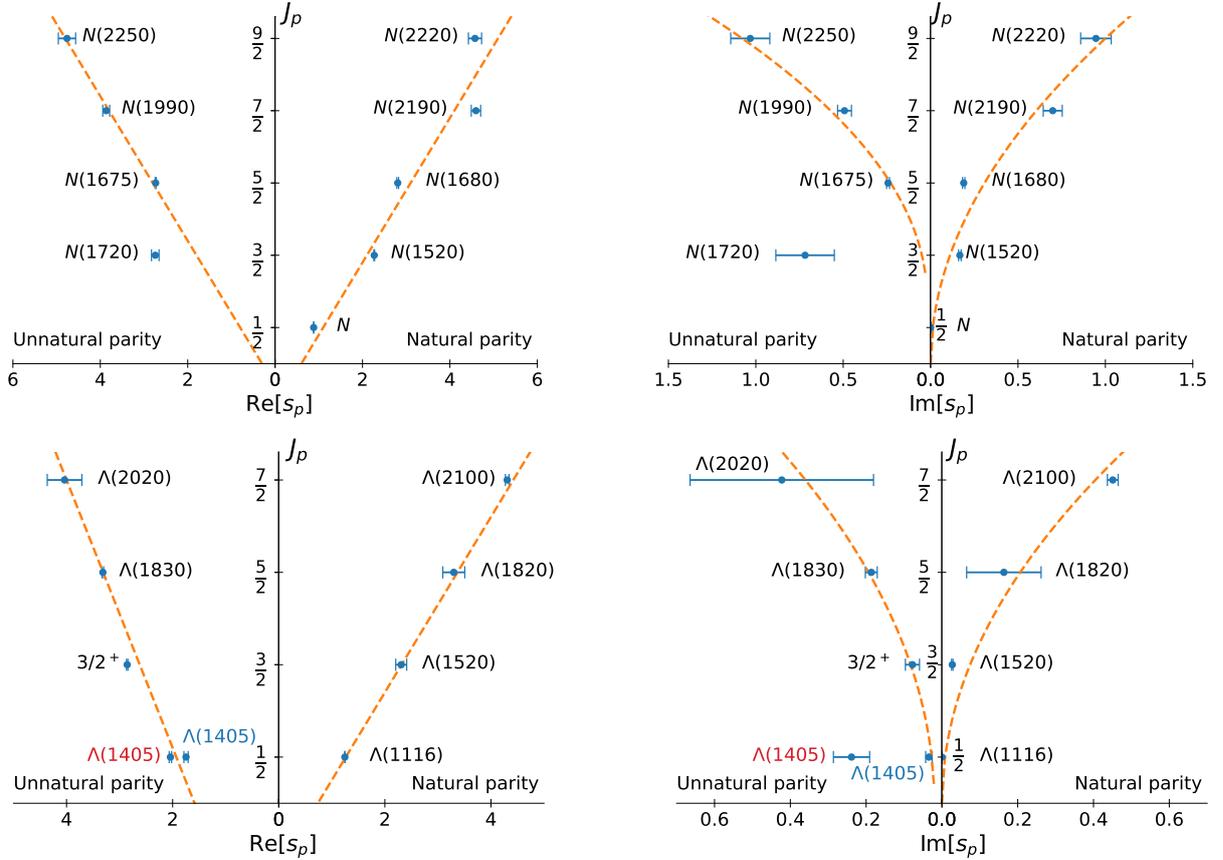

\centering
    \includegraphics[width=\textwidth]{{../figures/Lightbaryons/baryonpoles}.pdf} \\
    \includegraphics[width=\textwidth]{{../figures/Lightbaryons/hyperonpoles}.pdf}
 \caption{Leading Regge trajectories for $N^*$ (top row) and $\Lambda$ (bottom row) resonances. Left column
 shows the Chew-Frautschi plots, while the right column plots the spin as a function of the imaginary part of the pole position (mass times width). 
The $\Lambda$ poles are taken from~\cite{Mai:2014xna,Fernandez-Ramirez:2015tfa}. 
The $N^*$ poles are taken from~\cite{Anisovich:2011fc,Sokhoyan:2015fra}
We note that the $N^*(1535)$ ($J^P_p=1/2^-$, $\tau=-$, $\eta=+$) is not shown, as it 
belongs to a daughter trajectory.
}\label{fig:baryonpoles}
\end{figure}

Regge trajectories computed in the Chew-Frautschi plot 
ignore entirely the resonance widths, \ie the fact that they
are poles in the complex $s$-plane.
This is partially inconsistent, as 
unitarity demands the Regge
trajectory $\alpha(s)$ to be a complex function as well~\cite{Gribov:1962fx}. 
Therefore, one should plot 
the imaginary part of the pole (\ie the width times the mass of the resonance) as a function of the spin,
as proposed in~\cite{Fernandez-Ramirez:2015fbq}. More details about Regge theory and their implications to study production of hadrons will be given in Section~\ref{sec:regge}, while here we focus on using  the trajectories as a tool to organize the existing spectrum. 

For baryons, Regge trajectories are
classified according to 
isospin $I$, 
naturality $\eta$, and signature $\tau$.\footnote{For baryons, $\tau = (-1)^{J_p-1/2}$, for antibaryons  $\tau = (-1)^{J_p+1/2}$. The naturality is $\eta = P \tau$, with $P$ the parity.}
The quantum numbers 
identify  a given $I^\eta_{(\tau)}$ trajectory. For example, 
the nucleon trajectory corresponds to  
$I^\eta_{(\tau)}=\frac{1}{2}^+_{(+)}$.
The $\Lambda$ and $\Sigma$ poles were extracted in~\cite{Fernandez-Ramirez:2015tfa},
fitting the single energy partial waves from~\cite{Zhang:2013sva}
of $\bar{K}N$ scattering data 
with a $K$-matrix model that incorporates
analyticity in the angular momentum. The results are summarized in \figurename{~\ref{fig:hyperonpoles}}. 
Together with the
two $\Lambda(1405)$ poles from~\cite{Mai:2014xna}, their 
leading Regge trajectories were studied in~\cite{Fernandez-Ramirez:2015fbq}.

\figurename{~\ref{fig:baryonpoles}}
plots the spin of the resonance as a function of the real or imaginary part of the pole position. 
We note that only one of the $\Lambda(1405)$ poles lies on the same trajectory as the higher $\Lambda$s.
The linearity of the Chew-Frautschi plot is apparent, which suggest an interpretation as dominantly three-quark states.
The second plot provides
additional insight, specially regarding the two $\Lambda(1405)$ poles. 
In~\cite{Fernandez-Ramirez:2015fbq,JPAC:2018zjz}
it is argued that linearity
in the Chew-Frautschi plot is not enough
for a three-quark interpretation,
but since most of its width should be due to the
phase space contribution, a square-root-like behavior should emerge when plotting spin \vs imaginary part of the pole position. 
However, only one of the $\Lambda(1405)$ follows this pattern.
This suggests that the heavier $\Lambda(1405)$
might be mostly a compact state, while the lightest would have a different nature, most likely a molecule~\cite{Fernandez-Ramirez:2015fbq}, although no consensus on the topic has been reached yet.

The nonstrange light baryon spectrum can be studied in the same way~\cite{JPAC:2018zjz}.
The pole extraction can be taken from several partial wave analyses of meson scattering and photoproduction data available in the literature~\cite{Cutkosky:1979fy,Cutkosky:1980rh,Ronchen:2018ury,Anisovich:2011fc,Sokhoyan:2015fra,Svarc:2014aga,Svarc:2014zja}.
In \figurename{~\ref{fig:baryonpoles}} we show an example of $N^*$ trajectory.
The states nicely accommodate the Regge expectation, except for the
$N(1720)$, which in~\cite{Anisovich:2011fc,Sokhoyan:2015fra} has a large width $\Gamma_p \sim 300$--$430\mev$
that would place this state close to the daughter trajectory.
Hence, Regge phenomenology demands the existence
of another narrower state.

Such a state was actually claimed to be narrower in other analyses~\cite{Cutkosky:1979fy,Cutkosky:1980rh}
 with $\Gamma_p=120\mev$,
but no consensus was reached~\cite{LandoltBornstein1983:sm_lbs_978-3-540-39059-6_90,Svarc:2014zja,Ronchen:2018ury}. A recent CLAS analysis finds actually two $N(1720)$ with similar mass and widths, but different $Q^2$ behavior in electroproduction~\cite{Mokeev:2020hhu}.
The
ANL-Osaka analysis finds two poles
with masses $1703$ and $1763\mev$
and widths $70$ and $159\mev$,
respectively~\cite{Kamano:2013iva}.
Since quark models predict several $3/2^+$ states in this energy
region~\cite{Capstick:1985xss,Bijker:1994yr,Bijker:2000gq,Loring:2001kx}, it is possible that the data analyses are not able to resolve each pole individually.
Further research is necessary
to establish the number and properties of resonances in this energy region, before discussing their nature.

\subsection{Heavy quark spectroscopy}
\label{sec:heavy}
\begin{figure}[t]
\includegraphics[width=\textwidth]{{../figures/exotics_review}.pdf}
\caption{Summary of ordinary charmonia, \XYZ and pentaquarks listed by the PDG~\cite{pdg}.}\label{fig:charmonia_et_exotica}
\end{figure}

The unexpected discovery of the $X(3872)$ in 2003 ushered in a new era in hadron spectroscopy~\cite{Belle:2003nnu}. Experiments have claimed a long list of states, collectively called \XYZ, that appear mostly in the charmonium sector, but do not respect the expectations for ordinary $Q\bar Q$ states, summarized in \figurename{~\ref{fig:charmonia_et_exotica}}. 
An exotic composition is thus likely required~\cite{Olsen:2017bmm,Brambilla:2019esw}. Several of these states appear as relatively narrow peaks in proximity of open charm threshold, suggesting that hadron-hadron dynamics can play a role in their formation~\cite{Guo:2017jvc,Dong:2020hxe}. Alternatively, quark-level models also predict the existence of supernumerary states, by increasing the number of quark/gluon constituents~\cite{Esposito:2016noz}. The recent discovery of a doubly-heavy $T_{cc}^+$~\cite{LHCb:2021vvq,LHCb:2021auc} and of a fully-heavy $X(6900)$~\cite{LHCb:2020bwg} states make the whole picture extremely rich. 
  Having a comprehensive description of these states  
  will improve our
understanding of the nonperturbative features of QCD. Most of the analyses from \belle and \babar suffered from limited statistics, and strong claims were sometimes made with simplistic models on a handful of events. Currently running experiments like \lhcb and \bes have overcome this issue, providing extremely precise datasets that also require more sophisticated analysis methods and theory inputs. The status of ordinary and exotic charmonia is summarized in \figurename{~\ref{fig:charmonia_et_exotica}}. 
Depending on their width and the production mechanism, the states can roughly be classified into the following categories:
\begin{enumerate}
 \item Narrow ($\lesssim 50\mev$) states that appear in $b$-hadron decays: $X(3872)$, $P_c(4312)$, $P_c(4440)$, $P_c(4457)$, \dots, and at $e^+e^-$ colliders: $X(3872)$, $Y(4230)$, $Z_{c,b}^{(\prime)}$, \dots
 \item Broad ($\gtrsim 50\mev$) states that appear in $b$-hadron decays: $\chi_{c0}(4140)$, $Z(4430)$, $Z_{cs}(4000)$, $P_c(4380)$, \dots
 \item States produced promptly at hadron machines: $X(3872)$, $T_{cc}^+$, $X(6900)$, \dots
\end{enumerate}
The narrow signals do not require a thorough understanding of interferences with the background. Since they often appear close to some open flavor threshold, they call for analysis methods that incorporate such information. To some extent, it is possible to give model-independent statements.
The $X(3872)$ is very special. It has $J^{PC}=1^{++}$, violates isospin substantially decaying into $\jpsi \rho$ and $\jpsi \omega$ with similar rates, and lies exactly at the $\bar D^0 D^{*0}$ threshold. Its lineshape was recently studied by \lhcb, which triggered several discussions~\cite{LHCB:2020xds,Esposito:2021vhu,Baru:2021ldu}. The $Z_c(3900)$ (with $J^{PC}=1^{+-}$) was seen as a peak in the $\jpsi\,\pi$ invariant mass in the $e^+ e^- \to  \jpsi\,\pi \pi$ process, and as an enhancement at the $D\bar D^*$ threshold in $e^+ e^- \to  \pi D \bar D^*$. Similarly, a $Z_c'(4020)$ with same quantum numbers peaks in $h_c\,\pi$ invariant mass in the $e^+ e^- \to  h_c\,\pi \pi$ process, and enhances the cross section at the $D^*\bar D^*$ threshold in $e^+ e^- \to  \pi D \bar D^*$. The system of two $1^{+-}$ at the two thresholds seems replicated in the bottomonium sector, by the $Z_b(10610)$ and $Z_b^\prime(10650)$. The proximity to threshold motivated their identification as hadron molecules~\cite{Tornqvist:1993ng,Braaten:2003he,Voloshin:2003nt,Close:2003sg,Swanson:2006st,Cleven:2011gp,Guo:2013sya,Wang:2013cya}, but tetraquark interpretations are also viable~\cite{Maiani:2004vq,Ali:2011ug,Ali:2014dva,Maiani:2019cwl}.

The discovery of pentaquark candidates in $\Lambda_b^0 \to J/\psi p K^-$ decay in 2015 also boosted the field substantially. The \lhcb collaboration reported a narrow and a broad state, the $P_c(4450)$ and the $P_c(4380)$, with likely opposite parities~\cite{Aaij:2015tga}. The subsequent 1D analysis in 2019, with ten-times higher statistics, reported a composite structure of the narrow peak, that split into $P_c(4440)$ and $P_c(4457)$, and found a new isolated peak, the $P_c(4312)$~\cite{LHCb:2019kea}. In light of this new information, the quantum numbers reported previously are no longer reliable. Again, the signals can be interpreted as compact five-quark states~\cite{Maiani:2015vwa,Lebed:2015tna,Anisovich:2015cia,Ali:2019npk}, weakly bound meson-baryon molecules~\cite{Chen:2015loa,Meissner:2015mza,Chen:2015moa,Roca:2015dva,Guo:2019fdo,Guo:2019kdc,Xiao:2019aya,Liu:2019tjn}, or triangle singularities~\cite{Szczepaniak:2015hya,Guo:2015umn,Mikhasenko:2015vca,Guo:2016bkl,Bayar:2016ftu,Guo:2019twa,Nakamura:2021qvy}. 

We will discuss the examples of the $Z_c(3900)$ in Section~\ref{sec:zc3900} and of the $P_c(4312)$ in Section~\ref{sec:pc4312}. In some reactions, it is possible that 3-body dynamics, and in particular triangle singularities can play a role. An example of this will be discussed in Section~\ref{sec:pc4337}.

Broad states are much less evident in data, and can be extracted only with sophisticated amplitude analyses, where unitarity is usually neglected. Having a full exploration of the systematic model variations on the lines of what was presented in the previous sections has not been done yet due to limits in computational and human resources.
In the open charm sector, several U$\chi$PT-like analyses suggest the $D^*_0(2300)$, $D^*_1(2430)$, $D_{s0}(2317)$ and $D_{s1}(2460)$ to have a molecular nature. For the $D^*_0(2300)$, a double pole structure similar to the $\Lambda(1405)$ is suggested~\cite{Guo:2009ct,Albaladejo:2016lbb,Du:2017zvv,Du:2020pui}.

In prompt production, the initial kinematics is uncontrolled. While there is no doubt that strong signals exist, there is a long debate on whether or not one can infer the nature of such states from the production properties at high energies~\cite{Bignamini:2009sk,Artoisenet:2009wk,Bignamini:2009fn,Artoisenet:2010uu,Albaladejo:2017blx,Esposito:2017qef,Wang:2017gay,Esposito:2020ywk,Braaten:2020iqw}.
It is worth mentioning that, with the exception of the $X(3872)$, the \XYZ have been observed in one specific  
production channel.\footnote{Actually, {D\O} claimed to observe the $Z_c(3900)$ also in inclusive $b$-hadron decays~\cite{D0:2018wyb}. However, the statistics is still low, and the inclusive analysis prone to sizeable systematic effects. For this reason, we do not consider this claim to as convincing as the other ones, which justify our statement above. } Exploring alternative production mechanisms would provide complementary information, that can further shed light on their nature. The study of \XYZ photoproduction will be discussed later in Section~\ref{sec:xyz_exclusive}.

\subsubsection{\texorpdfstring{The $Z_c(3900)$}{The Zc(3900)}}
\label{sec:zc3900}

\begin{figure}
\centering
    \includegraphics[width=0.5\textwidth]{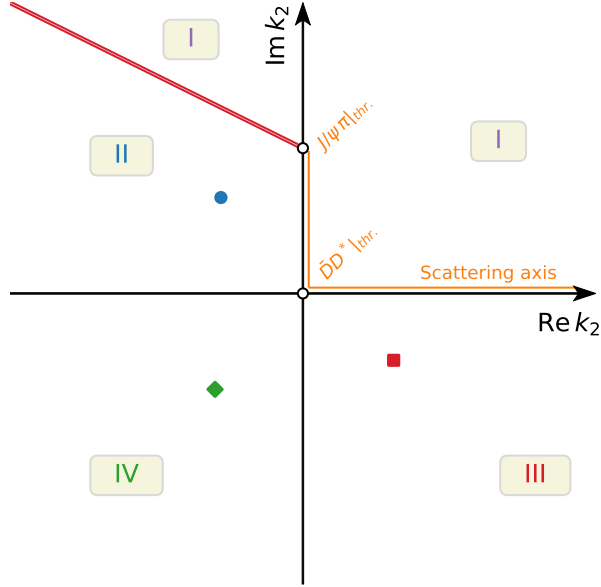} 
      \caption{Analytic structure of the $Z_c(3900)$ amplitude near the 
    $\bar D D^*$ threshold. The adjacent Riemann sheets are continuously connected along the axes. Several possibilities for the pole to appear. A pole on the III sheet above the $DD^*$ threshold (red square) generates a usual Breit-Wigner-like lineshape, and is likely due to a genuine QCD resonance. A pole on the II sheet below threshold (blue circle) is likely due to a bound state of $\bar DD^*$. Similarly, a pole on the IV sheet is not immediately visible on the physical region (orange), but enhances the threshold cusp. This is likely due to a virtual state. See the text for more details.}\label{fig:complexplaneZc}
\end{figure}

 \begin{figure}[b]
 \centering
 \includegraphics[width=.25\textwidth]{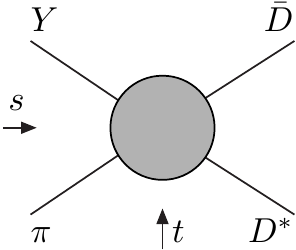}\hspace{2cm}
 \includegraphics[width=.25\textwidth]{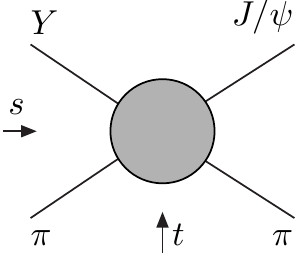}
 \caption{Channel definitions. In channel $1$ we consider the exchange of a $D_1(2420)$ in $t$ and of a $\bar D_0(2300)$ in $u$ in addition to the possible $Z_c$ in $s$.  In channel $2$ we consider the exchange of a $f_0(980)$ and a $\sigma$ in $t$, in addition to the possible $Z_c$ in $s$ and $u$. }
 \label{fig:zcdiagrams}
  \end{figure}
     
As we previously stated, the $Z_c(3900)$ peaks in $\jpsi\,\pi$~\cite{Ablikim:2013mio,Liu:2013dau,Xiao:2013iha,BESIII:2015cld}, and enhances the $D\bar D^*$ cross section at threshold~\cite{Ablikim:2013xfr,Ablikim:2015swa,Ablikim:2015gda}. 
Several possibilities are viable: It might be a bound or virtual state of $D \bar D^*$, that moves into the complex plane due to the coupling to $\jpsi\,\pi$; it might be a genuine QCD resonance; it might be a mere threshold cusp enhanced by the presence of a triangle singularity closeby.  
 The best candidate to produce a triangle cusp is the $D_1(2420)$  resonance in $D^* \pi$. 

Each microscopic interpretation reflects into the analytic properties of the amplitude: Bound and virtual states would likely appear on the II and IV sheet, compact states generally lie on the III sheet. This is schematically represented in \figurename{~\ref{fig:complexplaneZc}}. A more refined classification based also on the sign of the scattering length will be discussed in Section~\ref{sec:pc4312}, or can be given looking at the pole residues according to Weinberg's criterion, see \eg~\cite{Weinberg:1965zz,Baru:2003qq,Matuschek:2020gqe}. Triangle singularities produce a branch point that can also enhance a peak.
Such peaks cancel in the elastic case. Therefore, studying the inelasticities with a proper coupled channel analysis can constrain the strength of the triangle. Moreover, the details of the line shape are related to the sheet the physical pole is located, and eventually offer a tool to study the nature of the $Z_c(3900)$.

The amplitude can be parametrized in the isobar model:
\begin{equation} 
f_i(s,t,u) = 16\pi\,\left[a^{(t)}_{i}(t) +a^{(u)}_{i}(u) + \sum_j t_{ij}(s)\left(c_j + \frac{s}{\pi}  \int_{s_j}^\infty  ds^\prime  \frac{\rho_j(s^\prime) b_{j}(s^\prime) }{s^\prime\left(s^\prime - s\right)}\right)\right], \label{eq:KTZc} 
\end{equation}
with $i$ running over the two channels $\bar D D^*$ and $\jpsi \pi$, with Mandelstam variables $s,t,u$ as represented in \figurename{~\ref{fig:zcdiagrams}}. We are mostly interested in the distribution is $s$, where the $Z_c(3900)$ is directly observed.
Isospin symmetry is assumed, so that the neutral and charged datasets are studied together. Little information is available about the angular distributions, so there is no point in considering spin. Since we are not interested in  channels with nonexotic quantum numbers, we fill the $a^{(t,u)}_{i}$ isobars with simple Breit-Wigners for the $D_1(2420)$, $D^*_0(2300)$, and for two effective $\pi\pi$ resonances whose mass and width are let free in the fit.
The $s$-channel is unitarized {\it \`a la} Khuri-Treiman (see Section~\ref{sec:KT}), which gives the dispersive integral in Eq.~\eqref{eq:KTZc}, in terms of projections of the isobars in the other channels:
\begin{equation}
 b_{i}(s) = \frac{1}{32\pi} \int_{-1}^1 d z_s\, \left[a^{(t)}_{i}\!\left(t(s,z_s)\right)  +a^{(u)}_{i}\left(u(s,z_s)\right) \right] \, .
\end{equation}
The cross channels are not unitarized, so no integral equation has to be solved.
The scattering amplitude $t_{ij}$ describes the final state interactions, and can be parametrized with different functional forms, that allow for different singularities in the complex plane. We use the $K$-matrix parametrization that explicitly encodes unitarity $t_{ij} = \left[K^{-1}(s) - i \rho(s)\right]^{-1}_{ij}$, and consider four scenarios:
\begin{enumerate}
\item \scenIII: We use Flatt\'e, $K_{ij} = g_i g_j / (M^2 - s)$, and force $b_i(s)\equiv 0$. Although unphysical, this choice is the closest to the parametrization used in the experimental analyses, and eases the comparison; 
\item \scenIIItr: Same, restoring the correct $b_i(s)$. These two scenarios naturally produce either a bound state pole below the $\bar D D^*$ threshold, or a resonant pole above it, depending on the value of $M$.
\item \scenIV: $K$ is a symmetric constant matrix, which produces either bound or virtual states. 
\item \scentr: Same, but forcing the pole to be far from threshold penalizing its position in the fit, to assess whether the triangle singularity alone is able to generate the observed structure.
\end{enumerate}

 \begin{figure}[t]
 \centering
\includegraphics[width=.48\textwidth]{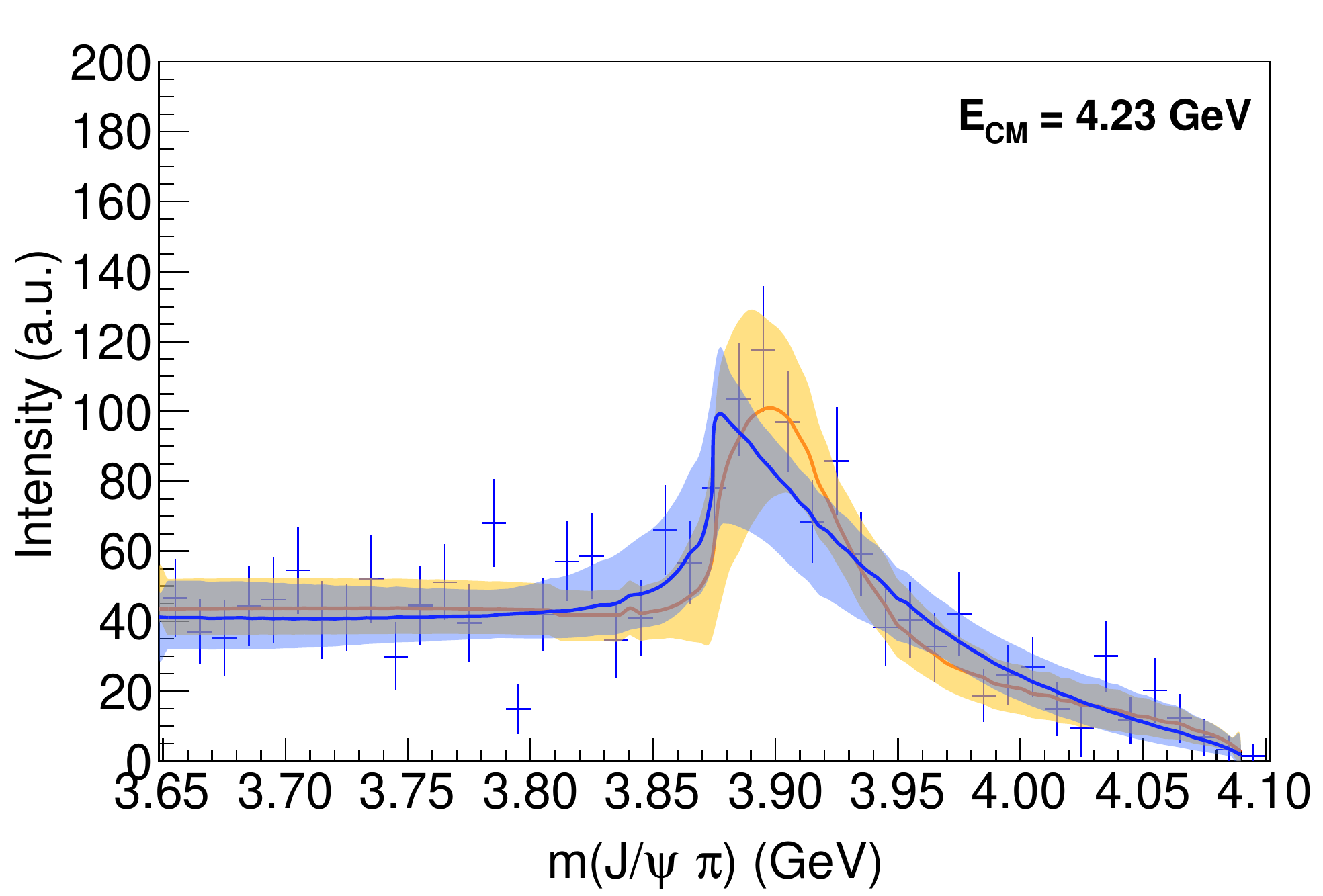}
\includegraphics[width=.48\textwidth]{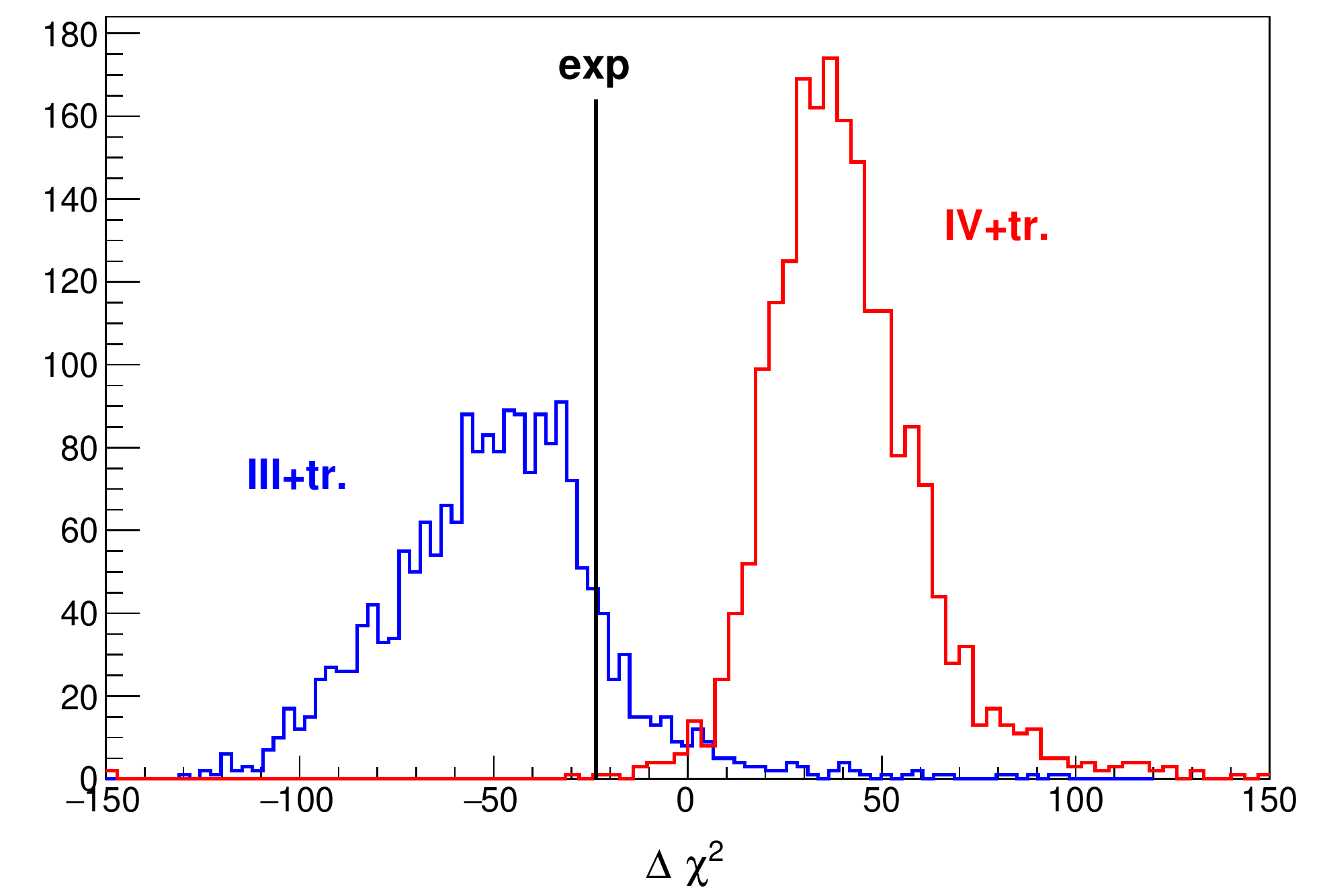}
\caption{(left) Result of the fit for the scenario \scenIII and \scentr. The colored lines and bands show the fit result with the relative $1\sigma$ error, calculated with bootstrap. Data are the $\jpsi \pi^0$ projection of the $e^+e^-\to \jpsi \pi^0\pi^0$ at $E_\text{CM} = 4.23\gev$, by \bes~\cite{BESIII:2015cld}. The errors shown are statistical only. (right) Loglikelihood ratio test. We histogram the $\chi^2$ difference of the \scenIII and \scenIV models, assuming that \scenIII (blue) or \scenIV is the truth. The black line highlights the  value of $\Delta\chi^2$ obtained from data. One gets a $2.7\sigma$ rejection of \scenIV over \scenIII.}
 \label{fig:zclineshapes}
 \end{figure}

We perform a minimum $\chi^2$ fit of these models to the $e^+e^-\to \jpsi \,\pi\pi$~\cite{Ablikim:2013mio,BESIII:2015cld} and $\to \bar D D^* \pi$~\cite{Ablikim:2015swa,Ablikim:2015gda}, for two values of total energy. 
In \figurename{~\ref{fig:zclineshapes}} we show an example of how the various models result in different lineshapes.

 All the models fit reasonably well the data, with $\chi^2/\dof$ ranging from $1.2$ to $1.3$. A likelihood ratio test does not give rejections larger than $3\sigma$~\cite{Demortier:2007zz,Faccini:2012zv}. We conclude that present statistics prevents us from drawing any strong statements.
 
\begin{figure}[t]
\includegraphics[width=.32\textwidth]{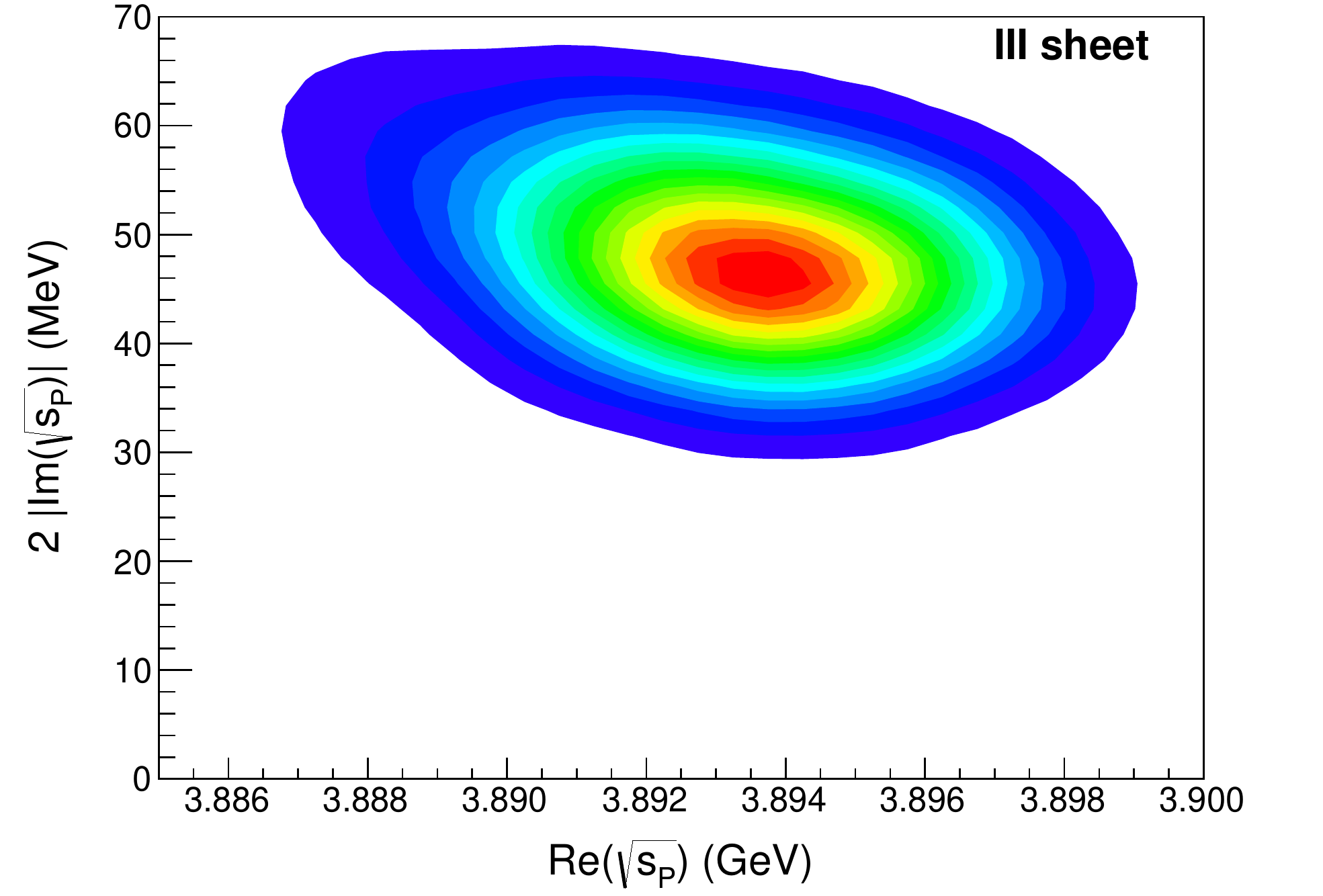}
\includegraphics[width=.32\textwidth]{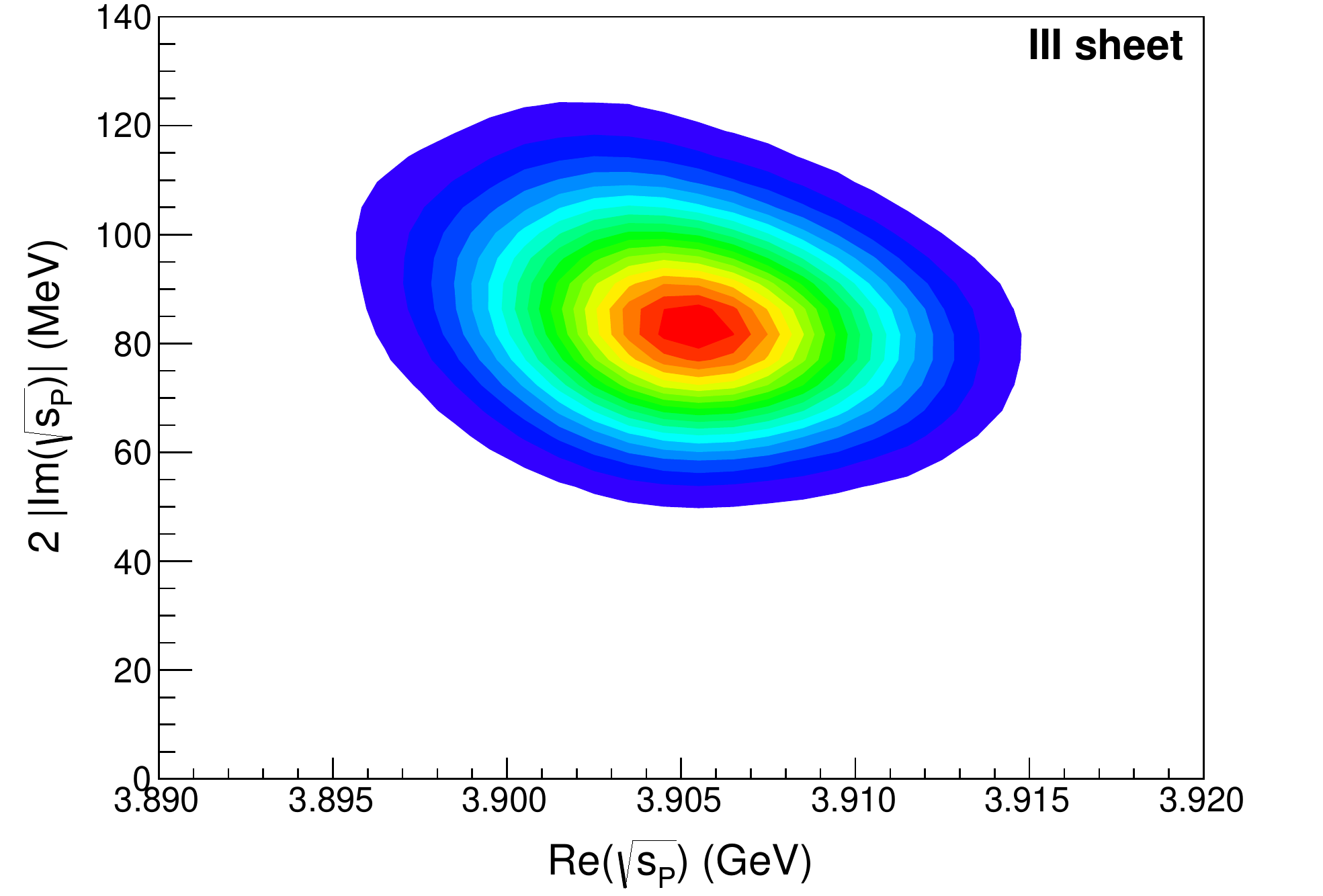}
 \includegraphics[width=.32\textwidth]{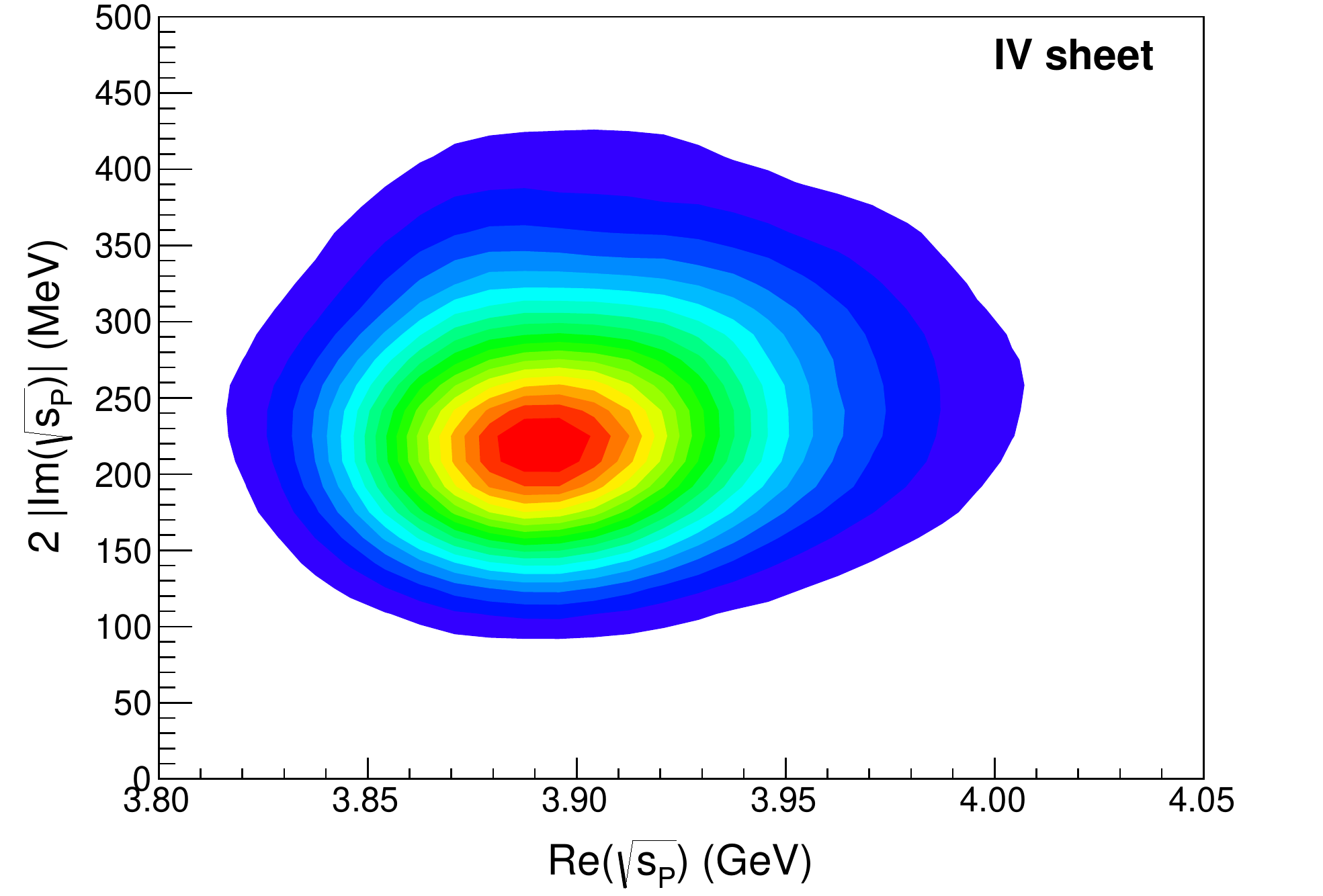} 
 \caption{Pole position according to the scenarios which allow for the presence of a pole in the scattering matrix close to the physical region. The colored regions correspond to the $1\sigma$ confidence level.
Figures from~\cite{Pilloni:2016obd}.}
 \label{fig:poles}
 \end{figure}

 In the models where a $Z_c$ pole appears, we can quote its position, and estimate the uncertainties using bootstrap. The results are summarized  
in \figurename{~\ref{fig:poles}} and \tablename{~\ref{tab:poles}}. We observe that:
 \begin{enumerate}
 \item \scenIII: The pole appears above the $\bar D D^*$ threshold, on the III sheet, and the width is $\Gamma \simeq 50\mev$, marginally compatible with the value quoted by the PDG, $M = 3886.6 \pm 2.4\mev$, $\Gamma = 28.1 \pm 2.6 \mev$~\cite{pdg}.  
\item \scenIIItr:  The presence of the logarithmic branching point close to the physical region allows for the pole to be slightly deeper in the complex plane, $\Gamma \simeq 90\mev$. The mass is still safely above threshold. 
\item \scenIV: In this case the peak is generated by the combination of the logarithmic branching point with the virtual state pole on the IV sheet. 
Given that this sheet is not directly connected with the physical region, and that the triangle singularity contributes to the strength of the signal, the pole position is not well constrained.
  \end{enumerate}
 
 To summarize, we presented a coupled-channel study of the $Z_c(3900)$. We write a unitarized model that takes into account the possible rescattering with the bachelor particle in both channels. We consider several scenarios that, depending on the parametrization chosen, produce singularities that favor different physical interpretations, but present statistics is not able to distinguish between those. Similar conclusions were reached in~\cite{Albaladejo:2015lob,Albaladejo:2016jsg}. In the following Section, we will present a similar analysis where quality of data is actually able to discriminate between the various hypotheses.
 
\begin{table}[b]
\centering
\caption{Mass and width of the $Z_c(3900)$ according to the scenarios which allow for the presence of a pole. Table from~\cite{Pilloni:2016obd}.} 
\label{tab:poles}
\begin{tabular}{c|ccc}
\hline
\hline
 &  \scenIII & \scenIIItr & \scenIV \\ \hline
$M \equiv \Re \sqrt{s_P}$ (\mevp) &  $3893.2^{+5.5}_{-7.7}$ & $3905^{+11}_{-9}$ & $3900^{+140}_{-90}$\\
$\Gamma \equiv 2 \left|\Im \sqrt{s_P}\right|$ (\mevp)&  $48^{+19}_{-14}$  &$85^{+45}_{-26}$  & $240^{+230}_{-130}$ \\
\hline
\hline
\end{tabular}

\end{table}

\subsubsection{The $P_c(4312)$}\label{sec:pc4312}
As discussed in Section~\ref{sec:heavy},
the discovery of two pentaquark resonances,
$P_c(4380)$ and $P_c(4450)$ in the $\Lambda^0\to J/\psi K^- p$
decay by \lhcb in 2015~\cite{Aaij:2015tga} triggered a frenzy of theory work to determine their nature. Later, with ten times more events~\cite{LHCb:2019kea}, the $P_c(4450)$ signal
was resolved into two peaks, $P_c(4440)$ and $P_c(4457)$,
and a new $P_c(4312)$ was discovered.
The latter is particularly interesting as it is a very clean isolated structure
that peaks approximately $5\mev$ below the \SigmaD\ threshold,
making it a prime candidate
for a hadron molecule composed 
of the two particles~\cite{Chen:2019bip,Chen:2019asm,He:2019ify,Guo:2019kdc,Xiao:2019mvs,Xiao:2019aya,Du:2019pij,Du:2021fmf}. 
Such a \SigmaD\ molecule with $J^P = 1/2^-$ 
 was predicted in various 
 models~\cite{Wu:2010jy,Wu:2010vk,Wang:2011rga,Yang:2011wz,Xiao:2013yca,Yamaguchi:2017zmn}.
However, the opening of a threshold and the
\SigmaD\ interaction can also generate a virtual state~\cite{Eden:1964zz},
where the interaction is attractive and generates a signal
in the cross section, but is not strong enough to bind a state.
A well known example is in neutron-neutron scattering, where the cross section is enhanced at threshold, even though no dineutron bound state exists~\cite{Hammer:2014rba}.
The fact that such a narrow ($\sim 10\mev$) peak 
appears on top of
what seems to be a smooth background permits a simplified analysis of the one-dimensional \jpsip invariant mass distribution.
This was done in~\cite{Fernandez-Ramirez:2019koa} following
a bottom-up approach,
favoring the virtual state interpretation.

\begin{figure}[t]
    \centering
    \includegraphics[width=0.5\textwidth]{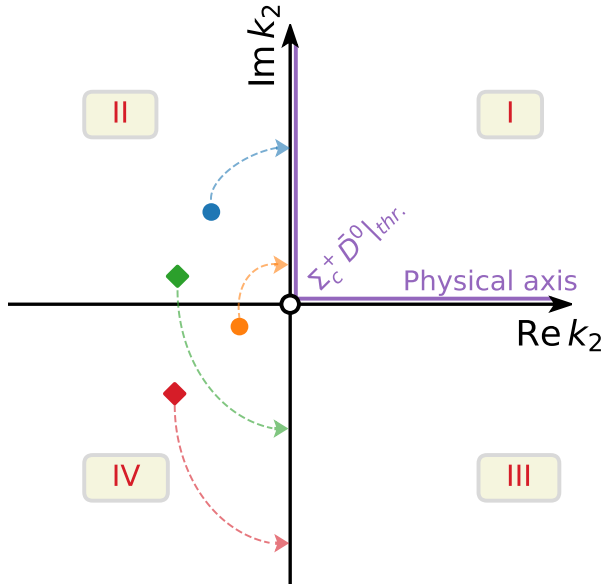} 
    \caption{Analytic structure of the $P_c(4312)$ amplitude near the 
    \SigmaD\ threshold. The adjacent Riemann sheets are continuously connected along the axes. The four possibilities for a resonant pole structure are depicted. When the \jpsip and \SigmaD\ channels decouple, the poles move to the imaginary $k_2$ axis along paths by the arrows. Poles moving to the positive (negative) axis correspond to bound (virtual) states.
}\label{fig:complexplanepc4312}
\end{figure}

\begin{figure*}[t]
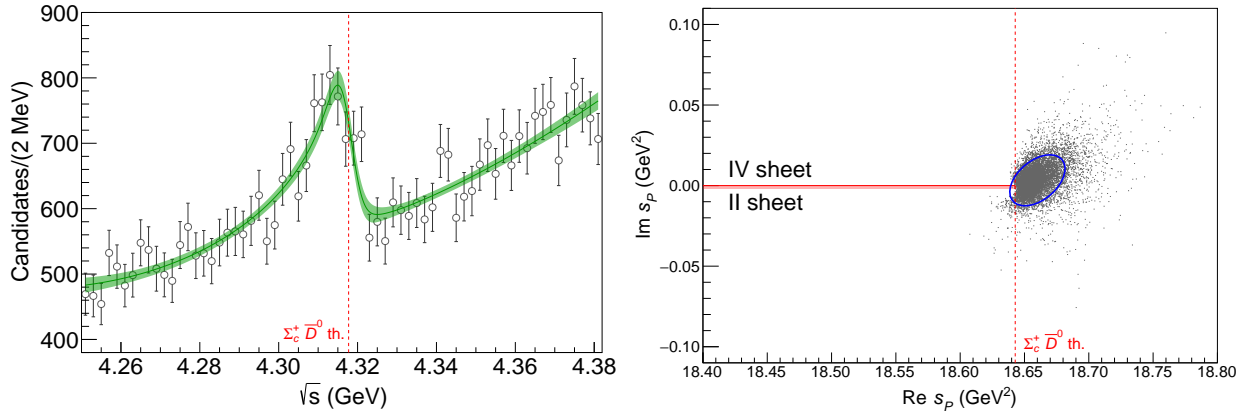

\centering
\includegraphics[width=.49\textwidth]{{../figures/pentaquarks/plot_nc}.pdf}
\includegraphics[width=.49\textwidth]{{../figures/pentaquarks/general_nc}.pdf}
\caption{
(left) Fit to the $\cos\theta_{P_c}$-weighted \jpsip mass distribution from \lhcb in the \Pc\ region~\cite{LHCb:2019kea}. The theory curve is convoluted with experimental resolution. 
The solid line and green 
band show the fit result and the $1\sigma$ 
confidence level. 
(right) Pole obtained from the $10^4$ bootstrap fits.
The physical region is highlighted with a pink band.
The poles lie on
the II and IV Riemann sheets
(which are continuously connected above the $\SigmaD$ threshold as shown in
\figurename{~\ref{fig:complexplanepc4312}}).
The blue ellipse accounts the 68\% of the cluster. 
Figures from~\cite{Fernandez-Ramirez:2019koa}.
} \label{fig:fitspc4312}
\end{figure*}

We consider a two-channel production process, $\Lambda^0_b \to K^- (\jpsip)$ and $\to K^- (\SigmaD)$. The presence of the bachelor antikaon does not create peaking structures, and the three-body effects described in Section~\ref{sec:three} can be neglected. Since we focus on events around the \Pc peak only, far away from the \jpsip threshold, the phase space is basically an imaginary constant, and we can claim that this absorbs all the inelastic channels lighter than \SigmaD. Similarly, the contributions from heavier channels can be absorbed by the real parameters of the scattering amplitude.
The events distribution is given by~\cite{Frazer:1964zz,Fernandez-Ramirez:2019koa} 
\begin{equation}
\frac{dN}{d\sqrt{s}}= \rho(s) \left[ \abs{F(s)}^2   
 + B(s) 
\right],\label{eq:pc4312amplitude}
\end{equation} 
where $\rho(s)$ is a phase space factor. 
The \Pc\ signal is assumed to have a well-defined spin and, hence, it appears in a single partial wave $F(s)$. The smooth background $B(s)$ is a linear polynomial that parametrizes all the other  partial waves that can be added incoherently.
The amplitude $F(s) = P_1(s)\, T_{11} (s)$ is a product of a 
smooth production function $P_1(s)$ 
that encapsulates
both the $\jpsip\,K^-$
production and  the cross channel $\Lambda^*$ resonances projected into the same partial wave as \Pc, 
and of the  $\jpsip\to\jpsip$ scattering amplitude  $T_{11}(s)$ that contains the details of the \Pc. In a $P$-vector formalism~\cite{Aitchison:1972ay}, another term $P_2(s) T_{12}(s)$ would also appear, but since it contains the same singularities as $P_1(s) T_{11}(s)$, it can be reabsorbed there.
Close to the
\SigmaD\ threshold, can be expanded as
\begin{equation} 
\quad \left(T^{-1}\right)_{ij} = M_{ij} - i k_i \,\delta_{ij}, \label{eq:caseAB}
\end{equation} 
with $i,j =1,2$. The $k_i$ momenta of the two channels are given by 
$k_1 = \sqrt{s - (m_{\psi} + m_{p})^2}$
and
$k_2 = \sqrt{s - (m_{\Sigma^+_c} + m_{\bar{D}^0})^2}$.
The $M_{ij}$ are given by
\begin{equation} 
M_{ij}(s) = m_{ij}  - c_{ij} s + \text{higher order terms}, \label{eq:mpc4312}
\end{equation} 
where $c_{ij}=0$ under the scattering length approximation.
The $m_{11}$, $m_{12}$ and $m_{22}$,
are fitted to the data.
The analytic structure of $T_{11} (s)$
is shown in \figurename{~\ref{fig:complexplanepc4312}}.
The amplitude has four poles in the complex $s$ plane. Two of them are a conjugated pair that appears either on the II or IV sheet near 
the \SigmaD\ threshold where the scattering length expansion
is reasonable. 
The other two poles lie far away from the region of interest and 
are irrelevant.
If $m_{12}\to 0$, the \SigmaD\ channel decouples from \jpsip. In this limit, the $P_c(4312)$ pole would become either a stable bound state on the I sheet, or a virtual state on the II sheet, depending on whether the pole would approach the positive or negative $\Im k_2$ axis, as represented in \figurename{~\ref{fig:complexplanepc4312}}. This is controlled by the sign of $m_{22}$, the inverse scattering length of the \SigmaD channel: If it is positive (negative) the resonance corresponds to a virtual 
(bound) state. 

The fit result 
is shown in \figurename{~\ref{fig:fitspc4312}} together with the
pole position from $10^4$ bootstrap fits.
For each bootstrap fit only one pole appears in this region. 
The resulting interpretation was a virtual state with
$M_P = 4319.7 \pm 1.6\mev$ and $\Gamma_P = -0.8 \pm 2.4\mev$, 
where the negative value of the width corresponds to a IV sheet pole. Consistent results are obtained with the three datasets published by LHCb.
Since a virtual pole does not lie in one of the proximal sheets, one would expect it to enhance the threshold cusp, so that the peak position should be exactly at threshold. However, convoluting with experimental resolution (roughly $3\mev$ in the region of interest~\cite{LHCb:2019kea}) can slightly move the position of the peak. This is why the fit shown in \figurename{~\ref{fig:fitspc4312}} peaks below threshold, even if it corresponds to a virtual state.

Still in the scattering length approximation,
the nature of the \Pc\ was also studied using 
a deep neural network, 
employed as described in Section~\ref{sec:ai}. The network was trained
against four classes of lineshapes: $b|2$, $b|4$, $v|2$, and $v|4$; where the letter
stands for the nature of the state, \ie  bound or virtual, 
and the number for the Riemann sheet where the pole is placed.
The result for the classification process is shown in~\figurename{~\ref{fig:shap.violin}}. The analysis heavily favors
the virtual state interpretation~\cite{Ng:2021ibr}.
The model can be extended to the effective range approximation, allowing $c_{ij}\neq 0$ with a similar fit quality. The \Pc\ pole
is pushed at $M_P = 4319.8 \pm 1.5\mev$ and $\Gamma_P = 9.2 \pm 2.9\mev$
on the II sheet, but jumps on the IV sheet as soon as $m_{12}$ is made smaller, also favoring a virtual state interpretation. This model by construction contains two more pairs of conjugate poles. One of the poles is systematically captured on the III Riemann sheet by the bump at about $4380\mev$, at the edge of the fitted window. Since the fit quality was not meaningfully improved with respect to the scattering length approximation and it was not possible to claim enough statistical significance, this was not claimed as a discovery. However, the unitary model of~\cite{Du:2019pij}, which enforces $\chi$PT and heavy quark spin symmetry, finds evidence of a narrow pole at the same mass. More systematic studies are thus required to assess whether this pole corresponds to a physical state or not. 

\subsubsection{\texorpdfstring{An example of triangle singularity: The $P_c(4337)$}{An example of triangle singularity: The Pc(4337)}}\label{sec:pc4337}

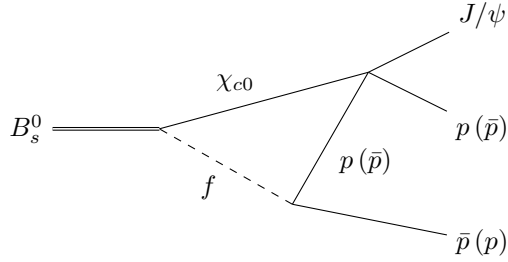
\begin{figure}[t]
    \centering
    \begin{tikzpicture}
        \begin{feynman}
        \vertex (x) {\(B_s^0\)};
        \vertex (a) at (1.75, 0);
        \vertex (b) at (4.5,  0.75);
        \vertex (c) at (3.5, -1.0);
        \vertex (p1) at (6, 1.5)   {\(  J/\psi \)};
        \vertex (p2) at (6, 0.)    {\( p \,(\bar{p}) \)};
        \vertex (p3) at (6, -1.5)  {\( \bar{p} \,(p) \)};
        \diagram* {
            (x) -- [double] (a),
            (a) -- [plain, edge label = \( \chi_{c0} \)] (b) -- [plain, edge label = \( p\, (\bar{p}) \)] (c) -- [scalar, edge label = \( f \)] (a),
            (b) -- [plain] (p1),
            (b) -- [plain] (p2),
            (c) -- [plain] (p3),
        };
        \end{feynman}
    \end{tikzpicture}
    \caption{Considered triangle diagram for the
    $B^0_s \to \jpsi\, p \bar{p}$ in the region where
    the $P_c(4337)$ signal appears.}
    \label{fig:trianglepc4337}
\end{figure}

Triangle singularities have been proposed as a possible explanation
of several resonances, and in particular of some of the pentaquark signals~\cite{Mikhasenko:2015vca,Bayar:2016ftu,Guo:2019twa,Nakamura:2021qvy}.
Here we show a simple example of a triangle calculation 
applied to the new $P_c(4337)$ pentaquark recently reported by \lhcb
in the $B^0_s \to J/\psi p \bar{p}$ decay close to the $\chi_{c0}p$ threshold~\cite{LHCb:2021chn}.
The signal was found analyzing the Dalitz plot
with a significance smaller than $5\sigma$,
so discovery was not claimed.
A hint of a peak is visible in the $J/\psi p (\bar{p})$ projections, as seen in \figurename{~\ref{fig:trianglepc4337}}, while no clear resonance is seen in $p\bar p$.
Hence, one is tempted to perform an analysis of these mass distributions similar to that of the $P_c(4312)$.
The low statistics makes an analysis
of that kind not worth the effort. Also, a proper analysis should be implemented at the Dalitz plot level rather than on the projections.
Nevertheless, we will use these
invariant mass
distributions to illustrate how 
a signal can be studied assuming it is generated by a scalar triangle singularity.
The possible triangle is shown in \figurename{~\ref{fig:trianglepc4337}}, where the exchanged $f$ is, in principle, unknown.
Here there are two options. One is to search for a suitable state to be exchanged as $f$ among the known ones. In this case the
$f_2(1950)$ seem like an adequate candidate. The second 
option is to let the data decide the mass and width of the $f$ particle.

\begin{figure}[b]
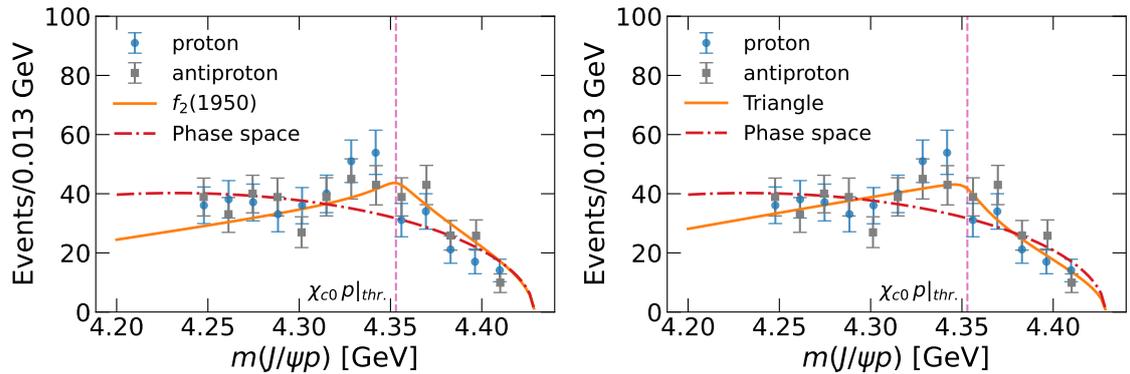

\centering
\includegraphics[width=.45\textwidth]{{../figures/pentaquarks/f2.pc4337}.pdf} 
\includegraphics[width=.45\textwidth]{{../figures/pentaquarks/f0.pc4337}.pdf}
\caption{Fits to the
$J/\psi p$ and $J/\psi\bar{p}$ projections 
from \lhcb~\cite{LHCb:2021chn}
in the energy
region of the $\chi_{c0}\,p$ threshold 
where the $P_c(4337)$ signal appears. (left) fits using phase space only and a triangle with an exchanged $f_2(1950)$. (right) fits using phase space only and a triangle with an exchanged $f$ whose mass and width are fitted to the data.} \label{fig:pc4337}
\end{figure}

The intensity distribution is given by
\begin{align}
\frac{dN}{d\sqrt{s}} = N_0
\, \rho(s)\, |M(s)|^2 \, ,
\end{align}
where 
$N_0$ is the normalization parameter,
the phase space is given by
$\rho(s)=\lambda^{1/2}(s,m_B^2,m_p^2)\, \lambda^{1/2}(s,m_p^2,m_{\psi}^2)\slash\sqrt{s}$,
and $M(s)$ is the scalar triangle amplitude
in \figurename{~\ref{fig:trianglepc4337}} given by~\cite{Mikhasenko:2015vca}
\begin{align}
M(s) = \int^1_0 \, \frac{dx}{y_+-y_-} \frac{1}{s}
& \left[ \log \frac{(1-x-y_+)}{-y_+}
- \log \frac{(1-x-y_-)}{-y_-} \right] \, ,
\end{align}
where
\begin{subequations}
\begin{align}
y_\mp  & = \frac{1}{2s}\left( -\beta \mp \sqrt{\beta^2-4\alpha s} \right) \, ,\\
\alpha & = x\, m_f^2 + (1-x)^2\, m_p^2 \, , \\
\beta  & = m_\chi^2 - (1-x)\, (s+m_p^2) - x\, m_B^2 \, .
\end{align}
\end{subequations}
Including particle spins and barrier factors will modify the numerator of the integrand, but will not affect the position of the triangle singularity. Since the precise form of the numerator is model-dependent anyway, we will not discuss it here in this illustrative example.
To account for the width of the $\chi_0$, we use a complex mass $ m_\chi \to m_\chi - i\Gamma_\chi /2$. Hence we perform three different fits:
phase space only, $f=f_2(1950)$, and
$f$ with mass and width as free parameters. These fits are
to three projection sets in the 
$\left(4.25, 4.40\right)\gev$ range: $J/\psi p$, $J/\psi\bar{p}$,
and both combined. The fits are summarized in \tablename{~\ref{tab:pc4337}}.
\figurename{~\ref{fig:pc4337}} shows the fits of the three models to
the combined $J/\psi p$ and $J/\psi\bar{p}$ projections.
If we take these results at face value, the phase space alone
cannot explain the apparent bump, and the inclusion
of the $f_2(1950)$ as a triangle improves the result,
although not much. However, the triangle with a
fitted mass and with ($m_f=1774$\mev and $\Gamma_f=217$\mev) does improve the agreement between theory and experiment.
Actually,
this exchanged $f$ lies very close
to the $f_0(1710)$ shown in Section~\ref{sec:scalars}.
However, this result has to be taken as a simple exercise,
given that the statistical significance of the signal is very low.

\begin{table}[t]
\caption{Summary of $\chi^2$/\dof for the 
fits to the $P_c(4337)$ signal
in the region of the $\chi_{c0}\,p$ threshold.
For the free $f$ fit we obtain $m_f=1774$\mev and $\Gamma_f=217$\mev.} \label{tab:pc4337}
\begin{center}
\begin{tabular}{c|ccc}
\hline
\hline
 Fitted projection & Phase space & $f_2(1950)$  &
 Free $f$\\
\hline
$J/\psi p$ & 1.52 & 1.73 & 0.76\\
$J/\psi \bar{p}$ & 1.81 & 1.31& 1.25 \\
$J/\psi p$ \& $J/\psi \bar{p}$ & 1.6 & 1.47 & 0.95\\
\hline
\hline
\end{tabular}
\end{center}
\end{table}

%============================================
% 3-body scattering and decays
%============================================
\section{Three-body scattering and decays}
\label{sec:three}
\subsection{Three-body decay and Khuri-Treiman equations} \label{sec:KT}

One of the main issues posed by the presence of hadrons in any reaction is their final-state interactions, which are formally expressed in terms of the unitarity of the amplitude. In two-body scattering, unitarity is usually imposed in the direct channel only, as one is not sensitive to the details of the crossed channels. This is certainly not the case for a three-body decay, where the three possible two-hadron channels are physical, and one ideally wants to impose unitarity in all channels at once.
The Khuri-Treiman (KT) formalism is a dispersive approach which indeed allows one to do so. KT equations were first written for $K \to 3\pi$ decays~\cite{Khuri:1960zz}. Soon after several papers appeared discussing different aspects of the formalism~\cite{Bronzan:1963mby,Aitchison:1965zz,Aitchison:1966lpz,Pasquier:1968zz,Pasquier:1969dt}. 
For the lightest mesons and lowest waves, KT equations can be justified in chiral perturbation theory at lowest orders via the so-called reconstruction theorem~\cite{Stern:1993rg,Knecht:1995tr,Ananthanarayan:2000cp,Zdrahal:2008bd,Bijnens:2007pr}. 
In Ref.~\cite{Albaladejo:2018gif} the formalism was applied to $\pi\pi$ scattering, and it was found to be equivalent to Roy equations~\cite{Roy:1971tc,Ananthanarayan:2000ht} when both formalisms are restricted to $S$- and $P$-waves. When higher waves are included in KT equations, still good agreement was found with other dispersive approaches~\cite{Garcia-Martin:2011iqs}. We also point out that the KT decomposition of $\pi\pi$ scattering in~\cite{Albaladejo:2018gif} is compatible with the amplitude decomposition obtained in~\cite{Yamagishi:1995kr} imposing crossing and chiral symmetries. 
In Ref.~\cite{Albaladejo:2019huw} we extended it to arbitrary quantum numbers of the decaying particle, by relating the isobar expansions in the three possible final states with the appropriate helicity crossing matrices (see also the discussion in Section~\ref{sec:smatrix}). 
The decays of vector mesons to three pions have been studied with this formalism~\cite{Niecknig:2012sj,Danilkin:2014cra,Ananthanarayan:2014pta,Caprini:2015wja,Dax:2018rvs,JPAC:2020umo}, and will be discussed in Section~\ref{sec:omegaKT}. 
However, the most important application is the study of the $\eta \to 3\pi$ decay~\cite{Neveu:1970tn,Anisovich:1993kn,Kambor:1995yc,Anisovich:1996tx,Walker:1998zz,Kampf:2011wr,Lanz:2013ku,Descotes-Genon:2014tla,Guo:2014vya,Guo:2015zqa,Guo:2016wsi,Colangelo:2016jmc,Colangelo:2018jxw, Albaladejo:2017hhj, Gasser:2018qtg}, as we will see in Section~\ref{sec:eta3pi}. 
 Other recent applications include Refs.~\cite{Isken:2017dkw,Niecknig:2015ija,Pilloni:2016obd,Niecknig:2017ylb}.

As in Section~\ref{sec:smatrix}, the amplitude can be decomposed in
\begin{equation}\label{eq:pwdecomp}
T(s,t,u) = \sum_{j=0}^{\infty} (2j+1) P_j(z_s) t_j(s)~,
\end{equation}
Equation~\eqref{eq:pwdecomp} is an \textit{infinite} sum of partial waves, each carrying both \textit{left-} and \textit{right-hand} cuts. The essence of the KT approach consists in performing instead an expansion of the amplitude into three (one for each two-meson subsystem) \textit{finite} sums of isobars or single-variable functions, carrying only a \textit{right-hand} cut. Explicitly,
\begin{equation}\label{eq:KTexp}
T(s,t,u) 
= \sum_{j=0}^{j_\text{max}} (2j+1)P_j(z_s) f_j(s) 
+ \sum_{j=0}^{j_\text{max}} (2j+1)P_j(z_t) f_j(t) 
+ \sum_{j=0}^{j_\text{max}} (2j+1)P_j(z_u) f_j(u)~.
\end{equation}

The ``original'' partial-wave expansion in Eq.~\eqref{eq:pwdecomp} is performed in a single channel, namely the $s$-channel. In other words, the partial-waves $t_j(s)$ depend solely on the $s$ Mandelstam variable. The dependence on $t$ and $u$ of $T(s,t,u)$ enters only through the Legendre polynomials, which are analytic functions of said variables. However, in practice one can model only a finite number of partial waves. As a consequence the analytic structure in the $t$- and $u$-variables is lost, because the sum of a finite number of analytic functions is again an analytic function, and the only way in which singularities (such as poles or cuts) in $t$ and $u$ could appear is if the infinite sum in $j$ diverges. Moreover, crossing symmetry is lost in truncation.
The KT expansion in Eq.~\eqref{eq:KTexp} solves these issues, by adding isobars $f_j$ in the three variables, so the analytic structure in $t$ and $u$ can be partially recovered. 

Furthermore, the application of dispersion relations to the single-variable functions $f_j(s)$ allows us to impose exact (elastic) unitarity to the two-meson subsystems, which can be essential in a three-body decay. Projecting this {\it model} decomposition for the amplitude into partial waves through~\eqref{eq:pwprojection}, we find
\begin{equation}\label{eq:fj_tj}
t_j(s) = f_j(s) + \hat{f}_j(s)~,
\end{equation}
where $\hat{f}_j(s)$ is called the {\it inhomogeneity},\footnote{The name inhomogeneity stems from the fact that if $\hat{f}_j=0$ then the equation for the discontinuity of the isobar $f_j(s)$, Eq.~\eqref{eq:KTdiscontinuity} is homogeneous.} given by
\begin{equation}\label{eq:fhatj}
\hat{f}_j(s) = \sum_{j'} \int_{-1}^{+1} \mathrm{d}z\,P_{j}(z)\, P_{j'}\left( z_t\left(s,t_z(s,z),u_z(s,z)\right) \right)\, f_{j'}(t_z(s,z))~.
\end{equation}
The structure of Eq.~\eqref{eq:fj_tj} is thus clear: The partial wave $t_j(s)$ receives a {\it direct} contribution from the isobar $f_j(s)$, plus an indirect contribution coming from the angular averages of all isobars of the crossed channels. To apply dispersion relations to the $f_j(s)$ functions, one writes its discontinuity,
\begin{equation}\label{eq:KTdiscontinuity}
\Delta f_j(s) = \Delta t_j(s) = \rho(s)\, \tau^\ast_j(s) \left( f_j(s) + \hat{f}_j(s) \right)~,
\end{equation}
where $\tau_j(s)$ is the two-meson elastic partial-wave amplitude, and $\rho(s)$ a phase space factor. The solution to the integral equation stemming from the dispersive representation of $f_j(s)$ reads
\begin{subequations}\label{eqs:solutionsKT}\begin{align}
f_j(s)       & = \Omega_j(s) \left( P_{j}^{(n)}(s) + I_j^{(n)}(s)\right)~,\\
I_j^{(n)}(s) & = \frac{1}{\pi} \int_{s_\text{th}}^{\infty} \mathrm{d}s' \left( \frac{s}{s'} \right)^{n+1} \frac{ \sin\delta_j(s')\, \hat{f}_j(s')}{\left\lvert \Omega_j(s') \right\rvert (s'-s)}~,\\
\Omega_j(s)  & = \exp \left[ \frac{s}{\pi} \int_{s_\text{th}}^{\infty} \mathrm{d}s'\, \frac{\delta_j(s')}{s'(s'-s)} \right]~.
\end{align}\end{subequations}
with $\Omega_j(s)$ and $\delta_j(s)$ the Omn\`es function and phase shift associated with the amplitude $\tau_j(s)$, respectively, and $P_j^{(n)}(s)$ a polynomial of $n$-th order. Above, $n$ represents the number of subtractions performed to the dispersion relation. 
Determining this number  is a rather delicate matter. From a purely mathematical point of view, subtracting the dispersion integral is simply a rearrangement of the equation. As the integral along the infinite circle should vanish, we just have to subtract sufficiently often to make this happen, as discussed in Section~\ref{sec:smatrix}. Oversubtracted dispersion relations still satisfy the same unitarity equation, with the advantage that the more subtractions are used, the more suppressed the dependence on the scattering shift at high energies becomes.  The price to pay here is that in doing so, one modifies the asymptotic behavior of the solution if no sum rule is imposed on the extra coefficients. However, this gives us more freedom than might be required by data.
From a physical point of view, the Froissart bound~\cite{Froissart:1961ux,Martin:1962rt} is often invoked to control the asymptotic behavior of the partial waves, hence the number of subtractions.

Equations~\eqref{eqs:solutionsKT} and~\eqref{eq:fhatj} represent a coupled system that can be solved, for example, iteratively. Things simplify considerably if one takes into account that the solutions of the dispersion relations are linear in the subtraction constants. This means that one can calculate a set of basis solutions that are independent of the numerical values of the latter. 

The equations obtained, Eqs.~\eqref{eqs:solutionsKT}, are valid for the scattering regime, and they have to be analytically continued for masses of the decaying particle $M>3m$, where $m$ is the mass of the light particle in the final state. The proper prescription was obtained in Refs.~\cite{Mandelstam:1960zz,Kacser:1963zz}. After analytically continuing $M_\eta^2$ to its physical value, extra singularities appear, which must be treated carefully. Also, depending on the solution method, $\hat f_j(s)$ could be needed for values of $s$ outside the physical domain, so singularities in the relation of $t$ with $s$ and  $\rm{cos}\theta$ also need to be taken care of. In this case, the integration path has to be chosen to avoid these extra singularities, see discussions in Refs.~\cite{Albaladejo:2017hhj,Gasser:2018qtg,Albaladejo:2019huw}.

In the example we have discussed the case for all-scalar particles. In more realistic applications of KT equations details on the amplitudes or isobar are different, but the essence of the method remains. 
Further discussions and references can be found in Refs.~\cite{Aitchison:2015jxa,Oller:2019opk,Fang:2021wes}.

\subsubsection{\texorpdfstring{$\omega \to 3\pi$ and $\psi \to 3\pi$ decays}{omega -> 3pi and psi to 3pi decays}}\label{sec:omegaKT}
The $\omega \to 3\pi$ decay has been previously studied with KT~\cite{Niecknig:2012sj,Danilkin:2014cra,Ananthanarayan:2014pta,Caprini:2015wja,Dax:2018rvs}, and other dispersive approaches~\cite{Terschlusen:2013iqa}. In particular, Refs.~\cite{Niecknig:2012sj,Danilkin:2014cra} predicted the Dalitz plot parameters (to be defined below) of the decay, either considering or neglecting KT effects, but using in both cases unsubtracted dispersion relations to solve KT equations. The \bes collaboration reported the measurement of the Dalitz plot parameters~\cite{Ablikim:2018yen}, and found a better agreement with the theoretical predictions of Refs.~\cite{Niecknig:2012sj,Danilkin:2014cra} when the rescattering effects were neglected, as can be seen in \tablename{~\ref{tab:ktomega}}. 
In view of this seeming disagreement, in Ref.~\cite{JPAC:2020umo} we have reviewed the application of the KT formalism to this decay.

\begin{table}\centering
\caption{Dalitz plot parameters $\alpha$, $\beta$, and $\gamma$, obtained by previous theoretical~\cite{Terschlusen:2013iqa,Niecknig:2012sj,Danilkin:2014cra} and experimental~\cite{Adlarson:2016wkw,Ablikim:2018yen} analyses. For the dispersive analyses~\cite{Niecknig:2012sj,Danilkin:2014cra}, we show the results obtained with and without KT equations (\ie, with $F_1(s)$ proportional to an Omn\`es function). Also shown are our results in Ref.~\cite{JPAC:2020umo} for the two solutions found in that work. The quoted uncertainty for Ref.~\cite{Niecknig:2012sj} corresponds to the range explored in that work. The uncertainties quoted for  Refs.~\cite{Adlarson:2016wkw,Ablikim:2018yen} correspond to the experimental statistical one. The first and second uncertainty quoted for Ref.~\cite{JPAC:2020umo} are statistical and systematic ones, respectively.\label{tab:ktomega}}
\begin{tabular}{c|c|ccc} \hline \hline
                        &  Reference & $\alpha$ ($\times 10^{-3}$)       & $\beta$ ($\times 10^{-3}$)    & $\gamma$ ($\times 10^{-3}$) \\ \hline
\multirow{6}{*}{\begin{tabular}{c} 2 par.\\ $(\alpha,\beta)$ \end{tabular}}
& Ref.~\cite{Danilkin:2014cra}, w KT      & $ 84$ & $28$ & -- \\
& Ref.~\cite{Danilkin:2014cra}, w/o KT    & $125$ & $30$ & -- \\  %\cline{2-5}
& Ref.~\cite{Niecknig:2012sj}, w KT       & $ 79(5)$ & $26(2)$ & -- \\
& Ref.~\cite{Niecknig:2012sj}, w/o KT     & $130(5)$ & $31(2)$ & -- \\  %\cline{2-5}
& WASA-at-COSY~\cite{Adlarson:2016wkw}    & $133(41)$ & $37(54)$ & -- \\
& BESIII~\cite{Ablikim:2018yen}           & $120.2(8.1)$ & $29.5(9.6)$ & -- \\ %\cline{2-5}          
& Ref.~\cite{JPAC:2020umo}, low $\phi_{\omega\pi^0}(0)$  & $121.2(7.7)(0.8)$ & $25.7(3.3)(3.3)$ & --  \\
& Ref.~\cite{JPAC:2020umo}, high $\phi_{\omega\pi^0}(0)$ & $120.1(7.7)(0.7)$ & $30.2(4.3)(2.5)$ & --  \\  \hline
\multirow{6}{*}{\begin{tabular}{c} 3 par.\\ $(\alpha,\beta,\gamma)$ \end{tabular}}
& Ref.~\cite{Danilkin:2014cra}, w KT      & $ 80$ & $27$ & $8$ \\
& Ref.~\cite{Danilkin:2014cra}, w/o KT    & $113$ & $27$ & $24$ \\   %\cline{2-5}
& Ref.~\cite{Niecknig:2012sj}, w KT       &  $77(4)$  & $26(2)$  &  $5(2)$ \\
& Ref.~\cite{Niecknig:2012sj}, w/o KT     & $116(4)$  & $28(2)$  & $16(2)$ \\   %\cline{2-5}
& BESIII~\cite{Ablikim:2018yen}           & $111(18)$ & $25(10)$ & $22(29)$ \\ %\cline{2-5}
& Ref.~\cite{JPAC:2020umo}, low $\phi_{\omega\pi^0}(0)$  & $112(15)(2)$ & $23(6)(2) $ & $29(6)(8)$ \\  
& Ref.~\cite{JPAC:2020umo}, high $\phi_{\omega\pi^0}(0)$ & $109(14)(2)$ & $26(6)(2) $ & $19(5)(4)$ \\ \hline
\hline
\end{tabular}%
\end{table}

For a vector decaying into three pions, the differential decay width is proportional to $\left \lvert T(s,t,u) \right \rvert^2 = \phi(s,t,u) \left \lvert F(s,t,u) \right \rvert^2$, where $\phi(s,t,u) = s\,t\,u - m_\pi^2 (m_V^2-m_\pi^2)^2$ is the Kibble function~\cite{Kibble:1960zz}, and $F(s,t,u)$ is an invariant amplitude. The Dalitz-plot parameters are obtained from a polynomial expansion of $\left\lvert F(s,t,u) \right\rvert^2$,
\begin{equation}\label{eq:dalitzplotexpansion}
\left\lvert F(s,t,u) \right\rvert^2 = \left\lvert N \right\rvert^{2}\left[1+2\alpha Z+2\beta Z^{3/2}\sin3\varphi+2\gamma Z^{2}+\mathcal{O}(Z^{5/2})\right]\,.
\end{equation}
\begin{equation}\label{eq:dalitzxy}
\sqrt{Z}\,\cos\varphi=\frac{t-u}{\sqrt{3}R_{\omega}}\,,\quad \sqrt{Z}\,\sin\varphi=\frac{s_{c}-s}{R_{\omega}}\,,
\end{equation}
where $s_{c}=\frac{1}{3}(m_{\omega}^{2}+3m_{\pi}^{2})$ and $R_{\omega}=\frac{2}{3}m_{\omega}(m_{\omega}-3m_{\pi})$. In Eq.~\eqref{eq:dalitzplotexpansion}, $\alpha,\beta$ and $\gamma$ are the real-valued Dalitz-plot parameters and $N$ is an overall normalization. The experimental determination of these parameters by the WASA-at-COSY~\cite{Adlarson:2016wkw} and \bes~\cite{Ablikim:2018yen} collaborations are shown in \tablename{~\ref{tab:ktomega}}, together with the theoretical predictions of Refs.~\cite{Danilkin:2014cra}~and~\cite{Isken:2017dkw}.

The partial wave expansion of the amplitude reads
\begin{equation}
F(s,t,u) = \sum_{j\,\text{odd}} \left( p(s) q(s) \right)^{j-1} P'_j(z_s) f_j(s)\, . 
\end{equation}
Similarly as explained above, the KT formalism is applied to the amplitude $F(s,t,u)$, and truncating the KT expansion to $j_\text{max}=1$ (only $\pi\pi$ $I=J=1$ wave), we have:
\begin{subequations}\begin{align}
F(s,t,u) & = F_1(s) + F_1(t) + F_1(u)~,\\
f_1(s)   & =  F_1(s) + \hat{F}_1(s)~,\\
\hat{F}_1(s) & = 3 \int_{-1}^{+1} \mathrm{d}z\, \frac{1-z^2}{2} F_1(t(s,z_s))~,\\
F_1(s) & = \Omega_{1}^{1}(s) \left( a + b\,s + \frac{s^2}{\pi} \int_{4m_\pi^2}^{\infty}\mathrm{d}s' \frac{\sin\delta_{1}^{1}(s')\, \hat{F}_1(s')}{(s')^2 \left\lvert \Omega_1^{1}(s') \right\rvert\, (s'-s)} \right)~. \label{eq:F1Omega_ab}
\end{align}\end{subequations}
The above expression for $F_1(s)$ is the solution of the integral equation corresponding to a once-subtracted dispersion relation.

Together with the $\omega \to 3\pi$ Dalitz plot parameters, we also analyze the $\omega \pi^0$ transition form factor, $f_{\omega\pi^{0}}(s)$, that controls the $\omega \to \pi^0 \gamma^\ast$ amplitude. A once-subtracted dispersion relation for this TFF gives
\begin{equation}
f_{\omega\pi^{0}}(s)=|f_{\omega\pi^{0}}(0)|\,e^{i\phi_{\omega\pi^{0}}(0)}+\frac{s}{12\pi^{2}}\int_{4m_{\pi}^{2}}^{\infty}\frac{ds^{\prime}}{(s^{\prime})^{3/2}}\frac{p^{3}(s^{\prime}) \; {F_{\pi}^{V}}^*(s^{\prime}) \; f_{1}(s^{\prime})}{(s^{\prime}-s)}~,
\label{eq:OmegaPiFF1sub}
\end{equation}
where $F_\pi^{V}(s)$ is the pion vector form factor, for which we take $F_\pi^{V}(s) = \Omega_{1}^{1}(s)$. This approximation does not reproduce all the details of the pion vector form factor~\cite{Hanhart:2013vba}. Nevertheless, it works for the $\pi\pi$ energy region explored here. The Omn\`es function $\Omega_{1}^{1}(s)$ is computed from the phase shift parametrization in~\cite{Garcia-Martin:2011iqs}, that is valid roughly up to $\Lambda\equiv\sqrt{s}=1.3\gev$. 
Beyond $1.3\gev$ we smoothly guide the phase to $\pi$ through~\cite{Gonzalez-Solis:2019iod,JPAC:2020umo}
\begin{eqnarray}
\delta_{\infty}(s)\equiv\lim_{s\to\infty}\delta_{1}^{1}(s)=\pi-\frac{a}{b+\left(s/\Lambda^{2}\right)^{3/2}}\,,
\label{Eq:PhaseExtrapolation}
\end{eqnarray}
where $a$ and $b$ are parameters taken such the phase $\delta(s)$ and its first derivative $\delta^{\prime}(s)$ are continuous at $s=\Lambda^{2}$
\begin{eqnarray}
a=\frac{3\left(\pi-\delta(\Lambda^{2})\right)^{2}}{2\Lambda^{2}\delta^{\prime}(\Lambda^{2})}\,,\quad b=-1+\frac{3\left(\pi-\delta(\Lambda^{2})\right)}{2\Lambda^{2}\delta^{\prime}(\Lambda^{2})}\,.
\end{eqnarray}

\begin{figure}[t]
    \centering
    \includegraphics[scale=1,keepaspectratio]{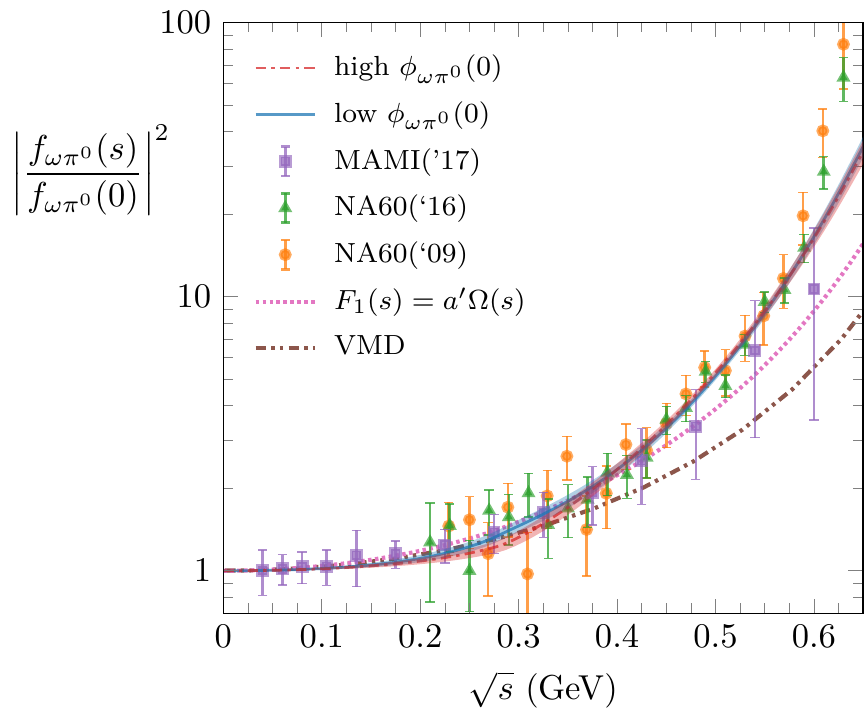}
    \caption{Transition form factor squared, $\left\lvert f_{\omega\pi^0}(s) \big / f_{\omega\pi^0}(0) \right\rvert^2$. The experimental data are from the A2 collaboration at MAMI~\cite{Adlarson:2016hpp} and the NA60 collaboration at SPS~\cite{Arnaldi:2009aa,Arnaldi:2016pzu}. The results of Ref.~\cite{JPAC:2020umo} are shown by the red and blue bands, corresponding to the high and low $\phi_{\omega\pi^0}(0)$ phase fits. The results obtained with the Vector Meson Dominance model (including an explicit $\rho$ pole as an amplitude) are shown with a double-dotted-dashed brown line, whereas  the results obtained when the KT effects are neglected are shown with a dotted pink line.
    Figure from~\cite{JPAC:2020umo}.
    \label{fig:omegapiTFF}}
\end{figure}

This ensures the expected asymptotic $1/s$ fall-off behavior of the pion vector form factor.\footnote{We have checked that continuation prescriptions different than Eq.~\eqref{Eq:PhaseExtrapolation}, \eg see Ref.~\cite{Moussallam:1999aq}, have small effects at low-energies, specially in the decay region of $\omega\to3\pi$ studied here.} Data from the A2 collaboration at MAMI~\cite{Adlarson:2016hpp} and by the NA60 collaboration at SPS~\cite{Arnaldi:2009aa,Arnaldi:2016pzu} for $\left\lvert f_{\omega\pi^0}(s) \right\rvert^2$ (normalized at $s=0$) for low $\omega\pi^0$ invariant mass are shown in \figurename{~\ref{fig:omegapiTFF}}. From the NA60 data, we will only consider in our fits the most up to date analysis~\cite{Arnaldi:2016pzu}. The free parameters in Eqs.~\eqref{eq:F1Omega_ab}~and~\eqref{eq:OmegaPiFF1sub} are the complex constant $b$, the absolute values $\left\lvert f_{\omega\pi^0}(0) \right\rvert$ and $\left\lvert a \right\rvert$, and the relative phase $\phi_{\omega\pi^0}(0) - \phi_a$, where $\phi_a$ is the phase of the constant $a$. We fit these parameters to the experimental data on $\left\lvert f_{\omega\pi^0}(s) \right\rvert^2$, the experimental Dalitz plot parameters, and the $\omega \to 3\pi$ and $\omega \to \pi^0 \gamma$ widths. Two different best fits are obtained, corresponding to a lower or higher value of the phase $\phi_{\omega\pi^0}(0)$~\cite{JPAC:2020umo}. In \tablename{~\ref{tab:ktomega}} we show the Dalitz plot parameters obtained as an output of the fit, in good agreement with the experimental ones. In \figurename{~\ref{fig:omegapiTFF}} we show our calculation of $\left\lvert f_{\omega\pi^0}(s) \right\rvert^2$ resulting from the fit, also in good agreement with data. Therefore we conclude that the KT equations are capable of describing the low-energy experimental information concerning $\omega \to 3\pi$ and $\omega \to \gamma^\ast \pi^0$, although a further subtraction has to be performed.

The KT description of $J/\psi\to3\pi$, proceeds in an identical fashion as the one discussed above.  
Despite the phase space here is much larger, so the production of excited states other than the $\rho(770)$ is in principle allowed, this decay is vastly dominated by the $\rho\,\pi$ intermediate state. The $\rho$ bands are clearly visible in the Dalitz plot from \bes~\cite{BESIII:2012vmy}, while almost no events appear in the center.
We perform fits to the \bes $m_{\pi\pi}$ invariant mass distribution after solving the KT equations for $J/\psi\to3\pi$~\cite{jpacinpreparation} and using the phase parametrization of~\cite{Pelaez:2019eqa}, which is valid up to $2\gev$.
The unsubtracted KT equation does not provide a good description of the data. 
In contrast, a satisfactory result can be achieved performing one subtraction in $F(s)$, with the subtraction constant fitted to data.
The result of the fit yields $b=0.20(1)e^{i2.68(1)}\gev^{-2}$.
While this fit provides an excellent description of the data up to $\sim1\gev$, contributions of higher waves seem to be required to describe the intermediate energy region around $\sim1.5\gev$.
The next allowed wave is the $F$-wave, which can be modeled by a resonance $\rho_{3}(1690)$.
The isobar decomposition of the amplitude including $F$-waves becomes~\cite{Niecknig:2012sj}
\begin{equation}
F(s,t,u)=F_1(s)+F_1(t)+F_1(u)+\kappa^{2}(s)P_{3}^{\prime}(z_{s})F_3(s)+\kappa^{2}(t)P_{3}^{\prime}(z_{t})F_3(t)+\kappa^{2}(u)P_{3}^{\prime}(z_{u})F_3(u)\,,
\end{equation}
where $P_{3}^{\prime}$ is the derivative of the Legendre polynomial.
The function $F_3(s)$ contains the $\rho_{3}(1690)$ contribution, which can be represented by a Breit-Wigner,
\begin{equation}
	F_3(s)=P(s)\frac{m_{\rho_{3}}^{2}}{m_{\rho_{3}}^{2}-s-im_{\rho_{3}}\Gamma_{\rho_{3}}(s)}\,,
	\label{Eq:FwaveAnsatz}
	\end{equation}
with the energy-dependent width given by
\begin{eqnarray}
	\Gamma_{\rho_{3}}(s)&=&\frac{\Gamma_{\rho_{3}}m_{\rho_{3}}}{\sqrt{s}}\left(\frac{p_{\pi}(s)}{p_{\pi}(m_{\rho_{3}}^{2})}\right)^{7}\left(F_{R}^{\ell}(s)\right)^{2}\,,\quad
	p_{\pi}(s)=\frac{\sqrt{s}}{2}\sigma_{\pi}(s)\,.
	\end{eqnarray}
The $F_{R}^{\ell=3}(s)$ denotes the Blatt-Weisskopf factor that limits the growth of the isobar~\cite{Blatt:1952ije},
\begin{eqnarray}
	F_{R}^{\ell=3}(s)&=&\sqrt{\frac{z_{0}(z_{0}-15)^{2}+9(2z_{0}-5)}{z(z-15)^{2}+9(2z-5)}}\,,\quad
	z=r_{R}^{2}p_{\pi}^{2}(s)\,,\quad z_{0}=r_{R}^{2}p_{\pi}^{2}(m_{\rho_{3}}^{2})\,,
	\end{eqnarray}
with the hadronic scale $r_{R}=2\gev^{-1}$.

The polynomial $P(s)$ in Eq.~\eqref{Eq:FwaveAnsatz} parametrizes some unknown energy dependence not directly related to the propagation of the $\rho_{3}(1690)$ resonance. 
Taking it linear, we add two additional (complex) parameters to the fit producing and improved description of the data.

The KT description of the partner reaction $\psi^{\prime}\to3\pi$ is formally identical to the one of $J/\psi\to3\pi$.
However, the experimental situation changes drastically for this decay: The $\rho\,\pi$ contribution is subleading and almost all events are found to be in the center of the Dalitz plot~\cite{BESIII:2012vmy}.
The significant differences between the $J/\psi$ and $\psi^{\prime}$ decays into three pions have attracted a lot of interest in the study of the transition mechanism $J/\psi$ and $\psi^{\prime}\to\rho\pi$. This is known as the ``$\rho\pi$ puzzle" and remains to be understood (see \eg~\cite{Chen:1998ma,Mo:2006cy,Wang:2012mf}, and references therein).
Other important aspects of $J/\psi \to 3\pi$ have been considered in~\cite{Guo:2010gx,Guo:2011aa}. A description orthogonal to KT, that 
takes into account the full tower of partial waves as given by the Veneziano amplitude, is found in~\cite{Szczepaniak:2014bsa}.

\subsubsection{\texorpdfstring{$\eta \to 3\pi$}{eta -> 3pi}}
\label{sec:eta3pi}
\begin{figure}
    \centering
    \includegraphics[width=.55\textwidth]{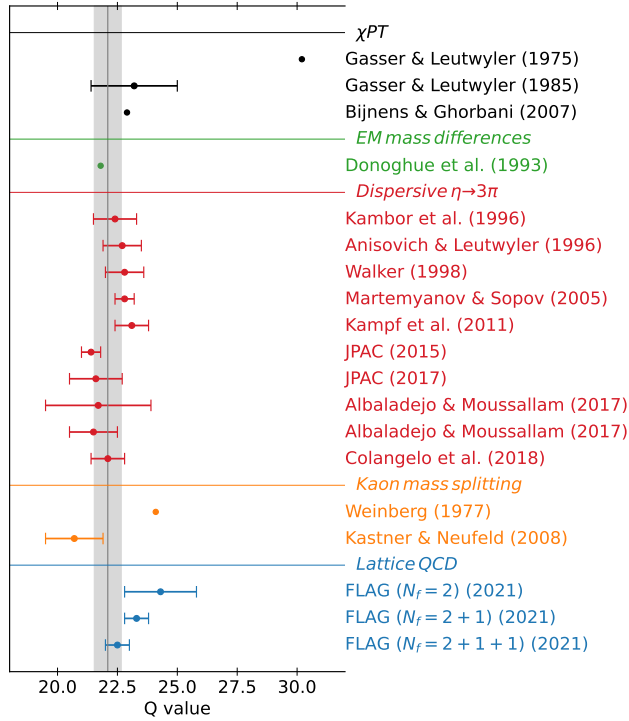} 
    \caption{Summary of $Q$ values found in the literature~\cite{Gasser:1974wd,Weinberg:1977hb,Gasser:1984pr,Donoghue:1993hj,Kambor:1995yc,Anisovich:1996tx,Walker:1998zz,Martemyanov:2005bt,Bijnens:2007pr,Kastner:2008ch,Kampf:2011wr,Guo:2015zqa,Guo:2016wsi,Albaladejo:2017hhj,Colangelo:2018jxw,Aoki:2021kgd}. The gray line is the
    weighted average of dispersive $\eta\to 3\pi$ extractions (red)
    and the gray band its $1\sigma$ uncertainty.
    \label{fig:results_etaQ}}
\end{figure}

The process $\eta \to 3\pi$ is very interesting because since this decay is forbidden by isospin symmetry--three pions cannot combine to a system with vanishing angular momentum, zero isospin, and even $C$-parity--it offers an unique experimental access to the light quark mass ratio
\begin{equation}
  Q^2 = \frac{m_s^2 - \hat{m}^2}{m_d^2 - m_u^2}\quad\text{and}\quad\hat{m}= \frac{m_u + m_d}{2}. 
\end{equation}
The origin of isospin breaking can be twofold: From electromagnetic corrections, and from the explicit breaking due to the mass difference $\Delta m= m_d-m_u$. In general, these two effects are of the same order (for example in the calculation of the mass difference between proton and neutron~\cite{Borsanyi:2014jba}). However, due to the Sutherland theorem~\cite{Sutherland:1966zz}, the electromagnetic contribution to $\eta \to 3\pi$ is suppressed~\cite{Baur:1995gc,Ditsche:2008cq}, so that the decay width gives immediate access to the $Q$ ratio. 
This can be extracted from data by comparing the experimental measured decay width with the reduced amplitude $M(s,t,u)$ integrated over the phase space:
\begin{equation}
\Gamma\big(\eta \to \pi^+ \pi^- \pi^0\big) = \frac{1}{Q^4} \frac{M_K^4 (M_K^2-M_\pi^2)^2}{6912 \pi^3 M_\eta^3 M_\pi^4 F_\pi^4} \int_{s_{\rm min}}^{s_{\rm max}} \diff s 
\int_{u_-(s)}^{u_+(s)} \diff u \left|M(s,t,u) \right|^2\,.
\label{eq:Qexp}
\end{equation}
The aim is to compute the amplitude $M(s,t,u)$ with the highest possible accuracy. This is not an easy task since there are strong rescattering effects among the final-state pions. 
These were initially calculated perturbatively in $\chi$PT.
The current algebra result is $\Gamma(\eta \to \pi^+ \pi^- \pi^0)_{\text{LO}} = 66\eV$~\cite{Osborn:1970nn}, and receives a substantial enhancement 
$\Gamma(\eta \to \pi^+ \pi^- \pi^0)_{\text{NLO}} = 160\pm 50\eV$ due to chiral one-loop corrections. The result is still far from the experimental value $\Gamma(\eta \to \pi^+ \pi^- \pi^0)= 300 \pm 12\eV$ suggesting a convergence problem. 
Moreover it has been shown that the two-loop calculation~\cite{Bijnens:2007pr} may lead to a precise numerical prediction 
only after the low-energy constants (LECs) appearing in the amplitude are determined reliably. 
In particular, the role played by the $\mathcal{O}(p^6)$ LECs is nonnegligible and they are largely unknown. 
A more accurate approach relies on dispersion relations to evaluate rescattering effects to all 
orders~\cite{Anisovich:1993kn,Anisovich:1996tx,Kambor:1995yc,Walker:1998zz}. 
This is not completely independent of $\chi$PT, because the dispersive representation requires the subtraction constants as input, and the latter can be matched to combinations of $\chi$PT LECs.

There has been a renewed interest in $\eta\to3\pi$ dispersive analysis due to new and more precise measurements of this decay. In particular recent measurements of the Dalitz plot of the charged ($\eta\to\pi^+\pi^-\pi^0$) channel by KLOE~\cite{KLOE:2008tdy,KLOE-2:2016zfv} 
and \bes~\cite{BESIII:2015fid} and of the neutral channel ($\eta\to\pi^0\pi^0\pi^0$) by A2~\cite{A2:2018pjo} have achieved an impressive level of precision. New measurements are planned by \bes and at \jlab by GlueX~\cite{Gan:2015nyc, JEF-PAC42} and CLAS~\cite{Amaryan:2013osa}, with completely different systematics and even better accuracy. 

In the application of KT equations to $\eta \to 3\pi$ decays, one truncates the expansion of the amplitude by neglecting $D$-and higher single-variable functions, thus writing
\begin{align} 
\label{eq:Mdecomp}
	M(s, t, u) = M_0^0(s) + (s-u) M_1^1(t) + (s-t) M_1^1(u) + M_2^0(t) + M_2^0(u) - \frac{2}{3} M_2^0(s) \,.
\end{align}
The functions $M_I^\ell(s)$ have isospin $I$ and angular momentum $\ell$. 
As said before, in the context of light mesons this decomposition is commonly referred to as a \textit{reconstruction theorem}~\cite{Stern:1993rg,Knecht:1995tr,Ananthanarayan:2000cp,Zdrahal:2008bd}. The latter relies on the observation that up to corrections of order $\mathcal{O}(p^{8})$ (or three loops) in the chiral expansion, partial waves of any meson--meson scattering process with angular momentum $\ell \geqslant 2$ contain no imaginary parts.  Since in Eq.~\eqref{eq:Mdecomp} the angular momentum of an isobar is unambiguously given by its isospin, we will omit $\ell$ in the following and refer to $M_I^\ell$ by $M_I$. The splitting of the full amplitude into these single-variable functions is not unique: There is some ambiguity in the distribution of the polynomial terms over the various $M_I$ due to $s+t+u$ being constant.

As discussed above, using analyticity and unitarity allows one to construct dispersion relations for the single-variable functions $M_I(s)$, arriving at
\begin{subequations}\label{eq:MIMIhat}\begin{align} 
\label{eq:MIdisp}
	M_I(s) & = \Omega_I(s) \left\{ P_I(s) + \frac{s^{n_I}}{\pi} \int _{4 m_\pi^2}^{\infty} \frac{\diff s^\prime}{s^{\prime n_I}}
		\frac{\sin \delta_I(s^\prime) \hat{M}_I(s^\prime)}{|\Omega_I(s^\prime)| (s^\prime - s -i \epsilon)}
		\right\},\\
	\hat M_I(s) & =\sum_{n,I'} \int_{-1}^{1} \diff \cos\theta~\cos^{n} \theta \, c_{n I I'} M_{I'}\big(t(s,\cos\theta)\big)~,
\end{align}\end{subequations}
which is completely analogous to Eqs.~\eqref{eqs:solutionsKT}. The explicit forms of the coefficients $c_{nII'}$ can be found \eg in Refs.~\cite{Anisovich:1996tx}. 
To study the convergence behavior of
the integrand we have to make assumptions as regards the
asymptotic behavior of the phase shifts. It is usually assumed that 
\begin{equation}
\delta_{0}(s) \to \pi \,,\quad \delta_{1}(s) \to \pi\,, \quad \text{and}\quad \delta_{2}(s) \to 0 \,, \quad \text{as} \quad s \to \infty \,.
\label{eq:asymdelta}
\end{equation}
An asymptotic behavior of $\delta(s) \to k\pi$ implies that the corresponding Omn\`es function behaves like $s^{-k}$ for high $s$. 
If the Froissart bound~\cite{Froissart:1961ux,Martin:1962rt} is assumed as discussed earlier, this implies $M_0(s), M_2(s) \to s$ and 
$M_1(s) \to {\rm const.}$, thus four subtractions are required. 
Since $s+t+u = M_\eta^2 + 3M_\pi^2$,
there exists a five-parameter polynomial transformation of the single-variable functions $M_I$
that leaves the amplitude $M(s,t,u)$ in Eq.~\eqref{eq:Mdecomp} invariant.
Therefore there is some freedom to assign the subtraction constants to the functions $M_I(s)$. 
In Ref.~\cite{Colangelo:2018jxw} a parametrization is used for Eq.~\eqref{eq:MIdisp}, such that above $1.7\gev$ the phase shifts $\delta_{0}(s)$ and $\delta_{1}(s)$ are set equal to $\pi$, whereas $\delta_{2}(s)$ is set to zero. In other words, the integral is cut at $s' = \left(1.7\gev\right)^2$, and therefore convergence is no longer an issue.\footnote{Cutting the integral introduces an unphysical branch point. However, the uncertainties associated with the input phases in the region above $1\gev$ were examined, and it was found that this cut barely affects the results.}
We can relax the Froissart bound and oversubtract the dispersive integrals~\eqref{eq:MIdisp} with the aim of being insensitive, in the physical region, to the high-energy inelastic behavior of the phase, which is unknown. The price to pay for this is that one has more subtraction constants to be determined. 
In some recent dispersive analyses~\cite{Kampf:2011wr,Colangelo:2016jmc,Colangelo:2018jxw}, 6 subtraction constants have been considered.
In Ref.~\cite{Albaladejo:2017hhj} only 4 subtraction constants are considered in the single channel approximation. The subtraction constants are unknown and have to be determined using a combination of experimental information and theory input.  Since the overall normalization multiplies $1/Q^2$, the quantity that should be extracted from the analysis, it cannot be obtained from data alone and one has to match to $\chi$PT.  On the other hand, this matching has to be performed in such a way that the problematic convergence of the chiral expansion is not transferred directly to the dispersive representation. This can be achieved by matching the amplitude around the Adler zeros. As discussed in Section~\ref{sec:KT}, several dispersive analyses have been performed over the last few years. All these analyses rely on the same theoretical ingredients described above with some subtle  differences. For instance, the analysis of JPAC~\cite{Guo:2015zqa,Guo:2016wsi} uses a different technique to solve the dispersion relation, called the Pasquier inversion~\cite{Pasquier:1968zz,Aitchison:1978pw,Guo:2014vya}. Moreover the left-hand cut is approximated using a Taylor series in the physical region. This allows to reduce the number of subtraction constants from six to three. The result is then matched to NLO $\chi$PT near the Adler zero to extract a value for $Q$. The analysis of Ref.~\cite{Colangelo:2016jmc,Colangelo:2018jxw} is a modern update of the approach of Anisovich and Leutwyler~\cite{Anisovich:1996tx}. There a matching to NLO and NNLO $\chi$PT has been performed. Moreover electromagnetic and isospin breaking corrections have been taken into account. Fits to experimental data by KLOE~\cite{KLOE-2:2016zfv}, but also to the recent neutral-channel Dalitz plot by A2~\cite{A2:2018pjo} have been explored. Finally the analysis of Ref.~\cite{Albaladejo:2017hhj} studies the impact of inelasticities on the dispersive integrals. To this end, the inelastic channels $\eta \pi$ and $K \bar K$ have been included.
\figurename{~\ref{fig:results_etaQ}} summarizes the results on the extraction of $Q$ from the different analyses. In principle, it is also possible to calculate the $Q$ ratio from Dashen theorem. Since $Q^{-2}$ depends linearly on $\Delta m$, it is proportional to the difference of the squared masses of the kaons induced by $\Delta m$ only, $ \left. m_{K^+}^2 - m_{K^0}^2 \right\rvert_{\Delta m}$. For pions, it holds $\left. m_{\pi^+}^2 - m_{\pi^0}^2 \right\rvert_{\Delta m} = \mathcal{O}\!\left(\Delta m^2 \right)$ instead~\cite{Gasser:1983yg,Gasser:1984gg}. According to Dashen theorem~\cite{Dashen:1969eg}, the QED-induced squared-mass difference is the same for kaons and for pions at lowest order in $\chi$PT. Therefore, one can write $Q^{-2}$ in terms of the experimental squared-mass differences of kaons and pions, obtaining $Q_\text{DT} \simeq 24.3$. While this estimate deviates sizeably from the values extracted from $\eta\to 3\pi$, the dispersive analyses agree well, allowing $\eta \to 3\pi$ to be the golden plate channel to extract the light quark mass ratios. For further discussions, see also Refs.~\cite{Gasser:1984pr,Urech:1994hd,Donoghue:1993hj,Martemyanov:2005bt,Kampf:2011wr}.

\subsection{\texorpdfstring{\threetothree scattering}{3->3 scattering}}
In recent years, the problem of describing multihadron scattering processes has generated significant interest. It is well-established experimentally that many resonances couple strongly to three- or more particle channels~\cite{Ketzer:2019wmd}. Some of the most intriguing particles which do not fit the na\"ive quark model predictions, like the Roper resonance $N^{*}(1440)$, the $a_1(1420)$ seen by COMPASS~\cite{COMPASS:2015kdx,COMPASS:2020yhb}, and the exotic $\pi_1(1600)$~\cite{COMPASS:2009xrl, COMPASS:2021ogp}, $X(3872)$, and other \XYZ states, have significant three-particle decay modes~\cite{pdg}. Three-body couplings might lead to non-standard line shapes and complicated structure of the amplitudes~\cite{Guo:2019twa}, allowing for ambiguities in interpretations of the hadron of interest~\cite{Aitchison:1979fj,Szczepaniak:2015hya,Szczepaniak:2015eza,Olsen:2017bmm,Nakamura:2019btl}.
To parametrize three-body processes, and in consequence, build and compare phenomenological models properly describing properties of the QCD states, one needs to establish a general theoretical framework of the three-body processes relying on the $S$-matrix principles.

In addition to phenomenological studies, usually based on particular models and approximations, it is desired to determine the properties of the strongly interacting resonances directly from the underlying theory, using Lattice QCD.
The essential challenge to study the resonance physics on the lattice
is the fact that resonances are not eigenstates of the QCD Hamiltonian.
Moreover, one cannot define scattering processes in a finite volume in the usual sense, since there are no asymptotic states, and the continuum spectrum becomes a discrete set of bound states in the box. Fortunately, it was shown by L\"uscher~\cite{Luscher:1986pf,Luscher:1990ux} that the scattering information is hidden in the volume dependence of the lattice spectrum. In the case of two hadrons being scattered off of each other one can obtain the two-body scattering phase shifts from the so-called two-body \emph{quantization condition}~\cite{Kim:2005gf, Rummukainen:1995vs, Liu:2005kr, Li:2012bi, Leskovec:2012gb, Lage:2009zv, He:2005ey, Feng:2004ua, Christ:2005gi, Bedaque:2004kc, Bernard:2010fp, Hansen:2012tf, Guo:2012hv}. This has been applied to many systems of physical relevance~\cite{Briceno:2017max}. The three-body generalization of the L\"usher's idea has been developed, leading to different three-particle quantization conditions~\cite{Polejaeva:2012ut,Hansen:2015zga, Mai:2017bge, Hammer:2017uqm, Hammer:2017kms, Blanton:2020gha}. They allow one to obtain objects called three-body $K_\text{df}$ matrices, from the three-particle finite-volume spectrum. They are analogous to the two-body $K$ matrix, however, they do not have a simple interpretation of a phase shift. Because of the multi-variable nature of the three-body process, formalisms required to describe the scattering of three hadrons become involved and a three-body $K_\text{df}$ matrix is related to the genuine three-body infinite-volume amplitude through the set of complicated integral equations. Once they are solved, one obtains the on-shell three-body scattering amplitude computed directly from QCD. In the last step, it has to be continued to the complex energies, to identify complex poles corresponding to three-body resonances.

Two main relativistic on-shell $\threetothree$ scattering formalisms have been developed and applied to a range of physical problems: (a) The relativistic EFT (RFET) established by Hansen, Sharpe, and  Brice\~no in Refs.~\cite{Hansen:2014eka,Hansen:2015zga,Briceno:2017tce,Briceno:2018aml}, and (b) The $S$-matrix unitarity, also referred to as the $B$-matrix approach, built by Mai \textit{et al.}~\cite{Mai:2017vot,Mai:2017bge,Doring:2018xxx} and the JPAC group~\cite{Jackura:2018xnx,Mikhasenko:2019vhk,Dawid:2020uhn}. All of these works have been shown to be equivalent both in their infinite-volume~\cite{Jackura:2019bmu} and finite-volume~\cite{Blanton:2020jnm} versions. In the following, we summarize both approaches and review the relevant results. Supplementary reviews can be found in Refs.~\cite{Hansen:2019nir, Mai:2021lwb}.

\subsubsection{Relativistic three-body formalisms}

Description of the three-body unitarity for the \threetothree scattering amplitude is considerably more involved than in the two-body case. Parametrizations satisfying unitarity in the three-body systems have been studied by various authors in the '60s and '70s~\cite{Grisaru:1966xyz, Mandelstam:1962xyz, Harrington:1962pr, Fleming:1964zz, Frazer:1962pr, Holman:1965mxx, Cook:1962zz, Ball:1962bwa, Hwa:1964xyz, Bjorken:1960zz, Blankenbecler:1965gx, Amado:1974za, Aaron:1973ca, Amado:1975zz, Aaron:1968aoz, Greben:1976xy}. The description of three-body states is usually based on the isobar representation, in which one writes the amplitude as a sum of  partial-wave expansions, one for each pair of particles in the external three-body state~\cite{Mandelstam:1962xyz, Fleming:1964zz, Holman:1965mxx, Herndon:1973yn}. Truncation of the partial wave decomposition leads to the so-called isobar approximation, which can provide a good description of three-particle final states in the kinematic region, where intermediate two-body resonances dominate over the scattering process. Moreover, the isobar approximation is capable of reproducing the threshold singularities in two-body subchannels by including only a finite number of partial waves. In the isobar representation, the \threetothree amplitude is decomposed into $\Ac_{\p'\p}$ isobar–spectator amplitudes, where the indices label one of the particles in the initial and final state. This particle is called the \emph{spectator}, whereas the other two form an \emph{isobar} (also called a \emph{pair}), corresponding to the given spectator. The isobar-spectator amplitudes can be pictured as describing a \twototwo scattering process of a quasi-particle and a stable spectator. 

%%%%%%%%%%%%%%%%%%%%%%%%%%%%%%%%%%%%
%	Figure :: 3-to-3 Unitarity
%%%%%%%%%%%%%%%%%%%%%%%%%%%%%%%%%%%%
\begin{figure}[t!]
    \centering
    \includegraphics[ width=0.95\textwidth, trim= 4 4 4 4,clip]{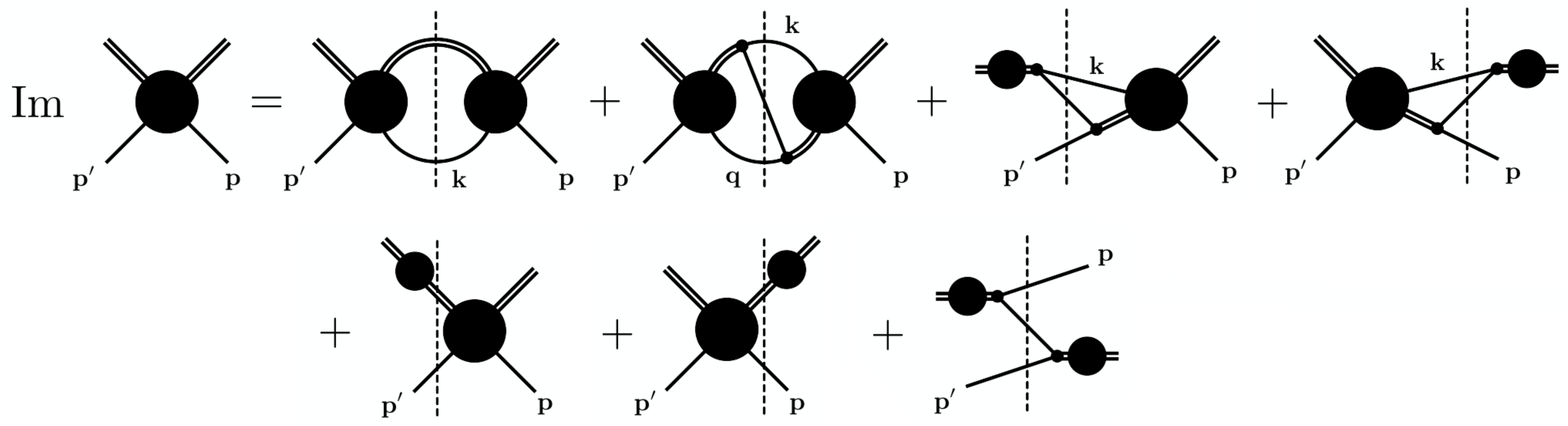}
    \caption{Diagrammatic representation for the \threetothree unitarity relation, Eq.~\eqref{eq:3to3unitarity}, for the connected isobar-spectator amplitude $\Ac_{\p'\p}$. A single external line represents a spectator, while a double external line---an isobar. Closed loops yield three-dimensional integrations over the labeled spectator momentum, and the dashed vertical lines represent placing all three intermediate state particles on their mass-shell. A solid circle with both external isobars and spectators is the amplitude $\Ac$, and a solid circle only with external isobars is the two-body amplitude $\Fc$. Momentum flow is from right to left, as before, and each amplitude on the left of the dashed line is hermitian conjugated. Figure adapted from~\cite{Jackura:2019bmu}.}
    \label{fig:diag_3to3_unitarity}
\end{figure}

To highlight the features of this parametrization we consider a simplified elastic scattering process in the center of momentum frame (CMF), in which the incoming and outgoing states consist of three spinless, indistinguishable particles of mass $m$ and total invariant mass squared $s$. Let $p=(\omega_{\p}, \p)$ be the four-momentum of the initial spectator in one isobar-spectator configuration, where $\omega_{\p} = \sqrt{\p^2 + m^2}$, and $\sigma_{\p}$ is the invariant mass squared of the corresponding initial isobar. We denote analogous variables for outgoing particles with a prime, e.g., the outgoing spectator's four-momentum is $p'=(\omega_{\p'},\p')$. Unitarity constrains the \threetothree amplitudes on the real energy axis, which restricts the imaginary parts of the partial-wave-projected isobar–spectator amplitudes. 
Using the notation of Refs.~\cite{Jackura:2019bmu, Dawid:2020uhn}, the elastic \threetothree scattering amplitude, $\Mc$, is defined as the three-body element of the $T$ matrix. It is convenient to work with the \emph{unsymmetrized} isobar-spectator amplitude $[\Mc_{\p'\p}]_{\ell' m'_\ell ; \, \ell m_\ell }$, written in the so called $(\p \ell m_{\ell})$ basis in which it can be treated as an infinite-dimensional matrix in the isobars angular momentum space. In the simplified case considered here, the three-body amplitude $\Mc$ becomes a symmetrized sum of 9 identical isobar-spectator amplitudes $\Mc_{\p'\p}$, corresponding to 9 identical divisions of the final and initial state particles into spectator-isobar configurations. The amplitude depends on eight kinematical variables: Initial and final isobar invariant masses squared, total invariant mass of the three-body system, the total angular momentum, and angular momenta of isobars $(\ell, m_\ell)$ and $(\ell', m'_{\ell})$. The multi-variable nature of the three-body scattering is the main factor making its description significantly more complicated than in the \twototwo case. The unsymmetrized partial-wave projected three-body amplitude $\Mc_{\p'\p}$ is further separated into a connected and disconnected part, $\Mc_{\p'\p} = \Ac_{\p'\p} + \Fc_{\p} \, \delta_{\p'\p} $ where $\delta_{\p'\p}$ is the properly normalized momentum-conserving $\delta$-function. The disconnected part is given by the two-body scattering amplitude in the isobar sub-channel. It depends on the isobar angular momentum and its invariant mass squared $\sigma_{\p}$. Above the isobar threshold, the disconnected amplitude satisfies the usual \twototwo unitarity relation, $\text{Im} \, \Fc_{\p} = \Fc_{\p}^{\dag} \, \bar \rho_{\p} \, \Fc_{\p}$, where $\bar\rho_{\p}$ is the two-body phase space multiplied by the threshold Heaviside function indicated with the bar. The unsymmetrized connected \threetothree amplitude satisfies the three-body unitarity,
    %%%%%
    \begin{align}
    \label{eq:3to3unitarity}
    \text{Im} \, \Ac_{\p'\p} & = 
     \int_{\k} \Ac_{\p'\k}^{\dag} \, \bar\rho_{\k} \, \Ac_{\k\p} 
    +  \int_{\q}\int_{\k} \Ac_{\p'\q}^{\dag} \,  \Cc_{\q\k} \,  \Ac_{\k\p} 
    + \int_{\k} \Fc_{\p'}^{\dag} \, \Cc_{\p'\k} \, \Ac_{\k \p}    %
     + \int_{\k} \Ac_{\p'\k}^{\dag} \,  \Cc_{\k\p} \, \Fc_{\p} 
     \nonumber \\
    \
    & + \Fc_{\p'}^{\dag} \, \bar\rho_{\p'} \, \Ac_{\p'\p} 
    +  \Ac_{\p'\p}^{\dag} \, \bar\rho_{\p} \, \Fc_{\p}
    + \Fc_{\p'}^{\dag} \, \Cc_{\p'\p} \, \Fc_{\p} \, .
    \end{align}
    %%%%%
where $\Cc_{\p'\p}$ is the \emph{recoupling} coefficient between a pair in one state to a different pair in the same state, which is defined as the imaginary part of the amputated one particle exchange (OPE) amplitude $\Gc_{\p'\p}$, see \figurename{~\ref{fig:diag_3to3_unitarity}}. The recoupling coefficients are a distinct feature of the three-body scattering unitarity relation. In the unitarity relation, the integrations are performed over momenta $\k, \q$ of spectators associated with intermediate three-particle states. Their energies are constrained by the on-shell condition as indicated by the dashed lines in \figurename{~\ref{fig:diag_3to3_unitarity}}. 

%%%%%%%%%%%%%%%%%%%%%%%%%%%%%%%%%%%%%%%%
%%%%%%%%%%%%%%%%%%%%%%%%%%%%%%%%%%%%%%%%

The most general parametrization satisfying these constraints is provided by the so-called $B$-matrix equation, which is a linear integral equation, analogous to the Bethe-Salpeter equation. It was introduced first in Ref.~\cite{Aaron:1968aoz} and later revisited in Refs.~\cite{Mai:2017vot, Mai:2017wdv, Jackura:2018xnx}, which corrected certain deficiencies of the original formulation related to the unitarity of the formalism above the breakup threshold. In the following, we give a concise overview of the $B$-matrix formalism, as described in Ref.~\cite{Jackura:2019bmu}. The $B$-matrix parametrization for the connected part $\Ac_{\p'\p}$ of the amplitude $\Mc_{\p'\p}$ is given by the matrix-integral linear equation,

%%%%%%%%%%%%%%%%%%%%%%%%%%%%%%%%%%%%
%	Figure :: B-matrix 
%%%%%%%%%%%%%%%%%%%%%%%%%%%%%%%%%%%%
\begin{figure}[t]
    \centering
    \includegraphics[ width=0.95\textwidth]{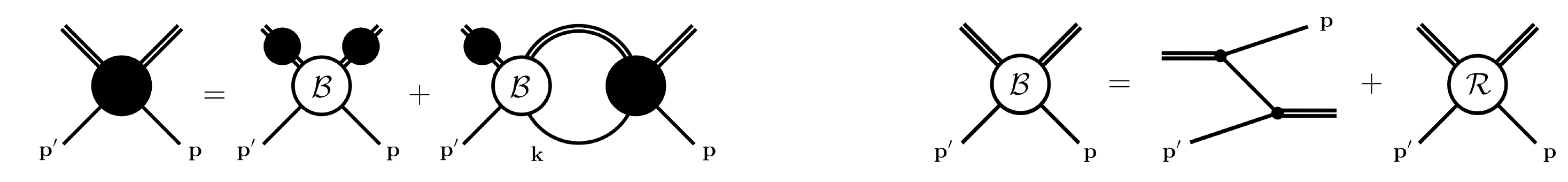}
    \put(-460,25){\colorbox{white}{(a)}}
    \put(-205,25){\colorbox{white}{(b)}}
    \caption{Diagrammatic representation of (a) the $B$-matrix representation for the on-shell amplitude, Eq.~\eqref{eq:b-matrix-param}, and (b) the $B$-matrix which is composed of the OPE $\Gc_{\p'\p}$, Eq.~\eqref{eq:B-matrix-decomp}, and the $R$-matrix. Figure adapted from~\cite{Jackura:2019bmu}}
    \label{fig:diag_3to3_B-matrix_amplitude}
\end{figure}

    %%%%%
    \begin{align}
    \label{eq:b-matrix-param}
    \Ac_{\p' \p} = \Fc_{\p'} \, \Bc_{\p' \p} \, \Fc_{\p} + \int_{\bm{k}} \Fc_{\p'} \, \Bc_{\p' \bm{k}} \, \Ac_{\bm{k} \p} \, ,
    \end{align}
    %%%%%
as demonstrated in \figurename{~\ref{fig:diag_3to3_B-matrix_amplitude}}. The $B$-matrix kernel is written as a sum of two terms,
    %%%%%
    \begin{align}
    \label{eq:B-matrix-decomp}
    \Bc_{\p'\p} = \Gc_{\p'\p} + \Rc_{\p'\p} \, ,
    \end{align}
    %%%%%
    
where the matrix $\Gc_{\p'\p}$ represents the long-range interaction due to one-particle exchange between the isobar and spectator required by unitarity. The amplitude $\Rc_{\p' \p}$ is a real matrix that embodies all short-range interactions. It is not constrained by unitarity, and it can be incorporated within a specific model allowing for the freedom to describe QCD resonances. Alternatively, it can be fixed from the lattice data as described below. 
Similarly to the unitarity relation, the intermediate particles are on the mass shell; therefore, the integration is performed over momentum $\k$ of a spectator associated with intermediate three-body state, compared to integration over four-momentum in the Bethe-Salpeter equation. Note that this is not a unique choice, as one may shift the remaining off-shell effects between kinematic functions and the three-body $\Rc$ function. As long as a given representation satisfies unitarity, it remains a valid approach \cite{Jackura:2019bmu}; the $B$-matrix equation being one such parametrization of the on-shell three-body amplitude.  The formalism can accommodate distinguishable spinless particles and was generalized to include two-to-three transitions~\cite{Dawid:2020uhn}.

The products of amplitudes present in Eq.~\eqref{eq:b-matrix-param} formally represent multiplications of infinite matrices in the angular momentum space. For practical use, they are truncated, leading to the finite matrix equation, in which one retains only contributions from dominating two-body sub-channels. Recall from the previous discussion, that the integration is restricted to the physical energy domain~\cite{Jackura:2018xnx, Dawid:2020uhn}. In principle, this requires one to include only the experimentally accessible sub-channel \twototwo amplitudes without the need for additional assumptions. Moreover, it might be beneficial for a description of states which lie close to their decay thresholds and often are interpreted as molecular systems bound by the nearly physical meson exchanges. However, restricting the integration to be over the physical intermediate state energies in the $B$-matrix construction can result in non-physical analytic properties of the amplitude in the unphysical region~\cite{Jackura:2018xnx,Dawid:2020uhn}. These arise from the chosen models for the amplitudes below threshold, e.g. smooth form-factors which regulate the high-momentum behavior~\cite{Aaron:1968aoz,Mai:2017vot, Mai:2017wdv, Sadasivan:2020syi}. While these effects do not alter the analytic behavior of the amplitude in the physical region, they may hinder the study of singularities associated with resonances since these occur on unphysical Riemann sheets in which one needs to discriminate between physical pole singularities and model-dependent effects. A potential cure for such effects is to write an appropriate dispersive representation for the amplitude~\cite{Dawid:2020uhn}.

In the REFT formulation, as developed in Refs.~\cite{Hansen:2014eka, Hansen:2015zga, Hansen:2015zta, Briceno:2017tce, Blanton:2019igq, Hansen:2020zhy, Blanton:2020gha, Blanton:2020gmf, Blanton:2021mih} the connected part of the unsymmetrized three-body amplitude is given as a solution of the equation analogous to Eq.~\eqref{eq:b-matrix-param}. It consists of the sum $\Dc + \Mc_{\text{df},3}$, where $\Dc$ is the \emph{ladder} amplitude driven by one-particle exchanges between \twototwo subprocesses, while $\Mc_{\text{df},3}$ is the separate short-range amplitude. The ladder amplitude is obtained by setting $\Rc_{\p'\p}=0$ in the $B$-matrix equation,
    %%%%%
    \begin{align}
    \label{eq:ladder-1}
    \Dc_{\p' \p} = \Fc_{\p'} \, \Gc_{\p' \p} \, \Fc_{\p} + \int_{\bm{k}} \Fc_{\p'} \, \Gc_{\p' \bm{k}} \, \Dc_{\bm{k} \p} \, .
    \end{align}
    %%%%%
The short-range part is given by an additional double-integral equation, driven by the three-body $K$ matrix called $\Kc_{\text{df},3}$, representing short-distance three-particle interactions,
    %%%%%
    \begin{align}
    \Mc_{\text{df},3; \p'\p} = \int_{\bm{k}} \int_{\bm{k}'} \mathcal{L}_{\p' \bm{k}'} \, \Tc_{\bm{k}' \bm{k}} \, \mathcal{L}^{\top}_{\bm{k} \p} \, ,
    \end{align}
    %%%%%
where $\Lc$ is the \emph{endcap} operator describing incoming and outgoing particles rescatterings, while $\Tc$ is given by the equation,
    %%%%%
    \begin{align}
    \Tc_{\p'\p} = \Kc_{\text{df},3; \p'\p} + \int_{\bm{k}} \int_{\bm{k}'} \Kc_{\text{df},3; \p'\k} \, i \rho_{\k} \mathcal{L}_{\bm{k} \bm{k}'} \Tc_{\bm{k}'\p} \, ,
    \end{align}
    %%%%%
The $\Kc_{\text{df},3}$ is the analog of the $\Rc$ matrix of the $B$-matrix formalism. In recent lattice studies both the $\Rc$ and $\Kc_{\text{df},3}$ have been determined for the realistic three-body systems (see the discussion below). For further details about the REFT formalism, we refer the reader to Ref.~\cite{Hansen:2019nir}.

The $B$-matrix parametrization can be analytically continued to the complex energy plane. The analytic properties of the formalism are discussed in Refs.~\cite{Jackura:2018xnx}. One of the most unique characteristics of the three-body equations appears from a kinematic singularity due to the exchange of a real particle. Analytic structure of the $S$-wave OPE was studied in Ref.~\cite{Jackura:2018xnx}. This process can be isolated from the full \threetothree scattering amplitude and affects the analytic structure of the interaction kernel. Through a single iteration of Eq.~\eqref{eq:b-matrix-param} it can be rewritten as a sum of: The bubble ($\Rc \times \Rc$), the triangle ($\Rc \times \Gc$) and the box ($\Gc \times \Gc$) diagrams. The authors of Ref.~\cite{Jackura:2018xnx} discuss explicitly the influence of the OPE on the $B$-matrix triangle amplitude, identifying spurious left-hand cuts. They propose a dispersion approach as a way to eliminate these spurious cuts and compare their result with the analytic structure of the covariant Feynman amplitude. In Ref.~\cite{Dawid:2020uhn} the authors perform a comparable analysis, using a model of relativistic three-body scattering with a bound state in the two-body subchannel. They focus on the contact interaction approximation in which the exact solution of the model can be achieved, being effectively a series of bubble diagrams. They show the emergence of similar singularities and eliminate them via the analogous dispersion scheme.

In Ref.~\cite{Sadasivan:2020syi} the $B$-matrix equation was applied to the coupled-channel case of the decay $a_1(1260) \to \pi^- \pi^- \pi^+$ with the dominant contribution provided by the $\rho \pi$ isobar-spectator channel in the $S$ and $D$ waves. The authors solved the equation by discretizing the particles' momenta on a complex contour, obtaining a matrix equation, which was handled numerically. The obtained amplitude was matched with the experimental data on the $\tau \to (3\pi) \nu_\tau$ to compute $a_1(1260) \to 3\pi$ Dalitz plots and lineshapes. Techniques for solving the $B$-matrix equation are also discussed in Ref.~\cite{Jackura:2020bsk}.
There, the authors study a three-body system with an $S$-wave bound state in the two-body subchannel. The setup can be considered as a simplified model of the nucleon-deuteron interaction.
They employ the \emph{ladder} approximation and reproduce results obtained using finite-volume spectra of the same model~\cite{Romero-Lopez:2019qrt}.

The three-body equations as presented are complicated to use in practical analyses, as it is necessary to parametrize the short-distance functions and to solve intricate integral equations. Moreover, for extracting the resonant spectrum from data, one is more interested in the short-distance physics than the long-distance rescattering terms which originate from two-body physics.
Moreover, real-axis singularities cannot be seen in physical process, since \threetothree scattering amplitudes are always convoluted with a production source.
A reformulation of the $B$ matrix method, similar to the REFT approach, separates these purely rescattering effects from the short-distance physics in order to provide a useful tool for practical data analysis~\cite{Mikhasenko:2019vhk}.
Resonance physics is a useful application of this approach, as
one can focus on constructing dynamical models for the short-distance term and consider the OPE as the ``dressing'' corrections.
Ref.~\cite{Mikhasenko:2019vhk} discusses further approximations which can be made to simplify subsequent analyses, such as factorization of the short-distance $R$ matrix or truncation of the partial-wave basis as it is done in Khuri-Treiman approaches.

The phase-space integral of the three-particles final state, which determines the imaginary part of the inverse amplitude, is the integral over the Dalitz plot. The OPE affects the Dalitz plot distribution in two ways: First, the interference of the different decay chains, e.g. a resonance in one pair of particle and in the other pair, is due to the real one-particle exchange. Second, the subchannel-resonance shape might deviate from the one measured in the two-body scattering due to the final-state interaction. Accounting for the interference is straightforward, while the second effect requires modeling and often a fitting of the model parameters to the data.

The Khuri-Treiman equations discussed in section~\ref{sec:KT} offer a good model to address the final-state interaction. The method is based on the analytic continuation of two-body unitarity. Hence, it implements a specific OPE ladder series of the \threetothree scattering~\cite{Aitchison:1966lpz}. More importantly, the effect of OPE computed with the KT can be used to complete the three-body unitarity~\cite{Mikhasenko:2019vhk}.

Similar to the three-body unitarity, the REFT formalism has been studied from various angles by different groups. A threshold expansion and isotropic approximation for the three-body $K$ matrix have been proposed~\cite{Hansen:2016fzj, Briceno:2018aml, Briceno:2018mlh} to simplify analyses of the lattice data. The formalism has also been applied to the study of the bound-state toy models in Refs.~\cite{Hansen:2016ync,Romero-Lopez:2019qrt}. In Ref.~\cite{Jackura:2019bmu}, it was shown that the REFT formulation in the infinite volume can be recovered from the $B$-matrix representation. This is an expected result in light of Ref.~\cite{Briceno:2019muc} which proved the unitarity of the REFT approach. The differences in both formalisms were proved to be consequences of different parametrization of two-body rescatterings in the initial and final states. Additionally, the authors presented the equivalence of the heavy-mass limit of both representations with the non-relativistic EFT approach by Bedaque, Hammer, and von Kolck~\cite{Bedaque:1998kg} and the Faddeev equations~\cite{Elster:2008hn}.

\subsubsection{Lattice studies of the three-body scattering}

The form of the $B$-matrix amplitude parametrization is suitable for investigating the three-body finite-volume spectrum in the finite-volume unitarity (FVU) formalism. The integral equation is modified because in a finite cubic volume with periodic boundary conditions, the particles' three-momenta become discretized. Thus, one replaces three-dimensional integrations by the summations over the available lattice momenta. Moreover, in the finite cubic box, the irreducible representations of the rotation group are divided into 10 irreducible representations of the octahedral group and become coupled due to the breaking of rotational symmetry~\cite{Bernard:2008ax, Leskovec:2012gb, Polejaeva:2012ut, Doring:2012eu, Hammer:2017kms}. The FVU approach is rooted in the fact that the three-body subprocesses which lead to the large, finite-volume power-law corrections in the values of observables, are described by amplitudes that contribute only imaginary parts to the three-body unitarity. In this sense, the three- and two-body unitarity imply a three-body quantization condition, that is derived from the finite-volume version of the $B$-matrix equation and takes the form of the determinant condition including the $B$ matrix and a known geometric function~\cite{Mai:2021lwb}.

The first study of the relativistic FVU quantization condition was completed in Ref.~\cite{Mai:2017bge} for the case of a single isobar and one irreducible representation of the lattice symmetry group, which can be considered an analog of a single partial-wave in the infinite volume limit. In practice, the energy levels extracted on the lattice correspond to a given representation of the group and are determined independently. In Ref.~\cite{Doring:2018xxx} the projection of the quantization condition to a given irreducible representation of the octahedral group was reported. It corresponds to a partial diagonalization of the quantization condition equation and thus greatly simplifies it for practical purposes. Finally, Ref.~\cite{Mai:2019fba} includes a prediction for the three-pion lattice spectrum from the unitarity quantization condition.

In Ref.~\cite{Blanton:2020jnm} it was shown that the quantization conditions corresponding to the REFT and FVU formalisms are equivalent. This was achieved by rewriting the REFT condition in terms of the $\Rc$ matrix, at the same time producing a generalization of the latter approach to arbitrary angular momenta of isobars, independently of Ref.~\cite{Mai:2021nul}.

The generic lattice-based computation of the three-body amplitude is implemented via the following, simplified procedure: First, one determines the two-body finite-volume amplitude $\Fc$ through a two-body convenient quantization condition for all relevant isobars in the three-body system. Secondly, one computes a set of three-body energy levels in a given octahedral representation and through the three-body quantization condition determines the three-body $\Rc$ matrix. In practice, a suitable model is needed to fit the short-range interaction to the finite volume $R$-matrix data. Finally, one inputs the obtained form of three-body forces into the infinite volume integral equation, Eq.~\eqref{eq:b-matrix-param}, to compute the three-body amplitude.

There is a growing number of results of few-body spectra from lattice QCD~\cite{Detmold:2008yn, Detmold:2008fn, Horz:2019rrn, Culver:2019vvu, Fischer:2020jzp, Alexandru:2020xqf, Hansen:2020otl, Mai:2021nul, Blanton:2021llb} that can be used to determine the nature of the three-hadron interactions in the QCD. The FVU formalism has been applied to extract three-body forces from various few-particle systems in lattice QCD, all of which were generated at a higher than physical pion mass. In Ref.~\cite{Mai:2018djl} the authors analyze the lattice $\pi^+\pi^+$ and $\pi^+\pi^+\pi^+$ data from Ref.~\cite{Detmold:2008fn}, extracting the matrix $\Rc$. Within the used parametrizations, the authors found the short-range forces to be consistent with zero in this system. This study was continued in Ref.~\cite{Brett:2021wyd}, based on the data of Ref.~\cite{Culver:2019vvu} and~\cite{Hansen:2020otl}, leading to a more precise determination of the three-body coupling. The authors found the result of their analysis to be small but non-zero, and consistent with the LO $\chi$PT at the heavy pion mass. In addition, the pion mass dependence of the three-pion amplitude was studied and compared to the LO $\chi$PT prediction. A clear conclusion of the consistency between the $\chi$PT and the Lattice QCD could not be made due to large systematic and statistical errors.

In Ref.~\cite{Fischer:2020jzp}, the REFT finite-volume approach was applied to the $3 \pi^+$ spectrum computed at three pion masses, including the physical one. The resulting $\Kc_{\text{df},3}$ term was analyzed in the isotropic approximation and found to be non-zero, showing a reasonable agreement with LO $\chi$PT. The three-body RFT formalism was also employed in Ref.~\cite{Hansen:2020otl}, for the same system at large pion mass, producing the three-body term in the isotropic approximation compatible with zero. It is worth noting that in the study the authors used the lattice output in the infinite-volume integral equations for the first time, producing scattering amplitudes and Dalitz plots. In Ref.~\cite{Mai:2021nul}, the authors extracted parameters of the $a_1(1260)$ from the Lattice QCD, at pion mass $244\mev$. They generalized the FVU three-body quantization condition to sub-systems with non-zero angular momenta and coupled channels, and performed analytic continuation of the $B$-matrix equations solution to determine the pole position of the resonance. Most recently, Ref.~\cite{Blanton:2021llb} presented a high-precision lattice computation of three-particle systems including either pions or kaons. The authors include the $D$-wave isobars in their work and determine the three-body $K$ matrix using three different pion masses in the REFT approach. They notice tensions between their results and previous studies and comment on the necessity of more accurate computations in the future.

%  _|_|_|_|_|                                                                  _|                            _|      _|            _|  
%      _|      _|      _|      _|    _|_|                _|_|_|      _|_|    _|_|_|_|    _|_|    _|_|_|    _|_|_|_|        _|_|_|  _|  
%      _|      _|      _|      _|  _|    _|  _|_|_|_|_|  _|    _|  _|    _|    _|      _|_|_|_|  _|    _|    _|      _|  _|    _|  _|  
%      _|        _|  _|  _|  _|    _|    _|              _|    _|  _|    _|    _|      _|        _|    _|    _|      _|  _|    _|  _|  
%      _|          _|      _|        _|_|                _|_|_|      _|_|        _|_|    _|_|_|  _|    _|      _|_|  _|    _|_|_|  _|  
%                                                        _|                                                                            
%                                                        _|                                                                            

\subsection{Application of three-body unitarity to resonance physics}
\label{sec:3bodyApp}

The construction of dynamical models for  three-body resonances
can proceed in a way similar to that of two-body amplitudes.
Unitarity determines the imaginary part of the inverse amplitude above
the particle production threshold. Following the analyticity requirement,
the self-energy function can be computed using dispersive techniques.
The remaining unknown part of the scattering amplitude is built through the parametrization of 
a real-valued function (or a matrix in the coupled-channel case) using the $K$-matrix approach. The OPE needs to be accounted for in the computation of the imaginary part (e.g. see diagram (b) in \figurename{~\ref{fig:a1.dalitz}}).
Firstly, it leads to the contribution of the interference of different chains for a three-body decay, and, secondly, it impacts the lineshape of the subchannel resonances.
The inclusion of only the OPE-related interference is referred to as \textit{approximate three-body unitarity}. The approach has been employed in several experimental analysis due to its relative simplicity and as a possibility to test data sensitivity to three-body effects~\cite{Mikhasenko:2019phe}.

\subsubsection{\texorpdfstring{Studies of $a_1(1260)$ resonance in the $3\pi$ system}{Studies of a1(1260) resonance in the 3pi system}}
\label{sec:a1}

\begin{figure}
    \centering
    \raisebox{-0.5\height}{
    \begin{overpic}[width=0.5\textwidth]{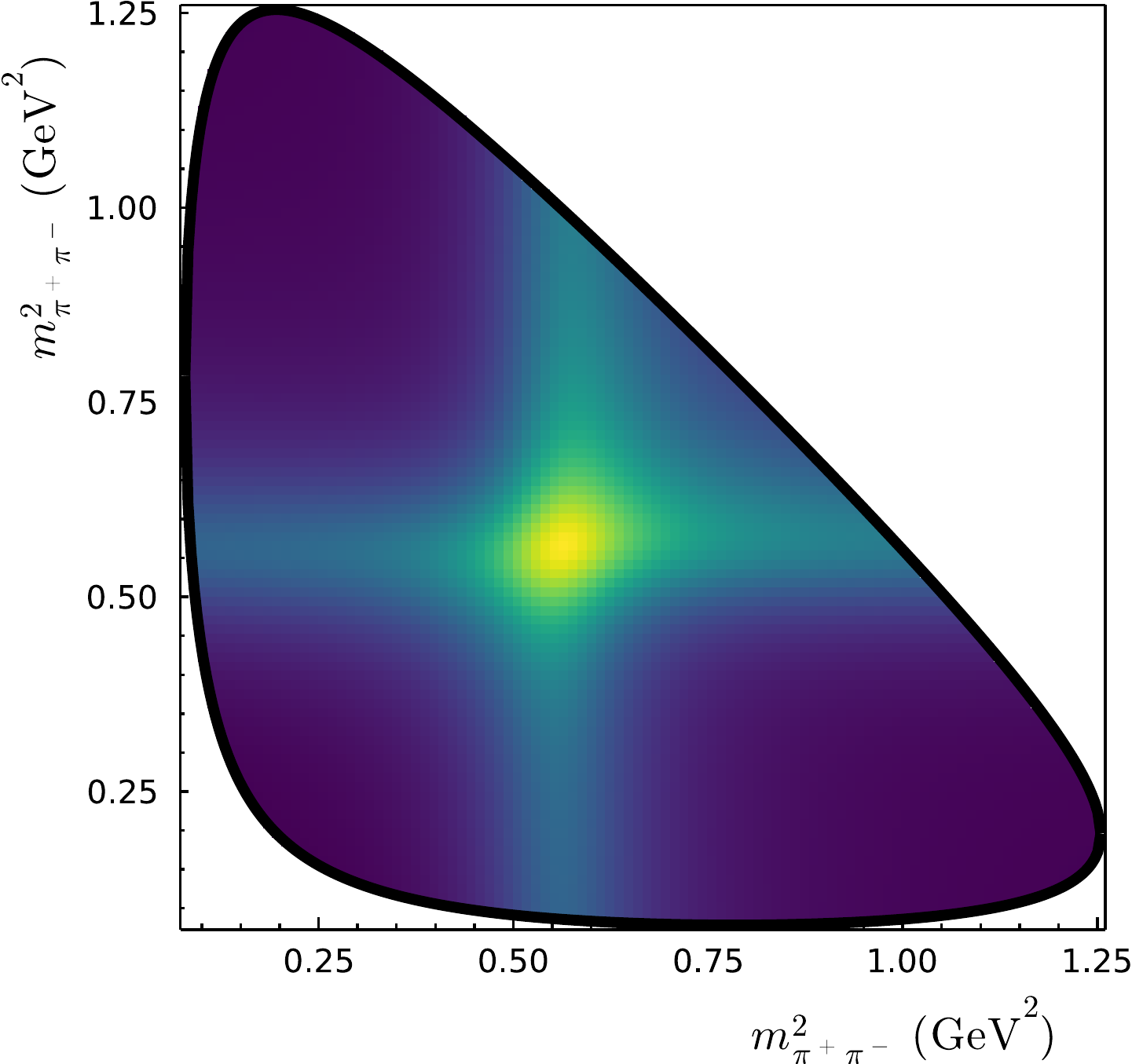}
      \put(25,52) {\color{white} (a)}
      \put(52,52) {\color{white} (b)}
      \put(53,18) {\color{white} (c)}
    \end{overpic}
    } \hspace{1.5cm}
    \raisebox{-0.5\height}{\includegraphics[width=0.25\textwidth]{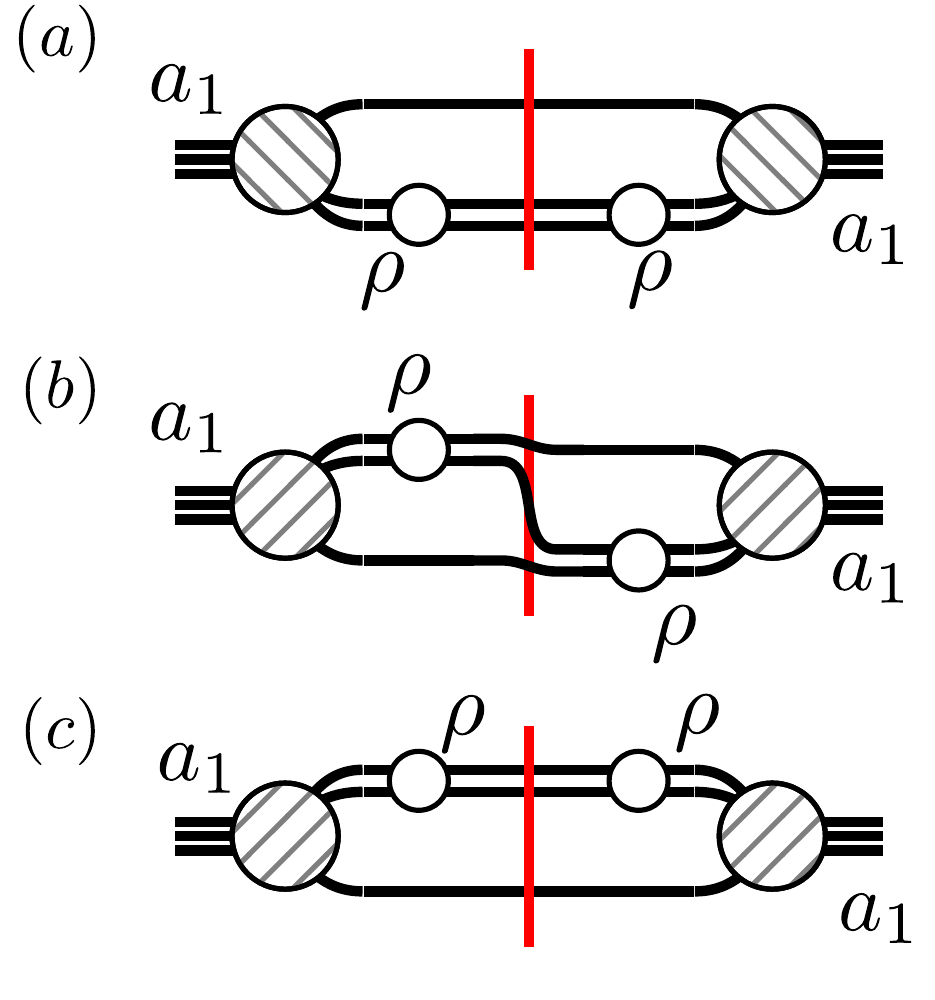}}
    \caption{Dalitz-plot distribution for the $a_1^-\to \pi^-\pi^-\pi^+$ decays modeled in Ref.~\cite{JPAC:2018zwp}.
    The mass of the system is fixed to the nominal  $a_1(1260)$ mass.
    The diagrams on the left panel represent the contributions to the $a_1$ self-energy.
    The kinematic regions where these are significant are indicated with labels on the Dalitz plot. In the quasi-two-body dispersive model, diagram (b) is neglected, and so is the interference of the two bands.}
    \label{fig:a1.dalitz}
\end{figure}

The $a_1(1260)$ resonance has a prominent role in the $\tau \to 3\pi\nu$ decay,
dominating the lineshape structure. Its mass is fairly known, but its width has large uncertainties and is just known to be large~\cite{pdg}.
The dominant decay channel is $a_1\to \rho\pi$ in the $S$-wave, where the $\rho$ subsequently
decays to two pions. 
The $a_1$ broad peak spans the range from $0.8$ to $1.6\gev$ of the three-pion invariant mass,
covering the nominal $\rho\pi$ threshold, which makes the explicit inclusion of the threshold essential for the proper analytic continuation of the amplitude to the complex energy plane and pole extraction.
The effect of the OPE is also significant since the two $\rho^0$ meson bands largely overlap in the Dalitz plot as shown in \figurename{~\ref{fig:a1.dalitz}}.

The reaction amplitude for the
resonant part, \aka  $a_1$, of the $3\pi\to 3\pi$ rescattering
is written as a Breit-Wigner with a nontrivial self-energy function
 that accounts for three-body effects.
The imaginary part of the amplitude is computed using the optical theorem
for the $a_1\to 3\pi$ decay; then the real part of the self-energy
is computed through dispersive integrals.
In this way we manage to implement the correct analytic structure.
We consider two models. The first one, \aka symmetrized-dispersive model
incorporates the OPE process via the interference the two coherent $a_1\to\rho\pi$ decay chains in the self-energy function of $a_1$.
Due to the presence of two same-charge pions in the decay of the $a_1^-$ meson, the imaginary part of the $\rho \pi \to \pi \rho$ bubble contains the term with the real pion exchange, as shown in \figurename{~\ref{fig:a1.dalitz}}, namely,
$\langle \pi^-_1 (\pi_2^- \pi^+)_{\rho^0} | \pi^-_2(\pi_1^-\pi^+)_{\rho^0} \rangle$.
The second model, \aka quasi-two-body dispersive  model,
neglects the OPE effect entirely and considers only one $\rho \pi$ decay chain for the $a_1\to 3\pi$ transition.
\begin{figure}
\centering
    \raisebox{-0.5\height}{\includegraphics[width=0.45\textwidth]{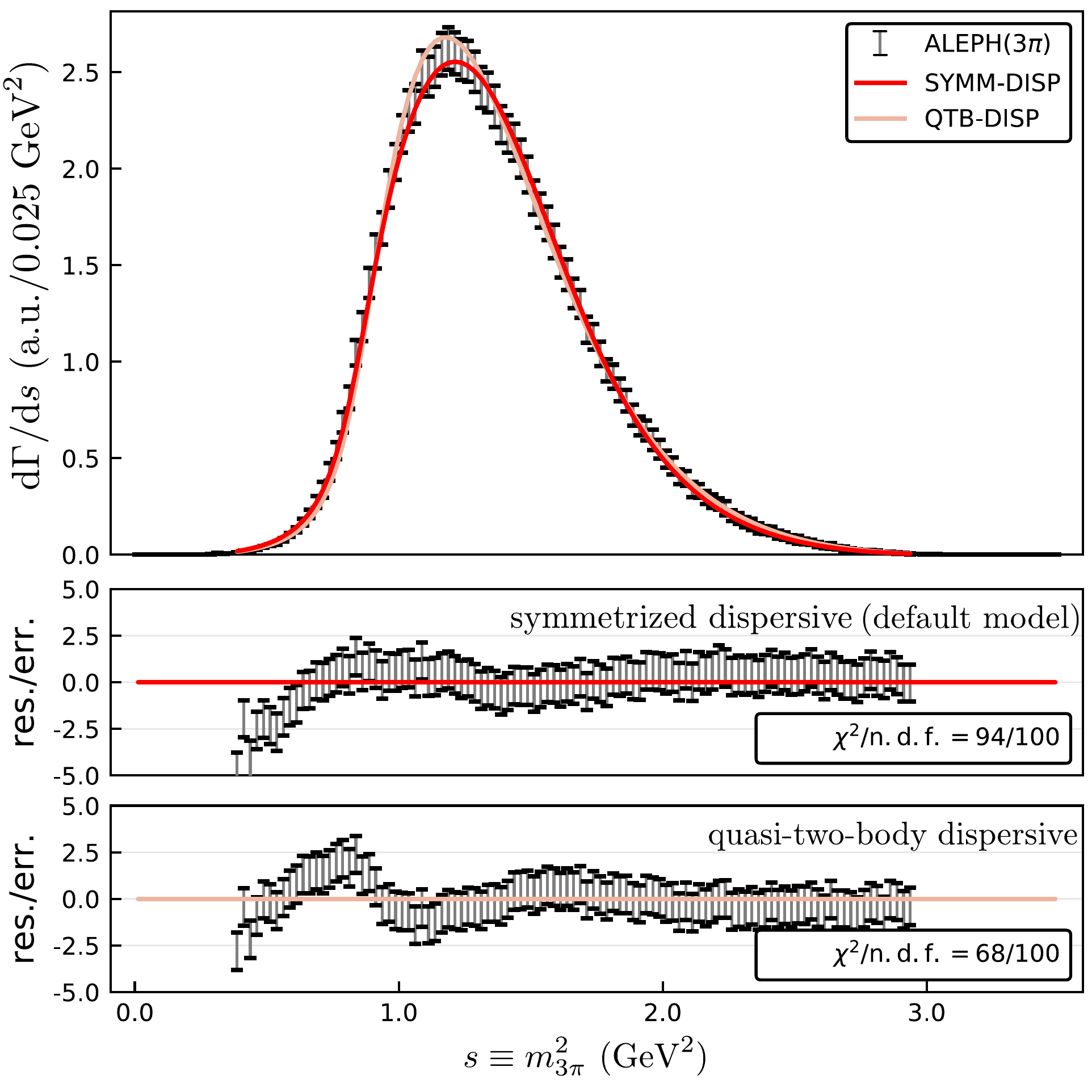}}\,\hspace{1cm}
    \raisebox{-0.5\height}{\includegraphics[width=0.4\textwidth]{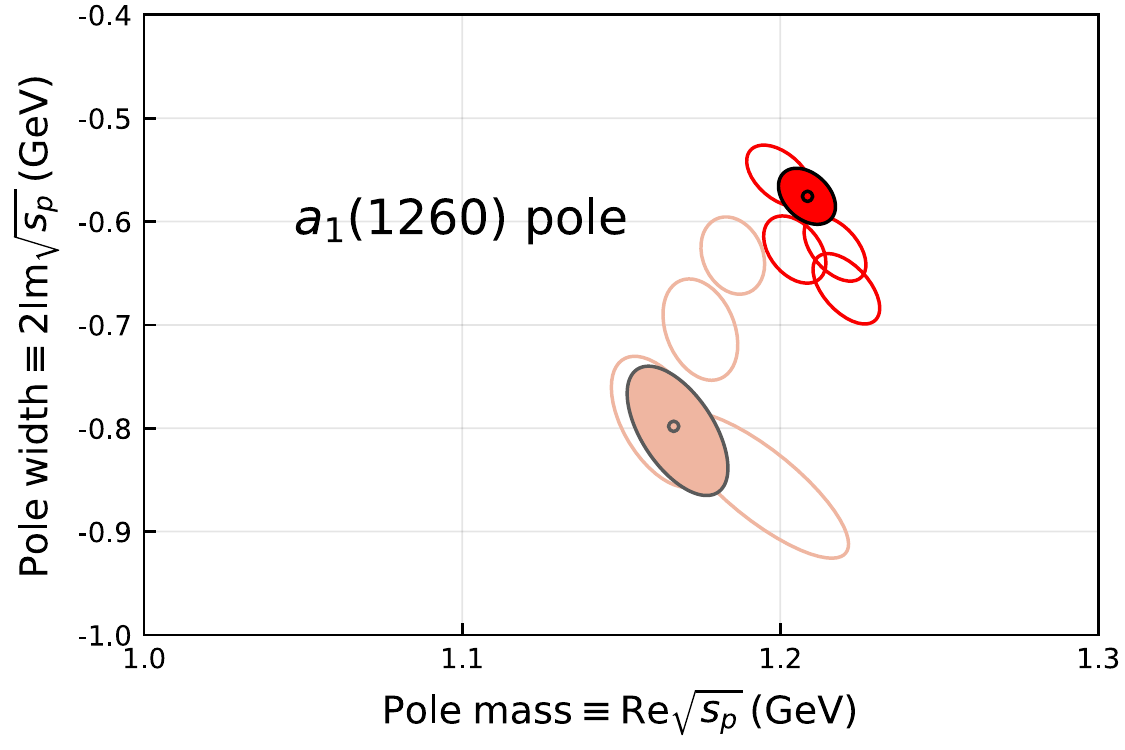}}
    \caption{(left) Three-pion spectrum of the  
    $\tau \to 3\pi \nu$ decay. Data are given by ALEPH~\cite{Davier:2013sfa, ALEPH:2005qgp}
    The model curves for the two dispersive models are overlayed.
    (right) Position of the $a_1(1260)$ pole in the complex energy plane for both
    models: The symmetrized-dispersive model (red) and the quasi-two-body dispersive model (orange). 
    The ellipses account for the $95\%$ confidence level.
    The results for the main fit are shown by filled ellipses,
    while the unfilled ellipses provide the systematic studies.
    Figures adapted from~\cite{JPAC:2018zwp}. 
    }
    \label{fig:a1.disp.fits.pole}
\end{figure}
\figurename{~\ref{fig:a1.disp.fits.pole}} shows the fits of both models to the
ALEPH dataset~\cite{Davier:2013sfa, ALEPH:2005qgp}. The data are clearly correlated, and the statistics results have be taken with a grain of salt. 
The models are consistent with the data and provide results of similar quality,
however, the parameters and hence the pole extractions are quite different.
In the figure we also show both pole extractions.
Not including the OPE makes the $a_1(1260)$ width larger and the mass lighter. The presence of the $\rho\pi$ cut and of spurious poles has been already discussed in Section~\ref{sec:spurious}, while here we focus on the extraction of physics.
The pole position obtained using the symmetrized-dispersive model reads:
\begin{align}
  m_p^{a_1(1260)} &= 1209 \pm 4 ^{+12}_{-9}\,\mev,&\quad
  \Gamma_p^{a_1(1260)} &= 580 \pm 10 ^{+80}_{-20}\, \mev,
\end{align}
where the first error is statistical 
and the second error comes from the systematic studies,
such as varying the $\rho$ lineshape.
There is an ongoing effect to take into account the rescattering in systematic manner using the Khuri-Treiman approach discussed in Section~\ref{sec:KT}, particularly see~\cite{Albaladejo:2019huw}.
The scalar $\pi\pi$ wave component is also of large interest for the future improvement of the model.
It might account for $20\%$ of the $a_1^-$ decay rate~\cite{CLEO:1999rzk}.

Moreover, an interplay of the $\pi\pi \to K\bar{K}$ in the $a_1^-$ decay leads to a spectacular manifestation of the triangle singularity.
An axial resonance-like $a_1(1420)$ signal with mass $1.42\gev$ and width of $150\mev$ was indeed reported by COMPASS~\cite{COMPASS:2015kdx}.
It was observed in the $P$-wave of the $f_0(980)\,\pi$ system
of the $\pi^-\,p \to 3\pi\,p$ reaction~\cite{COMPASS:2015gxz}.
The mass of the $a_1(1420)$ is slightly above the $K^*\bar{K}$ threshold.
In~\cite{Mikhasenko:2015oxp, Aceti:2016yeb} it was suggested that
the signal could be a consequence of final-state interactions 
in the $a_1(1260)$ decaying to $3\pi$ and $K\bar{K}\pi$,
in particular, a triangle singularity, finding an excellent agreement with the data~\cite{COMPASS:2020yhb}.

\subsubsection{\texorpdfstring{Studies of $\pi_2$ resonances in $3\pi$ system}{Studies of pi2 resonances in 3pi system}}
\label{sec:pi2}
The main puzzle of the $J^{PC} = 2^{-+}$ sector is an interplay of the two states called $\pi_2(1670)$ and $\pi_2(1880)$, which have been seen to decay predominantly into $3\pi$~\cite{pdg}.
The quark model does not explain two states with the same quantum numbers with masses so close together;
the $\pi_2(1880)$ is too light to be a radial excitation of the $\pi_2(1670)$, and is a prime candidate for a hybrid meson~\cite{Klempt:2007cp,Dudek:2011bn}.
Rather, the $\pi_2(2005)$ might be the radial excitation, and is
seen in the diffractive production of the $\omega \pi \pi$ system at BNL~\cite{E852:2004rfa}.

The $J^{PC} = 2^{-+}$ sector is well separated from the other quantum numbers in the COMPASS partial-wave analysis of
diffractive $\pi^-\,p\to 3\pi\,p$ reaction~\cite{COMPASS:2015gxz}, allowing to isolate the $\pi_2$ candidates.
The resonances decay to the $3\pi$ final states via $f_0$, $\rho$, and $f_2$.
The COMPASS mass-dependent analysis~\cite{COMPASS:2018uzl} indicates three $\pi_2$ states, $\pi_2(1670)$, $\pi_2(1880)$, and $\pi_2(2005)$.
The latter, however, significantly overlaps with the $\pi_2(1880)$, which limits the applicability of the Breit-Wigner model adopted.

\begin{figure}
    \centering
    \includegraphics[width=\textwidth]{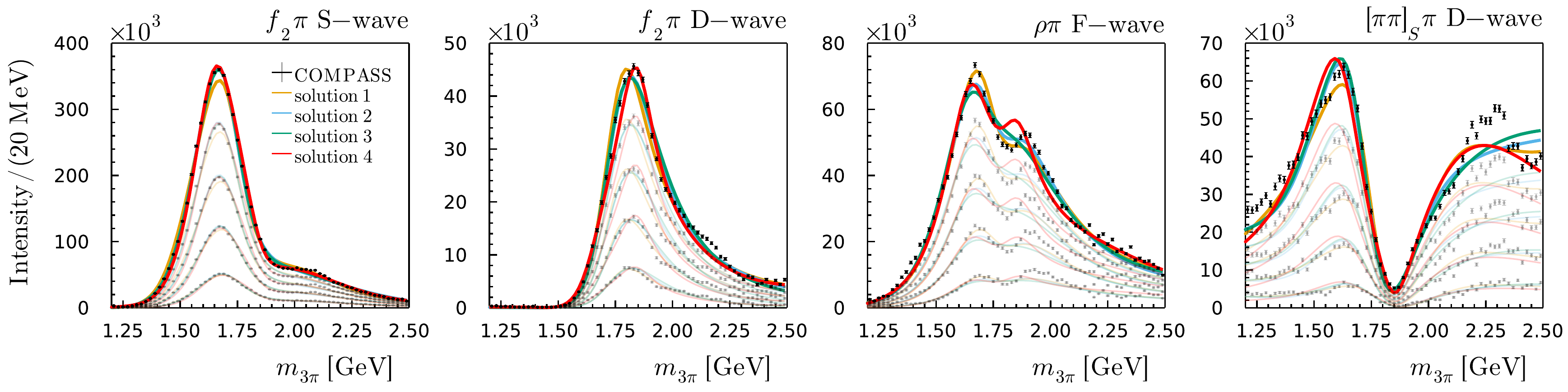}
    \caption{Intensities of the $J^{PC}M^{\epsilon}=2^{-+}0^+$ waves
    integrated over $t'$ bins in the $0.1\gevsq < -t' < 1\gevsq$ range.
    The sets of transparent points and curves correspond to increasing the lower limit of $-t'$ to $0.127$, $0.164$, $0.220$, and $0.326\gevsq$, respectively.
    Data are measured by COMPASS~\cite{COMPASS:2015gxz}. The curves are presented in~\cite{Mikhasenko:2019phe}.
    }
    \label{fig:2mp.fit}
\end{figure}

To study these states, we develop an exploratory coupled-channel model that incorporates 
both the three-body and the resonance-spectator thresholds in the complex plane~\cite{Mikhasenko:2017jtg,Jackura:2016llm,Mikhasenko:2019phe}. The OPE effects are neglected to simplify the setup.
The model accounts for the production mechanism using the $Q$-vector approach~\cite{Chung:1995dx, Cahn:1985wu, Au:1986vs}.
The production vector is modeled 
by a polynomial series of the conformal variable, $\omega(s) = (1-\sqrt{s}) / (1+\sqrt{s})$.
Then, the model is applied to the intensities and relative phases
of the four major $J^{PC}=2^{-+}$ waves
for eleven bins of $\pi\,p$ transferred momenta $t'$~\cite{COMPASS:2015gxz}.
A reasonable description of the set of the four waves is shown in \figurename{~\ref{fig:2mp.fit}}, and requires at least four $K$-matrix poles in the form of Eq.~\eqref{eq:Kmatrix}.
The production vector is modeled with a fourth-order polynomial in $\omega$,
independent for each wave and all $t'$ slices.

This optimization problem suffers from multimodality.
The colored lines in \figurename{~\ref{fig:2mp.fit}}
represent to the four solution for local minima with similar quality.
 
All the solutions suggest the presence of three poles in vicinity of the fit region.
The poles are ordered by their mass values and assigned to $\pi_2(1670)$, $\pi_2(1880)$, and $\pi_2(2005)$.
However, the parameters are significantly different across different solutions. Conservative estimate of masses and widths shown in \tablename{~\ref{tab:2mp.results}} are obtained by quoting the extreme values among all the selected solutions.
The pole positions of the states are correlated, to the extent that the mass intervals of $\pi_2(1880)$ and $\pi_2(2005)$ overlap.
However, all solutions prefer the heavier pole to be broader than the middle one.

\begin{table}[b]
  \centering
  \caption{
Summary of the parameters of $\pi_2$ resonances in the $K$-matrix model with four poles.
The second column gives the results of Ref.~\cite{COMPASS:2018uzl} using the Breit-Wigner model.
  }
  \label{tab:2mp.results}
  \begin{tabular}{c |c c | c c }
  \hline
  \hline
                  & $m_p$ (\mevp) & $\Gamma_p$ (\mevp)& $m_\mathrm{BW}$ (\mevp)& $\Gamma_\mathrm{BW}$ (\mevp)\\\hline
    $\pi_2(1670)$ & $1650$--$1750$ & $280$--$380$  & $1642^{+12}_{-1}$ & $311^{+12}_{-23}$ \\
    $\pi_2(1880)$ & $1770$--$1870$ & $200$--$450$  & $1847^{+20}_{-3}$ & $246^{+33}_{-28}$ \\
    $\pi_2(2005)$ & $1890$--$ 2190$ & $590$--$1340$  & $1962^{+17}_{-29}$ & $269^{+16}_{-120}$ \\
    \hline
    \hline
  \end{tabular}
\end{table}
The results are compared to the 
conventional approach of the Breit-Wigner model of Ref.~\cite{COMPASS:2018uzl}.
Our studies indicate that both $\pi_2(1670)$ and $\pi_2(1880)$ are required by the data, and their widths are below $450\mev$. The width of the $\pi_2(2005)$ is obtained in the interval from $590\mev$ to $1.34\gev$.
The $\pi_2(2005)$ pole is hinted in the data by the left shoulder of the $f_2\pi\,$S-wave in \figurename{~\ref{fig:2mp.fit}}.
However, the shortcomings of the models, e.g. the omission of OPE, in combination with the multi-channel complexity do not allow establishing the presence of resonance and its parameters reliably.

The main difficulty in describing the data is related
the nonresonant coherent background process named after Deck~\cite{Deck:1964hm}.
Studies of the Deck mechanism in Refs.~\cite{Ascoli:1974hi,Mikhasenko:2019phe} showed a large pollution in the $J^{PC}=2^{-+}$ waves.
The unitarization method proposed in Ref.~\cite{Basdevant:1977ya}
builds in the explicit form of the background while preserving
unitarity. The method requires dedicated studies of the partial wave projections of the Deck process. This will be the subject for future research that will lead to a better understanding of the sector.

%============================================
% Production
%============================================
\section{Production mechanisms}
\label{sec:production}
The mechanisms that produce hadron resonances in experiments offer another valuable piece of information for understanding their nature.
For example, most of the recent data on \XYZ states come from electroweak processes, as heavy meson/baryon decays or $e^+e^-$ annihilation. 
Matrix elements can most often be studied in terms of form factors~\cite{Hanhart:2012wi,Ropertz:2018stk,VonDetten:2021rax}. 

At high energy, (semi-)inclusive production processes enter the perturbative QCD regime. For example, deep inelastic scattering (DIS) of electrons off protons at large $Q^2$ has been the main experimental tool to scrutinize the inner structure of nucleons. Data on the corresponding cross sections and structure functions have been key ingredients in global QCD analyses of parton distributions~\cite{Collins:1989gx,Harland-Lang:2014zoa, Dulat:2015mca, Ball:2017nwa, Moffat:2021dji, Accardi:2016qay, Sato:2019yez, Alekhin:2017kpj,Abele:2021nyo,Liu:2021jfp}.
At lower energies and $Q^2$ inclusive data are saturated by a few exclusive channels, and perturbative calculations lose their validity. Having a comprehensive understanding of the low and high energy regimes at once is a highly nontrivial task that will be discussed in Section~\ref{sec:proton_structure}.

In peripheral production, where the momentum transferred is much smaller than the energy, 
forces and resonances themselves are constrained by the same strong interaction dynamics, and one can learn about one by studying the other. This duality is the cornerstone of Regge theory.
The most comprehensive study for establishing the role of Reggeons in quasi-elastic two body scattering will be discussed in Section~\ref{sec:regge}.
These studies are particularly effective in explaining single hadron (or resonance) photoproduction, as shown in Section~\ref{sec:photo} for the light sector. Quarkonia photoproduction will be discussed in Section~\ref{sec:jpsi_photo} in the context of pentaquark searches, and in Section~\ref{sec:xyz_exclusive} in the context of predicting \XYZ rates at electron-proton facilities.
At high invariant masses, one enters the so-called double-Regge regime, which we will describe in Section~\ref{sec:doubleregge}. 
Contributions to the  amplitude from 
resonances in the direct channel and Reggeons in the overlapping, crossed channels, cannot be added, 
as explicit for example in the Veneziano amplitude~\cite{Veneziano:1968yb,Veneziano:1974dr}, 
and one has to take specific care when involving  both in amplitude analysis.   
In \twototwo scattering, Reggeons dominate the high energy behavior of the cross section at forward (or backward) angles, while  resonances are visible at low energies in specific partial waves.  
Analyticity requires that these two regimes are connected, which allows us to write dispersion relations that can convert the Regge phenomenology at high energies into further constraints for the partial waves in the resonance region. This program of finite energy sum rules (FESR) will be discussed in Section~\ref{sec:FESR}.
This and other forms of duality have found some recent interest because of the discovery of several tetraquark and pentaquark candidates~\cite{Montanet:1980te,Rossi:2016szw}. Although establishing the presence of exotic states in the spectrum through their role as exchange forces might be a long shot,  the duality between Reggeons and resonances can play an important role in constraining models of exotics.

\subsection {Nucleon resonance contributions to inclusive electron scattering}
\label{sec:proton_structure}
Being able to describe the strong interaction physics across a broad range of energy and distance scales is crucial, but the nonperturbative regime is still far from being well understood. In inclusive electron-proton scattering, the transition from the low-energy resonance region to the high-energy regime (\ie DIS)
offers broad grounds for exploration~\cite{Feynman:1972xm, Farrar:1975yb, Close:1988br, Melnitchouk:1995fc, Brady:2011uy, Owens:2012bv,Nocera:2014uea,Accardi:2016qay, Alekhin:2017kpj, Moffat:2019qll}. Leading twist approximations\footnote{Operators contributing to DIS can be organized in terms of their twist, \ie their mass dimension minus the number of Lorentz indices. Higher twist operators are further suppressed by powers of the hard scale $Q^2$.} are found to be accurate at describing the region of invariant masses $W$ above the resonances, at sufficiently large photon virtualities, $Q^2 \gtrsim 1$--$2\gevsq$. Therefore, global QCD analyses~\cite{Accardi:2016qay, Alekhin:2017kpj, Sato:2019yez} usually involve cuts in both $W$ and $Q^2$~\cite{Jimenez-Delgado:2013sma,Harland-Lang:2014zoa, Dulat:2015mca, Ball:2017nwa, Gao:2017yyd, Ethier:2020way, Moffat:2021dji}. In order to bridge the gap between perturbative and nonperturbative regimes and to assess the parton distributions at large Bjorken-$x$, target mass corrections, higher twists and factorization-breaking corrections are called for. In addition, due to the resonance peaks appearing in the $W<2\gev$ region, the electroexcitation amplitudes of the resonances should be incorporated into the description of the structure functions~\cite{HillerBlin:2019jgp,Blin:2021twt}. High-precision measurements of inclusive electron scattering cross sections in the resonance region were made at \jlab's Halls~B and~C~\cite{Osipenko:2003bu, Liang:2004tj, Christy:2007ve,Malace:2009kw,Prok:2014ltt,Tvaskis:2016uxm,Golubenko:2019gxz}.

\begin{figure}[t]
\centering
\includegraphics[width=0.35\textwidth]{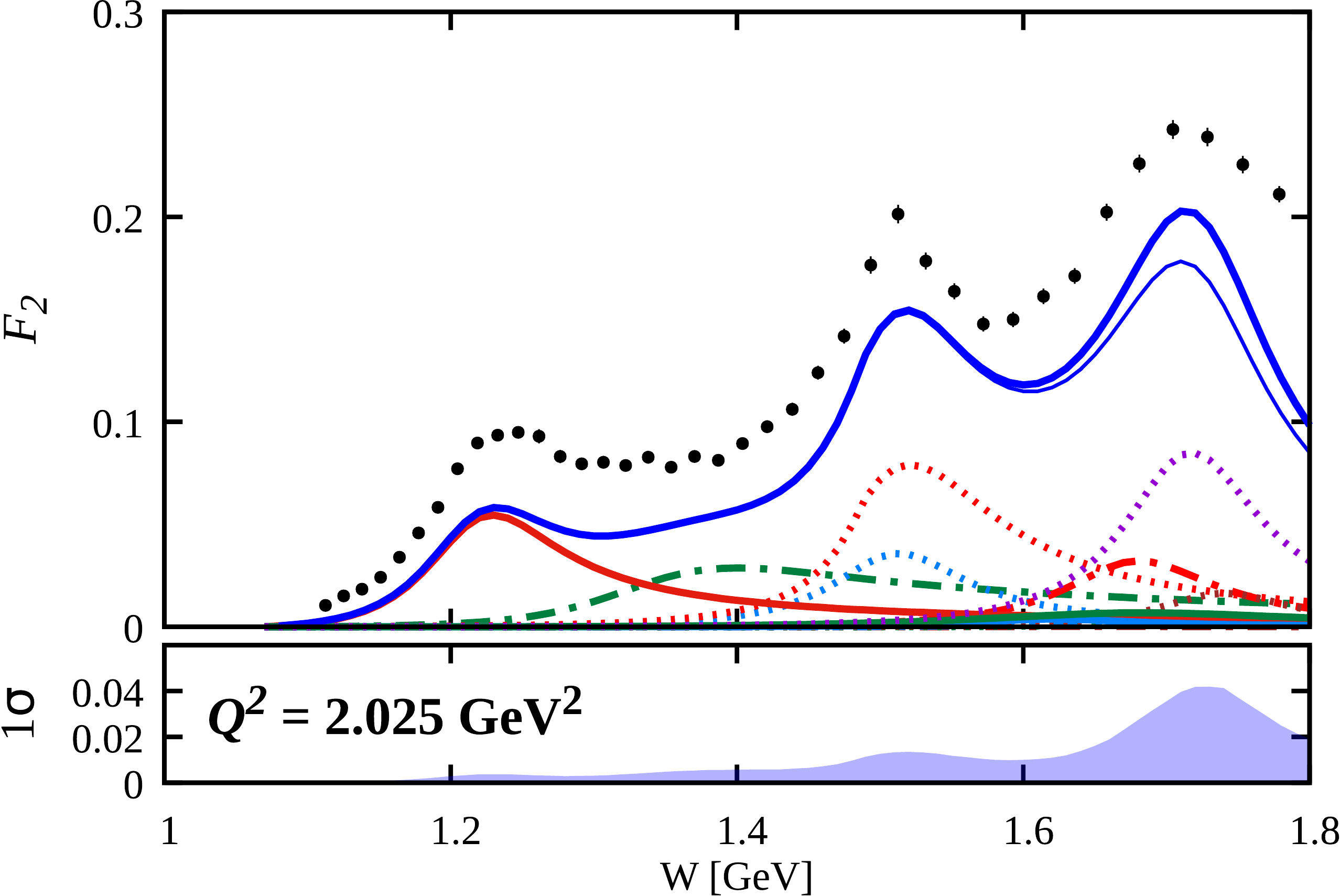}
\includegraphics[width=0.63\textwidth]{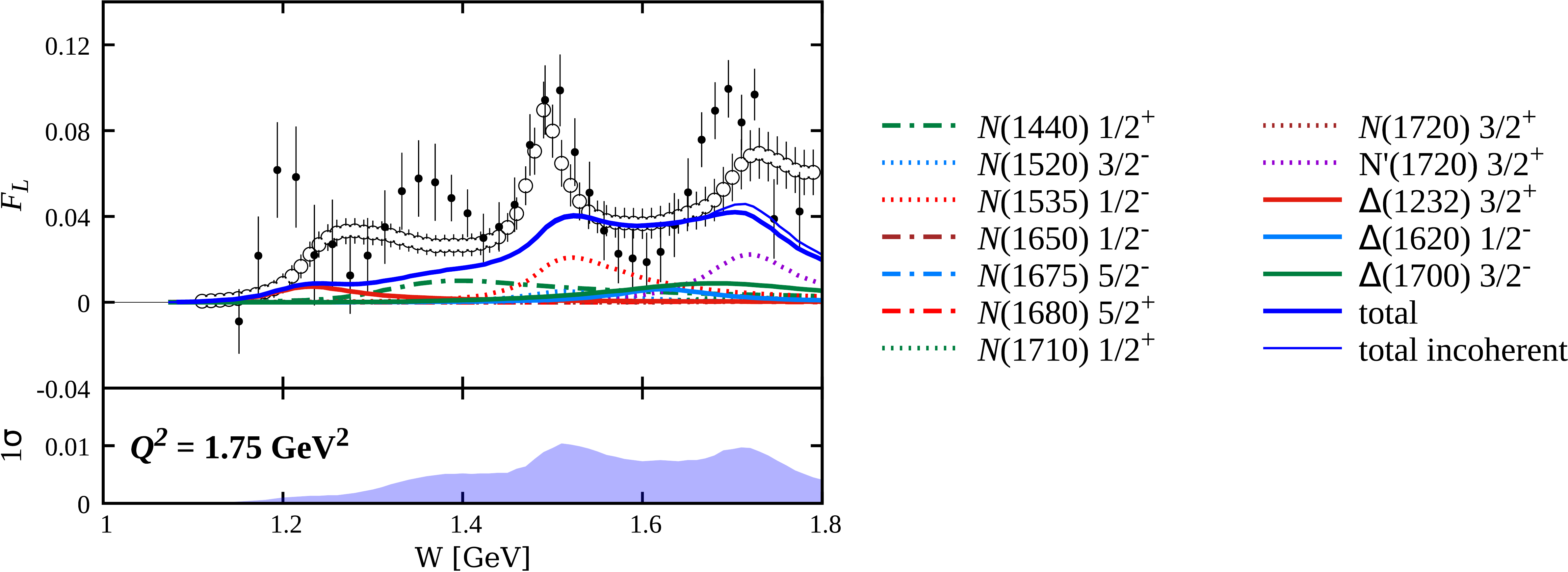}
\caption{Proton $F_2$ and $F_L$ structure function in the resonance region at different values of $Q^2$. The data  are compared with the full resonant structure functions computed by adding amplitudes (thick blue curves) and cross sections (thin blue curves) from the contributing resonances, using the central values of their electrocouplings. The contributions from individual resonances are shown separately, as indicated in the legends. Below each panel, we also show the uncertainty sizes of the thick blue curves (full coherent sum of resonant contributions), which are computed by propagating the electrocoupling uncertainties via bootstrap. 
The data in the left plot come from the interpolation of the CLAS database~\cite{CLAS:SFDB}, the data in the right plot come from~\cite{Tvaskis:2016uxm} (filled black circles) and~\cite{Liang:2004tj} (open black circles). 
Figure from~\cite{Blin:2021twt}.}
\label{fig:F2sing}
\end{figure}
A further phenomenological motivation for these studies is the observation of a duality between the structure functions in the nucleon resonance region, when averaged over resonances, and the scaling function extrapolated from the deep-inelastic scattering region~\cite{Bloom:1970xb}.
When integrating the structure functions over (finite intervals of) $x$, one obtains the (truncated) moments of the structure functions~\cite{Forte:1998nw, Forte:2000wh}.
The leading twist term is associated with incoherent scattering with individual partons in the nucleon~\cite{Piccione:2001vf, Kotlorz:2006dj, Kotlorz:2016icu}, while the higher twist corrections capture elements of long-distance, nonperturbative quark-gluon dynamics associated with color confinement in QCD~\cite{Ji:1994br,Psaker:2008ju}. Duality is interpreted as the dominance of the leading twist and the consequent suppression of higher twist contributions to the moments~\cite{DeRujula:1976baf}.

The inclusive structure functions are related to the total virtual photon-nucleon scattering cross sections $\sigma_T$ and $\sigma_L$, for transversely and longitudinally polarized photons, respectively~\cite{Drechsel:2002ar},
\begin{subequations}
\label{sf} 
\begin{align}
F_1(W,Q^2) &=\frac{{K m}}{4\pi^2\alpha}\, \sigma_T(W,Q^2),  \\
F_2(W,Q^2) &=\frac{{K m}}{4\pi^2\alpha}\, \frac{2x}{\rho^2}
             \left(\sigma_T(W,Q^2) +\sigma_L(W,Q^2)\right),
\end{align}
\end{subequations}
where $\alpha$ is the fine structure constant, $K= (W^2-m^2)\big/2m$ is the equivalent photon flux in the Hand convention~\cite{Hand:1963bb}, 
$\rho = \left(1 + 4m^2 x^2/Q^2\right)^{1/2}$ a kinematic parameter, and $m$ the proton mass.
The $F_2$ structure function can also be written in terms of the unpolarized virtual photoproduction cross section~$\sigma_{U}$,
\begin{align}
\label{eq:F2sig}
F_2(W,Q^2)
& = \frac{K m}{4\pi^2\alpha} 
    \frac{2x}{\rho^2} 
    \frac{1+R_{LT}}{1+\epsilon R_{LT}}\, \sigma_{U}(W,Q^2),
\end{align}
where
\begin{equation}
\label{xs}
\sigma_{U}(W,Q^2) = \sigma_T(W,Q^2) + \epsilon\, \sigma_L(W,Q^2),
\end{equation}
$\epsilon$ is the degree of transverse virtual photon polarization, determined by the scattered electron angle $\theta_e$,
\begin{equation}
\epsilon = \left( 1 + \frac{2 \rho^2}{\rho^2-1} \tan^2\frac{\theta_e}{2}
           \right)^{-1}\,,
\label{SigmaU1}
\end{equation}
and $R_{LT}=\sigma_{L}(W,Q^2)/\sigma_{T}(W,Q^2)$ is the ratio of longitudinal to transverse virtual photon cross sections. 
The longitudinal structure function is defined as
\begin{equation}
F_L(W,Q^2) =\frac{K m}{4\pi^2\alpha} 
    2x\,\sigma_L(W,Q^2)= \rho^2 F_2(W,Q^2) - 2 x F_1(W,Q^2).
\label{sfl}
\end{equation}

Here, we focus on the unpolarized structure functions $F_1$ and $F_2$, and their combination $F_L$. A compilation of the data for unpolarized structure functions and inclusive cross sections in the range $1.07 \leq W \leq 2\gev$ and $0.5 \leq Q^2 \leq 7\gevsq$, together with a tool for the interpolation between bins, is available online from the CLAS database~\cite{Golubenko:2019gxz, CLAS:DB, CLAS:SFDB}. At the same time, the experimental program of exclusive $\pi^+ n$, $\pi^0 p$, $\eta p$, and $\pi^+ \pi^-p$ electroproduction channels with CLAS at \jlab has provided the first and only available results on electroexcitation amplitudes, or electrocouplings of most nucleon resonances in the mass range $W < 1.8\gev$ and $Q^2< 5.0 \gevsq$~\cite{Carman:2020qmb,Mokeev:2020vab,Aznauryan:2011qj}. This makes it possible to evaluate the resonant contributions to inclusive electron scattering using parameters of the individual nucleon resonances extracted from data, expressing the amplitudes as a coherent sum over all relevant resonances in the mass range $W < 1.75\gev$~\cite{Carlson:1998gf, Melnitchouk:2005zr},
\begin{subequations}
\label{Eq:coherent}
\begin{align}
F_1^R
&= m^2 \sum_{IJ\eta} 
    \left[
    \left| \sum_{R^{IJ\eta}} G_+^{R^{IJ\eta}} \right|^2
  + \left| \sum_{R^{IJ\eta}} G_-^{R^{IJ\eta}} \right|^2
    \right]\,,
\\
\rho^2 F_2^R 
&= m \nu \sum_{IJ\eta}
    \left[
    \left| \sum_{R^{IJ\eta}} G_+^{R^{IJ\eta}} \right|^2
 +\ 2\, \left| \sum_{R^{IJ\eta}} G_0^{R^{IJ\eta}} \right|^2
 +  \left| \sum_{R^{IJ\eta}} G_-^{R^{IJ\eta}} \right|^2
    \right]\,,
\end{align}
\end{subequations}
where the outer sum runs over the possible values of spin $J$, isospin $I$ and intrinsic parity $\eta$, and the inner sums run over all those resonances $R^{IJ\eta}$ with same quantum numbers that are added coherently. 
The electrocouplings are encoded in the functions $G_{0,\pm}^{IJ\eta}$~\cite{Blin:2021twt}.

\begin{figure}[t]
\centering
\includegraphics[width=0.4\textwidth]{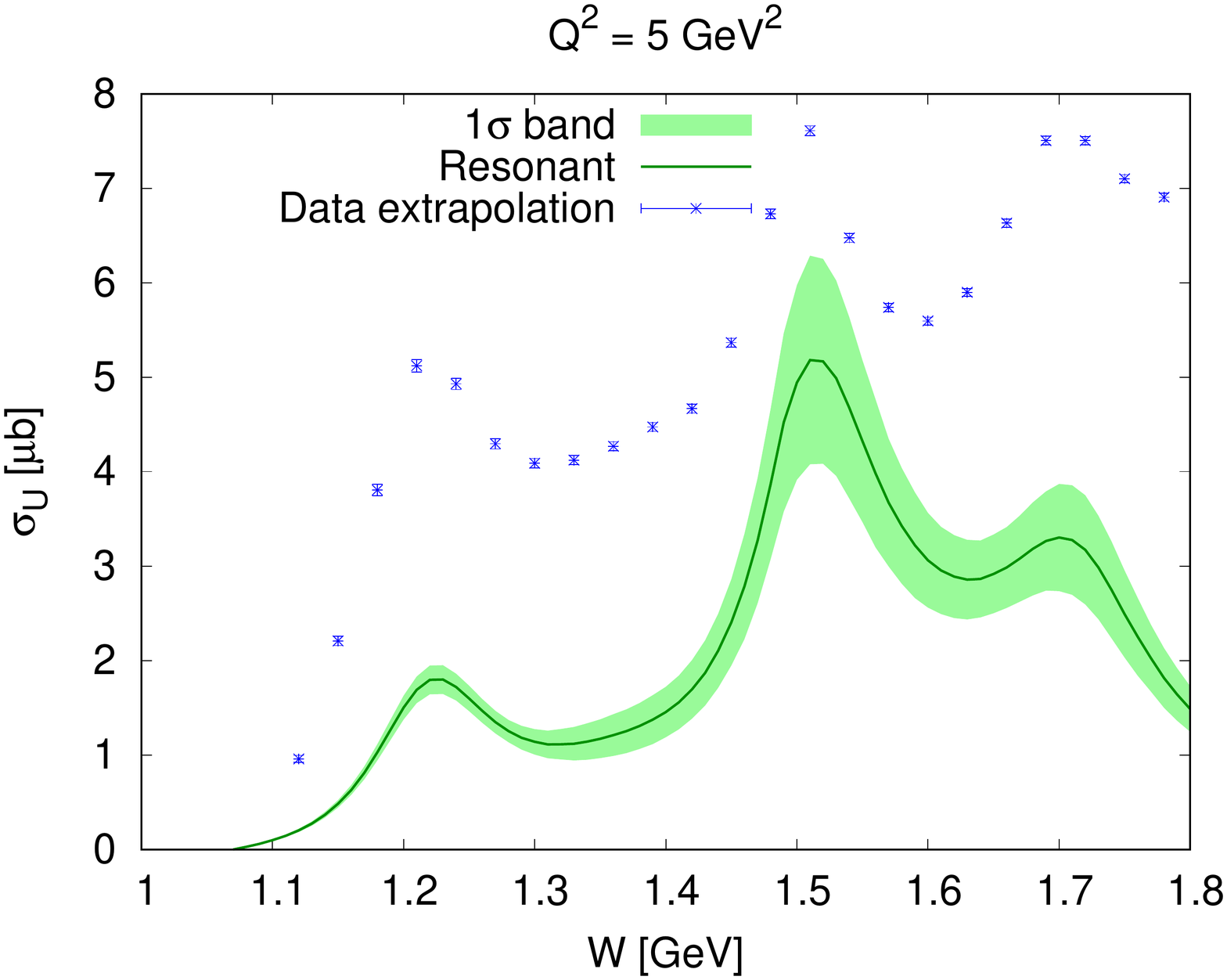}
\includegraphics[width=0.4\textwidth]{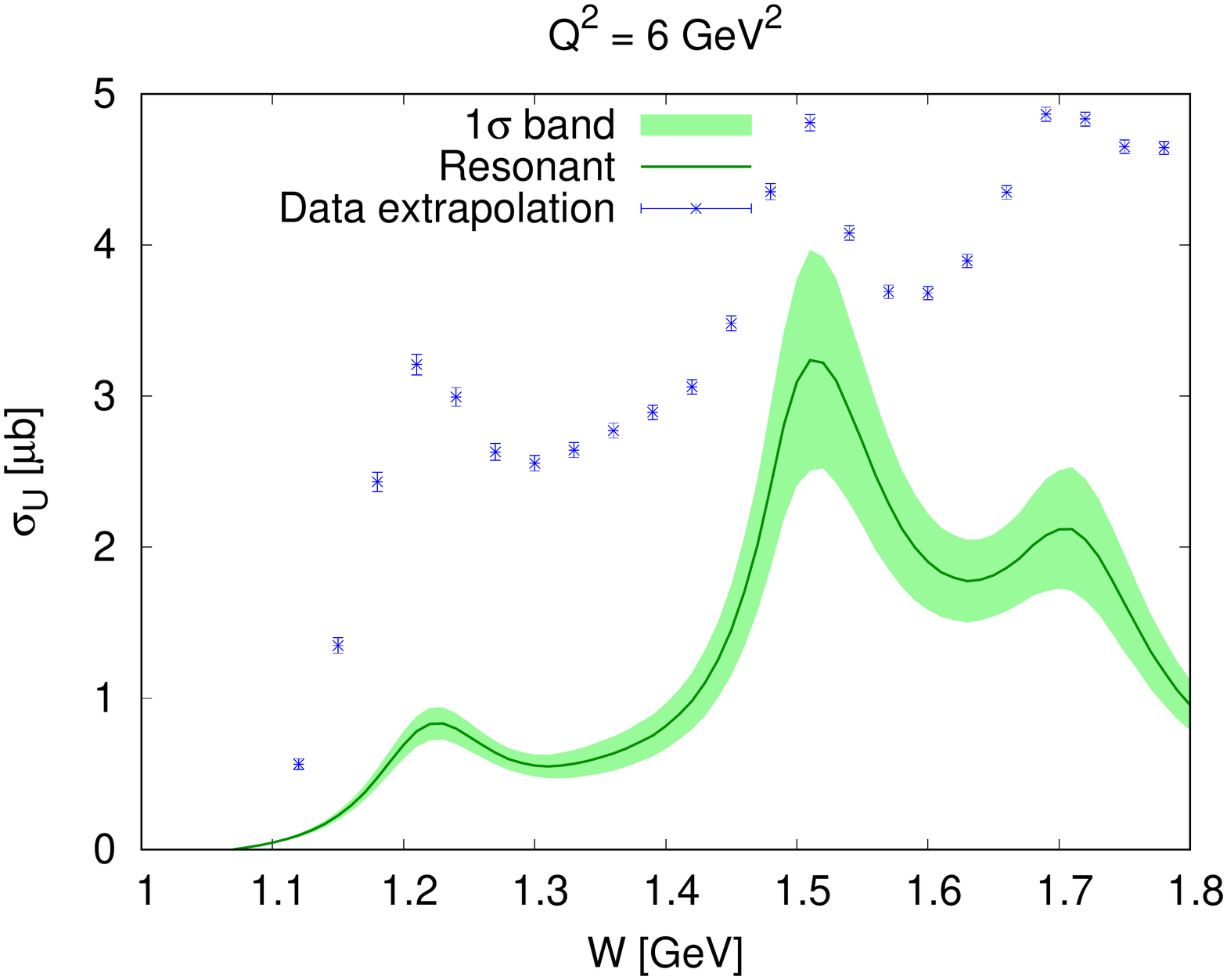}
\caption{Resonant contributions (green) to the unpolarized $\sigma_U$ virtual photon-proton cross sections, for an electron beam energy of $10.6\gev$, compared to the predicted inclusive virtual photon-proton cross sections in the kinematic area covered in the measurements with the CLAS12 detector~\cite{Burkert:2018nvj}. Figures from~\cite{HillerBlin:2019jgp}.}
\label{fig:sigU}
\end{figure}

Representative examples of $F_2(W,Q^2)$ and $F_L(W,Q^2)$ are shown in \figurename{~\ref{fig:F2sing}}.
Three distinct peaks are clearly seen in their $W$ dependencies and related to the resonant contributions.  In the first resonance region, the contribution from the $\Delta(1232)\,3/2^+$ decreases rapidly with $Q^2$, so that at $Q^2>2\gevsq$ the tail from the $N(1440)\,1/2^+$ state becomes essential. This is even more drastically so for the longitudinal $F_L$. In the second resonance region, the $N(1520)\,3/2^-$ and $N(1535)\,1/2^-$ give the largest contributions to $F_2$ and the contribution from the $N(1535)\,1/2^-$ becomes dominant as $Q^2$ increases. Additionally, the tail from $\Delta(1700)\,3/2^-$ becomes the main contribution to $F_L$ as $Q^2$ increases. Finally, the  peak in the third resonance region is generated by contributions from several resonances, one of the largest stemming from the $N'(1720)\,3/2^+$ state discovered recently in combined studies of $\pi^+ \pi^- p$ photo- and electroproduction at \jlab~\cite{Mokeev:2020hhu}.
Because of the intricate interplay with other resonances, the evolution with $Q^2$ of the third  peak in $F_2$ becomes rather involved, and the contribution from the $\Delta(1700)\,3/2^-$ dominates the resonant part at $Q^2 \sim 4\gevsq$.
This behavior suggests that further insight can be gained into its structure in the range of high $Q^2 > 4\gevsq$, which will be covered in future nucleon resonance studies with the CLAS12 detector~\cite{Carman:2020qmb, Brodsky:2020vco}. Therefore, in \figurename{~\ref{fig:sigU}} we also show the unpolarized inclusive cross section~\cite{Christy:2007ve}
 $\sigma_U(W,Q^2)$ as predicted for CLAS12 kinematics.
 
\begin{figure}[t]
\centering
\includegraphics[width=0.8\textwidth]{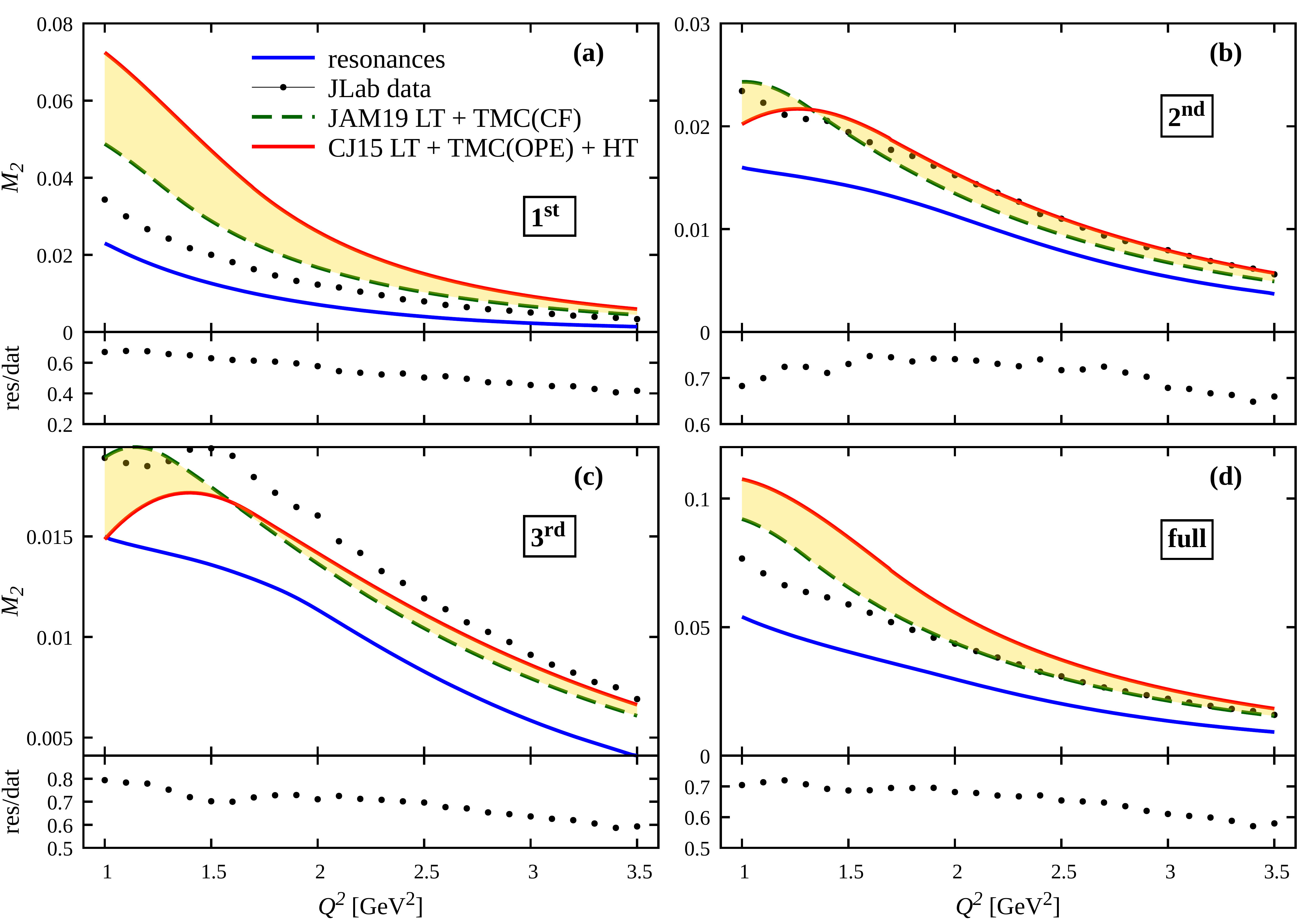}
\caption{Truncated moments $M_2$ of the $F_2$ structure function versus $Q^2$ for the 
three resonance regions, as well as 
for the full range $1.07<W<1.75\gev$. The moments from the experimental results~\cite{CLAS:SFDB} (black circles, with uncertainties smaller than the circle sizes) are compared with the resonant contributions (blue lines) and the structure function moments computed from the JAM19~\cite{Sato:2019yez} (green lines) and CJ15~\cite{Accardi:2016qay} (red lines) PDFs, with the latter including also higher twist terms. Also shown beneath each panel is the ratio of the resonant contributions to the data in each $W$ region. Figures from~\cite{Blin:2021twt}.}
\label{fig:F2trunc}
\end{figure}  

More generally, the pronounced differences seen in the $Q^2$-evolution of the three peaks are related to the different $Q^2$ evolutions of the electroexcitation amplitudes of the contributing resonances. Therefore, a credible evaluation of the resonant contributions relies essentially upon the knowledge of the electroexcitation amplitudes of all prominent resonances in the entire mass and $Q^2$ region under study. The extraction of reliable information from the experimental data on exclusive meson electroproduction will extend the capability of gaining insight into the nucleon parton distribution functions (PDFs) at large $x$ within the resonance excitation region. It also allows us to explore quark-hadron duality.  

We quantify the duality by considering the lowest truncated moment of $F_2$, which in an interval $\Delta x \equiv x_{\rm max} - x_{\rm min}$ at fixed $Q^2$ is given by
\begin{align}
\label{moments}
    M_{2}(x_{\rm min}, x_{\rm max}; Q^2)
    = \int_{x_{\rm min}}^{x_{\rm max}} \mathrm{d}x\, F_{2}(x,Q^2).
\end{align}
In \figurename{~\ref{fig:F2trunc}}, we compare the empirical moments with the moments of structure functions computed from the CJ15~\cite{Accardi:2016qay} and JAM19~\cite{Sato:2019yez} PDFs extrapolated from higher $W$.
The differences between parametrizations can be interpreted as systematic theoretical uncertainties associated with the extrapolations.\footnote{To run these extrapolations to such low values of $Q^2$, one has to include the leading $\mathcal{O}\!\left(\Lambda_\text{QCD}^2 / Q^2\right)$ corrections. The target mass correction is included in both extrapolations, while an estimate of next-to-leading twist operators is included in CJ15 only. An improvement of these estimates, as well as including higher twist terms, will improve the agreement between the extrapolations and data.} For the moments evaluated for the entire resonance region, from the pion threshold to $W = 1.75\gev$, there is reasonable agreement within uncertainties between the experimental data and the extrapolations from the DIS region for $Q^2 \gtrsim 2\gevsq$.
A similar agreement is observed in the second resonance region down to even smaller $Q^2$ values.
In the third region the extrapolated results generally underestimate the data by $\sim 10\%-30\%$, while in the first region the extrapolated results overestimate the data at all $Q^2$ considered.
Interestingly, the ratio of the resonance contributions to the truncated moments relative to the total remains fairly constant across the range of $Q^2$ considered. This suggests a similar $Q^2$ evolution of the resonant and nonresonant contributions to the structure function, thus pointing to a nonvanishing relative resonant \vs nonresonant size, even at larger $Q^2$.

Definitive conclusions about the longitudinal truncated moments are more difficult to draw on account of the greater systematic uncertainties associated with the experimental data extraction and the theoretical analysis prescriptions, motivating the need to complete the understanding of the leading and higher twist contributions to $F_L$, as well as of obtaining $L/T$ separated data in the resonance region. 

On the experimental side, our results motivate extensions of the inclusive electron scattering studies in the resonance region towards $Q^2 > 4\gevsq$, as well as the extraction of the $\gamma^* p N^*$ electrocouplings at high photon virtualities from the exclusive meson electroproduction data~\cite{Carman:2020qmb,Brodsky:2020vco}. Furthermore, the results suggest the intriguing future avenue of simultaneously describing the resonance and DIS regions, thus providing constraints for nucleon PDFs at large values of $x$~\cite{Melnitchouk:2005zr}.
A further extension is to explore the spin dependence of the exclusive-inclusive duality by analyzing the spin-dependent $g_1$ and $g_2$ structure functions.

\subsection{Regge theory and global fits} 
\label{sec:regge}

As mentioned, Reggeons and resonances are dual and not additive, so one has  to be careful when involving  both in amplitude analysis.   
Before exploring the applications of duality in the following sections, we first review the basic of the Regge theory and its recent applications in the analysis of modern experiments. 

Analyticity in angular momentum requires that singularities of partial waves are not independent, but rather connected by an analytic function called {\em trajectory}. One can show that indeed that poles in the complex angular momentum (Regge poles or {\em Reggeons}) correspond to the existence of an infinite tower of resonances of increasing mass and spin, as the ones described in Section~\ref{sec:baryons_and_hyperons} (see for example~\cite{Gribov:2003nw,Gribov:2009zz,Collins:1977jy}).
The contribution of a single Reggeon to a \twototwo process in the high-energy limit $s \gg -t$ can be written as 
\begin{align} \label{eq:Regge_pole}
A_{\substack{\lambda_b, \lambda_M \\ \lambda_N, \lambda'_N}}(s,t) = \beta_{\lambda_\gamma, \lambda_M}(t) \left[ \frac{\tau + e^{-i \pi \alpha(t)}}{2 \sin\pi\alpha(t)}  \left(\frac{s}{s_0}\right)^{\alpha(t)} \right]  \beta_{\lambda_N, \lambda'_N}(t)\,,
\end{align}
where $\lambda_{b}, \lambda_M, \lambda_N$ and $\lambda_N'$ are the helicities of the beam, produced meson, target and recoil, respectively. The trajectory $\alpha(t)$ gives the pole position of a resonance of given spin $J_R$ solving $\alpha(m_R^2) = J_R$. Indeed, in the vicinity of $t\simeq m_R^2$, the formula can be expanded
\begin{align} \label{eq:Regge_pole_expanded}
A_{\substack{\lambda_b, \lambda_M \\ \lambda_N, \lambda'_N}}(s,t \simeq m_R^2) = \beta_{\lambda_\gamma, \lambda_M}(m_R^2) \left[ \frac{1 + \tau(-1)^{J_R}}{2 \pi \alpha'(m_R^2)}  \left(\frac{s}{s_0}\right)^{J_R} \right] \frac{1}{t - m_R^2} \beta_{\lambda_N, \lambda'_N}(m_R^2)+\text{regular}\,,
\end{align}
which explains that Eq.~\eqref{eq:Regge_pole} contains a tower of increasing spin and mass, as given by the trajectory crossing integer numbers for mesons (or semi-integers for baryons).
The signature of the Reggeon is $\tau = (-1)^J$ with $J$ representing the spin of the lightest particle on the trajectory. Because of the factor $1 + \tau(-1)^{J_R}$, a trajectory can only contain odd or even spins. It is customary to refer to a trajectory with the name of the lightest particle laying on it. Vectors and tensors satisfying $P(-1)^J = +1$ are called ``natural exchanges'', while pseudoscalars and axial-vectors  with  $P(-1)^J = -1$ are denoted as ``unnatural exchanges.'' 
 In the following, we consider $\alpha(t)$ to be real functions, as their imaginary parts are connected to resonance widths and give subleading effects when describing meson production. 
The effects of the widths and their relation to the nature of light baryon states are studied in Section~\ref{sec:baryons_and_hyperons}.

\begin{figure}[t]
\centering
	\includegraphics[align=c,width=.5\textwidth]{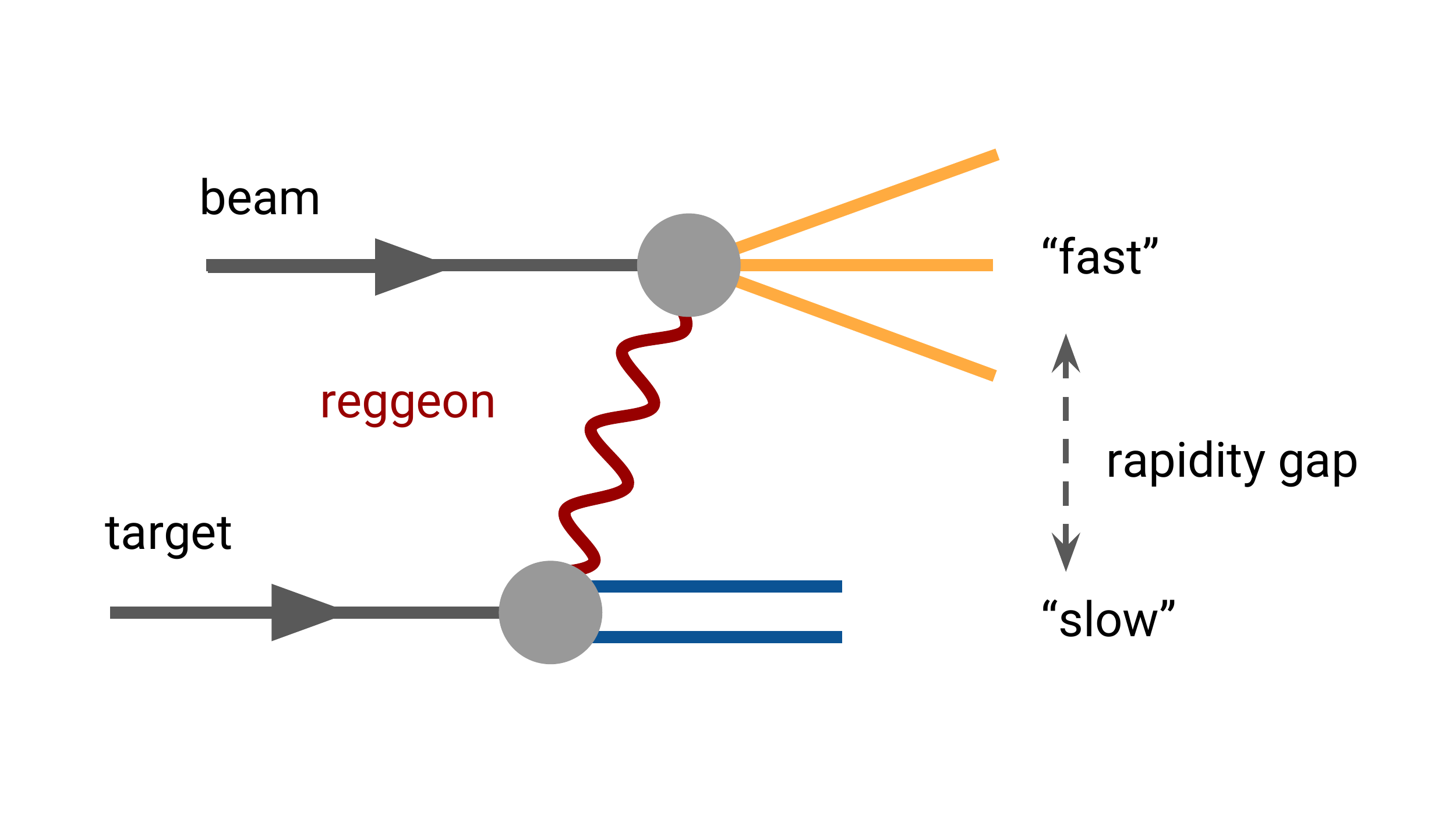}  \hspace*{.2in}
	\includegraphics[align=c,width=.4\textwidth]{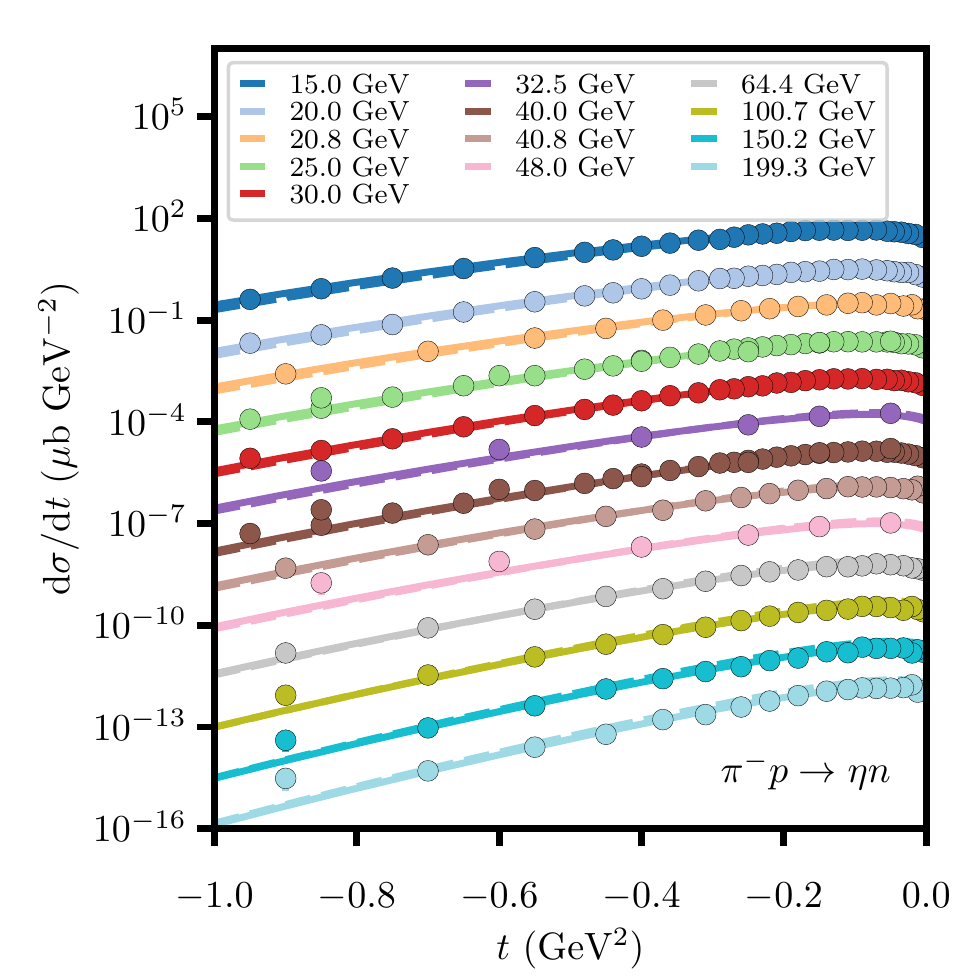}
	\caption{(left) Production of meson via Regge exchange (Reggeon). Since the momentum transferred is small, in a fixed-target experiment the fragments of beam and target will be fast and slow, respectively, in the lab frame. (right) $\pi^- p \to \eta n$ differential cross sections. Figures from~\cite{Nys:2018vck}.  \label{fig:Regge-prod} }
\end{figure}

Meson production at high energies can thus be explained in terms of Regge exchanges whose quantum numbers determine the energy dependence of the reaction from its trajectory $\alpha(t)$. As shown in Eq.~\eqref{eq:Regge_pole} and represented in \figurename{~\ref{fig:Regge-prod}}, the production amplitude factorizes into a top (beam-meson) and a bottom (target-recoil) vertex.

The total production amplitude will be a sum over Regge poles $R$~\eqref{eq:Regge_pole}, that at large energies behave like $s^{\alpha(t)}$. The sum is thus dominated by the exchanges $R$ having largest values of $\alpha_R(t)$ in the kinematic domain of interest. Since, in practice, trajectories are approximately linear $\alpha(t) = \alpha_0  + \alpha' t$, for small $t$ the production mechanism is dominated by the trajectory having the highest intercept $\alpha_0$.
In processes having exotic quantum numbers in the $s$-channel such as $p p$ or $\pi^+\pi^+$ elastic scattering, tensor and vector trajectories are forced to cancel each other since there is no direct resonance produced to be dual to. These vector and tensor exchanges are called degenerate. Exchange degeneracy (EXD) requires not only that vector and tensor trajectories $\alpha(t)$, but also couplings to the exotic channel $\beta(t)$ be equal. Similarly to the natural parity exchange (vector and tensor), one can show that unnatural parity exchange (pseudoscalar and axial-vector trajectories) are also degenerate. There are thus only two different Regge trajectories $\alpha(t)$ for Reggeons built out of mesons.

Natural exchanges have a larger intercept ($\alpha_0 \simeq 0.5$) than unnatural ones ($\alpha_0 \simeq 0$). These values lead to cross sections that decrease with energy, and cannot explain why the total cross section $pp\to \text{anything}$ slightly rises with energy. It has been postulated that a {\em Pomeron} trajectory ($\mathbb P$) with an intercept close to unity and having the quantum numbers of the vacuum is responsible for this phenomena. This trajectory is in principle related to the existence of purely gluonic particles, as glueballs  with $C=+$. Glueballs with $C=-$ would instead lie on the {\em Odderon} trajectory, which is responsible of the difference between $pp$ and $p\bar p$ total cross section at high energies, and whose existence is been recently debated~\cite{TOTEM:2020zzr,Donnachie:2019ciz,Cudell:2019mbe,Osterberg:2022qoy}.
The $t$-channel quantum number for various reactions are listed in \tablename{~\ref{tab:Regge_exchange}}, and their Regge trajectories $\alpha(t)$ are listed in \tablename{~\ref{tab:trajectories}}.

\begin{table}[b]
\caption{Regge trajectories of the Pomeron, and of other Regge exchanges.
Appropriate units of $\gevnospace$ are understood. 
\label{tab:trajectories}}
\begin{center}
        \begin{tabular}{c|c}
        \hline
        \hline
            Exchange & Regge trajectory $\alpha(t)$
            \\
            \hline
            $\mathbb P$ & $1.08 + 0.2 t$ \\
           $\rho$, $\omega$, $a_2$, $f_2$ & $0.9(t-m_\rho^2)+1$ \\
           $\pi$, $b_1$, $h_1$, $a_1$, $f_1$ & $0.7(t-m_\pi^2)$ \\
           \hline
           \hline
        \end{tabular}
        \end{center}
\end{table}

For a given process, all quantities appearing in Eq.~\eqref{eq:Regge_pole} are known except the couplings $\beta(t)$. These can be approximated to constants, once the $t$-dependence at small angles is explicitly factored out. 
Conservation of angular momentum implies that, in the forward direction,
\begin{align} \label{eq:small_t_general}
	A_{\substack{\lambda_b, \lambda_M \\ \lambda_N, \lambda'_N}}(s,t) \propto \left(\sqrt{-t'} \right)^{|\lambda_b - \lambda_M + \lambda'_N - \lambda_N|} \, .
\end{align}
Where we have defined $t' = t- t_\text{min}$ with $t_\text{min} = t(\theta = 0)$.   
The factorized form of the production amplitude implies a stronger constraint
\begin{align} \label{eq:small_t_factorized}
	A_{\substack{\lambda_b, \lambda_M \\ \lambda_N, \lambda'_N}}(s,t) \propto \left(\sqrt{-t'} \right)^{|\lambda_b - \lambda_M| + |\lambda'_N - \lambda_N|} \, .
\end{align}
By imposing Eq.~\eqref{eq:small_t_factorized} at the amplitude level, we can thus check whether mesons are photoproduced diffractively by comparing the $t$-dependence of the data and the model. As said, the constraints in Eq.~\eqref{eq:small_t_factorized} are only valid near the forward direction. Away from this limit, there are corrections of $\mathcal{O}(-t'/(m+m_\text{beam})^2)$, where $m$ is the mass of the produced meson and $m_\text{beam}$ is the mass of the beam.

The most recent and comprehensive study aimed at establishing the role of Reggeons in quasi-elastic two-body scattering with pion and kaon beams was performed in~\cite{Nys:2018vck}.
It was established that the leading Regge poles which give the high-energy  asymptotic behavior of scattering amplitudes (\ie poles with the largest intercept) indeed dominate the charge exchange reactions already for $p_\text{beam} > 5\gev$ . As can be seen in \figurename{~\ref{fig:Regge-prod}}, the model matches perfectly data across a wide energy range. 
Subleading effects include poles with lower intercept ({\em daughter trajectories}), or branch cuts in complex angular momentum, for example Reggeon-Pomeron boxes that model final-state interactions, \aka absorption. These effects are mainly visible when the leading amplitude vanishes, and can also contribute significantly to polarization observables, or in specific cases of pion exchange~\cite{Mathieu:2015eia, Mathieu:2015gxa,Nys:2016vjz,Mathieu:2018mjw}. The latter has indeed long range and can be significantly affected by final-state interactions.

\subsection{Single meson photoproduction}
\label{sec:photo}
\begin{figure}[t]
\begin{center}
	\includegraphics[width=0.9\textwidth]{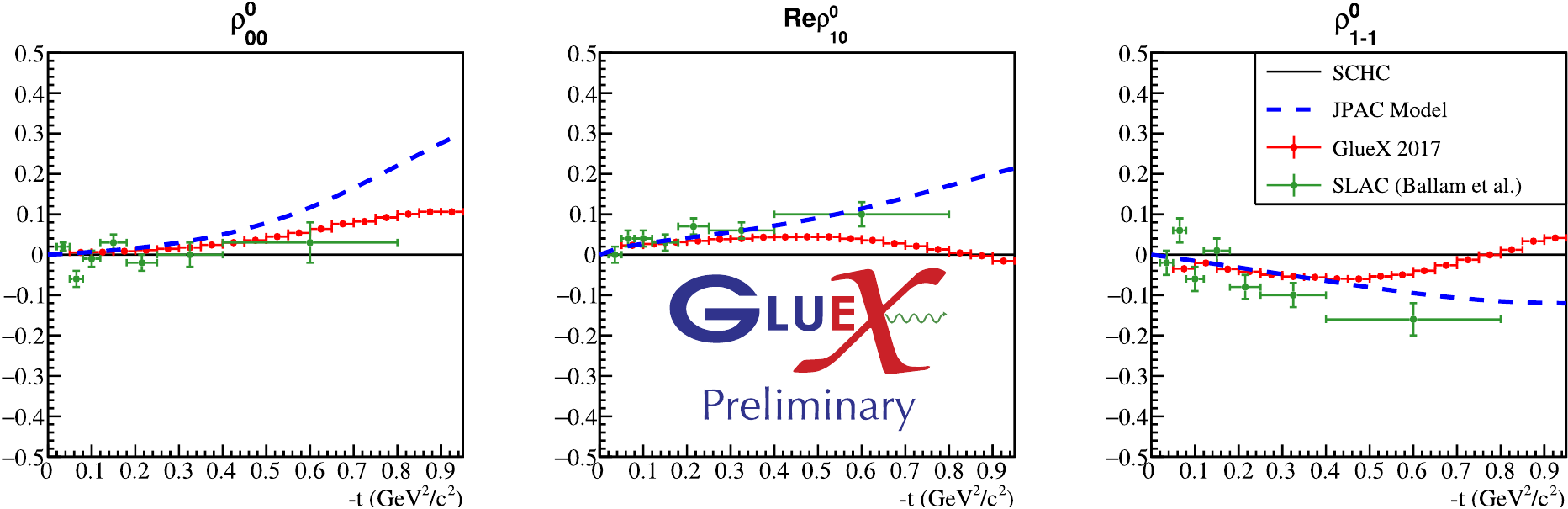}
\end{center}
\caption{Example of spin density matrix elements in $\rho^0 p$ photoproduction in the helicity frame. Comparison between GlueX preliminary data and the model of Ref.~\cite{Mathieu:2018xyc}. Figure from~\cite{Austregesilo:2019tld}. \label{fig:rho-sdme} }
\end{figure}

\begin{table}[t]
\caption{Regge exchanges in single meson photoproduction on a nucleon target. \label{tab:Regge_exchange}}
\begin{center}
        \begin{tabular}{c | c c c}
        \hline
        \hline
            Final state & Natural Ex. & Unnatural Ex. & Refs.
            \\
            \hline
            $\pi^0 p$, $\eta p$, $\eta' p$ & $\rho$, $\omega$ & $b_1$, $h_1$ & \cite{Mathieu:2015eia, Nys:2016vjz, Mathieu:2017jjs}  \\
            $\rho^0 p$, $\omega p$, $\phi p$ & $\mathbb{P}$, $a_2$, $f_2$ & $\pi^0$, $\eta$, $a_1$, $f_1$ & \cite{Mathieu:2018xyc} \\
            $a_2^0 p$, $f_2 p$ & $\rho$, $\omega$ & $b_1$, $h_1$ & \cite{Mathieu:2020zpm} \\
            $\pi^+ n$, $\pi^+ \Delta^{0}$, $\pi^- \Delta^{++}$ & $\rho$, $a_2$ & $\pi^\pm$, $b_1$ & \cite{JointPhysicsAnalysisCenter:2017del}\\
        \hline
        \hline
        \end{tabular}
        \end{center}
\end{table}

The upgrade of the \jlab facility has opened a new area for meson photoproduction~\cite{Dudek:2012vr}. With a $12\gev$ electron beam, mesons with a mass up to $\sim 4\gev$ are expected to be produced, as depicted on \figurename{~\ref{fig:Regge-prod}}. 

The GlueX and CLAS detectors are developing a rich meson spectroscopy program, including the study of exotic mesons and of other excited resonances, produced with a real  and quasi-real photon beam, respectively~\cite{battaglieri2005meson,battaglieri2011meson,GlueX:2015eeb}. However, before undertaking the search of new hadrons, we need to establish the production mechanisms of known mesons. In particular we need to assess whether at these energies they are produced diffractively, \ie the production amplitude factorizes into a photon-meson vertex and a nucleon vertex. This is achieved by comparing diffractive models for pseudoscalar, vector and tensor meson photoproduction to data. The Regge exchanges contributing to photoproduction of mesons on a nucleon target are summarized in \tablename{~\ref{tab:Regge_exchange}}. In these sections, we present several results on single meson photoproduction and their comparison to data. The results are discussed in order of model sophistication.

In this context, the production of vector mesons is an ideal place to start. The $\omega$, the $\phi$ and, to some extent, the $\rho$ are narrow resonances and so are easy to  reconstruct experimentally. The angular distributions of their decay products are given by the Spin Density Matrix Elements (SDME), which are known quadratic combinations of the production amplitudes. 
We can thus confirm the diffractive nature of vector meson photoproduction by constructing a model based on Regge exchanges that incorporates the small $t$ behavior in Eq.~\eqref{eq:small_t_factorized}, and comparing with \jlab data. 
In Ref.~\cite{Mathieu:2018xyc}, we developed a model for photoproduction of light neutral vector mesons ($V=\rho^0$, $\omega$, $\phi$). We considered the only relevant unnatural exchange to be $\pi^0$, and determine the coupling from the radiative decay $V\to \gamma \pi^0$. The natural exchanges (Pomeron and tensors), have three distinct helicity couplings to the $\gamma V$ vertex. These are denoted by the difference between the beam helicity and the vector meson one, and called non-flip, single-flip and double-flip couplings. 
For tensor exchange, the three couplings are related to the partial waves of the radiative decay $T\to V \gamma$ (where $T = f_2$, $a_2$), and in principle is accessible experimentally. In the absence of such information, we extracted their relative weights from the SLAC measurement of vector meson SDME from Ref.~\cite{Ballam:1972eq}. In the $s$-channel center-of-mass frame, the Pomeron is assumed to be helicity conserving. 
A year after the publication the model, GlueX has presented the preliminary version of the $\rho^0$ SDME~\cite{Austregesilo:2019tld}. In \figurename{~\ref{fig:rho-sdme}} we show that the comparison between prediction and data is excellent for $-t'< 0.5\gevnospace^2 \simeq m_{\rho}^2$, in agreement with the range of validity of the expansion \eqref{eq:small_t_factorized}. Data are compatible with the dominance of the helicity conserving Pomeron coupling, plus the addition of tensor exchanges with complete helicity structures.

In the case just discussed, unnatural exchanges turned out to be almost irrelevant. However, it is well known that pion exchange dominates at low $t$ in charge-exchange reactions such as $\gamma p \to \pi^- \Delta^{++}$. In single pseudoscalar photoproduction, the relative importance between natural and unnatural exchanges can be extracted from the beam asymmetry
\begin{align}
    \Sigma(t) & = \frac{\frac{d\sigma_\perp}{dt} - \frac{d\sigma_\parallel}{dt}}{\frac{d\sigma_\perp}{dt} + \frac{d\sigma_\parallel}{dt}}\, ,
\end{align}
where $\sigma_\perp$ ($\sigma_\parallel$) is the cross section for
photon beam with linear polarization, perpendicular (parallel)  polarization to the reaction plane.
At high energies, natural (unnatural) exchanges contribute only to $\sigma_\perp$ ($\sigma_\parallel$). Thus, positive (negative) $\Sigma$ implies the dominance of natural (unnatural) Reggeons. 

\begin{figure}[t]
\centering
	\includegraphics[width=.4\textwidth]{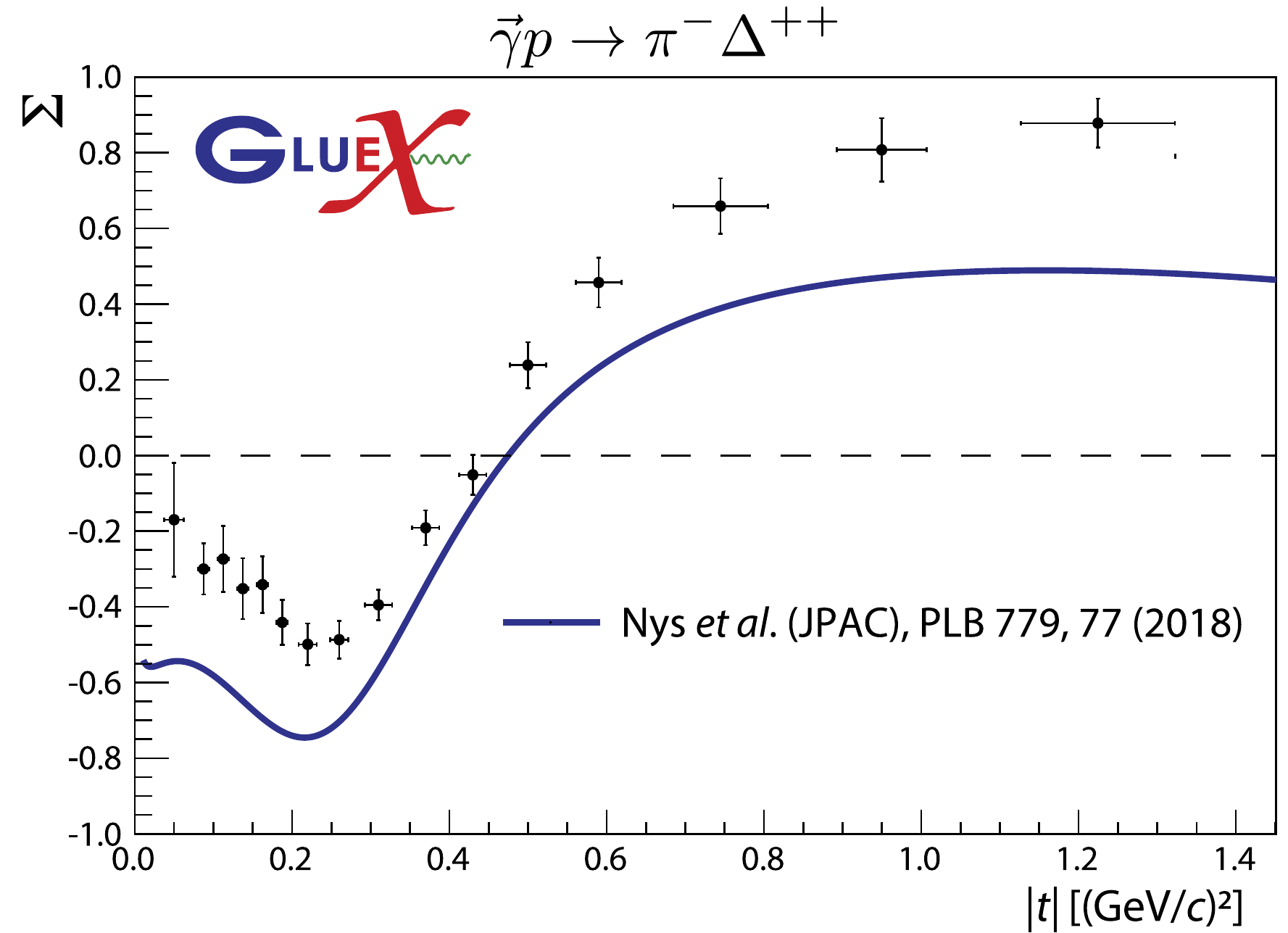} \hspace*{0.01cm}
	\raisebox{-0.04\height}{	\includegraphics[width=0.45\textwidth]{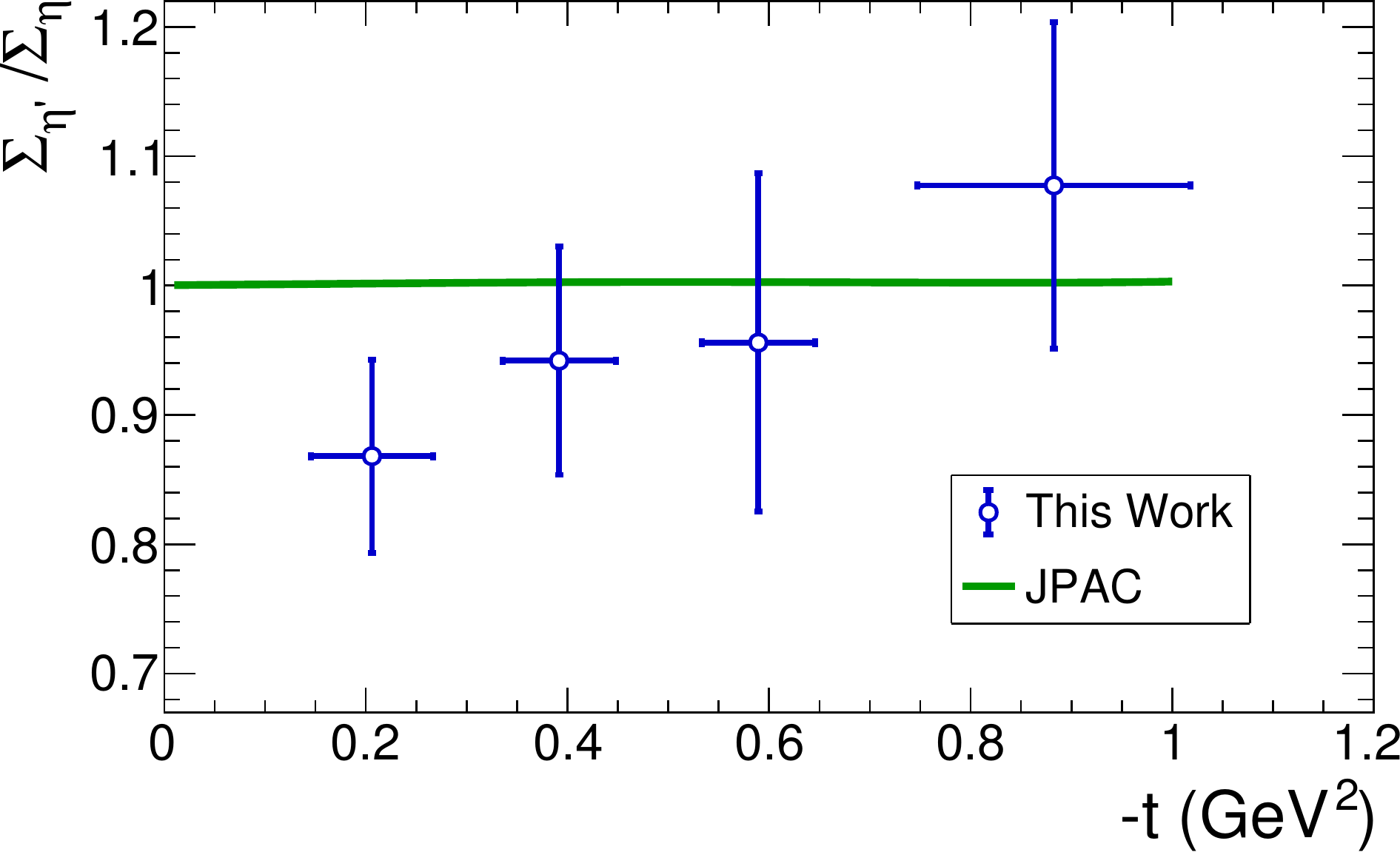}}
\caption{(left) Beam asymmetry in $\pi^- \Delta^{++}$ photoproduction. Comparison between GlueX data and the model in~\cite{JointPhysicsAnalysisCenter:2017del}. Figure from~\cite{GlueX:2020fam}. (right) $\eta/\eta'$ beam asymmetry ratio compared with the model of~\cite{Mathieu:2017jjs}. Figure from~\cite{GlueX:2019adl}. \label{fig:PiDelta-EtaBA}}
\end{figure}

One observes in \figurename{~\ref{fig:PiDelta-EtaBA}} that the beam asymmetry turns out to be negative at small $-t$, confirming the dominance of unnatural exchange (pion) in the forward region. Moreover the minimum of $\Sigma$ around $|t|\simeq 0.25\gevnospace^2$ has been predicted by the model in Ref.~\cite{JointPhysicsAnalysisCenter:2017del}, which includes $\rho$ and $a_2$ as dominant natural exchanges. As mentioned in Sec.~\ref{sec:regge},  pion exchange suffers from large absorption corrections. We used William's model also known as ``Poor's man absorption model", which provides a simple  prescription to take these corrections into account. Data on beam asymmetry tend to 1 faster than the model as $-t$ increases, indicating a stronger component of natural exchanges than expected. Nevertheless, the model describes correctly the gross features of the data. %The relative strength between the two different parity exchanges were determined on data in the range $|t|< 0.6\gevnospace^2$. 

\begin{figure}[!t]
\begin{center}
	\resizebox{\textwidth}{!}{\input{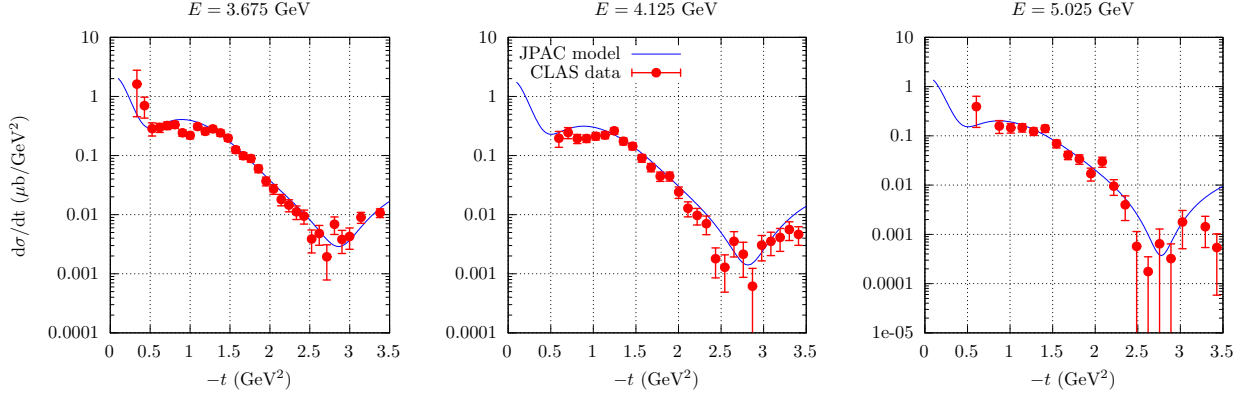}}
\end{center}
\caption{Differential cross section for Regge model 
at three energies compared to the 
$\gamma p \to \pi^0 p$
CLAS data~\cite{CLAS:2017kyf} 
Figure from~\cite{CLAS:2017kyf}. \label{fig:pi0-XS} }
\end{figure}

% pi0: Mathieu:2015eia
For large $-t$ values, corrections to the leading Regge poles, such as daughter trajectories or Regge cuts, might be important. %They were not included in the $\pi^- \Delta^{++}$ model. 
Their contribution are not easily derived theoretically, but can be estimated from data when available. In Ref.~\cite{Mathieu:2015eia}, we developed a model for $\gamma p \to \pi^0 p$ that includes Regge cuts, fitting to data measured at $E_\gamma = 6$--$15\gev$ in the range $-t< 1.5\gevnospace^2$.
The data display a dip at $t \simeq -0.5\gevnospace^2$ that is described in the model by including a zero in the vector Regge residue. 
This zero has a physical motivation, despite the curious name `nonsense wrong-signature zero' (NSWSZ). In tensor exchanges, Eq.~\eqref{eq:Regge_pole} has a scalar pole ($\alpha(t) = 0$) at negative mass squared $t \simeq -0.5\gevnospace^2$. Since this happens in the physical region, we have to add explicitly a zero to the residue to remove this unphysical pole. Because of the EXD discussed in Section~\ref{sec:regge}, the same zero appears in the vector trajectory as well.\footnote{The name `wrong signature' is due to the fact that a vector trajectory contains odd spins only, and usually does not bother about zeros and poles at even spin. Furthermore, the `nonsense' is due to the fact that, since the photon helicity is $\pm 1$, the minimum spin allowed in the $t$-channel is $J=1$. For more details see~\cite{Collins:1977jy}. Since this argument relies on exact EXD, it is not necessarily realized in nature. However, it gives a simple explanation for this kind of dips that occur in differential cross sections. Whether they exist or not, it must be seen by a comprehensive analysis of several reactions that allows us to disentangle the contributions of individual Regge exchanges. Alternatively, one can implement Finite Energy Sum Rules (FESR) to reconstruct the residues, as we will show in Section~\ref{sec:FESR}.} 
Regge cuts are parametrized in a similar way to Regge pole as in Eq.~\eqref{eq:Regge_pole}, but with a flatter trajectory $\alpha_c(t) \simeq 0.5 + 0.2 t$. In analogy with the Regge vector pole, we included a NSWSZ in the Regge cut couplings $\beta_c(t) \propto \alpha_c(t)$. Consequently, our model predicts a dip at $t \simeq -2.5\gevnospace^2$. New data from the CLAS detector~\cite{CLAS:2017kyf} in a wide $t$ range, displayed in \figurename{~\ref{fig:pi0-XS}}, confirm the presence of this dip at the same $t$ for several energies, and thus the presence of the NSWSZ in the Regge cut contribution to the production mechanism in $\pi^0$ photoproduction.

\begin{figure}%[htb]
\begin{center}
    \includegraphics[width=0.45\textwidth]{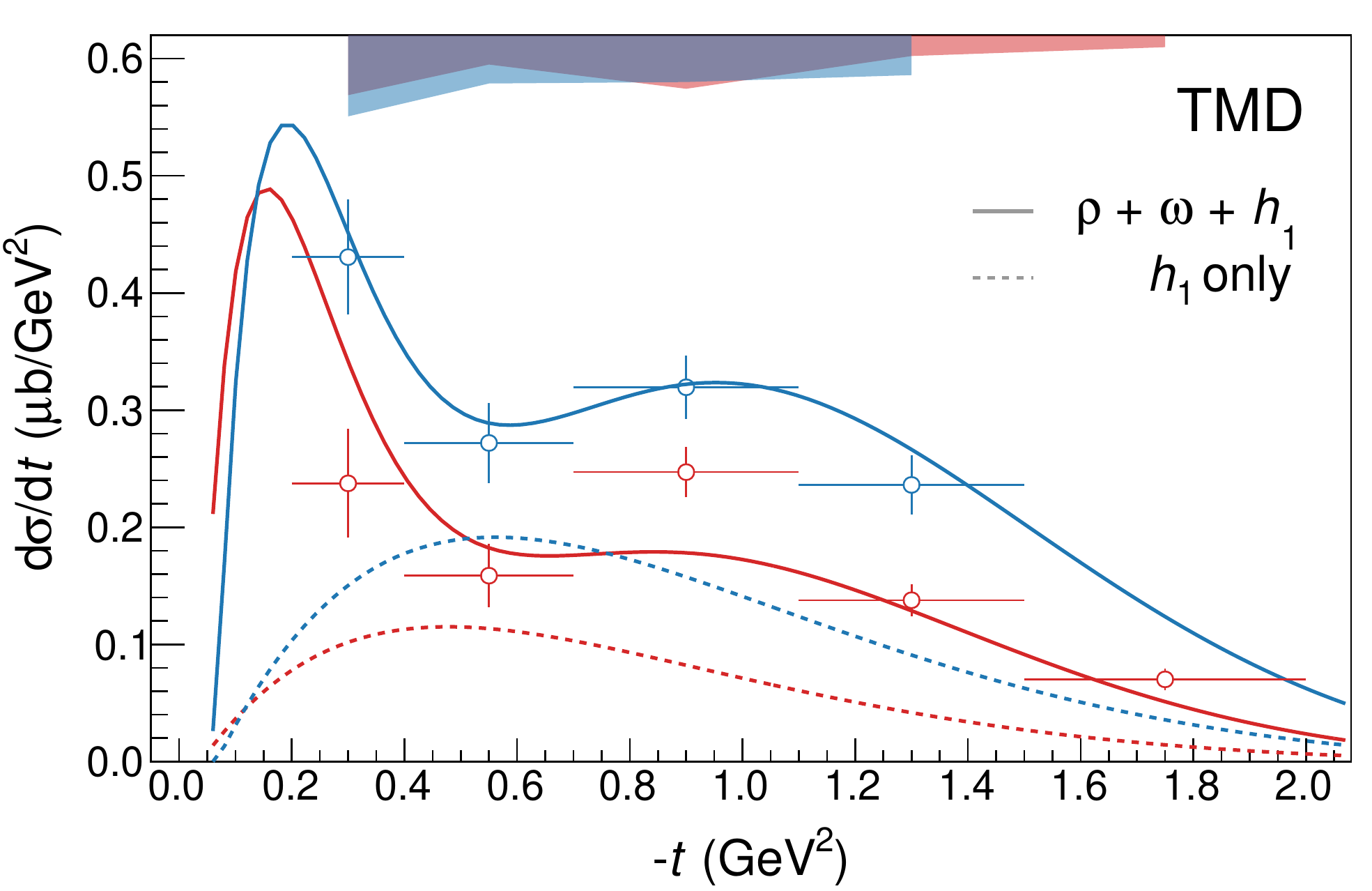} \hspace*{0.02cm}
	\raisebox{0.04\height}{\includegraphics[width=0.45\textwidth]{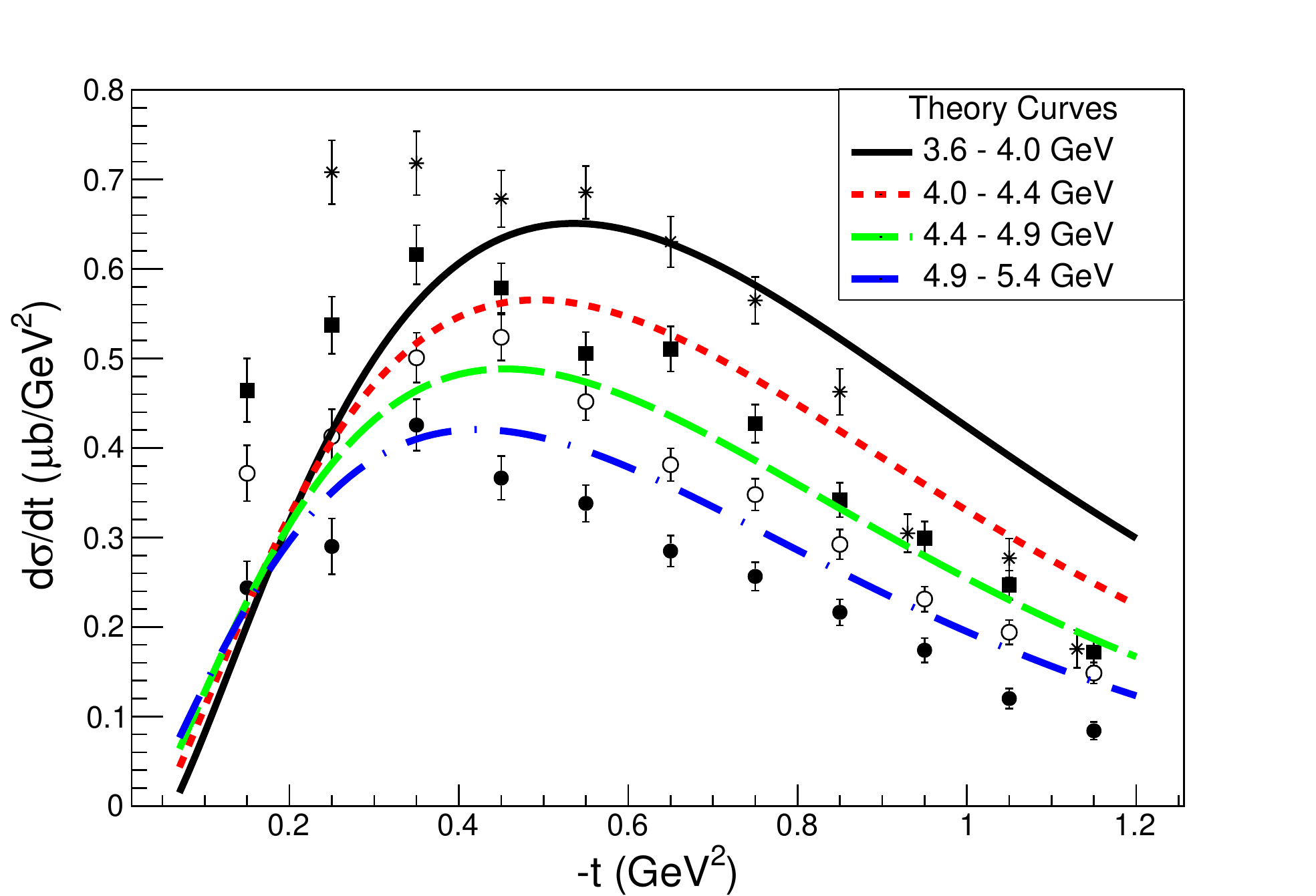}}
\end{center}
\caption{$a_2(1320)$ (left) and $f_2(1270)$ (right) differential cross section extracted by the CLAS collaboration~\cite{CLAS:2020ngl,CLAS:2020rdz} compared to the JPAC model of Ref.~\cite{Mathieu:2020zpm}. Figures from~\cite{Mathieu:2020zpm} (left) and~\cite{CLAS:2020ngl} (right). \label{fig:a2-f2-xs} }
\end{figure}

The presence of corrections to the leading pole approximation can be identified by comparing the beam asymmetries of $\eta$ and $\eta'$ photoproduction. Since Regge poles factorize, assuming only Regge exchanges and the absence of hidden strangeness terms as $\phi$ exchange leads to the equivalence of $\eta$ and $\eta'$ beam asymmetries. In Ref.~\cite{Mathieu:2017jjs}, we modeled the contribution of the $\phi$ exchange and obtained that the ratio of $\eta$ and $\eta'$ beam asymmetries would deviate from 1 by maximum $2\%$ in the range $-t<1\gevnospace^2$. This ratio, measured recently by the GlueX collaboration~\cite{GlueX:2019adl} and presented in \figurename{~\ref{fig:PiDelta-EtaBA}} (right), turned out to indeed be compatible with unity. However, the uncertainties are still quite large at about 10$\%$, and the data points are centered around a nominal value about $0.95$. One should then take this conclusions cautiously, as the data points could change significantly when more statistics are accumulated. 

Tensor mesons are photoproduced by the same $t$-channel quantum numbers as pseudoscalar mesons. Their differential cross sections indeed present the same pattern. That is, the isovector $\pi^0$ and $a_2$ cross section present a minimum around $t = -0.5\gevnospace^2$, while the isoscalar $\eta$ and $f_2$ cross sections do not, see \figurename{~\ref{fig:a2-f2-xs}}. Since the $\omega$ exchange dominates the isovector production, it is natural to associate the dip at $t = -0.5\gevnospace^2$ with a zero in the $\omega$ amplitudes. In our previous model of $\gamma p \to \pi^0 p$ we introduced a NSWSZ in both the $\omega$ and the $\rho$ amplitudes and explained that it was filled  by Regge cut contributions. In Ref.~\cite{Mathieu:2020zpm} we revised this hypothesis  and introduced the NSWSZ only in the $\omega$ production, so that the dip at $t=-0.5\gevnospace^2$ is filled by a nonvanishing $\rho$ contribution. In this model, the $\rho$ amplitude does not feature the NSWZ and lead to a nondipping shape of the $f_2$ differential cross section, in agreement with the recent measurements by the CLAS collaboration~\cite{CLAS:2020rdz,CLAS:2020ngl}.

\subsection{\texorpdfstring{Photoproduction of \jpsi and pentaquark searches}{Photoproduction of J/psi and pentaquark searches}}
\label{sec:jpsi_photo}
The use of photon beams to search for or confirm exotic hadrons is  appealing since it reduces the role of kinematic effects  
  and minimizes the model dependence of partial-wave analyses~\cite{Kubarovsky:2015aaa,Wang:2015jsa,Karliner:2015voa,HillerBlin:2016odx,Wang:2019krd}.

As discussed in Section~\ref{sec:heavy}, the LHCb data on $\Lambda^0_b \to \jpsi\,p\,K^-$ decay potentially indicate the existence of several baryon resonances in the $\jpsi\,p$ spectrum that do not fit predictions of the valence quark model~\cite{Aaij:2015tga,Aaij:2016phn,LHCb:2019kea}.
These states have the right mass to be produced directly at \jlab,  
   through a scan of the $J/\psi$ photoproduction  cross sections.
Searches proposed at Hall~C and CLAS12 are ongoing~\cite{Meziani:2016lhg,claspentaquark}, while the results by GlueX show no evidence of narrow peaks~\cite{Ali:2019lzf}.
With fits to these data existing so far, we could provide estimates for the upper limits of the pentaquark coupling sizes.
The quantum numbers are not reliably determined yet, so in order to provide an estimate we focus the discussion on the $P_c(4450)$ as determined in the older analysis~\cite{Aaij:2015tga}.

\begin{figure}[t]
\includegraphics[width=.50\textwidth]{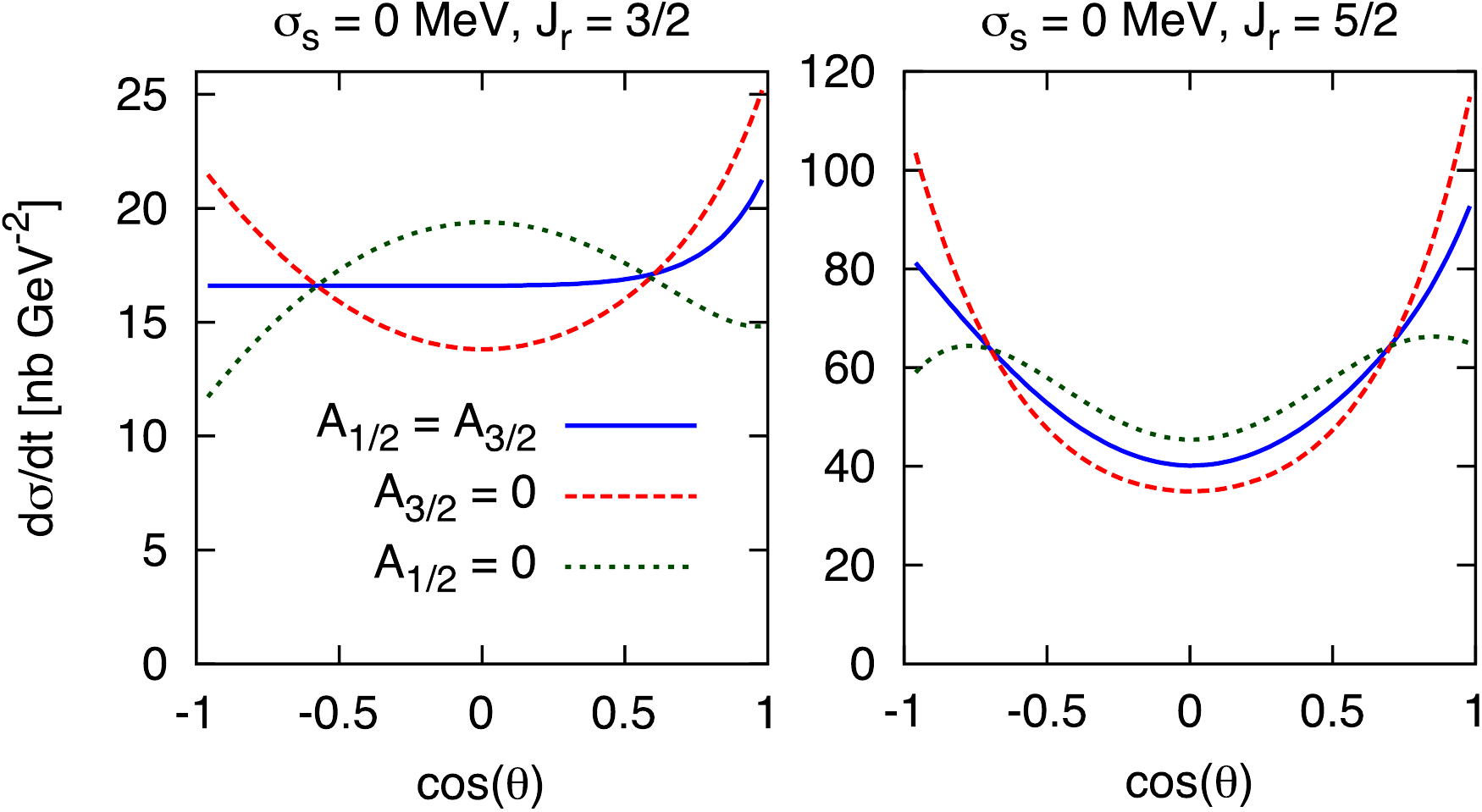} \hspace{.5cm}
\includegraphics[width=.44\textwidth]{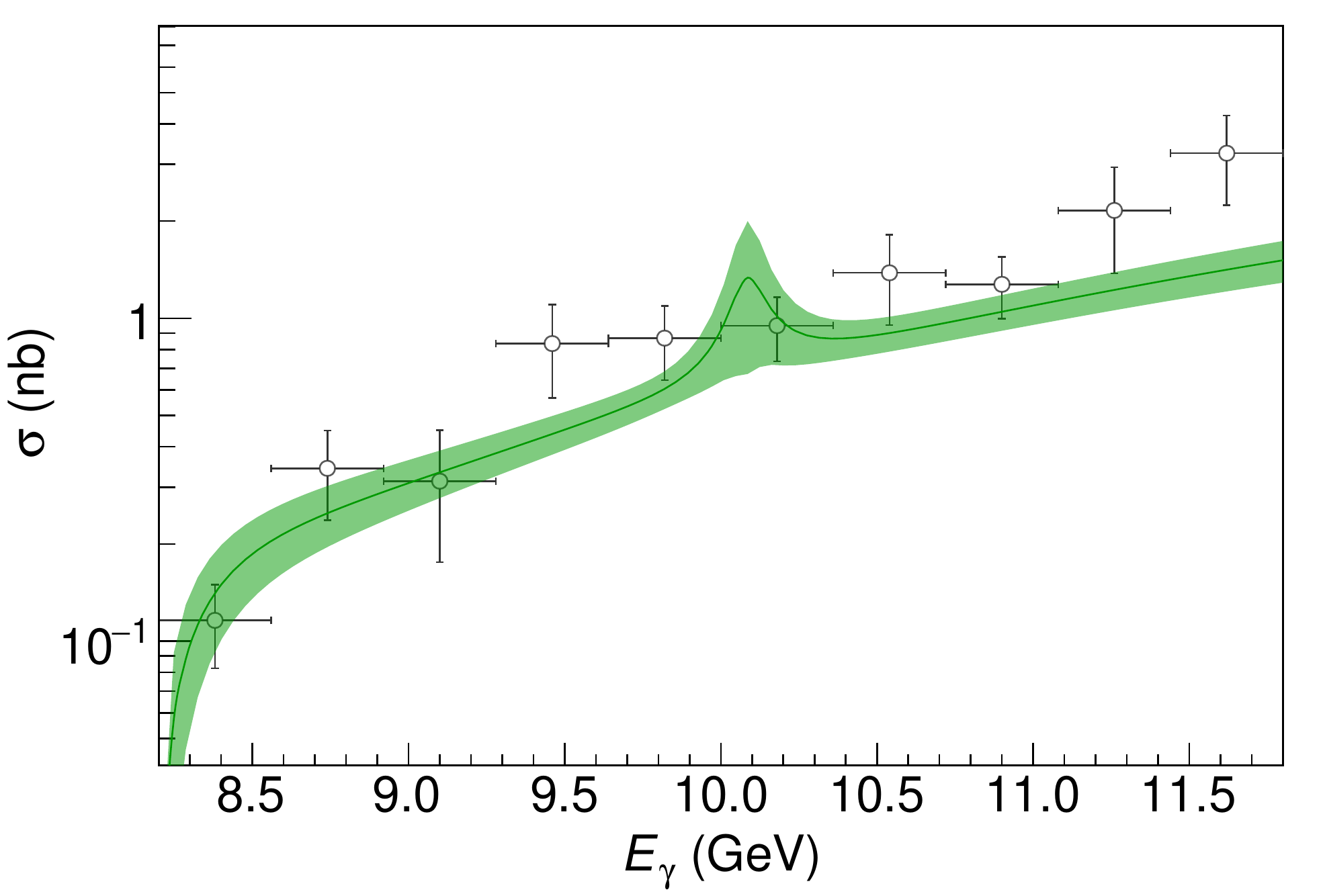}
\caption{(left) Examples of angular distributions of the pentaquark at the peak, depending on the spin-parity assignment and relative photocoupling sizes. (right) Fit to GlueX~\cite{Ali:2019lzf} data for a spin assignment of the $P_c(4450)$ $J^P=\frac{3}{2}^-$. Figures from~\cite{HillerBlin:2016odx} (left) and~\cite{Winney:2019edt} (right).}
\label{fig:dxsang}
\end{figure}

The data available in this channel is so scarce that there is no point in considering refined models. We describe the diffractive \jpsi production background with an effective Pomeron exchange~\cite{HillerBlin:2016odx,Winney:2019edt}.
We adopt the vector Pomeron model~\cite{WA102:1999poj,Lesniak:2003gf},
   \begin{align}
\left<\lambda_\psi \lambda_{p^\prime}|T_P|\lambda_\gamma \lambda_p\right> &= F(s,t) \, \bar{u}(p_f, \lambda_{p^\prime}) \gamma_\mu u(p_i, \lambda_p) [ \varepsilon^\mu(q, \lambda_\gamma)q^\nu - \varepsilon^\nu(q, \lambda_\gamma) q^\mu] \varepsilon^*_\nu(p_\psi, \lambda_\psi)\, ,
\end{align}
with 
\begin{equation}
   F(s,t) = iA \left(\frac{s - s_\text{th}}{s_0}\right)^{\alpha(t)} \frac{e^{b_0(t - t_\text{min})}}{s} \, ,
\end{equation}
that was successful in reproducing the azimuthal angular dependencies  (see Sections~\ref{sec:etapicompass} and~\ref{sec:doubleregge}). Since this is just an effective description, the Pomeron parameters are refitted to data.\footnote{An alternative microscopic description is given in~\cite{Du:2020bqj}.}
As said, since we do not need a refined description, instead of considering dual models that incorporate both resonances and Reggeons as in~\cite{Szczepaniak:2014bsa}, we simply add to the previous model a Breit-Wigner amplitude for the pentaquark,
   \begin{align}
   \left<\lambda_\psi\lambda_{p^\prime}|T_R|\lambda_\gamma \lambda_p\right>=f_{\textrm{th}}(s)&\frac{\left<\lambda_\psi \lambda_{p^\prime}|T_{\textrm{dec}}|\lambda_R\right>\left<\lambda_R|T_{\textrm{em}}^\dagger|\lambda_\gamma\lambda_p\right>}{M_R^2 - s - i \Gamma_RM_R}\,,
   \end{align}
 where $f_\text{th}(s)$ further suppresses the amplitude at threshold.  The strong decay $\left<\lambda_\psi \lambda_{p^\prime}|T_{\textrm{dec}}|\lambda_R\right>$ is determined by the spin-parity of the state, while $\left<\lambda_R|T_{\textrm{em}}^\dagger|\lambda_\gamma\lambda_p\right>$ depends on the unknown pentaquark photocouplings. According to 
Vector Meson Dominance (VMD), one can relate the two matrix elements assuming
 \begin{equation}
 \bra{\lambda_{\gamma}\lambda_p}T_\text{em} \ket{\lambda_R} =
 \frac{\sqrt{4\pi \alpha} f_\psi}{M_\psi} \bra{\lambda_{\psi}=\lambda_\gamma ,\lambda_p}T_\text{dec} \ket{\lambda_R},
 \label{eq:vmd}
\end{equation}
and with this provide an upper limit to the branching ratio of the $P_c(4450)$ to the final state where it is actually observed. A word of caution is in order: While in the light sector VMD gives reasonable results, having a photon so off-shell that it can oscillate into a heavy vector meson is questionable~\cite{Xu:2021mju}. Nevertheless, VMD predicts roughly the correct size of the ratio $\Gamma(\chi_{c2} \to \gamma \,J/\psi) / \Gamma(\chi_{c2} \to \gamma \gamma)$, so it seems an appropriate method to obtain at least order-of-magnitude estimates.
Before GlueX data were published, most of \jpsi photoproduction data were taken by HERA at higher energies~\cite{Chekanov:2002xi,Aktas:2005xu}, which we analyzed in the first publication~\cite{HillerBlin:2016odx}.
After GlueX data were made available, we noticed that our simple model is not able to fit consistently the low and high energy region. We therefore selected the data at $E_\gamma \lesssim 25\gev$ by GlueX and SLAC~\cite{Camerini:1975cy}.
The results  are summarized in \tablename{~\ref{tab:Pcfits}} for different spin-parity hypotheses~\cite{Winney:2019edt}, and a fit example is given in \figurename{~\ref{fig:dxsang}}.
   
\begin{figure}[t]
\includegraphics[width=.48\textwidth]{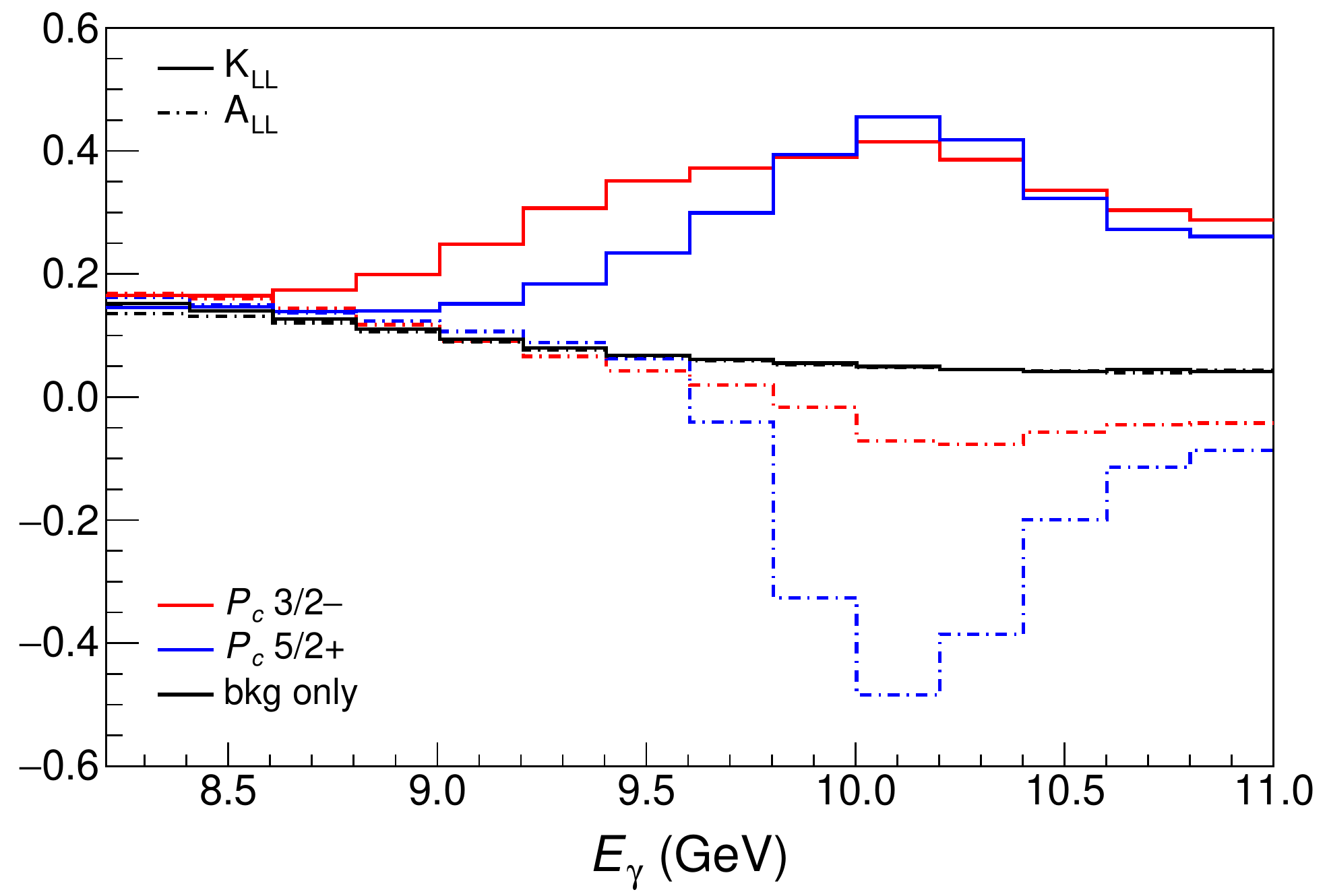}
\includegraphics[width=.48\textwidth]{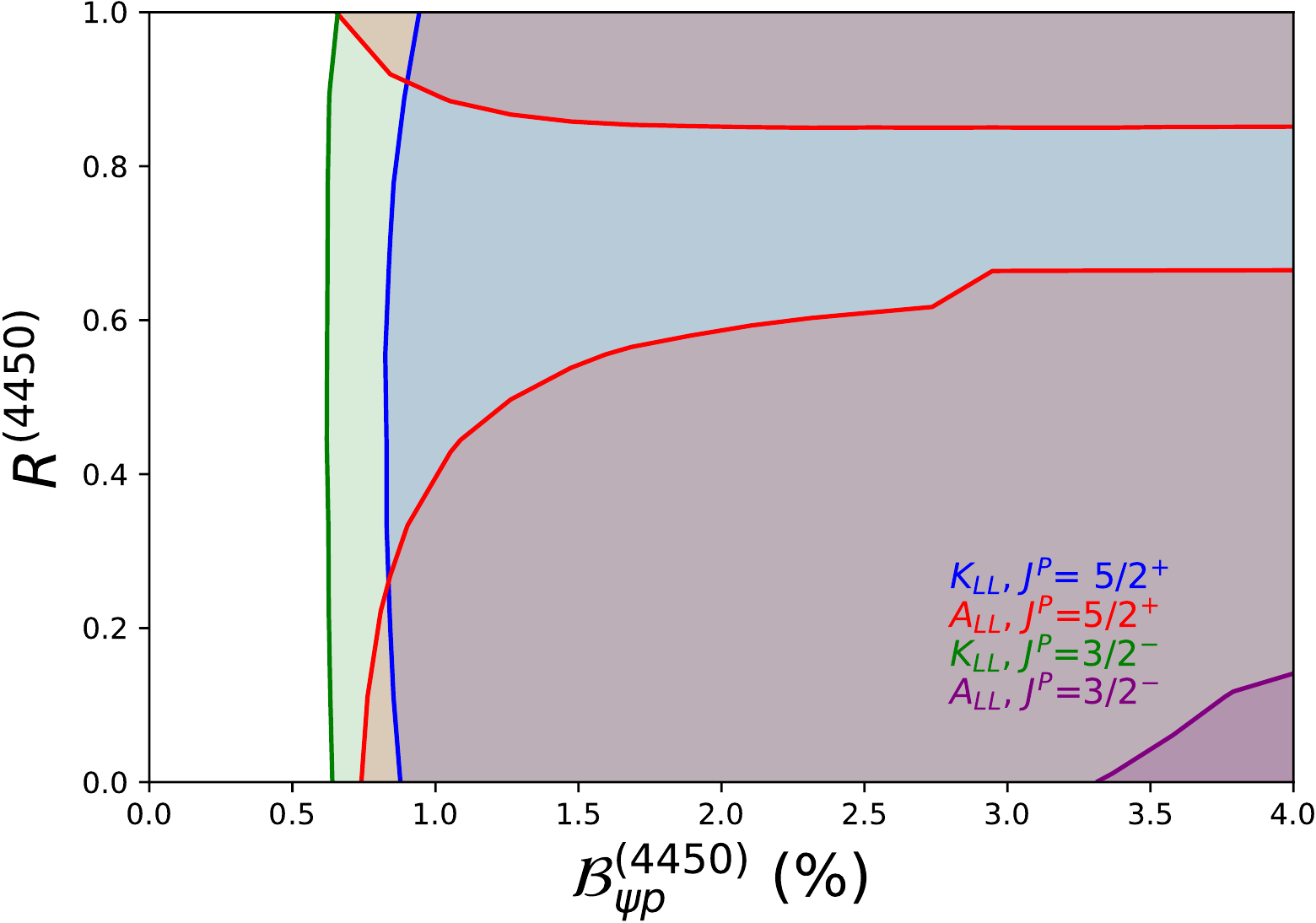}
\caption{(left) Predictions for \KLL (solid line) and \ALL (dash-dotted line) in the SBS acceptance in bins of energy in the presence of two pentaquarks, considering hadronic branching ratios $\mathcal{B}_{\psi p}^{(4450)} = \mathcal{B}_{\psi p}^{(4380)} = 1.3\%$, and photocouplings ratio $R^{(4450)} = 0.2$, $R^{(4380)} = 1/\sqrt{2}$ and an experimental resolution of $125\mev$. (right) $5\sigma$ sensitivity map of the dependence of the double polarization observables on the photocoupling ratio $R$ and the branching fraction $\mathcal{B}_{\psi p}$ of the $P_c(4450)$. Figures from~\cite{Winney:2019edt}.}
\label{fig:polarization}
\end{figure}
Furthermore, we could provide estimates of the angular distributions of the differential cross sections depending on the relative size of the photocouplings, see \figurename{~\ref{fig:dxsang}}. These kinds of studies will help pin down the quantum numbers of the pentaquarks if their signals are to be found in photoproduction experiments.

\begin{table}[b]
                \caption{Parameters of the fits for different $J^P$ assignments for the $P_c(4450)$ state. Uncertainties are at the 68\% confidence level,  except for the branching ratio, whose upper limit is quoted at 95\%. Table from~\cite{Winney:2019edt}.}\label{tab:Pcfits}
                \centering
        \begin{tabular}{c| c c c c }
        \hline
        \hline
            $J^P$                               & $\frac{3}{2}^-$               & $\frac{5}{2}^+$       & $\frac{3}{2}^+$               & $\frac{5}{2}^-$ \\
        \hline
        $A$                                     & $0.379\pm 0.051$              & $0.380 \pm0.053$      & $0.378 \pm 0.049$             & $0.381 \pm 0.053$             \\
        $\alpha_0$                              & $0.941\pm 0.047$              & $0.941 \pm0.049$      & $0.942 \pm 0.045$             & $0.941 \pm 0.048$             \\
        $\alpha^\prime$ ($\gevnospace^{-2}$)                & $0.364\pm0.037$               & $0.367 \pm0.039$      & $0.363 \pm 0.035$             & $0.365 \pm 0.037$             \\
        $b_0$ ($\gevnospace^{-2}$)                  & $0.12\pm 0.14$                & $0.13 \pm0.15$        & $0.12 \pm 0.14$               & $0.13 \pm 0.15$                \\
        $\mathcal{B}_{\psi p}^{(4450)}$ (95\%)                               & $\leq 4.3\%$                  & $ \leq 1.4\%$         & $\leq 1.8\%$                  & $\leq 0.71\%$ \\
        \hline
        \hline
        \end{tabular}
\end{table}

The use of polarization observables has been 
    proposed for an experiment
      at the Super BigBite Spectrometer (SBS) in Hall~A at \jlab~\cite{SBS:2018}. It has been argued that these may reach higher signal-to-background ratios than differential cross sections, which is particularly appealing due to the discovery of double-peak structures in the LHCb spectrum.
      Furthermore, the polarization data offer new and complementary information relevant in the evaluation of the resonance photo- and hadronic couplings. In Ref.~\cite{Winney:2019edt}, we provided sensitivity studies for the planned experiments on extracting the beam-target asymmetry \ALL, and the beam-recoil asymmetry \KLL, scanning the observable behavior with the relative coupling sizes, and mapping it as functions of the scattering angles and energies, to find the optimal experimental settings. The results are summarized in \figurename{~\ref{fig:polarization}}, where the presence of a broad $P_c(4380)$ with parity opposite to the $P_c(4450)$ was also considered. We found that 250 days of collected data with the SBS experiment would give more than $5\sigma$ sensitivity to the $P_c$ signals in large regions of the parameter space, in particular for \KLL.
      
      Another possibility is to produce $P_c$'s in backward $J/\psi$ photoproduction.
      The cross sections can be estimated by employing the techniques shown in the following section.
      Unfortunately, searches of hidden-charm pentaquarks in this way are hindered by large $N^*$ contributions~\cite{Albaladejo:2020tzt}.
      
      In view of the experiments in the electron-ion collider (EIC) era, it is instructive and timely to extend these studies to other polarization observables as well, and to investigate pentaquark photo- and electro-production also in semi-inclusive reactions. Furthermore, EIC colliders are unique factories for hidden-beauty $P_b$ searches~\cite{Cao:2019gqo}, hence it is important to provide theoretical studies for these states as well. Ultimately, for the signals to be found in photo- and electroproduction experiments, it shall be important to use the combined knowledge of different observables in order to draw conclusions about the quantum numbers, couplings, and nature of these exotic states.

\subsection{\texorpdfstring{\XYZ production in electron-proton collisions}{XYZ production in electron-proton collisions}}
\label{sec:xyz_exclusive}
Electromagnetic probes are expected to be
essential in the near future 
to provide insight on the nature of exotic hadrons.
Besides providing independent confirmation of exotica,
at high energies
these reactions are not affected by three-body dynamics and 
constitute efficient probes to 
determine exotic hadrons' quantum numbers and internal structure.
The use of real and quasi-real photon 
beams to search for exotic hadrons is currently being 
surveyed in the light sector at \jlab, with
some incursion into reachable low-lying charmonia,
in particular pentaquarks as was shown in the previous section.

\begin{figure}[t]
    \centering
    \includegraphics[width=.3\textwidth]{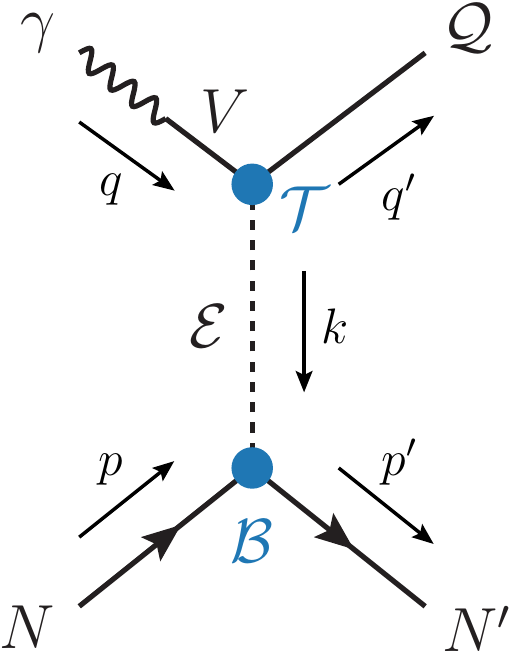} \hspace*{3cm}
    \includegraphics[width=.3\textwidth]{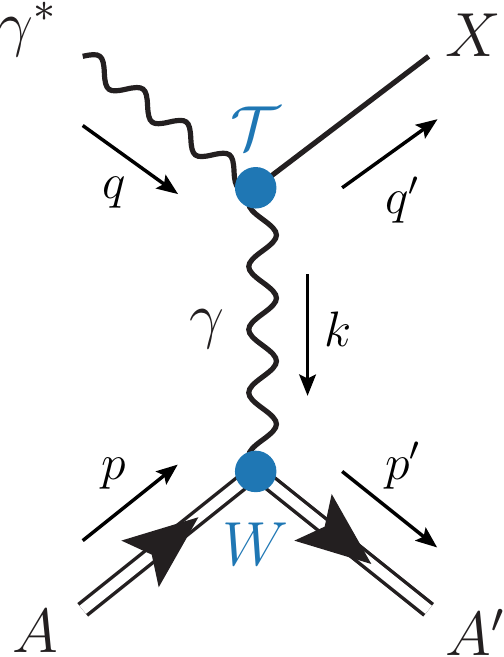}
    \caption{(left) Photoproduction of a quarkonium-like meson, $\calQ$ via an exchange $\calE$ in the $t$-channel. (right) Quasi-real photon production of $X(3872)$ 
    via Primakoff effect. Figures from~\cite{Albaladejo:2020tzt}.}
    \label{fig:photo_reaction_s}
\end{figure}
\begin{table}[t]
\caption{List of \XYZ states studied, with the corresponding branching ratios into vector quarkonia and light meson. The widths reported are obtained from Breit-Wigner extractions.\label{tab:exchanges}}
\begin{center}
        \begin{tabular}{c|cccc}
        \hline
        \hline
            $\calQ$ & $\Gamma_\calQ$ (\mevp) & $V$ & $\calE$ & $\calB(\calQ \to V \, \calE)$ (\%)
            \\ \hline
            \multirow{2}{*}{$X(3872)$} & \multirow{2}{*}{$1.19\pm0.19$} & \multirow{2}{*}{$\jpsi$} & $\rho$ & $4.9^{+1.9}_{-1.1}$ 
            \\
            & & & $\omega$ & $4.4^{+2.3}_{-1.3}$ 
            \\\hline
            $Z_c(3900)$ & $28.3\pm 2.5$ & $\jpsi$ & $\pi$ & $10.5 \pm3.5$ 
            \\\hline
            $Z_{cs}(4000)$ & $131\pm41$ & $\jpsi$ & $K$ & $\sim 10$ \\ \hline
            $X(6900)$ & $168 \pm 102$ & $\jpsi$ & $\omega$ & $\sim 1-4$ \\ \hline
            \multirow{3}{*}{$Z_b(10610)$} & \multirow{3}{*}{$18.4\pm2.4$}& $\Upsilon(1S)$ & \multirow{3}{*}{$\pi$} & $0.54^{+0.19}_{-0.15}$ \\
            & &$\Upsilon(2S)$ & & $3.6^{+1.1}_{-0.8}$ \\
            & &$\Upsilon(3S)$ & & $2.1^{+0.8}_{-0.6}$ \\\hline
            \multirow{3}{*}{$Z_b(10650)$} & \multirow{3}{*}{$11.5\pm2.2$} &  $\Upsilon(1S)$ & \multirow{3}{*}{$\pi$} & $0.17^{+0.08}_{-0.06}$ \\
            & &$\Upsilon(2S)$ & & $1.4^{+0.6}_{-0.4}$ \\
            & &$\Upsilon(3S)$ & & $1.6^{+0.7}_{-0.5}$ \\
        \hline
        \hline
        \end{tabular}
        \end{center}
\end{table}

The next generation of electron-hadron colliders, the EIC~\cite{Accardi:2012qut,AbdulKhalek:2021gbh} and the EicC~\cite{Anderle:2021wcy}, promise to open new possibilities of a spectroscopy program with higher energy and luminosity to study the plethora of the \XYZ states.
In preparation for these new facilities, it is necessary 
to provide theoretical estimates for the production of 
quarkonium(-like) states.
In particular, we are interested in  
exclusive processes where $\calQ$ is produced through photon fragmentation
from threshold to the expected EIC and EicC energies. We aim at providing predictions that are as based on data as possible, in order to minimize the model assumptions related to the microscopic nature of the \XYZ states. Moreover we also give predictions for the ordinary quarkonia generated by the production mechanism, in order to provide candles to assess the goodness of the model.

We write a helicity amplitude for the production process $\gamma N \to \calQ N^\prime$ (see \figurename{~\ref{fig:photo_reaction_s}}), 
    \begin{equation}
        \label{covariant_form}
        \mel{\lambda_\calQ, \lambda_{N^\prime}}{T_\calE}{\lambda_\gamma, \lambda_N} = 
        \calT^{\alpha_1\dots\alpha_J}_{\lambda_\gamma\lambda_\calQ} \;
        \calP^{(\calE)}_{\alpha_1\dots\alpha_J ;\beta_1\dots\beta_J} \; 
        \calB_{\lambda_N \, \lambda_{N^\prime}}^{\beta_1 \dots\beta_J} \, ,
    \end{equation}
where the rank-$J$ Lorentz tensors associated to the spin $J$ of the exchanged $\calE$, 
$\calT$ (top vertex) and $\calB$ (bottom vertex) are derived from assumed forms of the $\gamma\calQ\calE$ and $\calE N N^\prime$ interactions, say from effective Lagrangians consistent with expected symmetries of the reactions. The helicity structure is simplified when needed in order to make the amplitude depend on a single coupling.
Since most of the \XYZ states have been observed to decay into a vector quarkonium, one can assume VMD 
to calculate the photon-$\calQ$-$\calE$ coupling from the measurement of the branching ratio $\mathcal{B}\!\left(\mathcal{Q}\to V\,\mathcal{E}\right)$.\footnote{The applicability of VMD is discussed in Section~\ref{sec:jpsi_photo}.} The phenomenology of the bottom vertex is well constrained by photoproduction phenomenology.
\tablename{~\ref{tab:exchanges}} summarizes the exchanges and branching ratios for the considered exotics. 

We note that we are dealing with two energy regimes that require different treatments.
We expect that a model with exchange of a fixed-spin particle is valid from threshold to moderate values of $s$. However, it can be shown that this amplitude behaves as
\begin{align}
\label{eq:dtchannel}
	\mel{\mu_\mathcal{Q} \mu_\gamma}{T}{\mu'_{\bar N} \mu_N} &\propto \frac{d^{j}_{\mu'_{\bar N} - \mu_N,\mu_\mathcal{Q} - \mu_\gamma}(\theta_t)}{t-m_\mathcal{E}^2} \, ,
\end{align}
where $\cos \theta_t$ is the $t$-channel scattering angle, and depends linearly on $s$.  At high energies, this expression grows as $s^{j}$, which exceeds the unitarity bound. The reason for this is that a fixed-spin exchange amplitude is not analytic in angular momentum. Assuming that the large-$s$ behavior is dominated by a Regge pole rather than a fixed pole, 
we obtain the standard form of the Regge propagator of Eq.~\eqref{eq:Regge_pole}. This can be interpreted as originating from the resummation of the leading powers  of $s^{j}$ in the $t$-channel amplitude, which originate  from the exchange of a tower of particles with increasing spin. In the high-energy regime, the $\calP^{(\calE)}_{\alpha_1\dots\alpha_J ;\beta_1\dots\beta_J}$
is thus replaced by a Regge propagator. 

For the Pomeron-dominated $Y(4260)$ production, a fixed-spin description is no longer possible. However, we can use the results discussed in Section~\ref{sec:jpsi_photo} from Refs.~\cite{HillerBlin:2016odx,Winney:2019edt}, that were fitted to low-energy and high-energy $\jpsi$ photoproduction data separately. Together with a rescaling of couplings, and an upper limit on $Y(4260)$ production from HERA data~\cite{H1:2002yab}, one can obtain predictions also for this state.

\begin{figure}[t]
    \centering
    \includegraphics[width=0.45\textwidth]{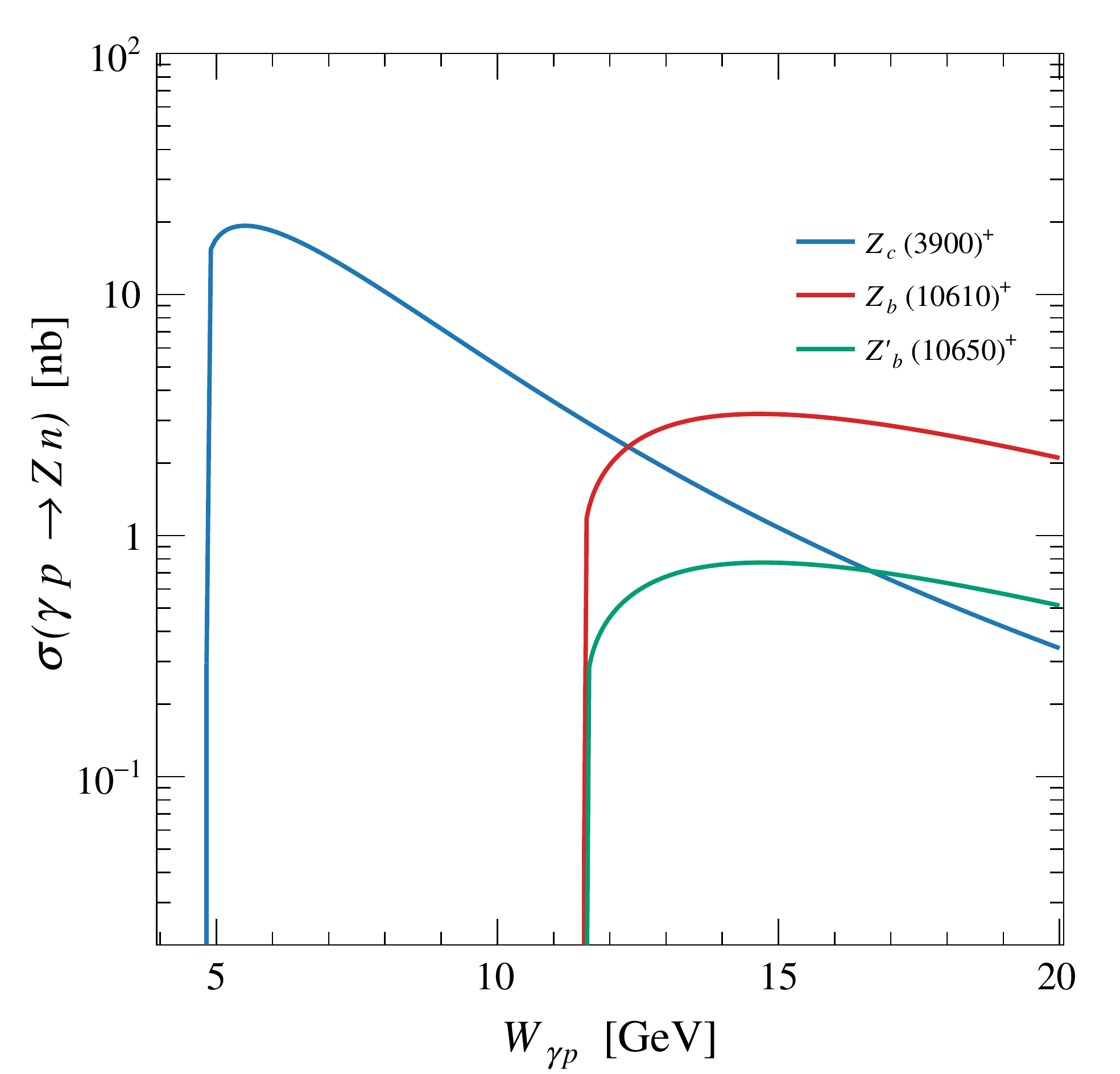}
    \includegraphics[width=0.45\textwidth]{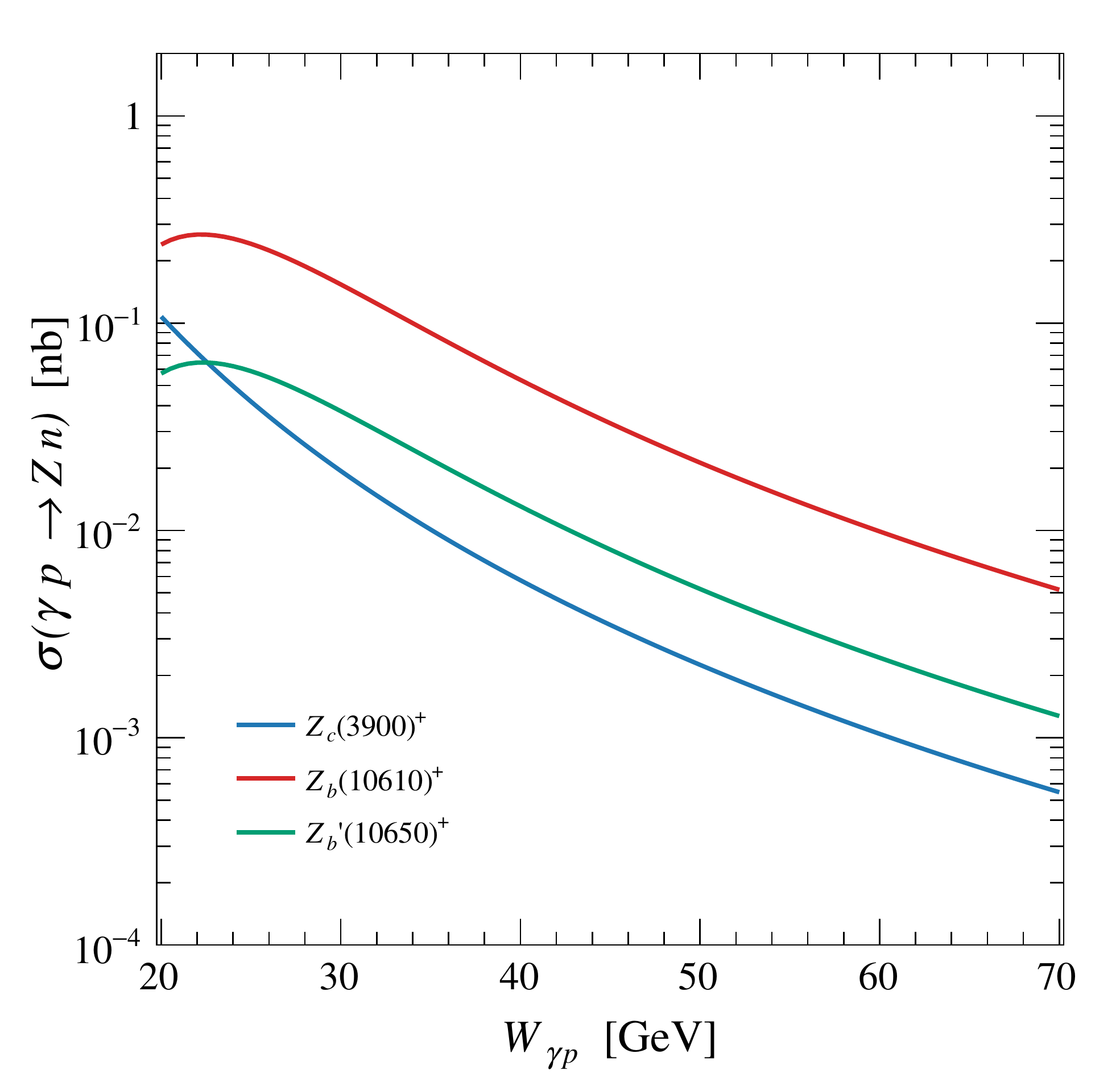}
    \caption{Integrated cross sections for the axial $Z_c$ and $Z_b^{(\prime)}$ states according to the low-energy fixed-spin model (left), and to the high-energy Regge exchange (right).
    Figures from~\cite{Albaladejo:2020tzt}.}
    \label{fig:exclusive_Z_xsections}
\end{figure}
\begin{figure}[!b]
    \includegraphics[width=.48\textwidth]{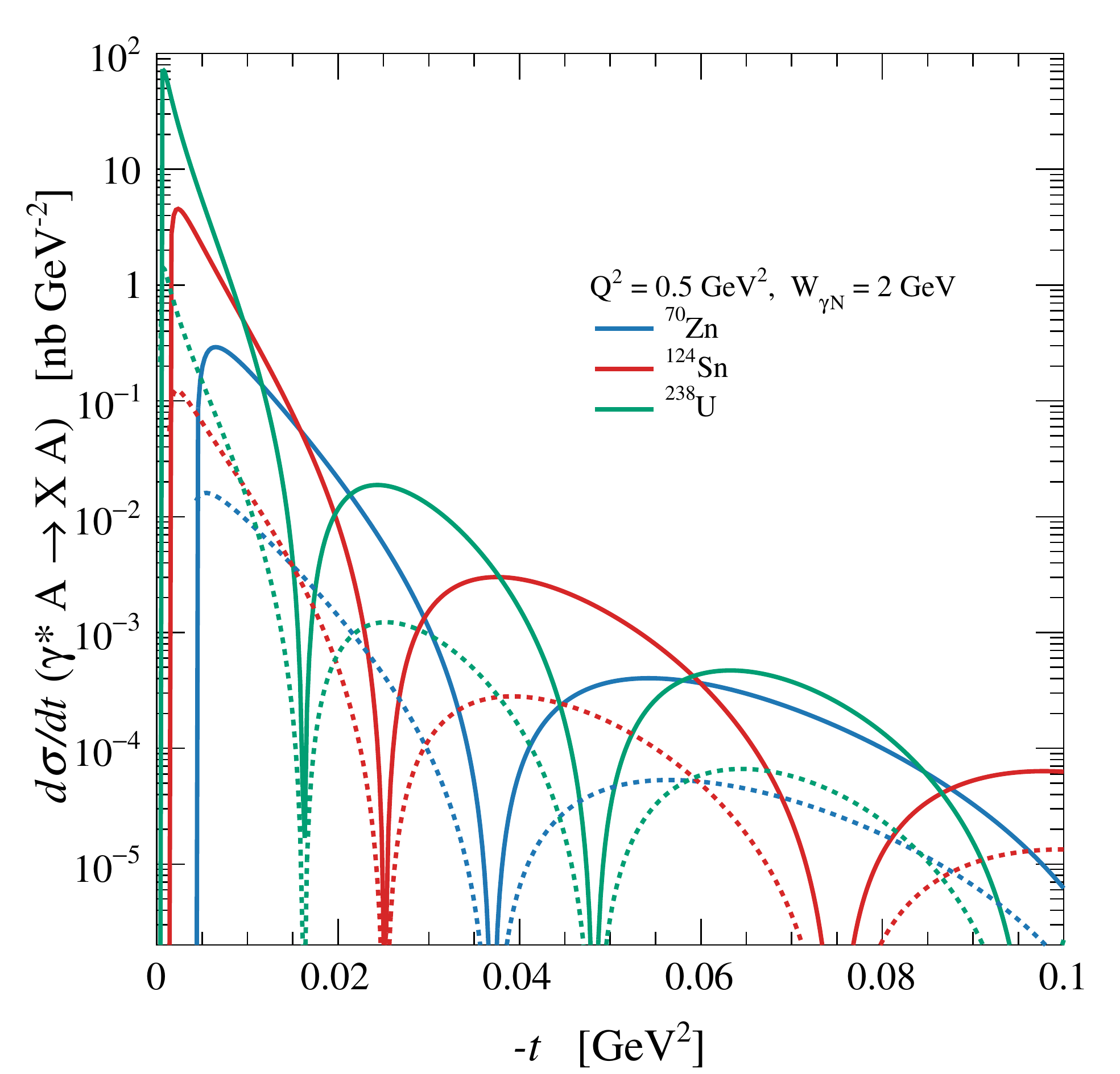}
     \includegraphics[width=.48\textwidth]{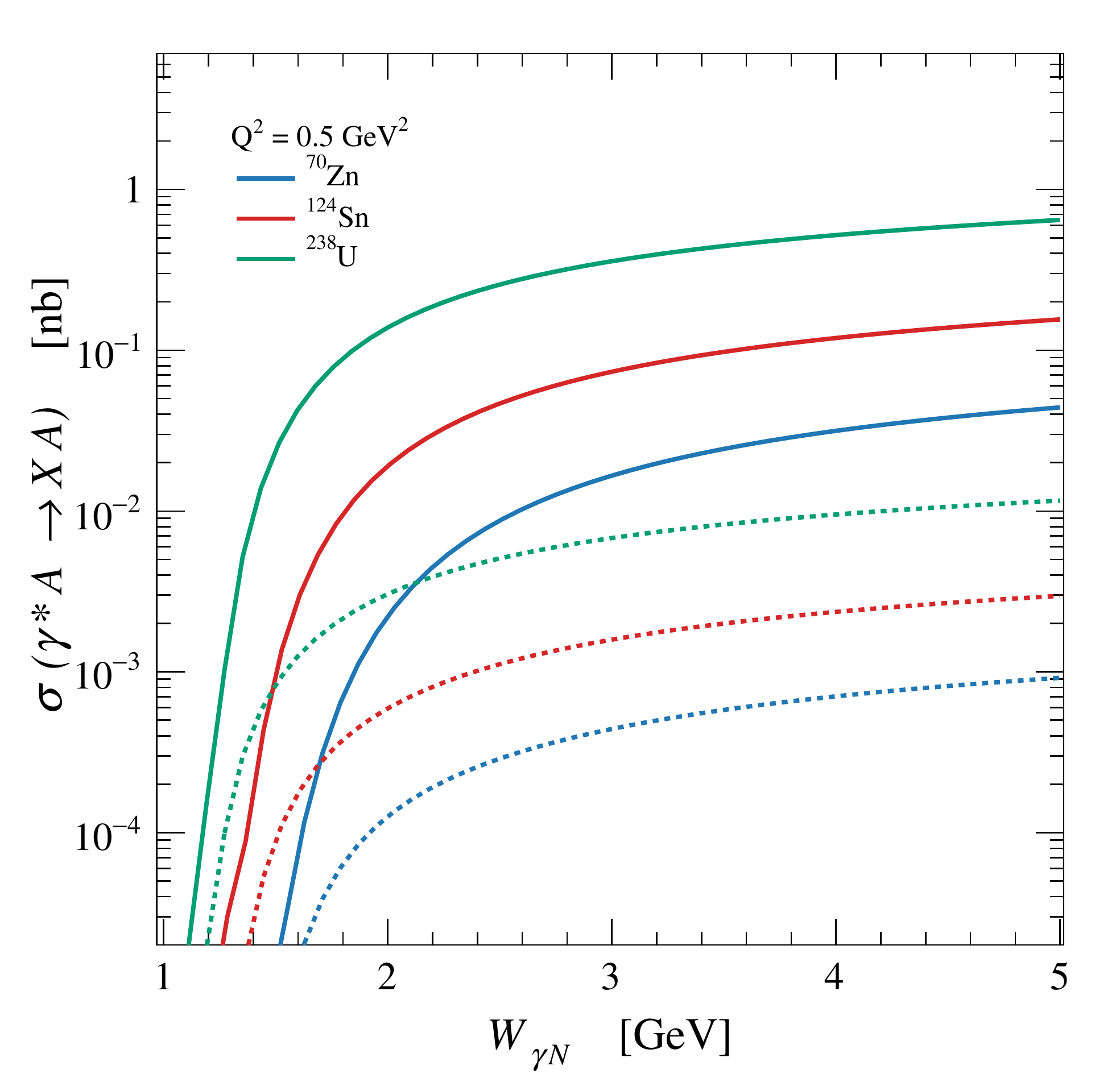}
    \caption{Differential cross sections for  $W_{\gamma N} = 2\gev$ (left) and integrated cross sections (right) for Primakoff production of $X(3872)$ off various nuclei. Solid and dashed curves correspond to longitudinal and transverse incoming photons, respectively.
    Figures from~\cite{Albaladejo:2020tzt}.}
    \label{fig:primakoff}
\end{figure}

An example of results is shown in \figurename{~\ref{fig:exclusive_Z_xsections}}.
We note that the strengths of the amplitudes do not necessarily match in the two regimes. 
The expectation is that the cross section decreases faster and matches the Regge prediction at $W_{\gamma p}\sim 20\gev$. COMPASS has measured upper limits for the $Z_c(3900)$ and $X(3872)$ photoproduction cross sections at an average energy of $\left\langle W_{\gamma p}\right\rangle = 13.7\gev$ of $\sim 0.35$ and $0.07\nb$ respectively, once branching ratios are taken into account. While the first result has roughly the same order of magnitude as our prediction, the second one is clearly smaller than our low energy prediction. This could be due to the Regge regime being reached earlier than expected, to the breaking of the VMD assumption, or to a dramatic dependence of the top coupling on the photon virtuality. In particular for the $X(3872)$, it is unlikely that all these fail at least close to threshold, so an independent confirmation of these measurements is needed.

Another possible mechanism to produce the $X(3872)$ at the EIC 
is through Primakoff effect exploiting the planned nuclear beams, 
ranging from the proton to uranium. The diagram is shown in \figurename{~\ref{fig:photo_reaction_s}}. The
Landau-Yang theorem~\cite{Landau:1948kw,Yang:1950rg}
prohibits the $X(3872)$ to couple to two real photons,
but nothing prevents it from coupling to a real and a virtual one.
Actually, a recent measurement by Belle found
$\tilde\Gamma^X_{\gamma \gamma} \times \calB(X(3872)\to \jpsi\,\pi^+\pi^-) = 5.5^{+4.1}_{-3.8}\pm 0.7\eV$~\cite{Teramoto:2020ezr}. 
The virtuality of the exchanged photon is suppressed for $-t \gg R^{-2} \sim \mathcal{O}(10^{-3})\gev^2$, $R$ being the nuclear radius, so that the exchanged photon is quasi-real. The $X\gamma \gamma^*$ coupling can be estimated from
Belle's width and the absolute branching ratios in~\cite{Li:2019kpj},
obtaining $g_{X \gamma \gamma^*} \sim 3.2\times 10^{-3}$.
The cross section is enhanced by the square of the atomic number of the nuclear beam, 
so we expect this production mechanism to be possible for high $Z$ beams. 
\figurename{~\ref{fig:primakoff}} shows differential and integrated
cross section predictions for a variety of nuclei for $Q^2 = 0.5\gev^2$
for an average photon nucleon energy $W_{\gamma N} = W_{\gamma A} / A=2\gev$, 
with $A$ being the mass number of the ion.

The integrated cross-section estimations show that
the near-threshold production of the $X(3872)$ and $Z$ states might be promising for the EIC or other electron-proton facilities. The $X(3872)$ may see production cross-sections of tens of nanobarn close to threshold, while charged quarkonium states near-threshold are predicted to be $\mathcal{O}(1\nb)$, and are well positioned for a high-luminosity spectroscopy program at the EIC. Additionally, diffractive vector production of vector states was also computed and shown to increase with energy, meaning the higher center-of-mass reach of the EIC is also beneficial for the production of $Y(4260)$ states.

\begin{figure*}
        \centering
        \includegraphics[width=\textwidth]{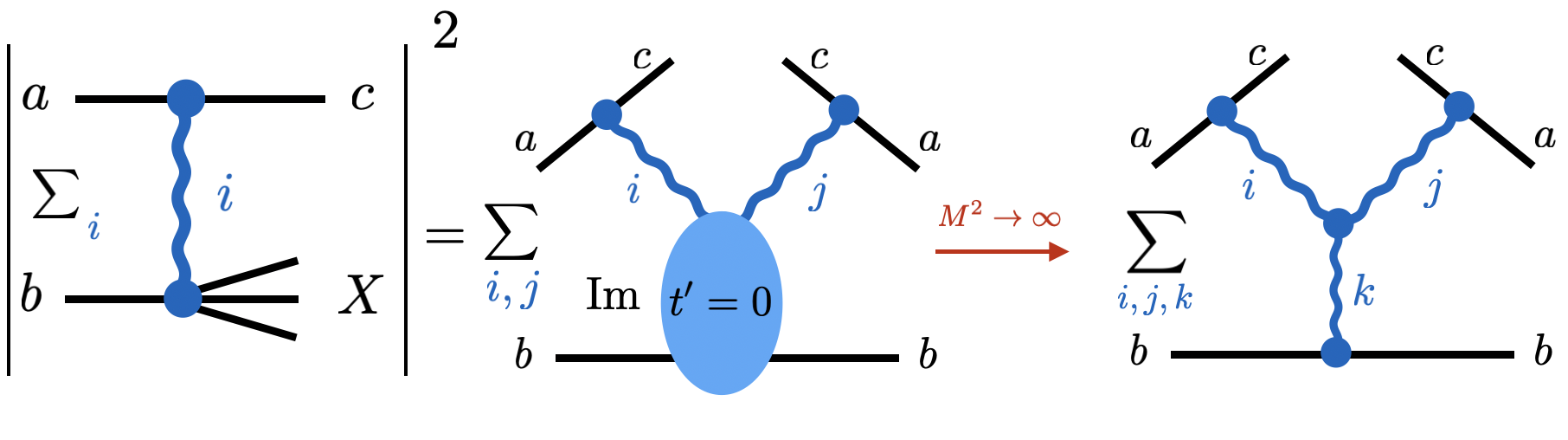}
        \caption{Diagrammatic representation of triple-Regge interactions in the $s,M^2\to\infty$ limit.}
        \label{fig:triple_regge}
    \end{figure*}

When compared to exclusive reactions, semi-inclusive production can, among others, offer the advantage of easier experimental accessibility. We therefore aim to extend the work described in the previous section to general $\gamma N\rightarrow \mathcal{Q}X$ processes, where $X$ represents any combination of final states with total invariant mass $M$ (\aka missing mass) that are produced in addition to the scrutinized $\mathcal{Q}$. We shall focus on the region characterized by large center-of-mass energy and missing masses, with $s \gg M^2 \gg m_p$. In this limit the $\mathcal{Q}$ meson is produced with high momentum in the near-forward region with $x \sim p_L/p \approx 1$. This region is dominated primarily by ``triple-Regge" interactions as shown in \figurename{~\ref{fig:triple_regge}},
where the sum over $i$, $j$ and $k$ refers to the possible exchanges contributing to the triple Regge vertex.
As can be seen, the top and bottom vertices can be taken from our previous work on exclusive reactions. 
The novel information to be given as input is the triple-Regge vertex itself. Here, production of neutral states is assumed to primarily proceed through a triple-Pomeron interaction~\cite{Field:1974fg}. For the charged $Z$ states, on the other hand, pion exchanges dominate the top vertices, and therefore in order to describe the triple-Regge exchange one needs to estimate the total $\pi\,N$ scattering cross section, for which one can take the asymptotic Regge approximation~\cite{COMPETE:2002jcr, Cudell:2001pn, Mathieu:2015gxa}. Other kinematic regions must be studied with other methods, see \eg~\cite{Yang:2021jof}.
This way, we aim to provide estimates and feasibility studies for semi-inclusive production of heavy quarkonia, which is particularly timely and promising in view of the EIC era.

\subsection{Two-meson production in the double-Regge region}
\label{sec:doubleregge}
As discussed in Section~\ref{sec:etapicompass} the $\eta\pi$ spectrum is of particular interest, since the odd partial waves have exotic quantum numbers and specifically, the $J^{PC}=1^{-+}$ partial wave hosts the $\pi_1$ hybrid candidate. To further understand the exotic meson production in this channel, one can invoke the Regge-resonance duality to relate the process $\pi p \to R\, p \to \eta^{(\prime)}\pi p$, where $R$ stands for an $\eta^{(\prime)} \pi$ resonance, to the double Regge region where the $\eta^{(\prime)} \pi$ invariant mass is large. 

In general the multi-Regge exchange formalism has been extensively studied theoretically in the past~\cite{Brower:1974yv, Drummond:1969ft,Bali:1967zz,Drummond:1969jv,Weis:1973vh,  Collins:1977jy,Shimada:1978sx};
more recently the double-Regge exchange was used to study
two-kaon photoproduction off the proton~\cite{Shi:2014nea}, and  to describe
the central meson production in the high energy proton-proton collisions~\cite{Lebiedowicz:2019jru,Lebiedowicz:2020yre,Cisek:2021mju}.
 
\begin{figure}[t]
    \centering
    \includegraphics[width=0.9\textwidth]{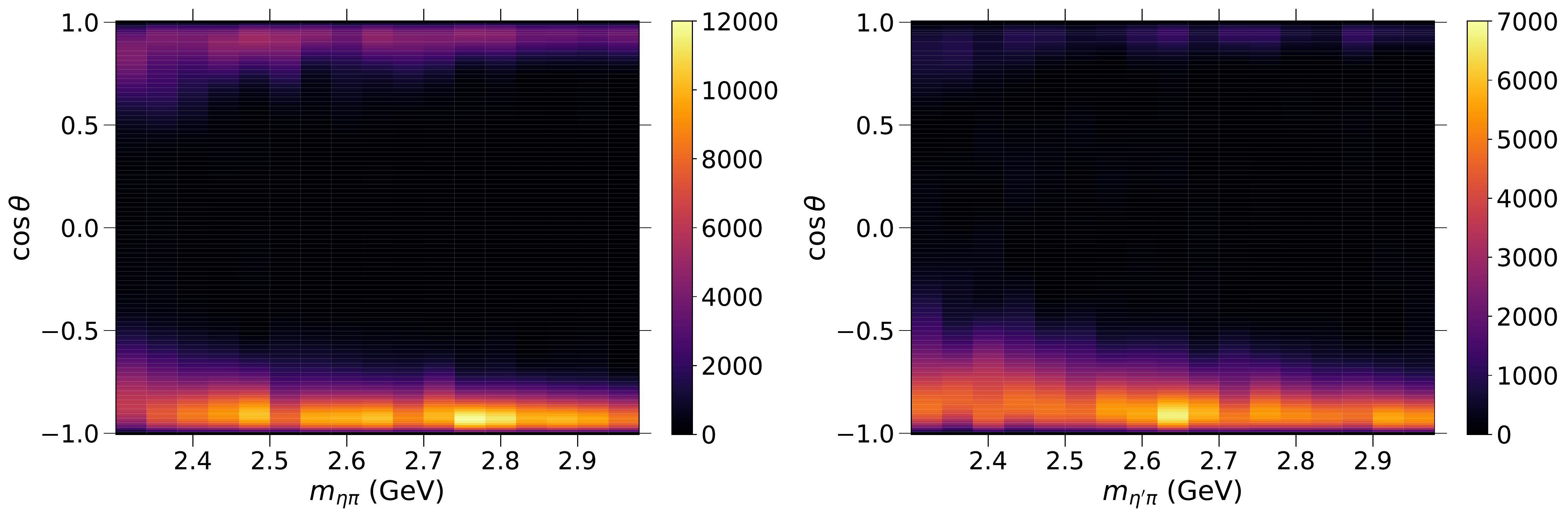}
      \caption{Intensity $I_\theta(\m ,\cos\theta)$ 
      density distribution computed from the 
      $\eta \pi$ (left) and $\eta' \pi$ (right)
      COMPASS partial waves. Figure from~\cite{Bibrzycki:2021rwh}. }\label{fig:compass_plot}
\end{figure} 

\begin{figure}[!b]
\centering
\includegraphics[width=0.40\textwidth]{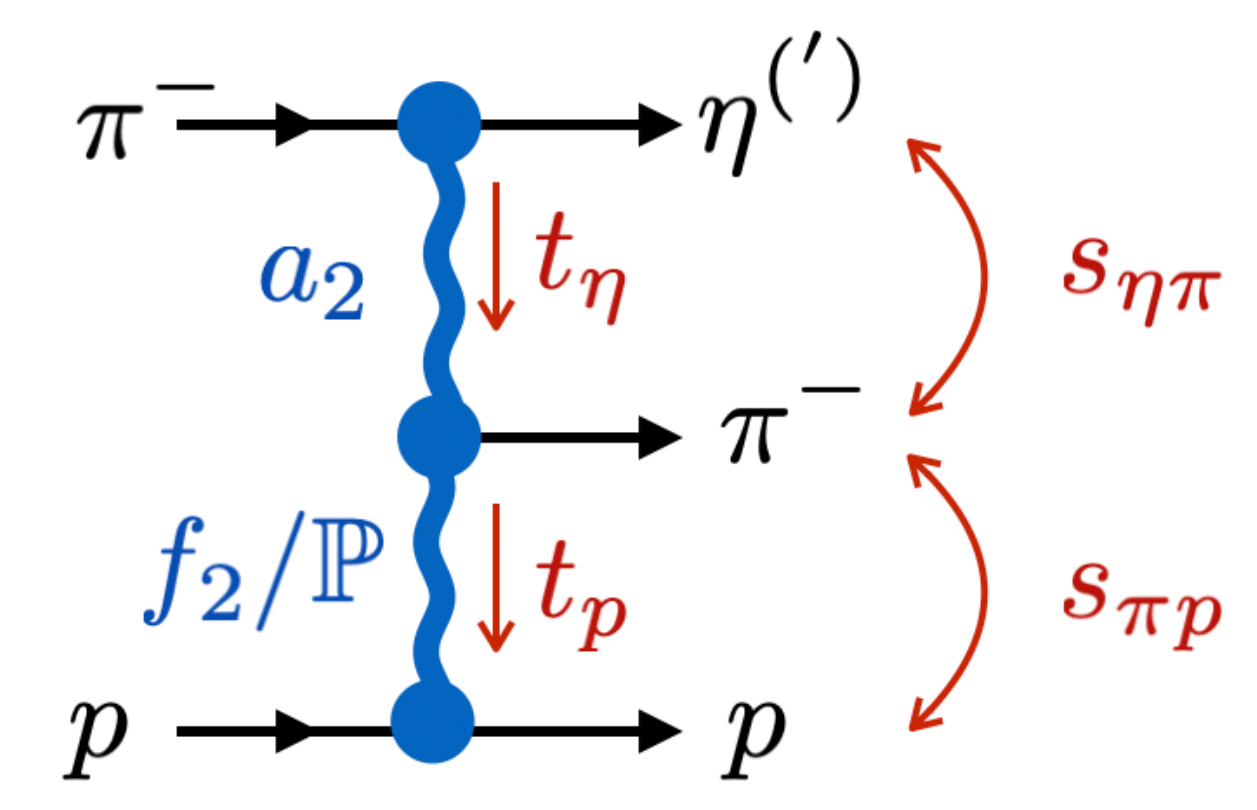} 
\includegraphics[width=0.40\textwidth]{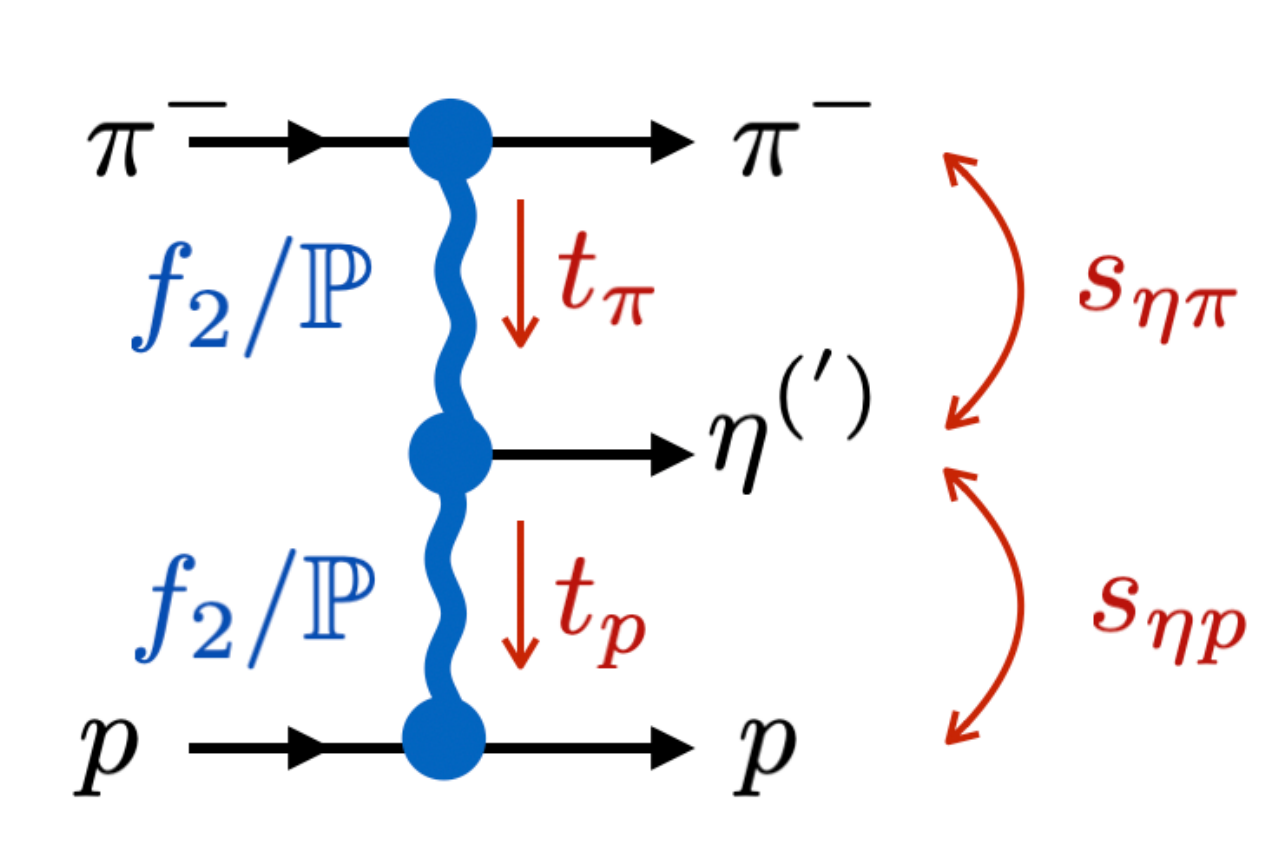}
\caption{Fast-$\eta$ (left) and fast-$\pi$ (right) amplitudes.
Figure from~\cite{Bibrzycki:2021rwh}.}
\label{fig:doublereggeexchanges}
\end{figure}

In Ref.~\cite{Bibrzycki:2021rwh}, we studied the $\pi^-p\to \eta^{(\prime)} \pi^- \,p$ data measured at COMPASS~\cite{COMPASS:2014vkj}.
The experimental $\eta^{(\prime)} \pi$  
$\m$-binned intensity distribution $I(\m , \cos \theta, \phi)$, where $\m$ is the $\etapi$ invariant mass,
can be computed from the published partial waves~\cite{COMPASS:2014vkj}. We focus on the $2.4 < \m < 3.0\gev$ region, where the double Regge is expected to dominate.
The angular variables determine the direction 
of the $\eta^{(\prime)}$ in the Gottfried-Jackson frame.
The $\phi$-integrated distributions 
\begin{align}
I_\theta (\m ,\cos \theta) = 
\int_0^{2\pi} \diff\phi \, I(\m ,\cos \theta, \phi)\, , \label{eq:Itheta}
\end{align}
are shown in \figurename{~\ref{fig:compass_plot}}
for a total of seventeen mass bins in each channel. 
We note that the intensity peaks in the forward $\cos\theta \sim 1$ and backward $\cos\theta \sim -1$ regions and both 
become narrower as the invariant mass $\m $ increases.
In the forward region, most of the beam momentum is carried by the $\eta^{(\prime)}$ (``fast-$\eta$'' region), 
and in the backward region by the pion (``fast-$\pi$'' region).
These features are typical of diffractive processes, 
pointing to the dominance of double-Regge exchanges 
for $\m \gtrsim 2.3\gev$.

The existence of a forward-backward asymmetry is apparent in \figurename{~\ref{fig:compass_plot}}
and by itself is proof of the existence of resonances
with exotic quantum numbers in that $\m$ range.
This asymmetry can be quantified through
\begin{align}
A(\m ) \equiv & \,   \frac{F(\m ) - B(\m )}{F(\m ) + B(\m ) } \, ,
\label{eq:intensities_asy}
\end{align}
where
\begin{align}
F(\m ) \equiv  \int_0^1 \diff\cos \theta \,
I_\theta(\m ,\cos\theta)& \, , \, 
B(\m ) \equiv  \int_{-1}^0 \diff\cos \theta \, I_\theta(\m ,\cos\theta) \, , 
\label{eq:fbintensities}
\end{align}
with $F(\m )$ and $B(\m )$ being the forward and backward intensities, respectively.

\begin{figure}[t]
    \centering
    \includegraphics[width=0.9\textwidth]{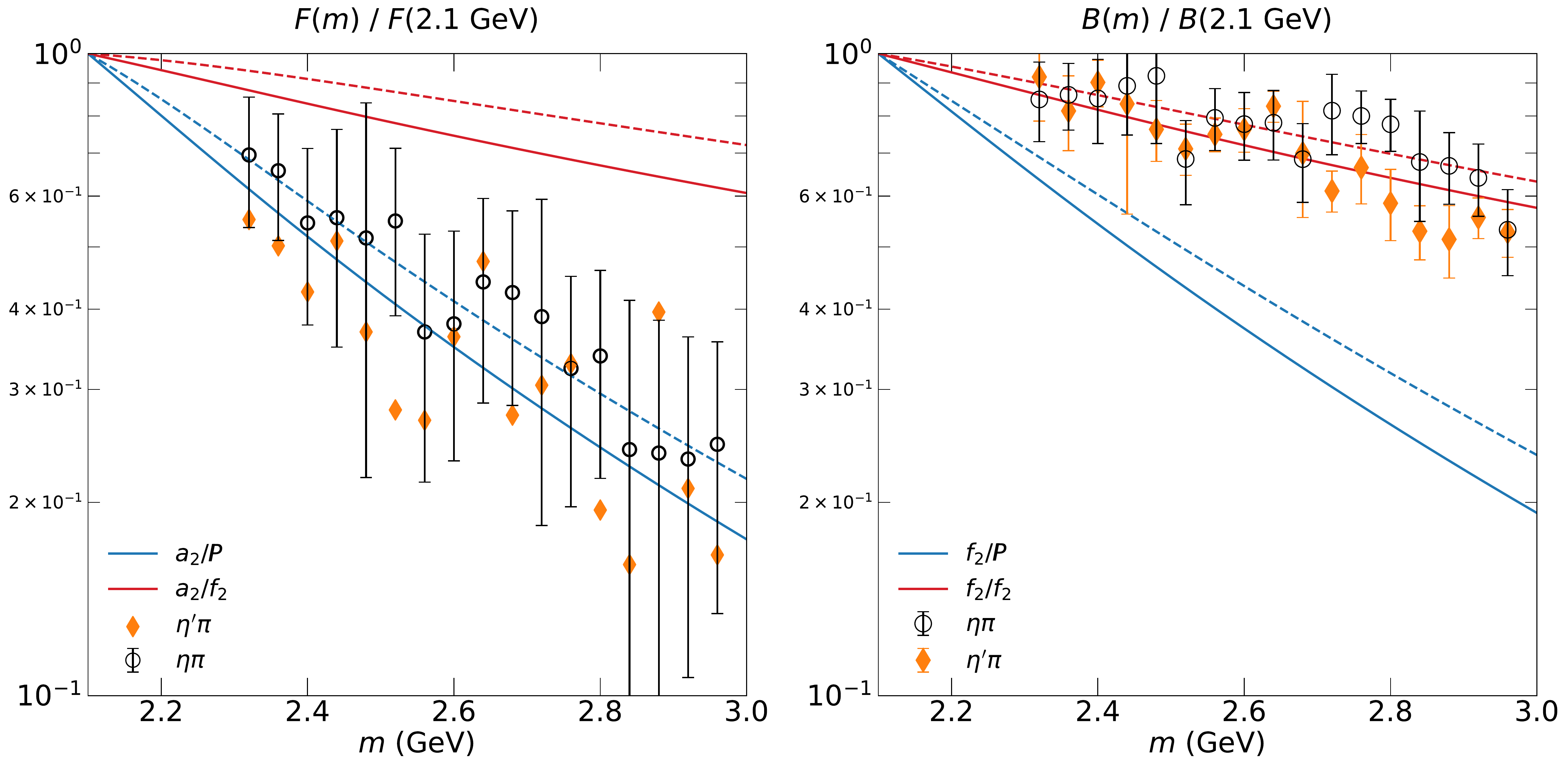}
      \caption{Forward (left) and backward (right) intensities as defined in Eq.~\eqref{eq:fbintensities}
      for the top-$a_2$ and top-$f_2$ amplitudes, respectively.
      Solid lines correspond to $\eta \pi$ and dashed
      to $\eta' \pi$.
      Each theoretical intensity is normalized to its value at $\m=2.1\gev$.
      In circles and diamonds we show the experimental data arbitrarily rescaled. Uncertainties for the forward $\eta'\pi$ intensity are very large, almost exhausting
the plot, and are therefore not shown.
      Figures adapted from~\cite{Bibrzycki:2021rwh}.
}\label{fig:mpieta_slope}
\end{figure}

COMPASS data were correctly
described by model of~\cite{Bibrzycki:2021rwh}. 
Specifically, the double-Regge exchange amplitudes depicted in \figurename{~\ref{fig:doublereggeexchanges}} were considered,
assuming the dominance of leading Regge poles.
The intensity is given by
\begin{align}
    I_\text{Th}(\m,\Omega) =  k(\m) \, |A_\text{Th}(\m,\Omega)|^2, \label{eq:model}
\end{align}
where $k(\m)= \lambda^\frac{1}{2}(\m^2,m^2_{\eta^{(\prime)}}, m_\pi^2)/(2\m)$ is the breakup momentum between the $\pi$ and the $\eta^{(\prime)}$
and the total amplitude $A_\text{Th}(\m,\Omega)$ is the sum of six  
double-Regge amplitudes,
\begin{align} 
    A_\text{Th}(\m,\Omega) &= c_{a_2\mathbb{P}} \,A_{a_2 \mathbb{P}} 
    + c_{a_2 f_2}\, A_{a_2 f_2}
    + c_{f_2\mathbb{P}}\, A_{f_2 \mathbb{P}} + c_{f_2f_2}\, A_{f_2 f_2}
    + c_{\mathbb{P}\mathbb{P}} \, A_{\mathbb{P}\mathbb{P}} 
    + c_{\mathbb{P} f_2} \, A_{\mathbb{P} f_2} \, ,
    \label{eq:ampTot}
\end{align}
where the $\{c\}$ are constants  fitted to the data.

An important property of multiperipheral amplitudes
is the absence of simultaneous singularities in overlapping channels.
For example, it is possible to identify that, for fast-$\eta$ production, 
the first two terms in Eq.~\eqref{eq:ampTot} are dual to resonances decaying to $\eta\pi$ and $\pi N$. It is then possible to write dispersion relations that enable us to independently study resonances in the beam and target fragmentation region.

\begin{figure}[t]
\centering
\includegraphics[width=0.46\textwidth]{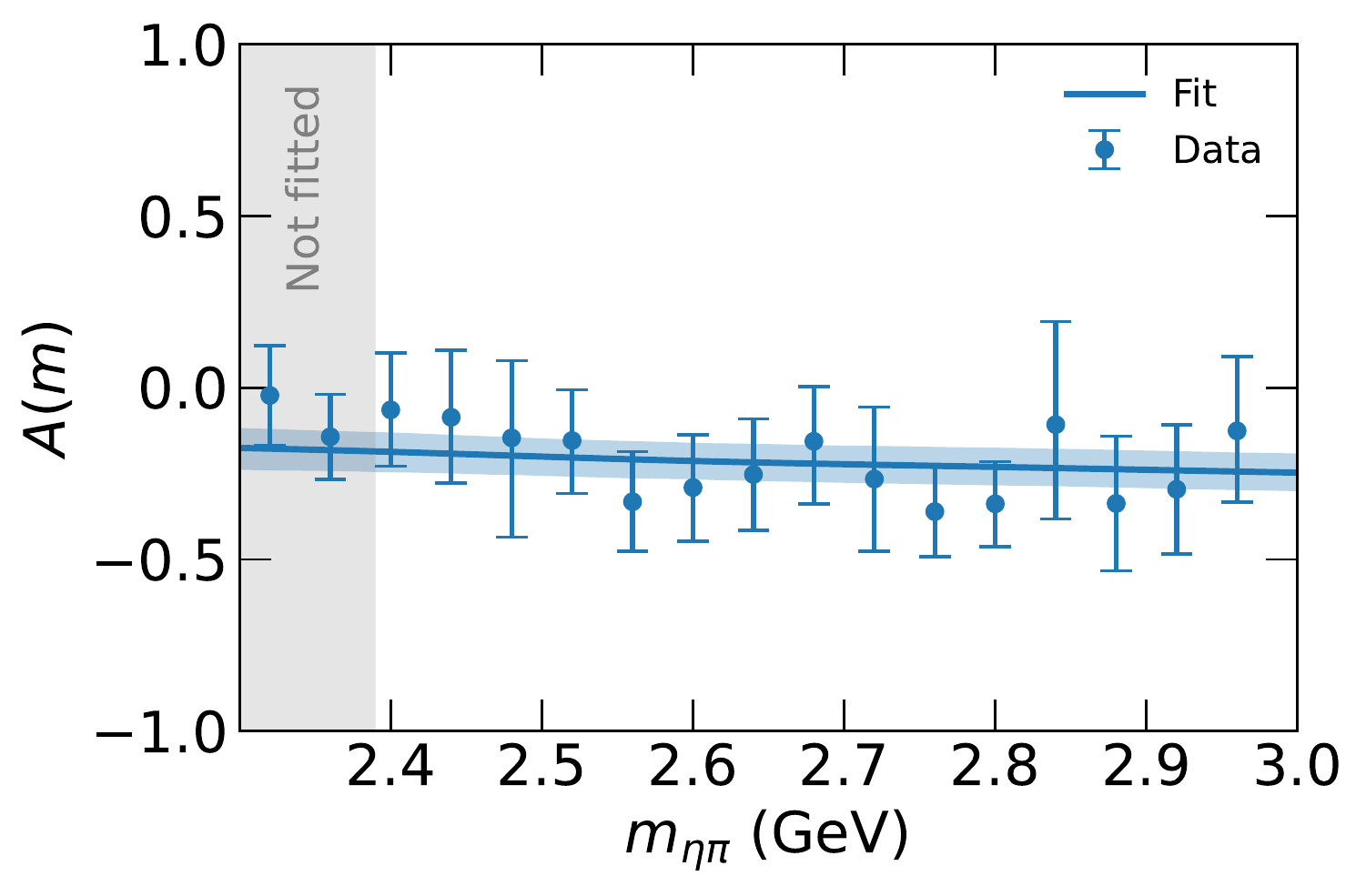} 
\includegraphics[width=0.46\textwidth]{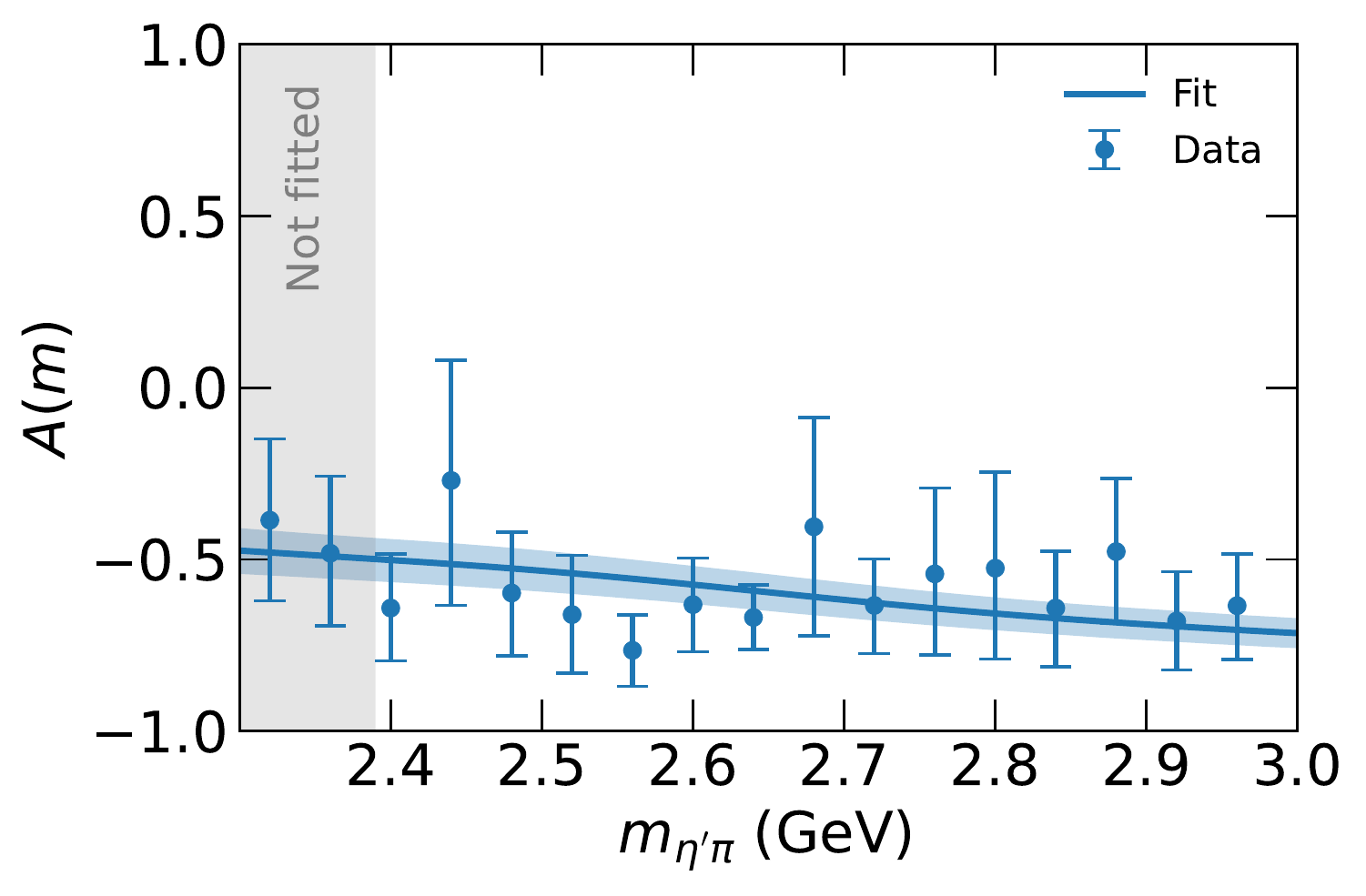}
\caption{Forward-backward intensity asymmetry as defined in Eq.~\eqref{eq:intensities_asy} for $\eta \pi$ (left) and $\eta' \pi$ (right).}
\label{fig:doublereggeasymmetry}
\end{figure}

The top exchange is dominated by the $a_2$ trajectory for fast-$\eta$, and by $f_2$ or $\mathbb{P}$ trajectories for fast-$\pi$. The bottom exchange is either 
$f_2$ or $\mathbb{P}$ for both amplitudes. Given the high energy of the COMPASS pion beam, the $\mathbb{P}$ was expected to be the relevant bottom exchange.
We found this to be true for the forward peak, where the slope of the
$F(\m)$ intensity is dominated by the $a_2/\mathbb{P}$ amplitude as shown in
\figurename{~\ref{fig:mpieta_slope}}. However, for the backward peak
we find that the slope of the $B(\m)$ intensity is dominated by the
bottom $f_2$ exchange, as also shown in \figurename{~\ref{fig:mpieta_slope}}.

The $\eta \pi$ intensity can be well
described with four amplitudes, either $a_2/\mathbb{P}$ or $a_2/f_2$, $f_2/f_2$, $f_2/\mathbb{P}$ or $\mathbb{P}/\mathbb{P}$. The inclusion of either bottom-$\mathbb{P}$ amplitude is necessary to describe the forward region, but the data do not show a clear preference between $f_2/\mathbb{P}$ and $\mathbb{P}/\mathbb{P}$.
Figure~\ref{fig:doublereggeasymmetry}
compares the asymmetry intensity $A(\m)$ to the fitted model. The existence
of a nonzero asymmetry is clear.
The $\eta'\pi$ data are consistently described by
the $a_2/\mathbb{P}$, $a_2/f_2$, $f_2/f_2$, and $\mathbb{P}/\mathbb{P}$ amplitudes. 
The $\mathbb{P}/\mathbb{P}$ contribution is necessary to describe
the data and is an indication of the large gluon impact on the $\eta'\pi$ system,
which relates to the existence of hybrid mesons.
This result is consistent with 
exchange degeneracy breaking between $a_2$ and $f_2$ in $\eta'\pi$ production. 

Additionally, the double-Regge amplitude model
contains an infinite number of partial waves 
and hence, these cannot be directly matched to the truncated waves from COMPASS.
However, both the truncated partial waves from COMPASS and those from the partial wave expansion of the Regge exchanges can be reconciled by performing an analysis on the theoretical amplitudes constrained to the
same number of partial waves employed by COMPASS.
Hence, once the double-Regge regime is reached, it is important to study the 
full amplitude rather than a truncated partial wave decomposition.

\subsection{Finite energy sum rules}
\label{sec:FESR}
In the previous sections, we provided models for Regge amplitudes and compared to experimental data. As was said, at low energies the \twototwo amplitude is saturated by a finite number of $s$-channel partial waves, dominated by resonances, while at high energies it can be represented by a sum over a finite number of leading Regge poles exchanged in the crossed channels.  Both are representations of the same analytic amplitudes, so they must be related by a dispersion relation, which can be used to provide strong constraints on resonance parameters by imposing that the sum of partial waves at  low energies matches the Regge amplitude at high energies. 
\begin{figure}
\begin{center}
	\includegraphics[width=0.45\textwidth]{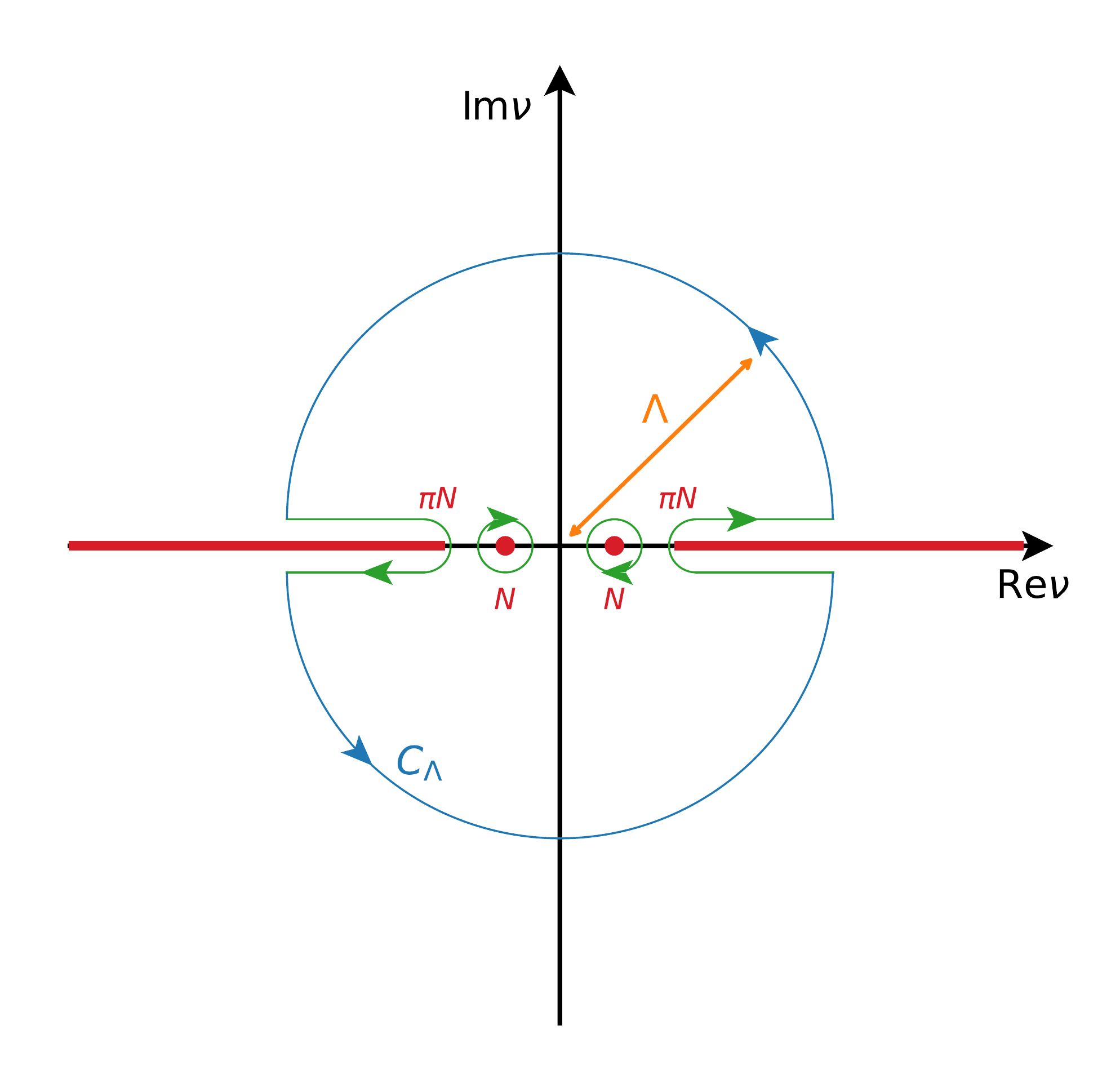}
\end{center}
\caption{Contour integration in the $\nu = (s - u)/2$ plane for the $\pi N$ FESR. The contour across the positive real axis accounts for the integral over  $\Im A_i(\nu,t)$ in the resonance region. At large $\Lambda$, the amplitude is saturated by a finite number of Regge poles, so that the integral over $C_\Lambda$ can be analytically computed. \label{fig:FESR-contour} }
\end{figure}
The dispersion relations are written for invariant amplitudes that are free of kinematic singularities. These are four in pseudoscalar photoproduction 
and two in $\pi N$ scattering. We will denote generically those amplitudes as 
$A_i$ and refer to~\cite{Nys:2016vjz, Mathieu:2017but} and~\cite{Mathieu:2015gxa} for their definition and their relation to observables. For $\pi N$ scattering, where the $s$- and $u$-channel both represent the $\pi N\to \pi N$ reaction, the $A_i$ have definite parity in the variable $\nu = (s - u)/2$.

For each $A_i$, one writes an integral over the contour depicted in \figurename{~\ref{fig:FESR-contour}}. This gives a relation between the imaginary part of the amplitude integrated over the resonance region, and an integral evaluated at complex high energies. In the latter, the amplitude is represented by Regge exchanges, and the integral can be computed analytically. Powers of $\nu$ can be multiplied to the amplitude, in order to calculate higher moments.
These relations are generally called finite-energy sum rules (FESR) and for pseudoscalar photoproduction they read
\begin{align}\label{eq:FESR}
	\frac{1}{\Lambda^{k+1}}\int_0^\Lambda \Im A_i(\nu,t) \nu^k \text{d}\nu &= \beta(t) \frac{(\Lambda/s_0)^{\alpha(t)-1}}{\alpha(t)+k} \, ,
\end{align}
where $k\in \mathbb{Z}$. The \rhs of Eq.~\eqref{eq:FESR} includes the nucleon pole and the discontinuity above the $\pi N$ threshold up to $\nu=\Lambda$ as depicted in \figurename{~\ref{fig:FESR-contour}}. The cutoff $\Lambda$ should be chosen large enough that for $\nu\gtrsim \Lambda$  the amplitude is saturated by Reggeons. In the \rhs of Eq.~\eqref{eq:FESR}, a sum over the leading Regge poles is understood.

These relations between the low and high energy regimes can be exploited in different ways. One such way is to provide further constraints to the resonance parametrization using high energy data. Another use would be the prediction of the cross section at high energies from the reactions at low energies.
 
\begin{figure}[t]
\begin{center}
    \resizebox{0.45\textwidth}{!}{\input{figures/fig-piN}}
	\raisebox{-0.03\height}{\includegraphics[width=0.42\textwidth]{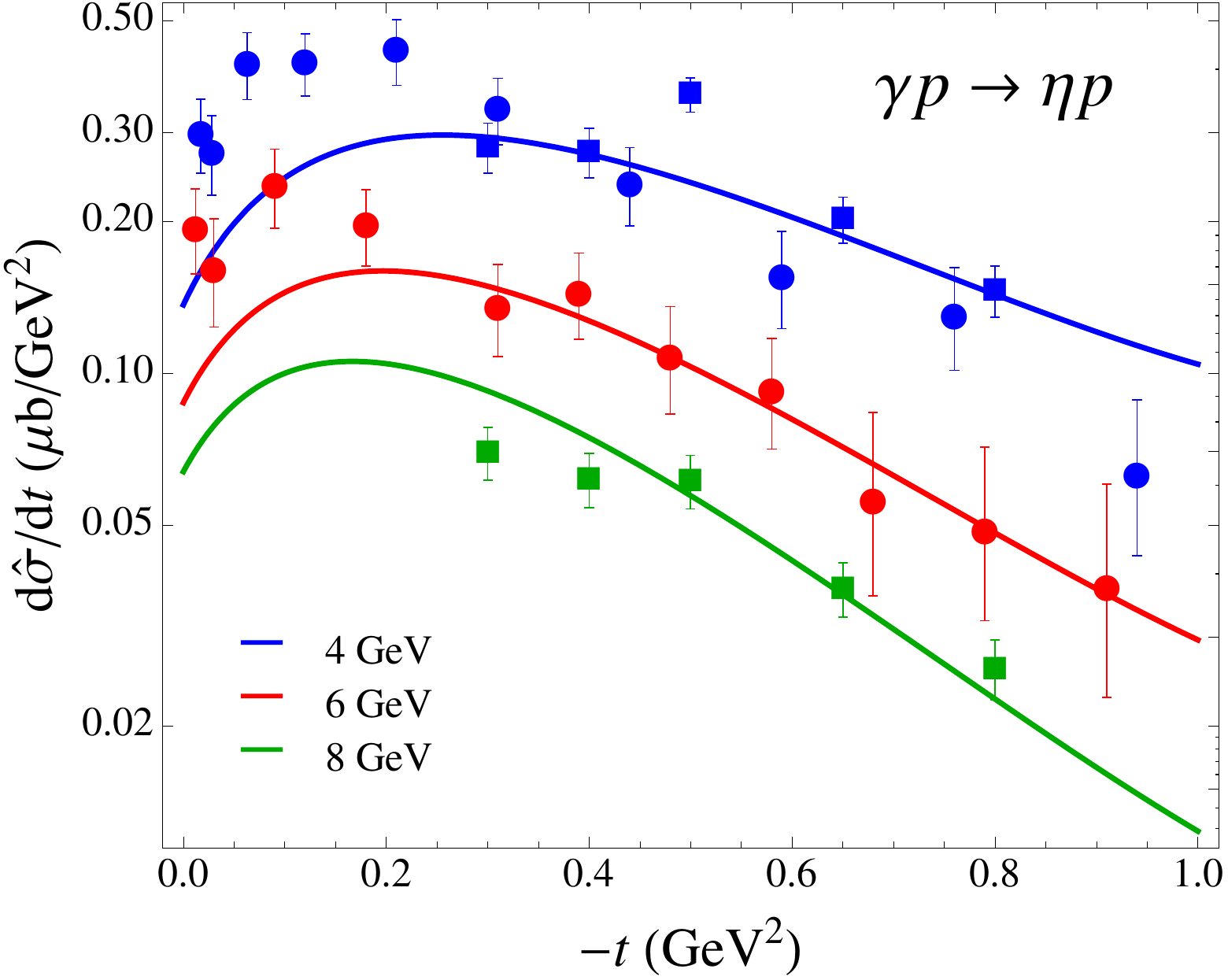}}
\end{center}
\caption{(left) Differential cross sections for $\pi^+ p$ (red) and $\pi^- p$ (blue) elastic scattering at $p_\text{lab}=5\gev$. 
Data from~\cite{Ambats:1972aar}.
Figure adapted from~\cite{Mathieu:2015gxa}. 
(right) $\eta$ photoproduction differential cross section computed from the low energy models using FESR. 
Data from~\cite{Braunschweig:1970jb} (circles)
and~\cite{Dewire:1971jln} (squares).
Figure from~\cite{Mathieu:2017but}. \label{fig:FESR1} }
\end{figure}

As an example, in \figurename{~\ref{fig:FESR1}} the differential cross section for $\pi^0$ photoproduction at high energies is compared with predictions based on FESR. The low-energy amplitude is calculated from the partial waves extracted by SAID~\cite{Workman:2012jf}. Analogous application of FESR to $\eta$ photoproduction~\cite{Nys:2016vjz}, using the partial wave from $\eta$-MAID~\cite{Chiang:2001as}. This study revealed a discrepancy between data and predictions in the forward $-t < 0.25\gevsq$ region, which was traced to the $A_4$ amplitude. In this region indeed the various PWA extractions available in the literature have strong disagreements, and other constraints like the one discussed here can be crucial to resolve the issue.

When precise data in the high energy regime are available, the Regge couplings $\beta_i(t)$ can be determined as explained in the previous section and both sides of Eq.~\eqref{eq:FESR} can be compared. In the case of $\pi N$ scattering, we observed an excellent agreement in all invariant amplitudes for both isospin channels~\cite{Mathieu:2015gxa}. The relation between the pattern of zeros in the low-energy and high-energy regimes is made apparent thanks to the FESR. For instance, the \lhs of Eq.~\eqref{eq:FESR} for the charged exchange amplitudes helicity nonflip and flip vanish at $t = -0.1$ and $-0.5\gevnospace^2$ respectively, which correspond to zeros of the $\rho$ exchange residues. The zero in the nonflip amplitude implies that the elastic $\pi^+ p$ and $\pi^- p$ cross sections coincide at that value of $t$ (\cf \figurename{~\ref{fig:FESR1}}, left panel), while the zero in the flip amplitude produces a dip in the $\pi^- p \to \pi^0 n$ cross section.
   
\begin{figure}[htb]
\begin{center}
	\includegraphics[width=0.43\textwidth]{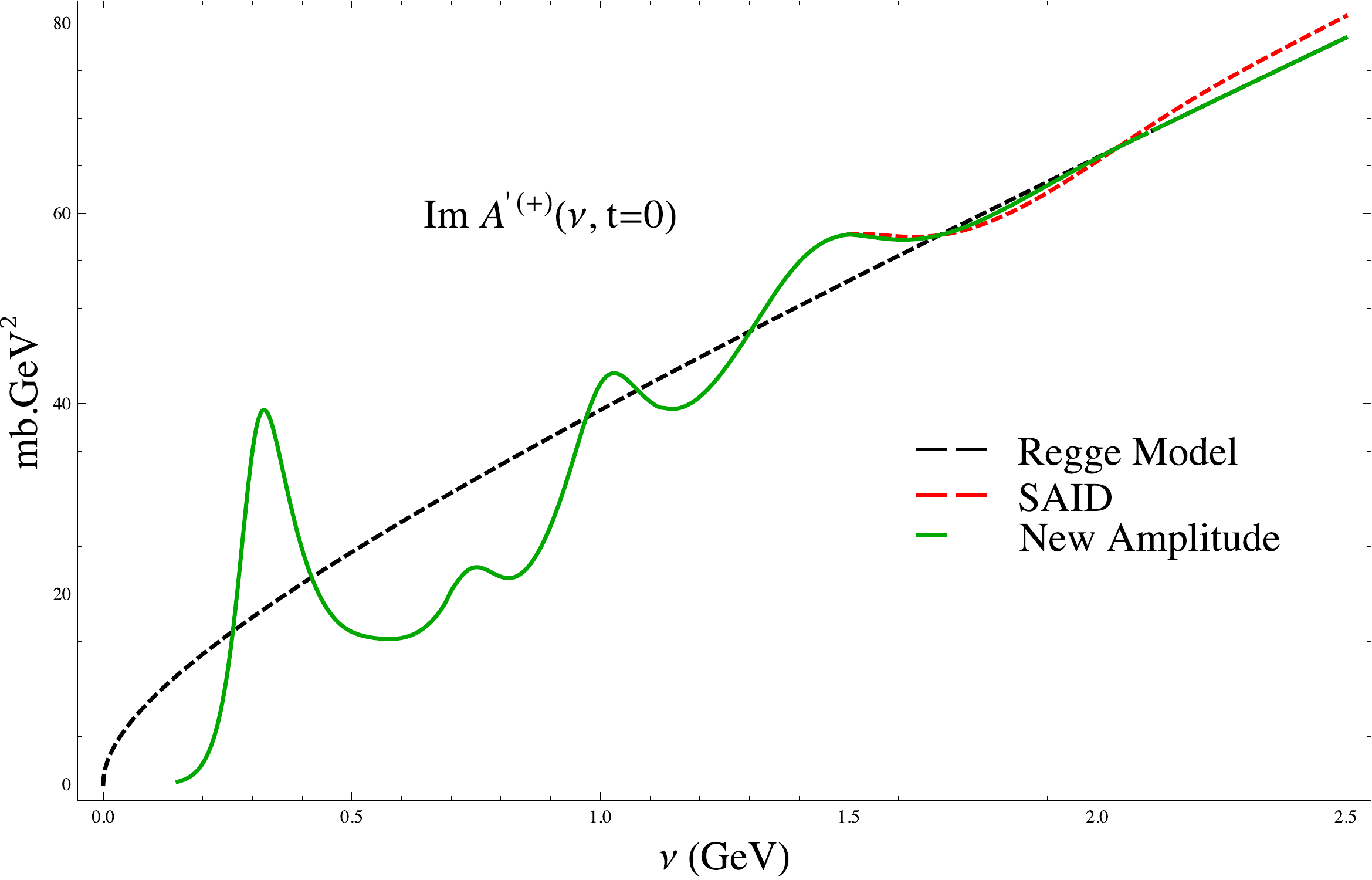} \hspace*{0.08cm}
	\includegraphics[width=0.43\textwidth]{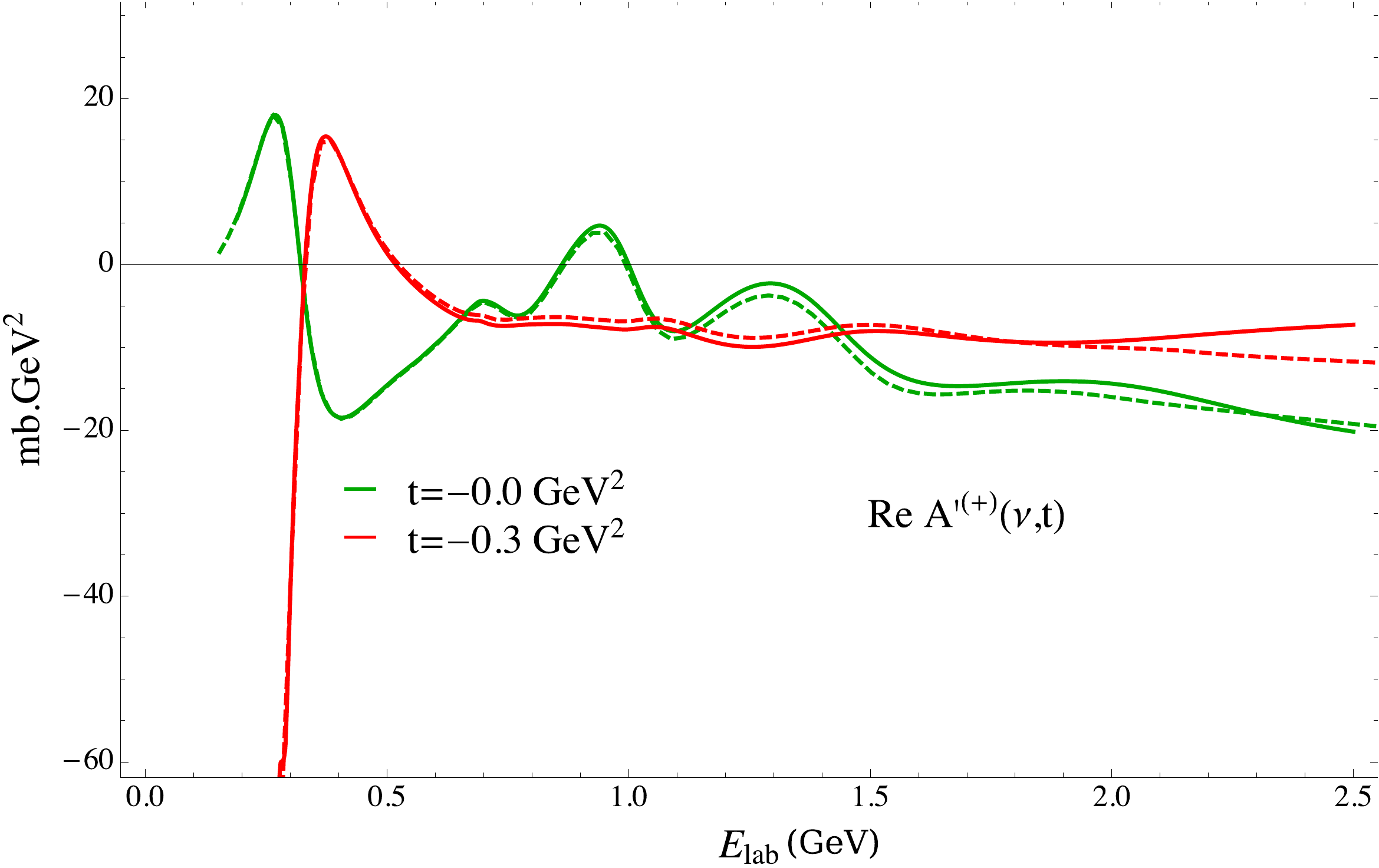}
\end{center}
\caption{(left) Illustration of FESR in Eq.~\eqref{eq:FESR}: The Regge parametrization is equivalent to the average of the imaginary part of the amplitude. (right) The real part of the amplitude reconstructed from the dispersion relation in Eq.~\eqref{eq:DR_FESR} (dashed) matches the original real part from SAID (solid)~\cite{Workman:2012jf}. 
Figures adapted from~\cite{Mathieu:2015gxa}. \label{fig:FESR2} }
\end{figure}

The very good agreement of both sides of the FESR allowed us to reconstruct the real part of the amplitudes via the dispersion relation
\begin{align} \label{eq:DR_FESR}
	A_i(\nu,t) & = \frac{1}{\pi} \int_0^\infty \text{d}\nu' \ \Im A_i(\nu',t) \left(\frac{1}{\nu'-\nu} \pm  \frac{1}{\nu'+\nu} \right)\, ,
\end{align}
where the relative sign depends on the parity properties of $A_i$.
In Ref.~\cite{Mathieu:2015gxa}, the imaginary parts of the amplitudes  in Eq.~\eqref{eq:DR_FESR} were taken from SAID in the low energy region and smoothly continued to the Regge parametrization matching the data and satisfying the FESR. The resulting real part of the amplitudes reconstructed from Eq.~\eqref{eq:DR_FESR} is in excellent agreement with the original real part from the SAID analysis, \cf \figurename{~\ref{fig:FESR2}}. This result exemplifies how one can determine the complete amplitudes in the complex plane by only fitting its imaginary part on the real axis, together with an appropriate description of the high energy region with Regge poles. 

In pion photoproduction, the situation is more complicated. 
The simultaneous inclusion of all three isospin channels leads to 12 invariant amplitudes. In Ref.~\cite{Mathieu:2017but} we computed the \lhs of the FESR using five independent partial wave analyses in the resonance region.
We then performed a global fit of the high energy data constrained by the FESR. The inclusion of Regge daughters was necessary to accommodate both the data and the features of the FESR. Finally our  solution  involves  the  minimum  Regge content  in  each  amplitude:  A leading Regge pole,  whose trajectory is constrained  around  the  expected  values,  and  a  second subleading term in the natural exchange amplitudes.  The latter allowed us to match the position of zeros   in  the two sides of  the  FESR,  and  to  describe  the  high-energy observables.

%============================================
% Summary
%============================================
\section{Summary}
\label{sec:conclusions}
Many new and unexpected hadrons have been discovered in the last twenty years.
To fully exploit the potential of present and future high-statistics datasets,
one must combine knowledge of reaction
theory, hadron phenomenology, 
and data analysis.
The ultimate goal of all this is to reduce model dependence as much as possible. In this respect, machine learning techniques have the potential to greatly contribute towards achieving this goal.

Since its foundation in 2013, the Joint Physics Analysis Center (JPAC) has focused its research on developing the necessary tools 
to tackle some of the many open challenges in hadron spectroscopy.
JPAC has contributed
to understand several aspects of the hadron spectrum and of resonance production, 
as well as of three-body dynamics.

Continuing this work in close cooperation with experimental collaborations will allow to improve the level of rigor on how to assess the properties of resonances in QCD, and eventually to understand why the microscopic constituents of matter arrange themselves into the rich picture one observes in nature.

%============================================
% Ack
%============================================
\section*{Acknowledgments} 

We dedicate this review to the memory of our colleague and  friend Mike Pennington, who was instrumental in the inception and development of the Joint Physics Analysis Center.
We thank our numerous colleagues within both the theory and experimental hadron physics communities for all their input, keen insight, and encouraging discussions that
helped unfold all the physics discussed in this review.

\vspace{.5cm}
This work was supported by the U.S.~Department of Energy under Grant No.~DE-AC05-06OR23177  under which Jefferson Science Associates, LLC, manages and operates Jefferson Lab, No.~DE-FG02-87ER40365 at Indiana University, and
No.~DE-SC0018416 at the College of William \& Mary,
National Science Foundation under Grant No.~PHY-2013184,
Polish Science Center (NCN) under Grant No.~2018/29/B/ST2/02576,
Spanish Ministerio de Econom\'ia y Competitividad and  
Ministerio de Ciencia e Innovaci\'on under Grants 
No.~PID2019–106080 GB-C21, 
No.~PID2019-105439G-C22,
No.~PID2020-118758GB-I00
and No.~PID2020-112777GB-I00 (Ref.~10.13039/501100011033),
UNAM-PAPIIT under Grant No.~IN106921,
CONACYT under Grant No.~A1-S-21389,
National Natural Science Foundation of China Grant No.~12035007
and the NSFC and the Deutsche Forschungsgemeinschaft (DFG, German Research Foundation) through 
% MM
the Germany's Excellence Strategy – EXC-2094 – 390783311 and through the 
Research Unit FOR 2926 (project number 40824754), as well as 
the funds provided to the Sino-German Collaborative Research Center TRR110 ``Symmetries and the Emergence of Structure in QCD" (NSFC Grant No.~12070131001, DFG Project-ID~196253076-TRR~110).
MA is supported by Generalitat Valenciana under Grant No.~CIDEGENT/2020/002.
CFR is supported by Spanish Ministerio de Educaci\'on y Formaci\'on Profesional under Grant No.~BG20/00133.
VM is a Serra H\'unter fellow.
JASC is supported by CONACYT under Grant No.~734789.
SGS is supported by the Laboratory Directed Research and Development program of Los Alamos National Laboratory under project No.~20210944PRD2, and by the U.S. Department of Energy through the Los Alamos National Laboratory. Los Alamos National Laboratory is operated by Triad National Security, LLC, for the National Nuclear Security Administration of U.S.~Department of Energy (Contract No.~89233218CNA000001).

%============================================
% Bibliography
%============================================
\bibliographystyle{elsarticle-num}
\bibliography{quattro}
\end{document}